\documentclass[11pt,preprint]{elsarticle}

\makeatletter
\def\ps@pprintTitle{%
 \let\@oddhead\@empty
 \let\@evenhead\@empty
 \def\@oddfoot{\centerline{\thepage}}%
 \let\@evenfoot\@oddfoot}
\makeatother

\journal{Joule}

\abstracttitle{Summary}

\biboptions{numbers,sort&compress,super}

\usepackage{libertine}
\usepackage{libertinust1math}
\renewcommand{\ttdefault}{\sfdefault}

\usepackage{geometry}
\usepackage{amsmath}
\usepackage{bbold}
\usepackage{graphicx}
\usepackage{eurosym}
\usepackage{mathtools}
\usepackage{url}
\usepackage{booktabs}
\usepackage{epstopdf}
\usepackage{xfrac}
\usepackage{tabularx}
\usepackage[version=4]{mhchem}
\usepackage{bm}
\usepackage[colorlinks]{hyperref}
\usepackage[nameinlink,sort&compress,capitalise,noabbrev]{cleveref}
\usepackage[leftcaption,raggedright]{sidecap}
\usepackage{subcaption}
\usepackage{blindtext}
\usepackage[parfill]{parskip}

\usepackage[prependcaption,textsize=scriptsize]{todonotes}

\usepackage{siunitx}
\DeclareSIUnit\year{a}
\DeclareSIUnit\tco{t_{\ce{CO2}}}
\DeclareSIUnit\sieuro{\mbox{\euro}}
\DeclareSIUnit\twh{TWh}
\DeclareSIUnit\mwh{MWh}
\DeclareSIUnit\kwh{kWh}

\newcommand{\co}{\ce{CO2}~}

\usepackage{longtable}
\usepackage{multirow}
\usepackage{threeparttable}
\usepackage{pdflscape}

\usepackage[export]{adjustbox}

\usepackage[resetlabels,labeled]{multibib}
\newcites{S}{Supplementary References}
\bibliographystyleS{elsarticle-num}

\graphicspath{
	{graphics/graphics-20221227/},
	{graphics/results/}
}

\newcommand{\costrun}{20221227-costs}
\newcommand{\imprun}{20221227-import}
\newcommand{\shprun}{20221227-shipping}
\newcommand{\hyrun}{20221227-main}

\newcommand{\abs}[1]{\left|#1\right|}

\def\cT{\mathcal{T}}
\newcommand{\R}{\mathbb{R}}

\def\el{${}_{\textrm{el}}$}

\usepackage{xcolor}
\usepackage{framed}
\definecolor{shadecolor}{rgb}{.95,.95,.95}

\usepackage{tocloft}
\cftsetindents{section}{1em}{2.75em}

\newcommand{\gridbenefitabs}{72}
\newcommand{\gridbenefitrel}{9.9}
\newcommand{\minacbenefitabs}{46}
\newcommand{\maxacbenefitabs}{61}
\newcommand{\minhybenefitabs}{12}
\newcommand{\maxhybenefitabs}{26}
\newcommand{\minacbenefitrel}{6.3}
\newcommand{\maxacbenefitrel}{8.1}
\newcommand{\minhybenefitrel}{1.6}
\newcommand{\maxhybenefitrel}{3.4}
\newcommand{\minsystemcost}{733}
\newcommand{\maxsystemcost}{805}
\newcommand{\acvshycost}{4.6}
\newcommand{\minaccost}{15.1}
\newcommand{\maxaccost}{37.9}
\newcommand{\minhycost}{3.2}
\newcommand{\maxhycost}{4.6}
\newcommand{\benefithyofac}{42.6}
\newcommand{\additivebenefitabs}{87}
\newcommand{\additivebenefitrel}{20.8}

\newcommand{\minoffwind}{206}
\newcommand{\maxoffwind}{245}
\newcommand{\minonwind}{1691}
\newcommand{\maxonwind}{1776}
\newcommand{\minsolar}{2666}
\newcommand{\maxsolar}{3598}
\newcommand{\meanrooftopshare}{16}
\newcommand{\meanutilityshare}{84}
\newcommand{\minoffshoreshare}{10}
\newcommand{\maxoffshoreshare}{12}
\newcommand{\minelectrolysis}{937}
\newcommand{\maxelectrolysis}{1250}
\newcommand{\mincfFT}{59}
\newcommand{\maxcfFT}{68}
\newcommand{\mincfelectrolysis}{35}
\newcommand{\maxcfelectrolysis}{41}
\newcommand{\utilisationAC}{36}
\newcommand{\utilisationHy}{78}
\newcommand{\hydrogenstorageacyhyy}{26}
\newcommand{\hydrogenstorageacyhyn}{22}
\newcommand{\hydrogenstorageacnhyy}{43}
\newcommand{\hydrogenstorageacnhyn}{21}

\newcommand{\ptlhydrogenusage}{1903}
\newcommand{\ptlwasteheat}{192}
\newcommand{\hydrogenindustrydemand}{195}
\newcommand{\hydrogentransportdemand}{275}
\newcommand{\hydrogenlosses}{139}
\newcommand{\hydrogenfuelcell}{287}
\newcommand{\hydrogenmethanation}{152}
\newcommand{\fossilgas}{366}
\newcommand{\biogas}{336}
\newcommand{\bluehydrogen}{78}
\newcommand{\mindac}{58}
\newcommand{\maxdac}{82}

\newcommand{\acoftotalbenefit}{84}
\newcommand{\hyoftotalbenefit}{36}
\newcommand{\maxretroshare}{69.1}
\newcommand{\minretroshare}{63.5}

\newcommand{\ewhkmelectricity}{1.04}
\newcommand{\ewhkmhydrogen}{2.16}
\newcommand{\ewhkmdiff}{0.9}

\begin{document}
 
\begin{frontmatter}

	\title{The Potential Role of a Hydrogen Network in Europe}

	\author[tubaddress]{Fabian Neumann\corref{correspondingauthor}}
	\ead{f.neumann@tu-berlin.de}
	\author[tubaddress]{Elisabeth Zeyen}
	\author[aarhus,aarhus2]{Marta Victoria}
	\author[tubaddress]{Tom Brown}
	\address[tubaddress]{Department of Digital Transformation in Energy Systems, Institute of Energy Technology, Technische Universität Berlin, Fakultät III, Einsteinufer 25 (TA 8), 10587 Berlin, Germany}
	\address[aarhus]{Department of Mechanical and Production Engineering, Aarhus University, Inge Lehmanns Gade 10, 8000 Aarhus, Denmark}
	\address[aarhus2]{Novo Nordisk Foundation CO$_2$ Research Center, Aarhus University, Aarhus, Denmark}

	\begin{abstract}
		Electricity transmission expansion has suffered many delays in Europe in recent
decades, despite its significance for integrating renewable electricity into the
energy system. A hydrogen network which reuses the existing fossil gas network
could not only help to supply demand for low-emission fuels, but could also to
balance variations in wind and solar energy across the continent and thus avoid
power grid expansion. We pursue this idea by varying the allowed expansion of
electricity and hydrogen grids in net-zero \co scenarios for a sector-coupled
and self-sufficient European energy system with high shares of renewables. We
cover the electricity, buildings, transport, agriculture, and industry sectors
across 181 regions and model every third hour of a year. With this high
spatio-temporal resolution, the model can capture bottlenecks in transmission
networks, the variability of demand and renewable supply, as well as regional
opportunities for the retrofitting of legacy gas infrastructure and development
of geological hydrogen storage. Our results show consistent system cost
reductions with a pan-continental hydrogen network that connects regions with
low-cost and abundant renewable potentials to demand centers, synthetic fuel
production and cavern storage sites. Developing a hydrogen network reduces
system costs by up to \maxhybenefitabs~bn\euro/a (\maxhybenefitrel\%), with
highest benefits when electricity grid reinforcements cannot be realised.
Between 64\% and 69\% of this network could be built from repurposed natural gas
pipelines. However, we find that hydrogen networks can only partially substitute
for power grid expansion. While the expansion of both networks together can
achieve the largest cost savings of \gridbenefitrel\%, the expansion of neither
is truly essential as long as higher costs can be accepted and regulatory
changes are made to manage grid bottlenecks.

	\end{abstract}


\end{frontmatter}

\section*{Graphical Abstract}

\includegraphics[width=\textwidth]{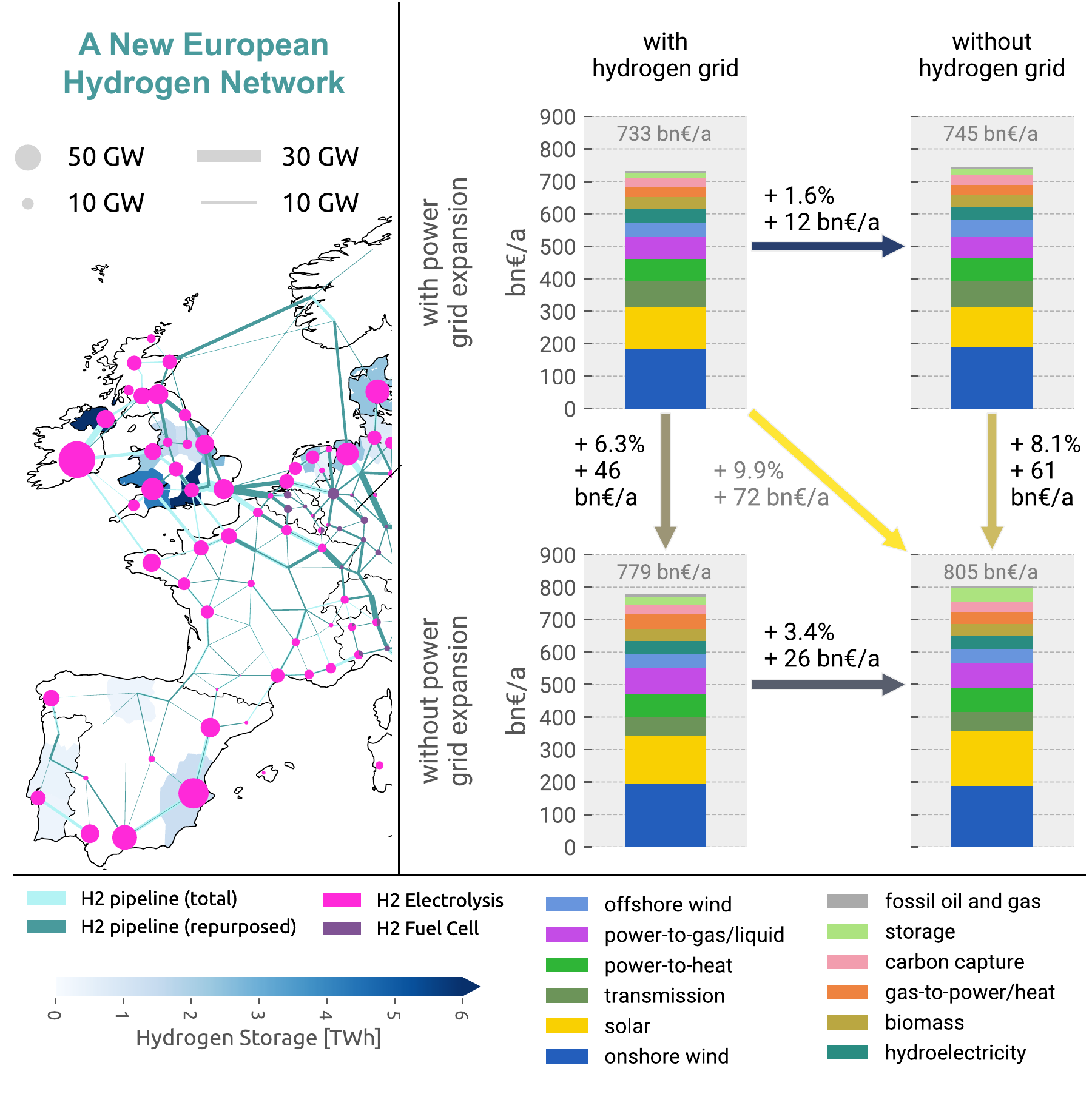}

\newpage
\par\noindent\rule{\textwidth}{0.4pt}

\section*{Highlights}

\begin{itemize}
	\item examines cost benefit of a hydrogen network including gas pipeline retrofitting in net-zero \co scenarios for Europe with high shares of renewables and no energy imports
	\item uses open energy system model PyPSA-Eur-Sec with 181 regions, 3-hourly resolution for a year and all energy sectors (electricity, buildings, transport, industry, agriculture) represented
	\item hydrogen network reduces energy system costs by up to \maxhybenefitrel\%, with highest cost reductions when power grid expansion is restricted
	\item between \minretroshare-\maxretroshare\% of hydrogen network uses retrofitted gas network pipelines
	\item cost benefit of electricity grid expansion is higher than of hydrogen network (\maxacbenefitrel\% versus \maxhybenefitrel\%), but both together reduce costs by up to \gridbenefitrel\%
\end{itemize}

\par\noindent\rule{\textwidth}{0.4pt}

\section*{Context \& Scale}

Many different combinations of infrastructure could make Europe climate-neutral
by mid-century, but not all solutions meet the same level of acceptance. For
example, power transmission reinforcements have experienced many delays, despite their
value for integrating renewable electricity. A hydrogen network which can reuse
gas pipelines could be a substitute for moving cheap but remote renewable
energy across the continent to where demand is.

We study trade-offs between new transmission lines and a hydrogen network in the
European energy system with all sectors represented and net-zero \co emissions.
We find that a hydrogen network consistently reduces system costs and that large
parts could use repurposed gas pipelines. Energy transport as electrons and
molecules offer complementary strengths, achieving highest cost savings
together. However, neither is essential as long as the system can be coordinated
around the resulting bottlenecks. This means that there are many affordable ways
to achieve net-zero emissions in Europe, giving policymakers different options
to choose from.

\par\noindent\rule{\textwidth}{0.4pt}

\newpage
\section*{Introduction}
\label{sec:intro}
\addcontentsline{toc}{section}{\nameref{sec:intro}}

There are many different combinations of infrastructure that would allow Europe
to reach net-zero greenhouse gas emissions by
mid-century.\cite{pickeringDiversityOptions2022} However, not all technologies
meet the same level of acceptance among the public. The last few decades have
seen public resistance to new and existing nuclear power plants, projects with
carbon capture and sequestration (CCS), onshore wind power plants, and overhead
transmission
lines.\cite{cohenRefocussingResearch2014a,bertschPublicAcceptance2016,boudetPublicPerceptions2019a}
The lack of public acceptance can both delay the deployment of a technology and
even stop its deployment altogether.\cite{batelSocialAcceptance2013a} This may
make it harder to reach greenhouse gas reduction targets in time or cause rising
costs through the substitution with other technologies. In particular,
electricity transmission network expansion has suffered many delays in Europe in
recent decades, despite its importance for integrating large amounts of
renewable electricity and electrifying the transport, buildings and industry
sectors.\cite{schlachtbergerBenefitsCooperation2017,trondleTradeOffsGeographic2020}

Hydrogen has the potential to become a pivotal energy carrier in such a climate-neutral
energy system.\cite{hanleyRoleHydrogen2018,staffellRoleHydrogen2019a} It is
needed in industry to produce ammonia for fertilisers and can be used for
direct reduced iron for
steelmaking.\cite{wangGreeningAmmonia2018,voglPhasingOut2021} It is also a
critical feedstock to produce synthetic methane and liquid hydrocarbons for use
as aviation and shipping fuels, and as a precursor to high-value chemicals in
industrial production.\cite{lechtenbohmerDecarbonisingEnergy2016} Hydrogen could
also be used for heavy-duty land transport and backup heat and power
supply.\cite{kluschkeMarketDiffusion2019,doddsHydrogenFuel2015}

The limited social acceptance for electricity grid reinforcement and the
advancing role of hydrogen raises the question of whether a new hydrogen network
could offer a replacement for balancing variable renewable electricity
generation and moving energy across the
continent.\cite{caglayanRobustDesign2021} Such a vision for a \textit{European
Hydrogen Backbone (EHB)} has recently been expressed by Europe's gas industry in
a series of reports.
\cite{gasforclimateEuropeanHydrogen2020,gasforclimateEuropeanHydrogen2021,gasforclimateExtendingEuropean2021,gasforclimateEuropeanHydrogen2022}
It would offer an alternative to connect remote regions with abundant
and cost-effective wind and solar potentials to densely-populated and
industry-heavy regions with high demand but limited supply options.

Since Europe's sizeable natural gas transmission network is set to become
increasingly redundant as the system transitions towards climate neutrality, the
option to repurpose parts of the network to transport hydrogen instead may
enhance the appeal of hydrogen networks further. This is because retrofitting
gas pipelines would greatly reduce the development costs of hydrogen
pipelines.\cite{cerniauskasOptionsNatural2020,tsikliosHydrogenTransport2022}
Moreover, repurposed and new pipelines may also meet higher levels of acceptance
among the local populations than transmission
lines.\cite{schonauerHydrogenFuture2022} Unlike transmission towers, pipelines
are less visible because they usually run below or near the ground. Particularly
where gas pipelines already exist, the perceivable impact would be minimal.

However, few studies have looked into how much building a hydrogen network in
Europe could reduce system costs. The industry-oriented EHB reports do not
include an assessment based on the co-optimisation of energy system components.
\cite{gasforclimateEuropeanHydrogen2020,gasforclimateEuropeanHydrogen2021,gasforclimateExtendingEuropean2021,gasforclimateEuropeanHydrogen2022}
Other sector-coupling studies have not included hydrogen networks at all,
\cite{brownSynergiesSector2018,pickeringDiversityOptions2022,childFlexibleElectricity2019a,kendziorskiCentralizedDecentral2022a}
or when they do, model Europe only at country-level resolution,
\cite{europeancommission.directorategeneralforenergy.METISStudy2021,victoriaSpeedTechnological2022}
have a country-specific focus with limited geographical scope or detail outside
the focus area,\cite{gilsInteractionHydrogen2021} investigate the mid-term
role rather than the long-term role of a hydrogen
network,\cite{europeancommission.directorategeneralforenergy.METISStudy2021} or
neglect some energy sectors or non-energy demands that involve hydrogen.
\cite{gilsInteractionHydrogen2021,Caglayan2019,caglayanRobustDesign2021}
However, high resolution at continental scope is needed to understand how a
hydrogen network can relieve power grid bottlenecks, where the costs of hydrogen
network development can be reduced by retrofitting gas pipelines, and where
geological sites for hydrogen storage are located. Previous one-node-per-country
studies could not have suitably assessed this.

This paper provides the first high-resolution examination of the trade-offs
between electricity grid expansion and a new hydrogen network in scenarios for a
European energy system with net-zero carbon dioxide emissions, no energy imports
and high shares of renewable electricity production. By leveraging recent
computational advances, we resolve 181 regions to study what role hydrogen
infrastructure can play in a future sector-coupled system. This enables us to
take account of network bottlenecks inside countries, see more precise locations
of demand and supply in the network and capture the variability of renewable
resources. For the first time, such an investigation also considers regional
potentials for the repurposing of legacy gas pipelines and the geological
storage of hydrogen in salt caverns.

Our analysis covers four main scenarios to examine if a hydrogen network
composed of new and retrofitted pipelines can compensate for a potential lack of
power grid expansion. These scenarios differ based on whether or not electricity
and hydrogen grids can be expanded. As supplementary sensitivity analyses, we
also evaluate the impact of restricted onshore wind potentials
(\cref{sec:si:onw}), more progressive technology assumptions
(\cref{sec:si:sensitivity-costs}), the impact of importing most hydrogen
derivatives from outside of Europe on network benefits (\cref{sec:si:sensitivity-imports}), and the use
of alternative shipping fuels (\cref{sec:si:sensitivity-shipping}).

For our analysis, we use an open capacity expansion model of the European energy
system, PyPSA-Eur-Sec, which, in contrast to many previous studies,
\cite{henningComprehensiveModel2014,mathiesenSmartEnergy2015,connollySmartEnergy2016,Loffler_2019,blancoPotentialHydrogen2018,brownSynergiesSector2018,in-depth_2018,victoria2020,ludererImpactDeclining2021,gea-bermudezRoleSector2021}
combines a fully sector-coupled approach with a high spatio-temporal resolution
and multi-carrier transmission infrastructure representation so that it can
capture the various transport bottlenecks that constrain the cost-effective
integration of variable renewable energy.\cite{frysztackiStrongEffect2021} The
model co-optimises the investment and operation of generation, storage,
conversion and transmission infrastructures for the least-cost outcome in a
single linear optimisation problem, covering 181 regions and a 3-hourly time
resolution for a full year. A sensitivity analysis varying the model's
spatio-temporal resolution is included in
\cref{sec:si:sensitivity-time,sec:si:sensitivity-space}. The regional scope
comprises the European Union without Cyprus and Malta as well as the United
Kingdom, Norway, Switzerland, Albania, Bosnia and Herzegovina, Montenegro, North
Macedonia, Serbia and Kosovo. It incorporates spatially distributed demands of
the electricity, industry, buildings, agriculture and transport sectors,
including dense fuels in shipping and aviation as well as non-energy feedstock
demands in the chemicals industry. Primary energy supply comes from wind, solar,
biomass, hydro, and limited amounts of fossil oil and gas. The energy flows
between the system's energy carriers are modelled by various technologies,
including heat pumps, combined heat and power (CHP) plants, thermal storage,
electric vehicles, batteries, power-to-X processes, hydrogen fuel cells, and
geological potentials of underground hydrogen storage. Data on existing
electricity and gas transmission infrastructure is also included to determine
grid expansion needs and retrofitting potentials. The model also features
detailed management of carbon flows between capture, usage, sequestration,
removal and emissions to the atmosphere to track carbon through the system. More
details on the model are presented in the \nameref{sec:methods} and
\nameref{sec:si}. The model is open-source and based on open data
(\href{https://github.com/pypsa/pypsa-eur-sec}{github.com/pypsa/pypsa-eur-sec}).

All investigations are conducted with a constraint that carbon dioxide emissions
into the atmosphere balance out to zero over the year, disregarding other
greenhouse gas emissions. The model can sequester up to 200 Mt\co per year,
allowing it to sequester industry process emissions that have a fossil origin,
such as the calcination in cement manufacturing, but restricting the use of
negative emission technologies compared to other works.
\cite{blancoPotentialHydrogen2018} In our scenarios, we also do not consider
clean energy imports to Europe, thus assuming that Europe is self-sufficient in
electricity and green fuels and feedstocks. We relax this constraint in
\cref{sec:si:sensitivity-imports}. Technology assumptions are taken widely from
the Danish Energy Agency for the year 2030.\cite{DEA} A sensitivity with
technology assumptions for the year 2050 is presented in \cref{sec:si:sensitivity-costs}.

\begin{figure}
    \centering
    \makebox[\textwidth][c]{
    \includegraphics[width=1.3\textwidth]{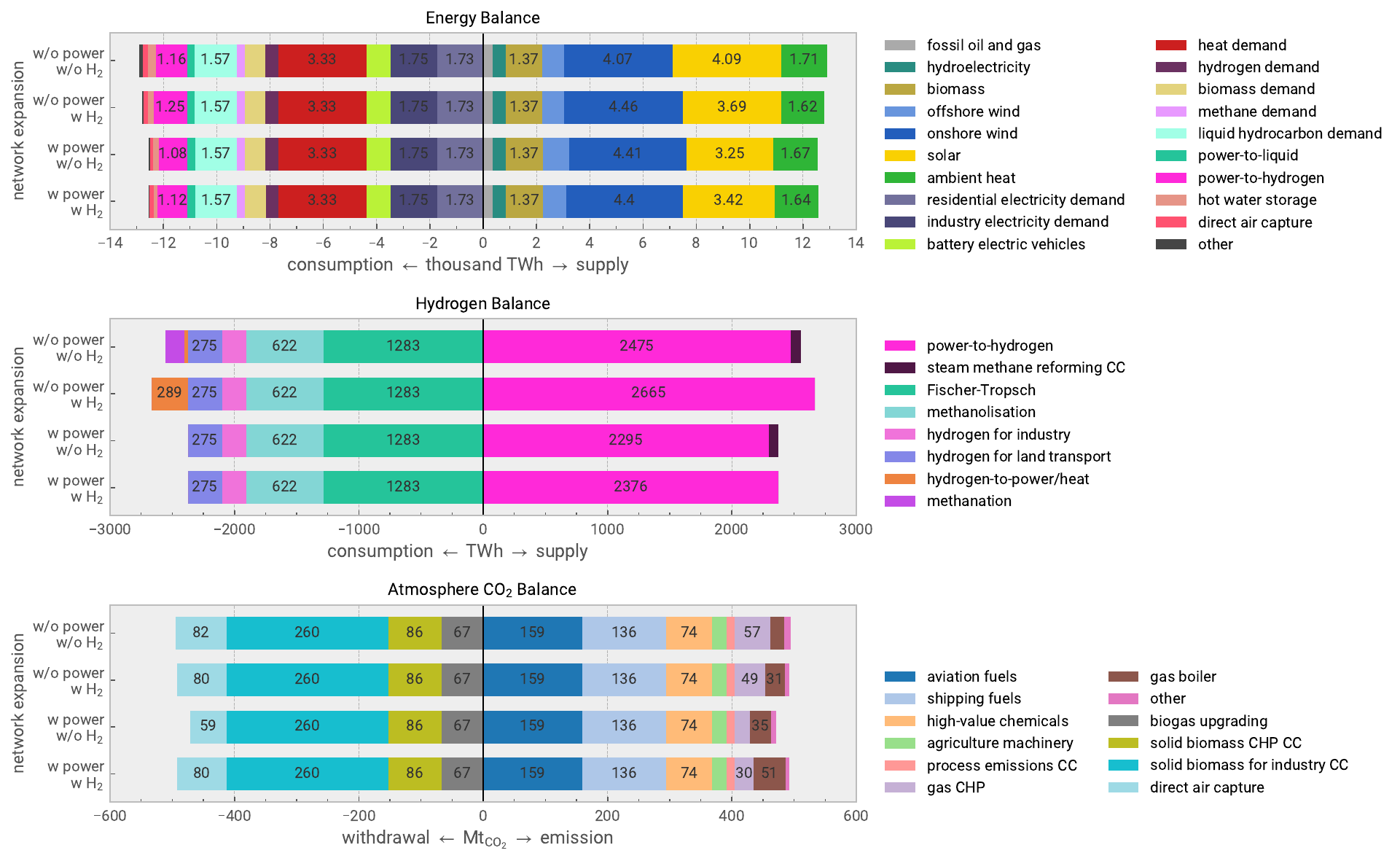}
    } \caption{ Energy, hydrogen and carbon dioxide balances across all
    scenarios. Energy consumption includes final energy and non-energy demands
    by carrier as well as conversion losses in thermal storage and electrofuel
    synthesis processes (e.g.~power-to-hydrogen, power-to-liquid). The ambient
    heat retrieved by heat pumps is counted as energy supply. A breakdown of
    final energy and non-energy demands by sector is shown by sector in
    \cref{fig:demand-by-sector-carrier}, by time in \cref{fig:demand-time}, and
    by region in \cref{fig:demand-space}. For technologies with carbon capture
    (CC) option, the carbon dioxide balance shows residual emissions due to
    imperfect capture rates. }
    \label{fig:balance}
\end{figure}

\section*{Energy, hydrogen and carbon balances show key technologies needed to satisfy European energy needs with net-zero emissions}
\label{sec:balances}
\addcontentsline{toc}{section}{\nameref{sec:balances}}

First of all, with the energy balance in \cref{fig:balance}, we underline the
central role of wind and solar electricity supply in all scenarios.
Hydroelectricity, biomass and the recovery of ambient heat through heat pumps
further support the energy supply, whereas fossil oil and gas only play a small
role, since carbon dioxide removal options to offset their unabated emissions
are limited by the assumed sequestration potentials. Electricity demand for
industrial processes, electrified transport and the residential sector,
alongside heat for hot water provision, space heating and industrial processes,
dominate the energy consumption. Conversion losses of power-to-X processes are
also shown in the energy balance and are most pronounced for electrolysis.
Overall, differences between the scenarios are small. With restricted network
expansion options, the energy supply shifts towards solar photovoltaics and the
total increases slightly. This rise compensates for the higher heat losses in
thermal energy storage and increased handling of added synthetic gas in these
scenarios.

\cref{fig:balance} also presents the balance of hydrogen consumption and supply.
The supply-side is dominated by the production of large amounts of green
electrolytic hydrogen between 2376~TWh/a and 2665~TWh/a depending on the
scenario. We only observe a limited production of blue hydrogen from steam
methane reforming with carbon capture in scenarios without hydrogen network
expansion (\SI{\bluehydrogen}{\twh\per\year}). A glance at the demand-side
reveals that, for the most part, hydrogen is only an intermediate product
between electricity and derivative products. There are only a few direct uses of
hydrogen, for instance, in the industry sector for producing ammonia and steel
with hydrogen-based direct reduction of iron, as well as for heavy-duty land
transport. Most hydrogen is used to produce derivatives like Fischer-Tropsch
fuels, methane, ammonia and methanol, which are used for dense aviation and
shipping fuels, fertilisers and as a feedstock for producing high-value
chemicals.

The production of liquid hydrocarbons consumes \ptlhydrogenusage~TWh/a of
hydrogen, of which \ptlwasteheat~TWh/a is useable in the form of waste heat for
district heating networks. Around \hydrogenlosses~TWh/a of hydrogen is lost
during synthetic fuel production. A total of \hydrogentransportdemand~TWh/a is
used in land transport, while the industry sector consumes
\hydrogenindustrydemand~TWh/a for ammonia and steel production, excluding the
consumption of hydrogen for other industry feedstocks (e.g.~for high-value
chemicals).  If the electricity grid expansion is restricted, but hydrogen can
be transported, some more hydrogen is produced to be re-electrified in fuel
cells during critical phases of system operation
(\hydrogenfuelcell~TWh$_{\ce{H2}}$). These fuel cells would mostly be built
inland in Central Europe (see later section \nameref{sec:es}), where the lack of
a strong grid connection requires local dispatchable heat and power supply as a
backup for periods of low renewables feed-in and cold weather. However, in terms
of energy consumed the reconversion of hydrogen to electricity only assumes a
secondary role. In all scenarios with network expansion, no synthetic methane
for process heat in some industrial applications and as a heating backup for
power-to-heat units is produced. This is because the model prefers to use the
full potential for biogas (\biogas~TWh/a) and limited amounts of fossil gas
(\fossilgas~TWh/a), which are offset by sequestering biogenic carbon dioxide,
over synthetic production.

Only when neither hydrogen nor power network expansion were allowed, do we see
methanation (\ce{H2}-to-\ce{CH4}, \SI{\hydrogenmethanation}{\twh} hydrogen). In
this case, despite the associated conversion losses, synthetic methane is used
as a transport medium for hydrogen to utilise the existing gas network to bypass
the restricted transport options for hydrogen and electricity. Apart from
imperfect capture rates of 90\% that requires supplementing some CO$_2$, the
combination of carbon-capturing steam methane reforming creates a carbon cycle
provided that the \co is returned to the methanation sites with an appropriate
\co transport infrastructure.

The atmospheric \co balance in \cref{fig:balance} shows that liquid hydrocarbons
in shipping, aviation and the incineration or eventual decay of plastics
constitute the major uncaptured carbon dioxide emissions in the system. Some
additional \co is emitted through using unabated methane (natural gas, biogas or
synthetic) in gas boilers and CHP plants in the heating sector during the
challenging cold winter periods with low renewable energy supply and high space
heating demand. Industrial process emissions are largely captured such that,
owing to imperfect capture rates, only residual emissions are released into the
atmosphere. Most carbon dioxide removal is achieved through biomass
technologies. For instance, biogenic \co is captured in biomass CHP plants or
industrial low-temperature heat applications. Direct air capture was used in all
scenarios, but takes a much smaller part supplementing the \co available from
biogenic or fossil sources once they are exhausted. Of the \co handled by the
system for the synthesis of electrofuels and long-term sequestration, the
largest share is of biogenic origin (62\%) followed by captured fossil \co
emissions from fuel combustion and process emissions (25\%). Direct air capture
has the smallest share with \SIrange{\mindac}{\maxdac}{\mega\tco\per\year}
(13\%). The broad availability of captured \co from industrial processes and
biofuel combustion is advantageous for the system, as it lowers the cost of fuel
synthesis by avoiding costly and energy-intensive direct air capture.

For a comprehensive overview of energy and carbon flows between carriers in each
scenario see \cref{fig:si:sankey,fig:si:carbon-sankey} which can be
interactively explored at
\href{https://h2-network.streamlit.app}{h2-network.streamlit.app}.

\begin{figure}
    \centering
    \begin{subfigure}[t]{\textwidth}
        \centering
        \caption{cost reductions induced by hydrogen and power grid expansion}
        \includegraphics[width=0.85\textwidth]{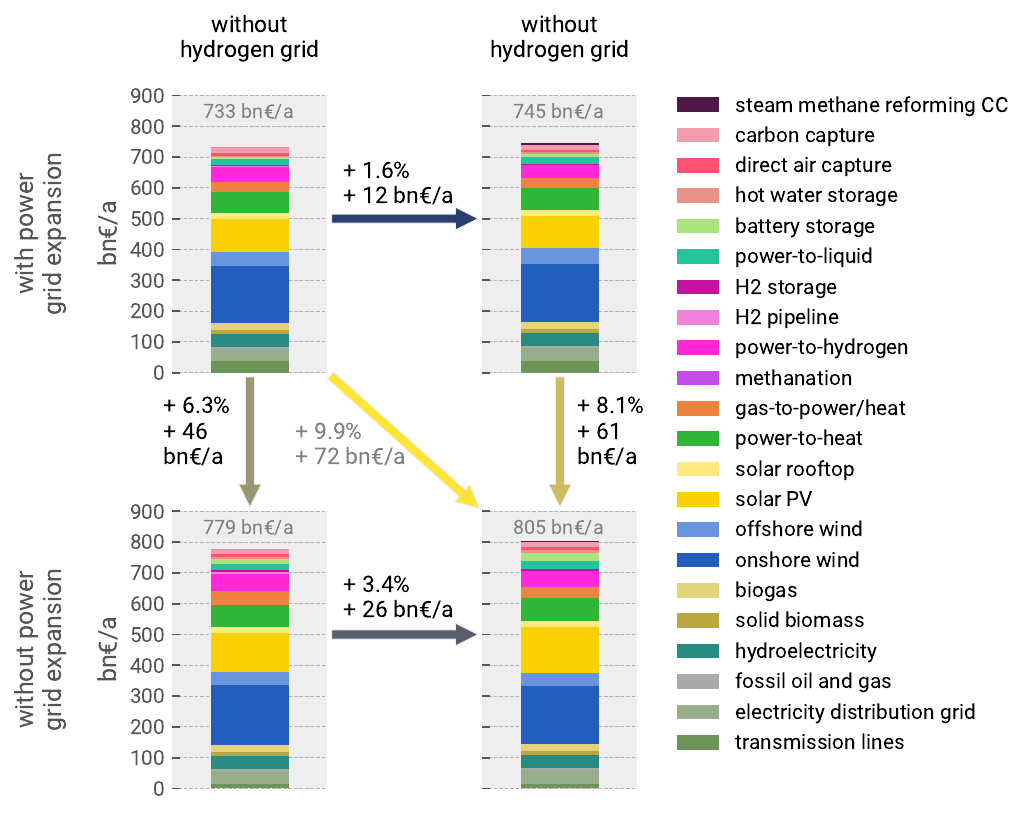}
        \label{fig:sensitivity-h2-a}
    \end{subfigure}
    \begin{subfigure}[t]{\textwidth}
        \centering
        \caption{system cost difference to full hydrogen and power grid expansion scenario}
        \includegraphics[width=.85\textwidth]{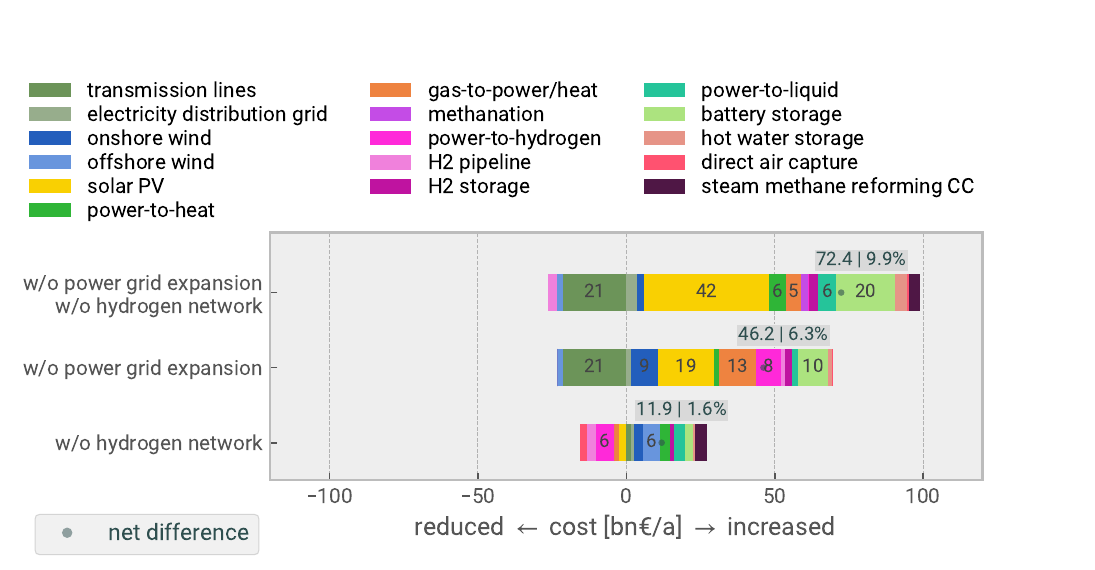}
        \label{fig:sensitivity-h2-b}
    \end{subfigure}
    \caption{Cost reductions achieved by developing electricity and hydrogen
    network infrastructure. \cref{fig:sensitivity-h2-a} compares four scenarios
    with and without expansion of a hydrogen network (left to right) and the
    electricity grid (top to bottom). Each bar depicts the total system cost of
    one scenario alongside its cost composition. Arrows between the bars
    indicate absolute and relative cost increases as network infrastructures are
    successively restricted. \cref{fig:sensitivity-h2-b} shows in monetary terms
    how the model reacts to grid expansion restrictions relative to the
    least-cost solution with full hydrogen and power grid expansion.}
    \label{fig:sensitivity-h2}
\end{figure}

\section*{Cost benefit of hydrogen network is consistent, and strongest without power grid expansion}
\label{sec:h2}
\addcontentsline{toc}{section}{\nameref{sec:h2}}

In \cref{fig:sensitivity-h2}, we first compare the total energy system costs and
their composition between the four main scenarios, which vary in whether or not
the power grid can be expanded beyond today's levels and whether a hydrogen
network based on new and retrofitted pipelines can be built. Across all
scenarios, the total costs are dominated by investments in wind and solar
capacities, power-to-heat applications (primarily heat pumps), electrolysers,
and electrofuel synthesis plants (for transport fuels and as a feedstock for the
chemicals industry). Total energy system costs vary between \minsystemcost~and
\maxsystemcost~bn\euro/a, depending on available network expansion options.

Overall, we find that energy system costs are not overly affected by
restrictions on the development of electricity or hydrogen transmission
infrastructure, and systems without grid expansion appear as equally feasible
alternatives. Nonetheless, realisable cost savings range in the order of tens of
billions of euros per year. The combined net benefit of hydrogen and electricity
grid expansion beyond today's levels is \gridbenefitabs~bn\euro/a; a system no
further network expansion would be around \gridbenefitrel\% more expensive. This
limited cost increase can be attributed to the high level of synthetic fuel
production for industry, transport, and backup electricity and heating
applications. The option for a flexible operation of conversion plants,
inexpensive energy storage and low-cost energy transport as hydrocarbons between
regions offer sufficient leeway to manage electricity and hydrogen transport
restrictions effectively (see \nameref{sec:es}). However, regulatory changes
would be needed in order to manage the network bottlenecks (see
\nameref{sec:policy}).

The total net benefit of power grid expansion is between
\minacbenefitabs-\maxacbenefitabs~bn\euro/a
(\minacbenefitrel-\maxacbenefitrel\%) compared to costs for transmission
reinforcements between \minaccost-\maxaccost~bn\euro/a. System costs decrease
despite the increasing investments in electricity transmission infrastructure.
Power grid reinforcements enable renewable resources with higher capacity
factors to be integrated from further away, resulting in lower capacity needs
for solar and wind. The electricity grid also allows renewable variations to be
smoothed in space and facilitates the integration of offshore wind, resulting in
lower hydrogen demand for balancing power and heat and less hydrogen
infrastructure (comprising electrolysis, cavern storage, re-conversion,
pipelines). Restrictions on power grid expansion conversely raise costs by
forcing more local production from solar photovoltaics and increased hydrogen
production. As a hydrogen network could compensate for the lack of grid capacity
to transport energy over long distances, the benefit of electricity grid
reinforcements is strongest if no hydrogen network can be developed.
\cref{sec:si:lv} presents in more detail the progression of system cost changes
in intermediate steps between a doubling of power grid capacity and no grid
expansion.

The presence of a new hydrogen network can reduce system costs by up to
\maxhybenefitrel\%. The net benefit between
\minhybenefitabs-\maxhybenefitabs~bn\euro/a
(\minhybenefitrel-\maxhybenefitrel\%) largely exceeds the cost of the hydrogen
network, which costs between \minhycost-\maxhycost~bn\euro/a. The hydrogen
network offers an alternative for bulk energy transport from the windiest and
sunniest regions in Europe's periphery to low-cost geological storage sites and
the industrial clusters in Central Europe with high energy demand but less
attractive and more constrained renewable potentials (see
\nameref{sec:energy-moved}). We find that its system cost benefit is strongest
when the electricity grid is not expanded. However, even with high levels of
power grid expansion, the hydrogen network is still beneficial infrastructure.

Although power grid reinforcements provide higher cost reductions, hydrogen and
electricity networks are stronger together. Around \hyoftotalbenefit\% of the
combined cost benefit of transmission infrastructure can be achieved solely with
a new hydrogen network. In contrast, \acoftotalbenefit\% of the combined cost
benefit can be reached by just reinforcing the electricity transmission system.
Compared to the combined net benefit of \gridbenefitabs~bn\euro/a, the
individual benefits sum up to a value that is only \additivebenefitrel\% higher
(\maxacbenefitabs{} + \maxhybenefitabs{} = \additivebenefitabs{} bn\euro/a).
Thus, offered cost reductions are mainly additive.

This also means that a hydrogen network cannot substitute perfectly for power
grid reinforcements. It can only partially compensate for the lack of grid
expansion, yielding \benefithyofac\% of the cost reductions achieved by
electricity grid expansion. This is because electricity has more versatile uses
in the newly electrified transport, buildings and industry sectors. Hydrogen can
only be used directly in a few specialised sectors, and if it has to be produced
only to be re-electrified later there are expensive efficiency losses.  A system
built exclusively around hydrogen network expansion is just \acvshycost\% more
expensive than an alternative system that only allows electricity grid
expansion. Overall, our results show that energy transport as electrons and
molecules offer complementary strengths. From a system-level perspective,
network expansion leads to small cost reductions.

\section*{Common design features in four net-zero carbon dioxide emission scenarios for Europe}
\label{sec:es}
\addcontentsline{toc}{section}{\nameref{sec:es}}

Across all scenarios, we see \SIrange{\minoffwind}{\maxoffwind}{\giga\watt}
offshore wind, \SIrange{\minonwind}{\maxonwind}{\giga\watt} onshore wind, and
\SIrange{\minsolar}{\maxsolar}{\giga\watt} solar photovoltaics
(\cref{fig:si:capacities}). The wide range of solar capacities is due to an
increased localisation of electricity generation through solar photovoltaics
when the expansion of transmission infrastructure is limited. As network
expansion options are constrained, we see demand for local daily storage with
batteries almost quadrupling (from 73 to 272 GW with a typical energy-to-power
ratio of 6 hours) and doubling for weekly and seasonal storage with hydrogen and
thermal storage (from 73 to 141~TWh, see \cref{fig:si:capacities}). For all
scenarios, the capacities of photovoltaics split on average into
\meanrooftopshare\% rooftop PV and \meanutilityshare\% utility-scale PV. The
offshore share of wind generation capacities varies between \minoffshoreshare\%
and \maxoffshoreshare\% and is highest when networks can be fully expanded.

\begin{figure}
    \centering
    \vspace{-2cm}
    \makebox[\textwidth][c]{
        \begin{subfigure}[t]{0.5\textwidth}
            \centering
            \caption{with power grid reinforcement, with hydrogen network}
            \includegraphics[width=\textwidth, trim=0cm .3cm 7cm 0cm, clip]{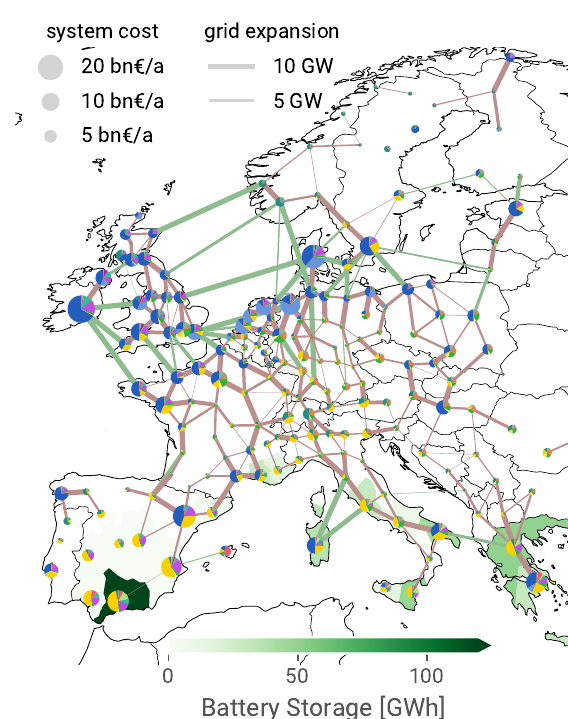}
            \label{fig:tsc:w-el-w-h2}
        \end{subfigure}
        \begin{subfigure}[t]{0.5\textwidth}
            \centering
            \caption{with power grid reinforcement, without hydrogen network}
            \includegraphics[width=\textwidth, trim=0cm .3cm 7cm 0cm, clip]{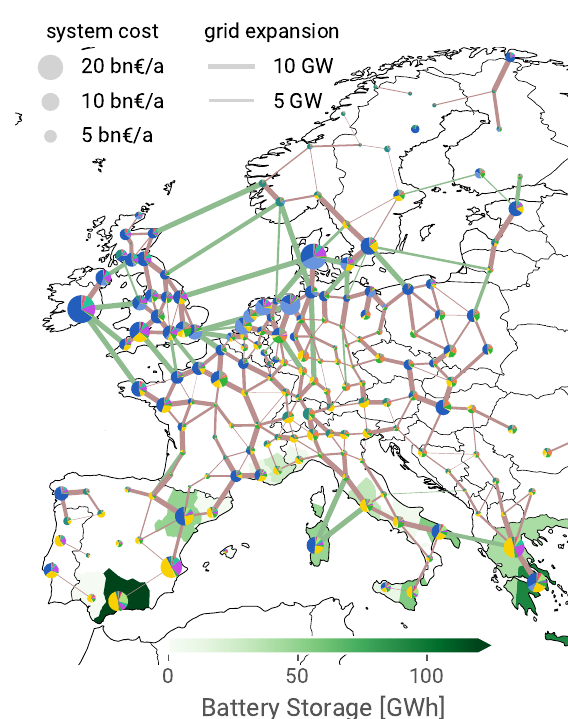}
            \label{fig:tsc:w-el-wo-h2}
        \end{subfigure}
    } \makebox[\textwidth][c]{
        \begin{subfigure}[t]{0.5\textwidth}
            \centering
            \caption{without power grid reinforcement, with hydrogen network}
            \includegraphics[width=\textwidth, trim=0cm .3cm 7cm 0cm, clip]{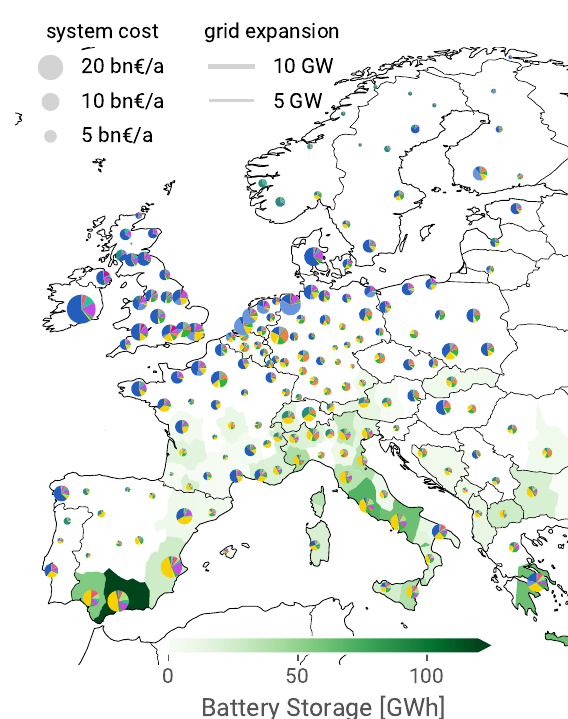}
            \label{fig:tsc:wo-el-w-h2}
        \end{subfigure}
        \begin{subfigure}[t]{0.5\textwidth}
            \centering
            \caption{without power grid reinforcement, without hydrogen network}
            \includegraphics[width=\textwidth, trim=0cm .3cm 7cm 0cm, clip]{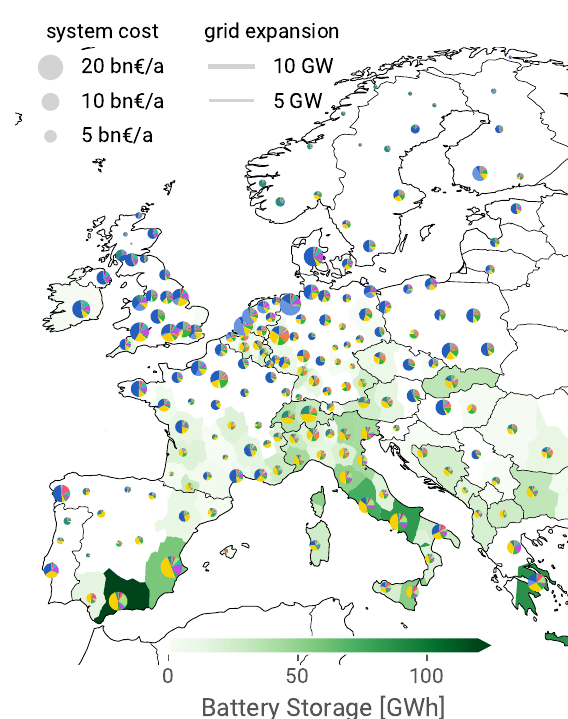}
            \label{fig:tsc:wo-el-wo-h2}
        \end{subfigure}
    } \vspace{-.5cm} \makebox[1\textwidth][c]{
        \centering
    \includegraphics[width=1.1\textwidth]{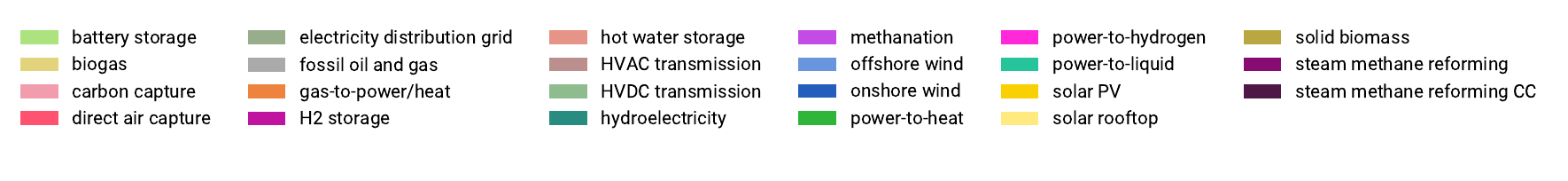}
    } \caption{ Regional distribution of system costs and electricity grid
    expansion for scenarios with and without electricity or hydrogen network
    expansion. The pie charts depict the annualised system cost alongside the
    shares of the various technologies for each region. The line widths depict
    the level of added grid capacity between two regions, which was capped at 10
    GW.}
    \label{fig:tsc}
\end{figure}

The spatial distribution of investments per scenario is shown in \cref{fig:tsc}.
While solar capacities are found throughout Europe, especially in the South,
onshore and offshore wind capacities are mostly found in the North Sea region
and the British Isles. When allowed, new electricity transmission capacity is
built where they help the integration of remote wind production and the
transport to inland demand centres. Consequently, most grid expansion is seen in
and between Northwestern and Central Europe. Battery storage pairs with solar
generation in Southern Europe, particularly when power grid reinforcement is
limited. Besides their wider use overall, battery deployment also progresses
northbound in this case.

Furthermore, electrolyser capacities for power-to-hydrogen conversion see a
massive scale-up ranging from
\SIrange{\minelectrolysis}{\maxelectrolysis}{\giga\watt} depending on the
scenario. The capacities are lowest when the electricity grid can be expanded.
In this case, their locations correlate strongly with wind and solar capacities
(Pearson correlation coefficient $R^2=0.64$ for each, \cref{fig:tsc}). If no
hydrogen or electricity transmission expansion is allowed, the electrolysis
correlates more strongly with wind ($R^2=0.74$) than solar ($R^2=0.46$). The
build-out of hydrogen production facilities is accompanied by a network of
pipelines and hydrogen underground storage in Europe to help balance generation
from renewables in time and space.

In space, a new pipeline network transports hydrogen from preferred production
sites to the rest of Europe, where hydrogen is consumed by industry (for
ammonia, high-value chemicals and steel production), aviation and shipping, as
well as fuel cell CHPs for combined power and heat backup. Varying in magnitude
per scenario, we see major net flows of hydrogen from Great Britain to the
Benelux countries, Germany and Norway, from Northern Germany to the South, and
from the East of Spain to Southern France. The favoured network topology
strongly depends on the potentials for cheap renewable electricity. If onshore
wind potentials were restricted, e.g. due to limited social acceptance in
Northern Europe, the network infrastructure would be tailored to deliver larger
amounts of solar-based hydrogen from Southern Europe to Central Europe. We
discuss this supplementary sensitivity analysis in
\cref{sec:si:onw,sec:si:onw-compromise}.

The development of a hydrogen network is driven by the fact that (i)
spatially-fixed hydrogen demand for steelmaking and ammonia industry as well as
heavy-duty land transport is located in areas with less attractive renewable
potentials (\cref{fig:demand-space:hydrogen}), (ii) the best wind and solar
potentials are located in the periphery of Europe (\cref{fig:energy-density}),
(iii) bottlenecks in the electricity transmission network give impetus to
alternative energy transport options and re-electrification capacities as backup
supply in weakly connected areas, and (iv) moving hydrogen from production sites
to where the geological conditions allow for cheap underground storage is
significantly more cost-effective than local storage in steel tanks
(\cref{fig:clustered-caverns}). Another subsidiary location factor for hydrogen
network infrastructure is linked to the siting of electrofuel production.
Because we assume that waste heat from these processes can be recovered for
district heating networks, urban areas with attractive renewable potentials
nearby are preferred sites for fuel synthesis to which the hydrogen needs to be
transferred. Since we assume no constraints for the transport of liquid
hydrocarbons, the spatial distribution of hydrogen consumption for fuel
synthesis is not a siting factor that is considered. Just like the positioning
of hydrogen fuel cells, the location of hydrogen consumption for electrofuel
production is endogenously optimised. Because we further assume sufficient
infrastructure for the transport of captured carbon dioxide, the location of
carbon sources and sinks neither influences the siting of fuel synthesis plants.

The flexible operation of electrolysers further supports the system integration
of variable renewables in time. Hydrogen production leverages periods with
exceptionally high wind speeds across Europe by running the electrolysis with
average utilisation rates between \mincfelectrolysis\% and \maxcfelectrolysis\%
(see \cref{fig:output-ts-1,fig:output-ts-3}). The produced hydrogen is buffered
in salt caverns which then allows for higher full load hours of fuel synthesis
processes. For Fischer-Tropsch and methanolisation plants, we see combined
average utilisation rates between \mincfFT\% and \maxcfFT\% which aligns with
the higher upfront investment costs of these processes. Their operation is very
steady in the summer months and mostly interrupted in winter periods with low
wind speeds and low ambient temperatures to give way to backup heat and power
supply options (see \cref{fig:output-ts-1,fig:output-ts-3}). By exploiting
periods of peak generation and curbing production in periods of scarcity, large
amounts of variable renewable power generation that serves the system's abundant
synthetic fuel demands can be incorporated into the system cost-effectively.
This ultimately leads to little curtailment of renewables between 2\% and 3\%
(\cref{fig:si:curtailment}) even without grid reinforcements, and low levels of
firm capacity. In relation to a peak electricity consumption of 2626~GW\el, we
observe OCGT and CHP plant capacities between 106 and 218~GW\el, most of which
are gas CHP plants. The lowest values were attained when additional power
transmission could be built.

Hydrogen storage is required to benefit from temporal balancing through flexible
electrolyser operation. We find cost-optimal storage capacities between
\SIrange{\hydrogenstorageacyhyy}{\hydrogenstorageacnhyy}{\twh} with a hydrogen
network and \SIrange{\hydrogenstorageacnhyn}{\hydrogenstorageacyhyn}{\twh}
without a hydrogen network while featuring similar filling level patterns
throughout the year (\cref{fig:si:soc}). Almost all hydrogen is stored in salt
caverns, exploiting vast geological potentials across Europe mostly in Northern
Ireland, England and Denmark. We observe no storage in steel tanks unless both
hydrogen and electricity networks cannot be expanded. In this case, we see up to
1~TWh of steel tank capacity, which represents 5\% of the total hydrogen storage
capacity. If the options for network development are restricted, more hydrogen
storage is built to balance renewables in time rather than in space.

\section*{Hydrogen network takes over role of bulk energy transport}
\label{sec:energy-moved}
\addcontentsline{toc}{section}{\nameref{sec:energy-moved}}

\begin{figure}
    \centering
        \begin{subfigure}[t]{0.49\textwidth}
            \centering
            \caption{transmission capacity built}
            \includegraphics[width=\textwidth]{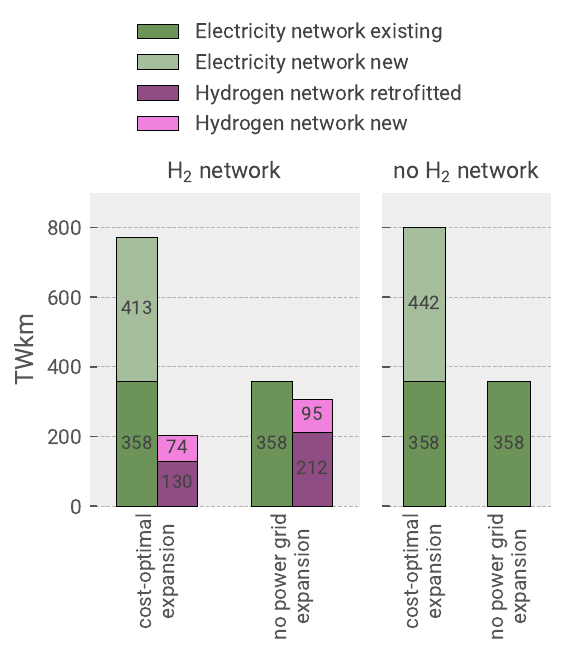}
            \label{fig:network-stats:twkm}
        \end{subfigure}
        \begin{subfigure}[t]{0.49\textwidth}
            \centering
            \caption{energy volume transported}
            \includegraphics[width=\textwidth]{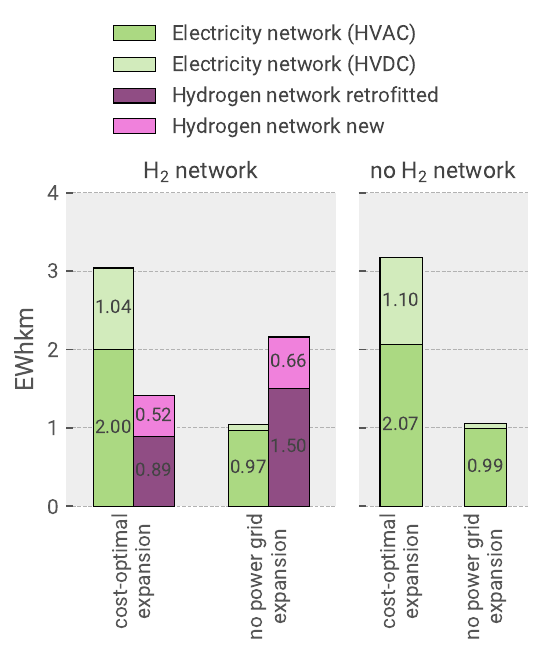}
            \label{fig:network-stats:ewhkm}
        \end{subfigure}
    \caption{Transmission capacity built and energy volume transported for
        various network expansion scenarios. For the hydrogen network, a
        distinction between retrofitted and new pipelines is made. For the
        electricity network, a distinction is made between existing and added
        capacity or how much energy is moved via HVAC or HVDC power lines. Both
        measures weight capacity (TW) or energy (EWh) by the length (km) of the
        network connection.}
    \label{fig:network-stats}
\end{figure}

Depending on the level of power grid expansion, between 204 and 307~TWkm of
hydrogen pipelines are built (\cref{fig:network-stats:twkm}). The higher value
is obtained when the hydrogen network partially offsets the lack of electricity
grid reinforcement. On the other hand, restricting hydrogen expansion only has a
small effect on cost-optimal levels of power grid expansion. The length-weighted
power grid capacity is more than doubled in the least-cost scenario; without a
hydrogen network, the cost-optimal power grid capacity is 7\% higher.

When both hydrogen and electricity grid expansion is allowed, the hydrogen
network transports approximately half the amount of energy transmitted via the
electricity network (\cref{fig:network-stats:ewhkm}). This is striking because
the hydrogen network capacity is little more than a quarter that of the power
grid (\cref{fig:network-stats:twkm}). In consequence, the utilisation rate of
\utilisationHy\% of the hydrogen network is much higher than the
\utilisationAC\% of the electricity grid (\cref{fig:si:grid-utilisation}). One
plausible explanation for this observation is that the buffering of produced
hydrogen in cavern storage allows more coordinated bulk energy tranport in
hydrogen networks, whereas the power grid directly balances the variability of
renewable electricity supply and is subject to linearised power flow physics
(Kirchhoff's circuit laws).

When electricity grid expansion is restricted, the hydrogen network plays a
dominant role in transporting energy around Europe. In this case, around twice
as much energy is moved in the hydrogen network (\ewhkmhydrogen~EWhkm) than in
the electricity network (\ewhkmelectricity~EWhkm). Between only power grid
expansion and only hydrogen network expansion, the difference in the total
volume of energy transported is only \ewhkmdiff\%.

\section*{New hydrogen network can leverage repurposed natural gas pipelines}
\label{sec:repurposed}
\addcontentsline{toc}{section}{\nameref{sec:repurposed}}

\begin{figure}
    \centering
    \makebox[\textwidth][c]{
        \begin{subfigure}[t]{0.65\textwidth}
            \centering
            \caption{hydrogen infrastructure with power grid reinforcement}
            \includegraphics[width=\textwidth]{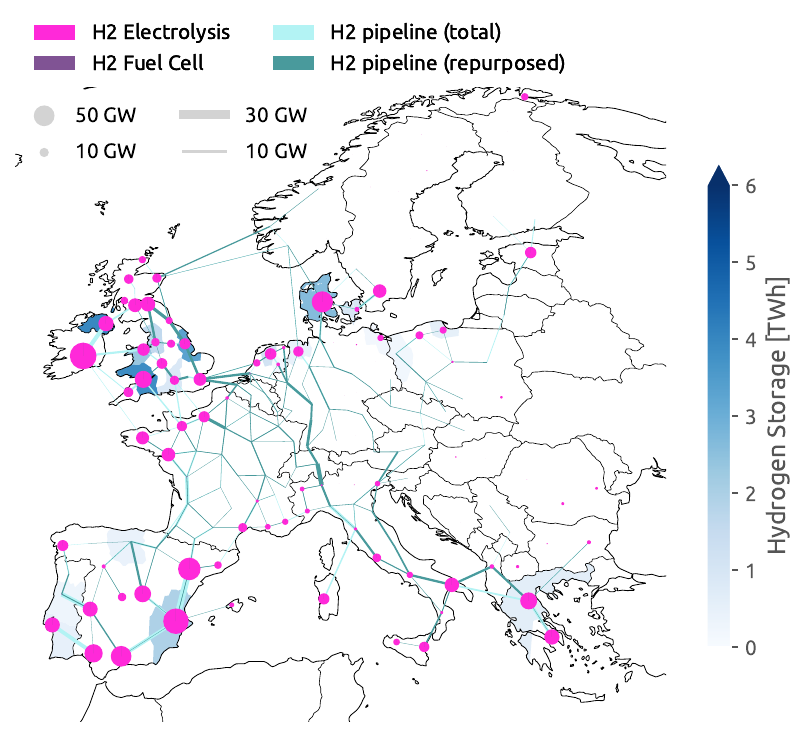}
            \label{fig:h2-network:w-el}
        \end{subfigure}
        \begin{subfigure}[t]{0.65\textwidth}
            \centering
            \caption{hydrogen infrastructure without power grid reinforcement}
            \includegraphics[width=\textwidth]{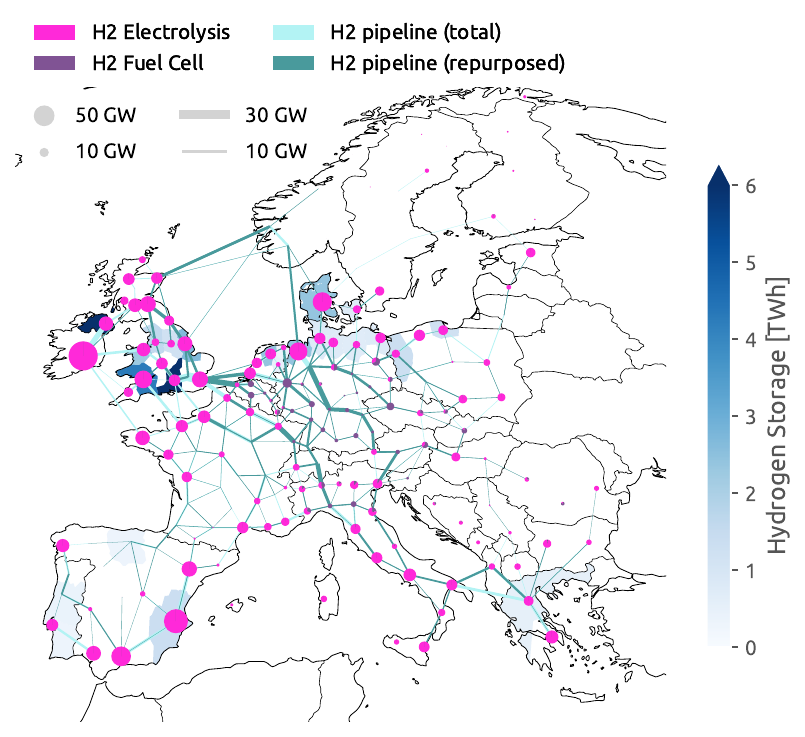}
            \label{fig:h2-network:wo-el}
        \end{subfigure}
    } \makebox[\textwidth][c]{
    \begin{subfigure}[t]{0.65\textwidth}
        \centering
        \caption{hydrogen flows with power grid reinforcement}
        \includegraphics[width=\textwidth]{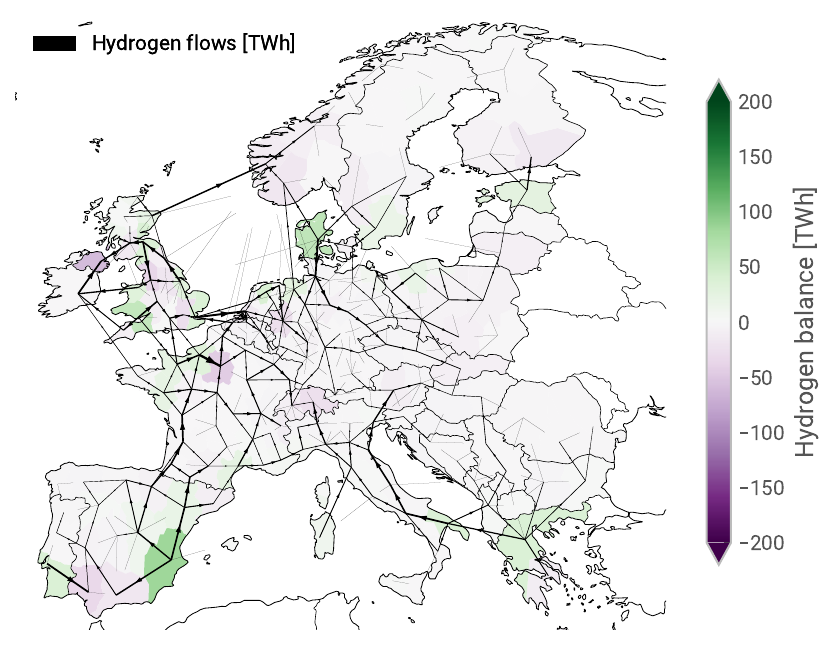}
    \end{subfigure}
    \begin{subfigure}[t]{0.65\textwidth}
        \centering
        \caption{hydrogen flows without power grid reinforcement}
        \includegraphics[width=\textwidth]{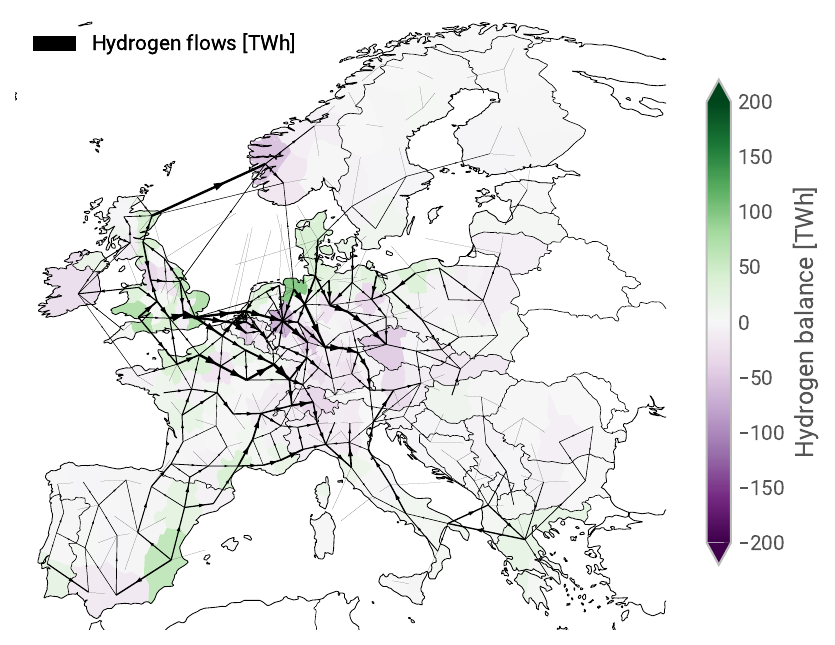}
    \end{subfigure}
    } \caption{Optimised hydrogen network, storage, reconversion and production
    sites with and without electricity grid reinforcement. The size of the
    circles depicts the electrolysis and fuel cell capacities in the respective
    region. The line widths depict the optimised hydrogen pipeline capacities.
    The darker shade depicts the share of capacity built from retrofitted gas
    pipelines. The coloring of the regions indicates installed hydrogen storage
    capacities. The second row shows net flow of hydrogen in the network and the
    respective energy balance. Flows larger than 2~TWh are shown with arrow
    sizes proportional to net flow volume.}
    \label{fig:h2-network}
\end{figure}

With our assumptions, developing electricity transmission lines is approximately
60\% more expensive than building new hydrogen pipelines. We assume costs for a
new hydrogen pipeline of 250~\euro/MW/km, whereas, for a new high-voltage
transmission line, we assume 400~\euro/MW/km (see~\cref{sec:si:costs}). Despite
higher costs, we observe that electricity grid reinforcements are preferred over
hydrogen pipelines. Part of the reason is that electricity has more versatile
end uses in transport, buildings and industry in our scenarios with high levels
of direct electrification. Hydrogen can only be used directly in a few
specialised sectors, and if hydrogen has to be produced only to be
re-electrified later, the efficiency losses mean additional generation capacity
would be needed to compensate. This makes energy transport in form of hydrogen
less competitive. However, hydrogen pipelines are particularly attractive where
the end-use is hydrogen-based.

The appeal of a hydrogen network is further spurred when existing natural gas
pipelines are available for retrofitting. Repurposing costs just around half
that of building a new hydrogen pipeline (117 versus 250 \euro/MW/km;
see~\cref{sec:si:costs}). For the capacity retrofit we include costs for
required compressor substitutions and assume that for every unit of gas pipeline
decommissioned, 60\% of its capacity becomes available for hydrogen transport.
The threefold lower volumetric energy density of hydrogen compared to natural
gas is offset by the possibility to attain higher volume flows with hydrogen. In
consequence, even detours of the hydrogen network topology may be cost-effective
if, through rerouting, more repurposing potentials can be tapped.

As \cref{fig:h2-network} illustrates, the optimised hydrogen network topology is
built around supporting flows into the industrial and population centres of
Central Europe. We see strong pipeline connections in Northwestern Europe to
integrate wind-based hydrogen hubs as well as connections for the transport of
solar hydrogen hubs from Spain, Italy and Greece. Individual pipeline
connections between regions have optimised capacities up to 30 GW. Of the total
hydrogen network volume, between 64\% and 69\% consists of repurposed gas
pipelines. The share is highest when the electricity grid is not permitted to be
reinforced. Up to a quarter of the existing natural gas network is retrofitted
to transport hydrogen instead, leaving large capacities that are used neither
for hydrogen nor methane transport. In our scenarios, 29-42\% of retrofittable
gas pipelines fully exhaust their conversion potential to hydrogen. The most
notable corridors for gas pipeline retrofitting are located offshore across the
North Sea and the English Channel and in Great Britain, Germany, Austria,
Switzerland, Northern France and Italy. The most prominent new hydrogen
pipelines are built in the British Isles particularly to connect Ireland,
Northern France, the Netherlands, and in Spain and Portugal. The sizeable
existing natural gas transmission capacities in Southern Italy and Eastern
Europe are largely not repurposed for hydrogen transport in this self-sufficient
scenario for Europe.

However, this picture would change if clean energy import options were
considered. Since most hydrogen is used to produce synthetic hydrocarbons and
ammonia, much of the hydrogen demand would fall away if these derivatives were
imported. In a sensitivity analysis in \cref{sec:si:sensitivity-imports}, we
show that the relative cost benefits of hydrogen network expansion are not
strongly affected by importing all liquid hydrocarbons, even though this action
would reduce the cost-optimal extent of hydrogen infrastructure by more than
50\%. Moreover, direct hydrogen imports into Europe by pipeline or ship could
alter cost-effective network topologies as new import locations need to be
connected rather than domestic production sites. For instance, the networks role
might change from distributing energy from North Sea hydrogen hubs to
integrating inbound pipelines from North Africa with increased network
capacities in Southern Europe.\cite{wetzelGreenEnergy2022}

\section*{Regional imbalance of supply and demand is reinforced by transmission}
\label{sec:imbalance}
\addcontentsline{toc}{section}{\nameref{sec:imbalance}}

\begin{figure}
    \centering
    \makebox[\textwidth][c]{
        \begin{subfigure}[t]{0.65\textwidth}
            \centering
            \caption{with power grid reinforcement, with hydrogen network}
            \includegraphics[width=\textwidth]{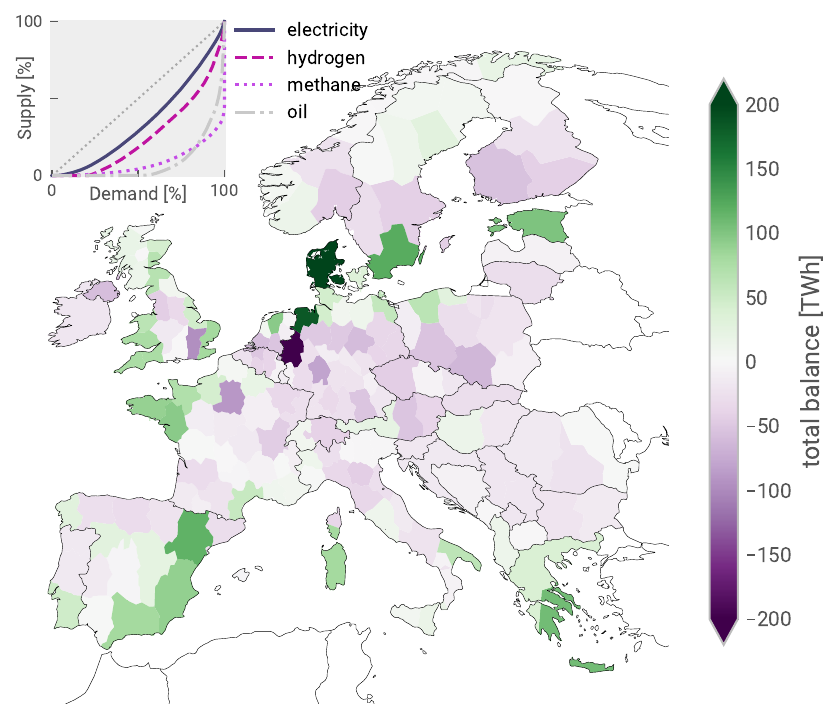}
            \label{fig:io:w-el-w-h2}
        \end{subfigure}
        \begin{subfigure}[t]{0.65\textwidth}
            \centering
            \caption{with power grid reinforcement, without hydrogen network}
            \includegraphics[width=\textwidth]{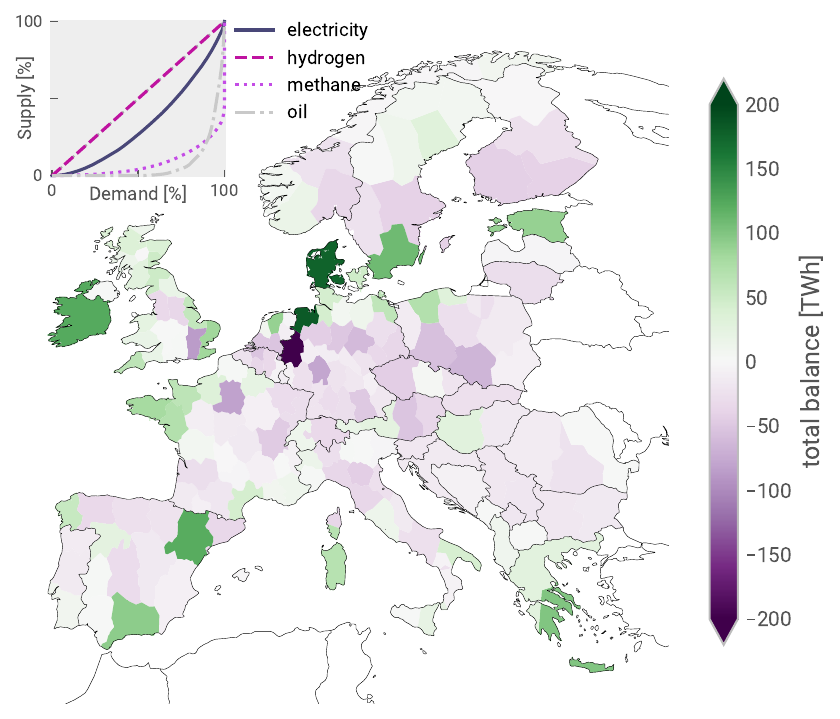}
            \label{fig:io:w-el-wo-h2}
        \end{subfigure}
    } \makebox[\textwidth][c]{
        \begin{subfigure}[t]{0.65\textwidth}
            \centering
            \caption{without power grid reinforcement, with hydrogen network}
            \includegraphics[width=\textwidth]{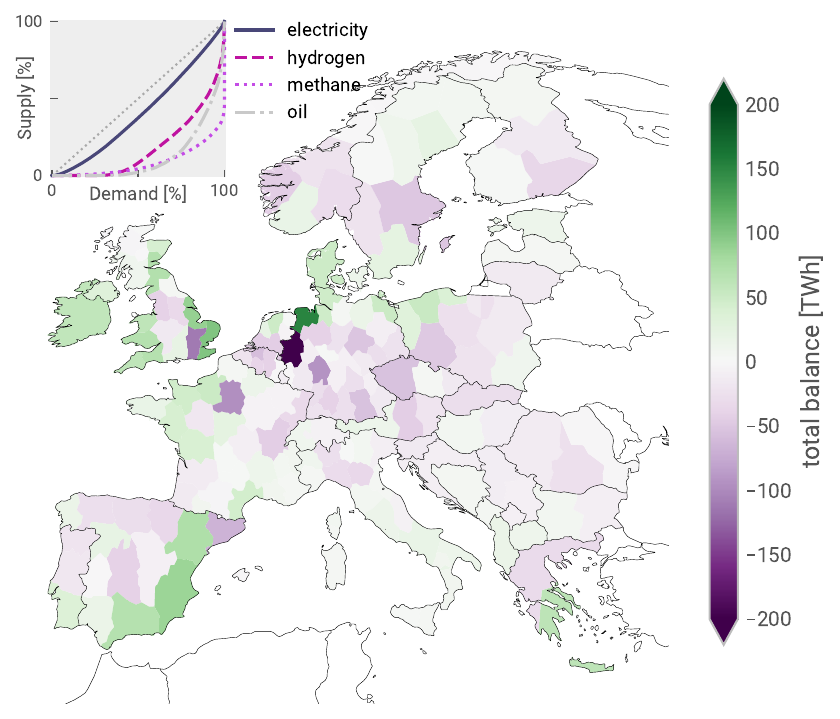}
            \label{fig:io:wo-el-w-h2}
        \end{subfigure}
        \begin{subfigure}[t]{0.65\textwidth}
            \centering
            \caption{without power grid reinforcement, without hydrogen network}
            \includegraphics[width=\textwidth]{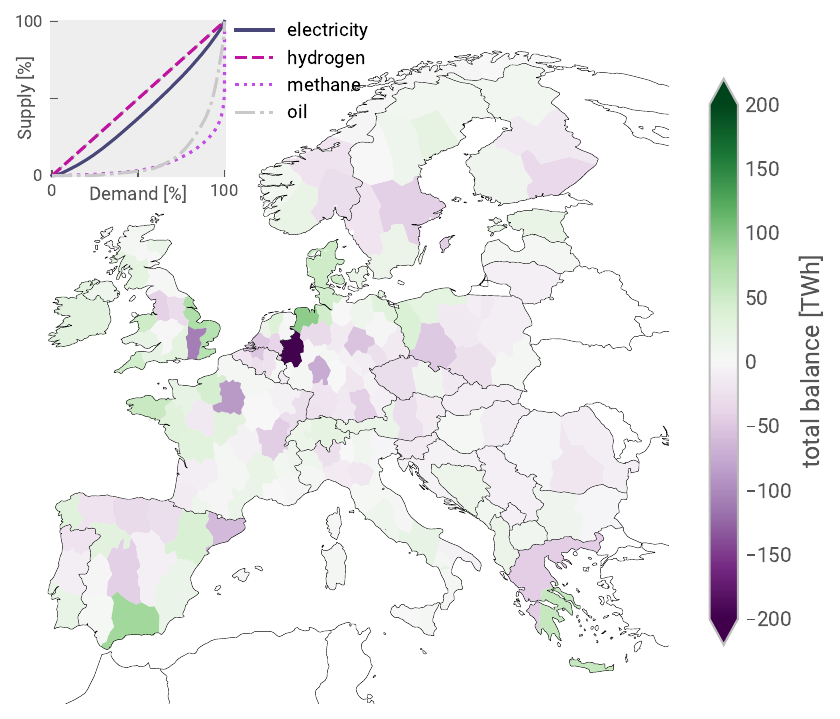}
            \label{fig:io:wo-el-wo-h2}
        \end{subfigure}
    } \caption{Total energy balances for scenarios with and without electricity
    or hydrogen network expansion for the 181 model regions, revealing regions
    with net energy surpluses and deficits. The Lorenz curves on the upper left
    of each map depict the regional imbalances of electricity, hydrogen, methane
    and liquid hydrocarbon supply relative to demand. Methane and liquid
    hydrocarbon supply can be of fossil, biogenic or synthetic origin. If the
    annual sums of supply and demand are equal in each region, the Lorenz curve
    resides on the identiy line. But the more imbalanced the regional supply is
    relative to demand, the further the curve dents into the bottom right corner
    of the graph.}
    \label{fig:io}
\end{figure}

In line with previously shown capacity expansion plans, energy surplus is found
largely in the windy coastal and sunny Southern regions that supply the inland
regions of Europe, which have high demands but less attractive renewable
potentials (\cref{fig:io}). The net energy surplus of individual regions amounts
to up to 260~TWh. Examples are Danish offshore wind power exports and large
wind-based production sites for synthetic fuels in Ireland. For Denmark, this
surplus is more than twice as high as its final energy demand, resulting in the
situation that three quarters of Denmark's energy production is exported. Net
deficits of single regions can have similarly high values, close to 200~TWh.
Examples are, in particular, the industrial cluster between Rotterdam and the
Ruhr valley as well as other European metropolises.

Energy transport infrastructure fuels the uneven regional distribution of supply
relative to demand. This is illustrated by the Lorenz curves in \cref{fig:io}
for different energy carriers. The Lorenz curves plot the carrier's cumulative
share of supply versus the cumulative share of demand, sorted by the ratio of
supply and demand in ascending order. If the annual sums of supply and demand
are equal in each region, the Lorenz curve resides on the identity line.
However, the more unequal the regional supply is relative to demand, the further
the curves dent into the graph's bottom right corner. For the least-cost
scenario, \cref{fig:io:w-el-w-h2} highlights that supply and demand of hydrogen
is slightly more regionally imbalanced than electricity. Reduced power grid
expansion causes more evenly distributed electricity supply
(\cref{fig:io:wo-el-w-h2,fig:io:wo-el-wo-h2}), and when hydrogen transport is
restricted (\cref{fig:io:w-el-wo-h2,fig:io:wo-el-wo-h2}), the production of
liquid hydrocarbons is increased in regions with attractive renewable potentials
because they can be transported at low cost.

\section*{Discussion}
\label{sec:discussion}
\addcontentsline{toc}{section}{\nameref{sec:discussion}}

To put our results into a broader perspective, for the discussion we compare
them to related literature and proposals presented in the gas industry's
European Hydrogen Backbone reports. This is followed by an appraisal of the
limitations of our study and a derivation of policy implications based on
spatial and operational insights.

\subsection*{Comparison to Related Literature}

Compared to the net-zero scenarios from the European
Commission,\cite{in-depth_2018} we see much larger wind and solar electricity
generation reaching beyond 8600~TWh compared to approximately 5700~TWh. This is
also reflected in the capacities built that exceed 2000~GW in our scenarios
compared to 1200~GW for wind and 3500~GW compared to 1000~GW for
solar.\cite{in-depth_2018} In terms of total electricity produced, our results
approximately show a tripling of today's generation compared to an increase by
145\% in the Commission's net-zero scenarios.\cite{in-depth_2018} Roughly one
third goes to regular electricity demand, one third goes to newly electrified
sectors in heating, transport and industry, and another third goes to hydrogen
production (dominated by demand for liquid hydrocarbons). The major difference
to the Commission's scenarios\cite{in-depth_2018} is caused by their lower
electrification rates, a 15\% share of nuclear power in the electricity mix,
higher biomass usage across all sectors (2900~TWh/a versus our 1400~TWh/a), and
a strong reliance on fossil fuels imports (2900~TWh/a) for non-energy uses (e.g.
plastics and other high-value chemicals). By considering landfill of plastics as
long-term carbon sequestration option, the Commission's scenarios see little
need to produce synthetic hydrocarbons for non-energy feedstocks. On the
contrary, our modelling, which assumes that all carbon in waste will be
incinerated or eventually decay into the atmosphere and limited sequestration
potentials, requires sustainable carbon sources for green electrofuels and
precludes the wide-ranging use of fossil oil.

Using pathway optimisation, Victoria et
al.~\cite{victoriaSpeedTechnological2022} investigate the timing of when certain
technologies become important for the European energy transition, and find a
hydrogen network consistently appearing after 2035. However, owing to a
one-node-per-country resolution in that study, little can be said about
subnational network infrastructure needs, retrofitting opportunities for gas
pipelines or regional geological storage potentials. Compared to our findings,
limited network expansion options affect total energy system costs less in
Victoria et al.~\cite{victoriaSpeedTechnological2022} A doubling of today's
transmission volume reduces cumulative system costs between 2020 and 2050 by 2\%
in Victoria et al.,~\cite{victoriaSpeedTechnological2022} compared to
\maxacbenefitrel\% in this study. Disabling hydrogen network expansion increases
cumulative costs by 0.5\% in Victoria et
al.,\cite{victoriaSpeedTechnological2022} compared to \maxhybenefitrel\% in this
study. This discrepancy arises because country-internal transmission bottlenecks
are not captured, whereby the integration costs of remote resources like
offshore wind within the countries are neglected.

Caglayan et al.~\cite{Caglayan2019} also consider European decarbonisation
scenarios with both electricity and hydrogen networks, but at lower spatial
resolution (96 regions) and without the industry, shipping, aviation,
agriculture and non-electrified heating sectors. A similar pattern of hydrogen
pipeline expansion towards the British Isles and North Sea is seen, but lower
overall electrolyser capacities (258~GW compared to our
\SIrange{\minelectrolysis}{\maxelectrolysis}{\giga\watt}) because not all
sectors are included. Caglayan et al.~\cite{Caglayan2019} also find cost-optimal
hydrogen storage of 130~TWh, whereas our scenarios involve just between
\hydrogenstorageacnhyn~and \hydrogenstorageacnhyy~TWh owing to the larger
flexible hydrogen demand diminishing the need for weekly and monthly balancing.

A large number of cost-effective designs for a climate-neutral European energy
system was also presented by Pickering et
al.~\cite{pickeringDiversityOptions2022} Their 98-region model with 2-hourly
resolution likewise includes all energy sectors including non-energy feedstocks
and also assumes energy self-sufficiency for Europe. However, hydrogen transport
options were not considered such that hydrogen must be produced locally.
Moreover, geological potentials for low-cost underground hydrogen storage and
the option to retrofit gas pipelines are not included. Owing to higher storage
cost in steel tanks and fewer assumed end-uses of hydrogen and its derivatives,
the scenarios involve less hydrogen storage (\SIrange{0}{6}{\twh} versus
\SIrange{\hydrogenstorageacnhyn}{\hydrogenstorageacnhyy}{\twh}) and lower
electrolyser capacities (\SIrange{290}{855}{\giga\watt} versus
\SIrange{\minelectrolysis}{\maxelectrolysis}{\giga\watt}) in our results.
Furthermore, whereas our model allows limited use of fossil fuels and with
options for carbon capture and sequestration, Pickering et
al.~\cite{pickeringDiversityOptions2022} eliminate the use of fossil energy and
only consider direct air capture as a carbon source. Overall, total energy
system costs lie in a similar range between 730 and 866 bn\euro/a compared to
costs between \minsystemcost~and \maxsystemcost~bn\euro/a in our study.

\subsection*{Comparison to the European Hydrogen Backbone}

Our results are aligned with the European Hydrogen Backbone
(EHB).\cite{gasforclimateEuropeanHydrogen2020,gasforclimateExtendingEuropean2021,gasforclimateEuropeanHydrogen2021,gasforclimateEuropeanHydrogen2022}
Whereas no detailed modelling lies behind the visions in the EHB reports, we
present analysis based on temporally resolved spatial co-planning of energy
infrastructures. We see cost-optimal hydrogen network investments in the range
of \minhycost-\maxhycost~bn\euro/a, while the EHB report covering 21 countries
finds slightly higher costs between
4-10~bn\euro/a.\cite{gasforclimateExtendingEuropean2021}\footnote{To calculate
the annuity of the overnight hydrogen network costs listed in the EHB reports, a
lifetime of 50 years and a discount rate of 7\% are assumed.} The extension to
28 countries reports costs between 7-14
bn\euro/a.\cite{gasforclimateEuropeanHydrogen2022} Compared to the hydrogen
backbone vision presented in the EHB from April
2021,\cite{gasforclimateExtendingEuropean2021}\footnote{The newer EHB report
from April 2022 \cite{gasforclimateEuropeanHydrogen2022} lacks sufficient data
to calculate length-weighted network capacities.} our scenarios show a
similarly sized hydrogen network with comparable retrofitting shares. Measured
by the length-weighted sum of pipeline capacities (TWkm), the 309 TWkm indicated
in the EHB report match the upper end of the range of 204-307~TWkm observed in
our scenarios. Likewise, the 69\% share of repurposed natural gas pipelines
\cite{gasforclimateExtendingEuropean2021} roughly agrees with our findings where
between \minretroshare\% and \maxretroshare\% of hydrogen pipelines are
retrofitted gas pipelines. In contrast to the EHB reports, we also explore
solutions without a hydrogen network, which we find to be feasible as well.

\subsection*{Limitations of the study and scope for future investigations}
\label{sec:limitations}

In our scenarios, Europe is largely energy self-sufficient. While limited
amounts of fossil gas and oil imports are allowed, no imports of renewable
electricity, chemical energy carriers or commodities from outside of Europe are
considered. However, including green imports may change system needs for
electricity and hydrogen transmission infrastructure substantially. New hydrogen
import hubs might require different bulk transmission routes. The import of
large amounts of carbon-based fuels and ammonia would furthermore diminish the
demand for hydrogen overall, and hence also the need to transport it. This
effect of wide-ranging imports of liquid hydrocarbon demand on infrastructure
needs is demonstrated in \cref{sec:si:sensitivity-imports} and should be
explored in more detail in future work.
\cite{fasihiTechnoeconomicAssessment2019,heuserTechnoeconomicAnalysis2019,hamppImportOptions2023}

Additionally, the very uneven distribution of energy supply in our results may
interfere with the level of social acceptance for new infrastructure to an
extent that may block a swift energy transition.
\cite{sasseDistributionalTradeoffs2019,sasseRegionalImpacts2020} Hence, future
investigations should weigh the cost surcharge of increased regionally
self-sufficient energy supply against the potential benefit of higher public
acceptance and increased resilience.

Previous research has shown that the system design can be changed in many ways
with only a small change in total
costs.\cite{Neumann2019,lombardiPolicyDecision2020,pedersenModelingAll2021,pickeringDiversityOptions2022}
This breadth of options makes robust statements about specific locational
infrastructure needs more vague. While we present selected design trade-offs
regarding transmission networks and some further sensitivities
(\cref{sec:si:sensitivity}), a more comprehensive exploration of near-optimal
solutions would be prudent, especially in the directions of carbon management
infrastructure, biomass usage, the level of energy imports, industry transition
and relocation options, more regionally balanced infrastructure, and increased
system resilience.

Owing to the absence of pathway optimisation, our results cannot offer insights
into the required transition steps and how the gradual transformation may
restrict certain options towards the final climate-neutral state. For example,
our results do not show which parts of gas network could be repurposed first or
where the benefit of a hydrogen network might be the highest initially. In the
context of multi-horizon planning, we also neglect the dynamics of technological
learning by
doing.\cite{heubergerPowerCapacity2017,fellingMultihorizonPlanning2022,zeyenEndogenousLearning2022}
The transformation to net-zero emissions requires vast and timely growth rates
of power-to-X and carbon dioxide removal technologies to realise anticipated
cost reductions,\cite{odenwellerProbabilisticFeasibility2022a} which we assume to
be given by assuming fixed technology cost.

Further limitations include that heat demands and the availability of renewables
vary considerably year by year such that our restriction to a single year may
limit the robustness to interannual weather variability; we do not consider new
nuclear power plants; we do not consider secondary benefits of grid expansion
for the provision of ancillary services; and for the transport and industry
sectors we make some exogenous assumptions about process switching, drive
trains, alternative fuels for industry heat and recycling rates which may have
turned out differently if they were endogenously optimised.

\subsection*{Derivation of policy implications from regional and operational insights}
\label{sec:policy}

Regardless of the energy carrier transported, our results highlight that
cooperation between European countries is important to reach net-zero \co
emissions most cost-effectively. This is because there are significant
differences in renewable resources across Europe. The cost differential between
supply in Europe's demand centers and periphery outweigh the cost of building
new transmission infrastructure. Thus, we see both substantial net importers
(e.g.~the industry clusters in Ruhr valley and Rotterdam area) and strong net
exporters of energy (e.g.~Denmark, Ireland, Spain, Greece). The option to
transport energy around Europe also counteracts incentives for industry
relocation. Expanding energy transport infrastructure may be less controversial
since it would affect regional development less than the migration of
industries.

Regarding hydrogen production, we see both solar-based hubs in Southern Europe
and wind-based hubs in Northern Europe using water electrolysis. The regional
and technological diversity in electrolytic hydrogen production is the preferred
solution, but the impetus for Southern solar-based hubs is greatly affected by
the evolution of other system components. Difficulties to install sufficient
onshore wind capacities around the North Sea would reinforce their relevance,
whilst the import of most liquid hydrocarbons from outside of Europe would
weaken the case for solar-based hubs. Our results also highlight that compared
to the amount of electrolytic hydrogen, blue hydrogen from steam methane
reforming with carbon capture only plays a marginal role and was only used in
our scenarios when no hydrogen network could be developed.

As the general hydrogen network benefit is not dependent on electricity grid
reinforcements, both networks could be developed in parallel. Thus, policymaking
could focus on options that are most easily achieved and widely accepted. While
the hydrogen network benefit is not affected by alternate technology cost
developments or import policies, the network topology is. Lower costs of solar
photovoltaics raise the appeal of hydrogen production hubs in Southern Europe,
altering the suitable hydrogen network layout. Likewise, wide-ranging hydrogen
imports from the MENA region would need to be supported with transmission
infrastructure in Southern Europe.

The flexible operation of electrolysers has several advantages for system
stability and integrating wind and solar generation cost-effectively and should
be incentivised. Fluctuating renewable generation is buffered in geological
hydrogen storage primarily in the UK, Denmark, Spain and Greece to achieve more
continuous production in capital-intensive fuel synthesis plants in accordance
with their operational restrictions. This leads to low curtailment rates of
renewables and a lower requirement for firm capacity, outlining the benefit of
cross-sectoral approaches for reducing \co emissions cost-effectively. Fuel cell
CHP plants in Germany can further support grid operation when the power grid
cannot be expanded. However, energetically the re-electrification of hydrogen
only plays a minor role in this sector-coupled system.

To reach the net-zero energy systems we have modelled with new transmission
networks and leveraging of various sector-coupling flexibilities, many changes
are needed in policy and regulation. Tight coordination between countries and
energy sectors is required to achieve low-cost solutions, similar to how the
process for the Ten Year Network Development Plan (TYNDP) has moved towards
joint planning.\cite{entso-eTYNDP20222022} To achieve the coordination of
dispatch and capacity expansion at the local level around grid bottlenecks,
particularly if electricity and hydrogen network expansion is limited, local
price signals are required corresponding to our 181 bidding zones
(\cref{fig:si:lmp-ac,fig:si:lmp-h2}). In our model, electric vehicles and heat
pumps operate flexibly, which requires the deployment of smart meters and
dynamic electricity tariffs to incentivise grid-supporting   behaviour. And
finally, a sustained rise in the price of \co emission certificates is needed.
The results we show are also contingent on adjusted regulations and rules for
building infrastructure and developing competitive markets for hydrogen and
carbon dioxide.

\section*{Conclusion}
\label{sec:conclusion}
\addcontentsline{toc}{section}{\nameref{sec:conclusion}}

In this work, we have investigated the potential role of a hydrogen network in
net-zero \co scenarios for Europe with high shares of renewables. The analysis
was performed using the open sector-coupled energy system model PyPSA-Eur-Sec
featuring high spatio-temporal coverage of all energy sectors (electricity,
buildings, transport, agriculture and industry across 181 regions and 3-hourly
resolution for a year). With this level of spatial, temporal, technological and
sectoral resolution, it is possible to represent grid bottlenecks as well as the
variability and regional distribution of demand and renewable supply. Thereby,
the system's infrastructure needs regarding generation, storage, transmission
and conversion can be assessed. This includes in particular trade-offs between
electricity grid reinforcement, which has limited public support, and developing
a hydrogen network, for which unused gas pipelines can be
repurposed.

Besides large-scale renewables expansion of wind turbines in Northern Europe and
solar photovoltaics in Southern Europe, the build-out of hydrogen infrastructure
is one of the biggest changes seen in our scenarios for the future European
energy system. Huge new electrolyser capacities enter the system and operate
flexibly to aid renewables integration. The siting of new hydrogen production
hubs is determined by access to excellent wind and solar resources in the
broader North Sea region and Spain in particular. Underground storage in salt
caverns is developed in the UK, Denmark, Spain and Greece for buffering, and a
new continent-spanning hydrogen pipeline network is built to connect cheap
supply and storage potentials in Europe's periphery with its industrial and
population centres. This new hydrogen network is supported by considerable
amounts of gas pipeline retrofitting: between \minretroshare\% and
\maxretroshare\% of the network uses repurposed pipes, especially in Central
European countries with existing gas infrastructure.

Our analysis reveals that a hydrogen network can reduce energy system costs by
up to \maxhybenefitrel\%. Cost reductions are shown to be highest when the
expansion of the power grid is restricted. However, hydrogen networks can only
partially substitute for grid expansion. We found that in fact both ways of
transporting energy and balancing renewable generation complement each other and
achieve the highest cost savings of up to \gridbenefitrel\% together. At the
same time, these findings also support the interpretation that neither
electricity nor hydrogen network expansion are essential for achieving a
cost-effective system design if such a cost premium can be accepted to achieve
alternative goals.

In conclusion, there appear to be many infrastructure trade-offs regarding how
and from where energy is transported across Europe, provided that energy
planning and operation can be tightly coordinated. More energy transport
capacity reduces costs, but some restrictions on grid expansion have only
limited impact on total energy system costs. This should enable policymakers to
choose from a wide range of compromise energy system designs with low cost but
higher acceptance.

\section*{Experimental Procedures}
\label{sec:methods}
\addcontentsline{toc}{section}{\nameref{sec:methods}}

\subsection*{Resource Availability}

\subsubsection*{Lead Contact}

Requests for further information, resources and materials should be directed to
the lead contact, Fabian Neumann
(\href{mailto:f.neumann@tu-berlin.de}{f.neumann@tu-berlin.de}).

\subsubsection*{Materials availability}

A dataset of the model results is available on Zenodo at
\href{https://doi.org/10.5281/zenodo.6821257}{doi:10.5281/zenodo.6821257}. The
code to reproduce the experiments is available on GitHub at
\href{https://github.com/fneum/spatial-sector}{github.com/fneum/spatial-sector}.
We also refer to the documentation of PyPSA
(\href{https://pypsa.readthedocs.io}{pypsa.readthedocs.io}), PyPSA-Eur
(\href{https://pypsa-eur.readthedocs.io}{pypsa-eur.readthedocs.io}), and
PyPSA-Eur-Sec
(\href{https://pypsa-eur-sec.readthedocs.io}{pypsa-eur-sec.readthedocs.io}).
Technology data was taken from
\href{https://github.com/pypsa/technology-data}{github.com/pypsa/technology-data}
(v0.4.0). An interactive scenario explorer can be found at
\href{https://h2-network.streamlit.app}{h2-network.streamlit.app}.

\subsubsection*{Data and Code Availability}

A dataset of the model inputs and results has been deposited to Zenodo at
\href{https://doi.org/10.5281/zenodo.6821257}{doi:10.5281/zenodo.6821257}. The
code to reproduce the experiments is available on GitHub at
\href{https://github.com/fneum/spatial-sector}{github.com/fneum/spatial-sector}.

\subsection*{Modelling Setup}

In this section the core characteristics and assumptions of the model
PyPSA-Eur-Sec are presented. More detailed descriptions of specific sectors,
energy carriers, renewable potentials, transmission infrastructure modelling,
and mathematical problem formulation are covered in the supplementary material
under \crefrange{sec:si:model-overview}{sec:si:math}.

The European sector-coupled energy system model PyPSA-Eur-Sec uses linear
\textbf{optimisation} to minimise total annual operational and investment costs
subject to technical and physical constraints, assuming perfect competition and
perfect foresight over one uninterrupted year of 3-hourly operation (see
\cref{sec:si:math} for mathematical formulation). In this study, we used the
historical year 2013 for weather-dependent inputs. Apart from existing
electricity and gas transmission infrastructure and hydroelectric power plants,
no other existing assets are assumed (\textit{greenfield optimisation} or
\textit{overnight scenario}), so that the model assumes a long-term equilibrium
in a market with perfect competition and foresight, and disregards pathway
dependencies. The model is implemented in the free and open software framework
Python for Power System Analysis (PyPSA).\cite{brownPyPSAPython2018}

PyPSA-Eur-Sec builds upon the model from Brown et
al.,~\cite{brownSynergiesSector2018} which covered electricity, heating in
buildings and ground transport in Europe with one node per country.
PyPSA-Eur-Sec adds biomass on the supply side, industry, agriculture, aviation
and shipping on the demand side, and higher spatial resolution to suitably
assess infrastructure requirements. In this study, the European continent is
divided into 181 regions. Unavoidable process emissions, feedstock demands in
the chemicals industry and the need for dense fuels for aviation and shipping,
also required the addition of a detailed representation of the carbon cycles,
including carbon capture from industry processes, biomass combustion and
directly from the air (DAC).

\cref{fig:multisector} gives an \textbf{overview} of the supply, transmission,
storage and demand sectors implemented in the model. To render interactions in
the sector-coupled energy system, we model the energy carriers electricity,
heat, methane, hydrogen, carbon dioxide and liquid hydrocarbons (oil, methanol,
naphtha) across the different energy sectors. Generator capacities (for onshore
wind, offshore wind, utility-scale and rooftop solar photovoltaic (PV), biomass,
hydroelectricity, oil and natural gas), heating capacities (for heat pumps,
resistive heaters, gas boilers, combined heat and power (CHP) plants and solar
thermal collector units), synthetic fuel production (electrolysers, methanation,
Fischer-Tropsch, steam methane reforming, fuel cells), storage capacities
(stationary and electric vehicle batteries, hydrogen storage in caverns and
steel tanks, pit thermal energy storage, pumped-hydro and reservoirs, and
carbon-based fuels like methane, methanol, and Fischer-Tropsch fuels), carbon
capture (from industry process emissions, steam methane reforming, CHP plants
and directly from the air), and transport capacities of electricity transmission
lines, new hydrogen and repurposed natural gas pipelines are all subject to
optimisation, as well as the operational dispatch of each unit in each
represented hour.

\textbf{Exogenous demand and supply assumptions} in the model include a fully
price inelastic and spatially-fixed demand for the different materials and
energy services in each sector, the extent of land transport electrification,
the use of methanol as shipping fuel and kerosene in aviation, process switching
in industry, the reuse and recycling rates of steel (70\%), aluminium (80\%) and
plastics (55\%) manufacturing, the ratio of district heating to decentralised
heating in densely populated regions, efficiency gains of 29\% due to building
retrofitting, hydroelectricity capacities (for reservoir and run-of-river
generators and pumped hydro storage).

For the \textbf{technology and cost assumptions}, we take estimates for the year
2030 for the main scenarios and run a sensitivity analysis with more progressive
cost projections for the year 2050 in \cref{sec:si:sensitivity-costs}. We take
technology projections for the year 2030 for the main scenarios to account for
expected technology cost reductions in the near-term while acknowledging that
the gradual transition to climate neutrality implies that much of the
infrastructure must be built well in advance of reaching net-zero emissions.
Many numbers come from the technology database published by the Danish Energy
Agency (DEA).\citeS{DEA} A complete referenced list of techno-economic
assumptions is compiled in \cref{tab:si:costs}. Among many other technologies,
for overnight costs we assume 636~\euro/kW$_e$ for rooftop PV, 487 \euro/kW$_e$
for utility-scale PV, 142~\euro/kWh and 160~\euro/kW for batteries,
1035~\euro/kW for onshore wind, 1524 \euro/kW for offshore wind, 450
\euro/kW$_e$ for electrolysers, 1100 \euro/kW$_e$ for fuel cell CHPs,
2~\euro/kWh for underground hydrogen storage, 0.54~\euro/kWh for central pit
thermal energy storage, 628-651~\euro/kW$_{out}$ for methanation,
methanolisation and Fischer-Tropsch processes, 572~\euro/kW$_{CH_4}$ for steam
methane reforming with carbon capture (i.e.~blue hydrogen), and 685~\euro/t for
direct air capture with uninterrupted operation.

The time series and potentials of variable renewable \textbf{energy supply}
(wind, solar, hydro, ambient heat) are computed from historical weather data
(ERA5 \cite{ecmwf} and SARAH-2 \cite{SARAH}). Potentials for wind and solar
generation take various land eligibility constraints into account, e.g.~suitable
land types and exclusion zones around populated and protected areas. As long as
emissions can be offset by negative emission technologies and sequestration
potentials are not exhausted, limited amounts of fossil oil and gas can still be
used as primary energy supply. While no assumption about the origin of fossil
energy is made, imports of renewables-based products into Europe are not
considered.

The full \textbf{transmission} network for European electricity transport is
taken from the electricity-only model version,
PyPSA-Eur,\cite{horschPyPSAEurOpen2018} and is clustered down to 181
representative regions based on the k-means network clustering methodology used
in Hörsch and Brown\cite{Hoersch2017} and Frysztacki et
al.~\cite{frysztackiStrongEffect2021}. This level of aggregation reflects, at
the upper end, the computational limit to solve a temporally resolved
sector-coupled energy system optimisation problem and, at the lower end, the
requirements to preserve the most important transmission corridors that cause
bottlenecks and limit the system integration of renewables. The impacts of
spatial aggregation are evaluated in \cref{sec:si:sensitivity-space}. Power
flows are modelled using a cycle-based load flow linearization from Hörsch et
al.~\cite{horschLinearOptimal2018} that significantly improves computational
performance. The power flow linearisation implies that no transmission losses
are considered. Hydrogen pipeline flows assume a simple transport model. This
means that while incoming and outbound flows must balance for each region and
pipes can transport hydrogen only within their capacity limits, no further
physical gaseous flow constraints are applied. The potential for gas pipeline
retrofitting is estimated based on consolidated network data from the
SciGRID\_gas project \cite{plutaSciGRIDGas2022a} such that for every unit of gas
pipeline decommissioned, 60\% of its capacity becomes available for hydrogen
transport.\cite{gasforclimateEuropeanHydrogen2020}

For \textbf{industry}, we assume that the demand for materials (such as steel,
cement, and high-value chemicals) remain constant, and disregard options for
industry relocation.\cite{toktarovaInteractionElectrified2022} The assumed
industry transformation is characterised by electrification, process switching
to low-emission alternatives (e.g. switching to hydrogen for direct reduction of
iron ore\cite{voglAssessmentHydrogen2018}), more recycling of steel, plastics
and aluminium\cite{circular_economy}, fuel switching for high- and
mid-temperature process heat to biomass and methane, use synthetic fuels for
ammonia and organic chemicals, and allow carbon capture. It is assumed that no
plastic or other non-energy product is sequestered in landfill, but that all
carbon in plastics eventually makes its way back to the atmosphere, either
through combustion or decay; this approach is stricter than other models.
\cite{in-depth_2018}

The \textbf{transport} sector comprises light and heavy road, rail, shipping and
aviation transport. For road and rail, electrification and fuel cell vehicles
for heavy-duty transport are available. For shipping, methanol is considered.
Aviation consumes kerosene whose origin (fossil or synthetic) is endogenously
determined. Half of the battery electric vehicle fleet for passenger transport
is assumed to engage in demand response schemes as well as vehicle-to-grid
operation.

The \textbf{buildings} sector includes decentral heat supply in individual
housing as well as centralised district heating for urban areas. Heating demand
can be met through air- and ground-sourced heat pumps, gas boilers,
gas/biomass/hydrogen CHPs, resitive heaters as well as waste heat from synthetic
fuel production in district heating networks. For district heating networks,
seasonal heat storage options are also available. Efficiency gains from building
retrofitting of 29\% are exogenous to the model based on Zeyen et
al.\cite{zeyenMitigatingHeat2021}

For \textbf{biomass}, only waste and residues from agriculture and forestry are
permitted, using the medium potential estimates from the JRC ENSPRESO database.
\cite{ruizENSPRESOOpen2019} This results in 336~TWh per year of biogas that can
be upgraded and 1038~TWh per year of solid biomass residues and waste for the
whole of Europe. Biomass can be used in combined electricity and heat generation
with and without carbon capture, as well as to provide low- to
medium-temperature process heat in industry.

\textbf{Carbon capture} is needed in the model both to capture and sequester
process emissions with a fossil origin, such as those from calcination of fossil
limestone in the cement industry, as well as to use carbon for the production of
hydrocarbons for dense transport fuels and as a chemical feedstock, for example
to produce plastic. \co can be captured from exhaust gases (industry process
emissions, steam methane reforming, CHP plants) or by direct air capture.
Captured \co can be used to produce synthetic hydrocarbons via Sabatier,
Fischer-Tropsch or methanolisation processes. Up to 200 Mt\co/a may be
sequestered underground, which is sufficient to capture process emissions but
limits the system's reliance on negative emission technologies. Landfill of
plastics is not considered as long-term sequestration option.

\section*{Acknowledgements}

We are grateful for helpful comments by Johannes Hampp. We thank the four
reviewers for their valuable feedback and suggestions.

\section*{Declaration of Interests}

The authors declare no competing interests.

\section*{Author Contributions}

\textbf{F.N.}:
Conceptualization --
Data curation --
Formal Analysis --
Investigation --
Methodology --
Software --
Supervision --
Validation --
Visualization --
Writing - original draft --
Writing - review \& editing
\textbf{E.Z.}:
Data curation --
Formal Analysis --
Investigation --
Software --
Validation --
Writing - review \& editing
\textbf{M.V.}:
Formal Analysis --
Investigation --
Methodology --
Software --
Writing - review \& editing
\textbf{T.B.}:
Conceptualization --
Data curation --
Formal Analysis --
Funding acquisition --
Investigation --
Methodology --
Project administration --
Resources --
Software --
Supervision --
Writing - original draft --
Writing - review \& editing

\addcontentsline{toc}{section}{References}
\renewcommand{\ttdefault}{\sfdefault}
\bibliography{/home/fneum/zotero}

\newpage

\makeatletter
\renewcommand \thesection{S\@arabic\c@section}
\renewcommand\thetable{S\@arabic\c@table}
\renewcommand \thefigure{S\@arabic\c@figure}
\makeatother

\renewcommand{\citenumfont}[1]{S#1}

\setcounter{equation}{0}
\setcounter{figure}{0}
\setcounter{table}{0}
\setcounter{section}{0}

\newgeometry{margin=2cm}

\addcontentsline{toc}{section}{Supplementary Information}
\section*{Supplementary Information}
\label{sec:si}
\normalsize

\begin{figure}[ht!]
    \centering
    \includegraphics[width=0.8\textwidth]{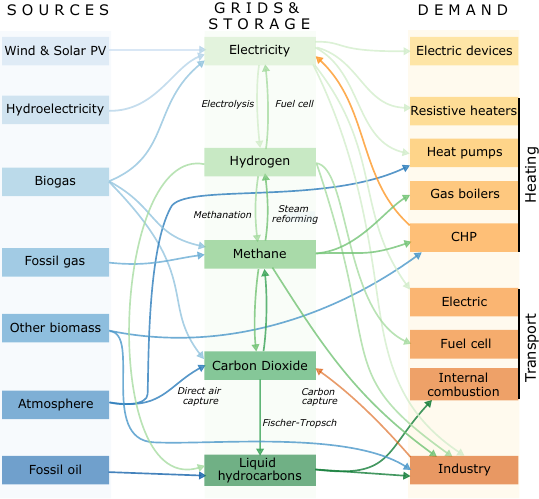}
    \caption{Overview of the circulation of energy and carbon in PyPSA-Eur-Sec.}
    \label{fig:multisector}
\end{figure}

\begin{figure}[ht!]
    \centering
    \includegraphics[height=0.3\textheight]{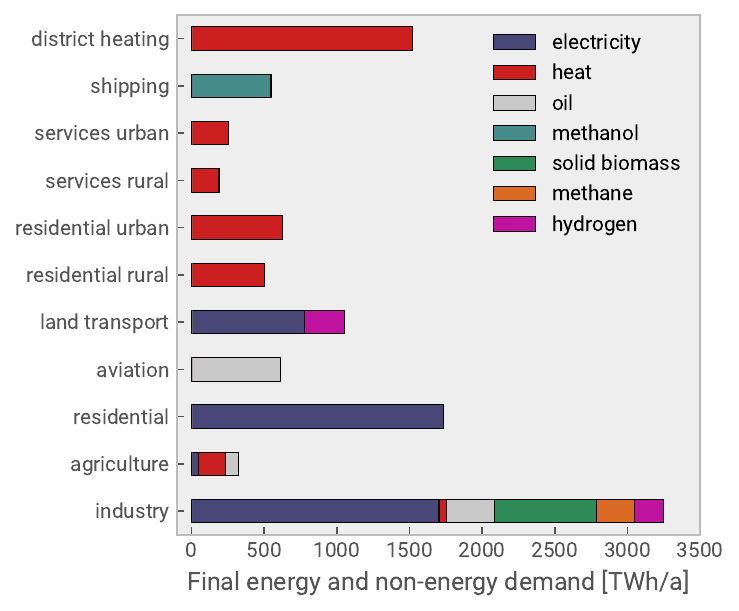}
    \includegraphics[height=0.3\textheight]{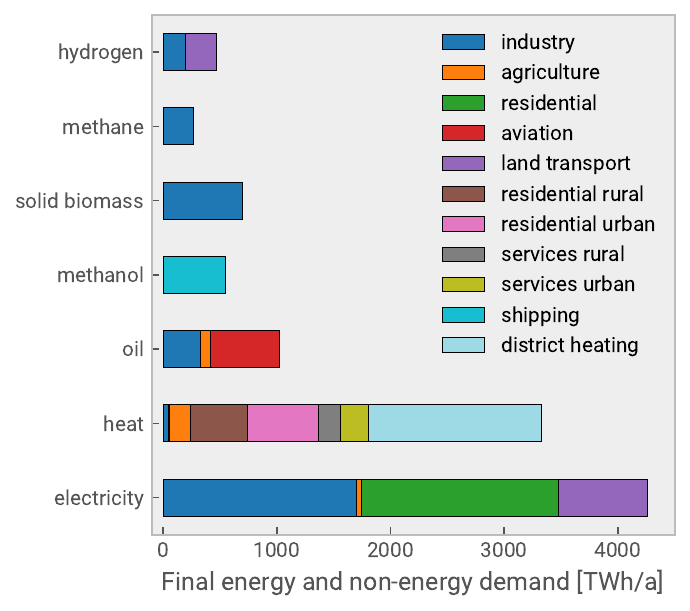}
    \caption{Annual final energy and non-energy demand by carrier and sector.}
    \label{fig:demand-by-sector-carrier}
\end{figure}

\begin{figure}[ht!]
    \centering
    \includegraphics[height=0.27\textheight]{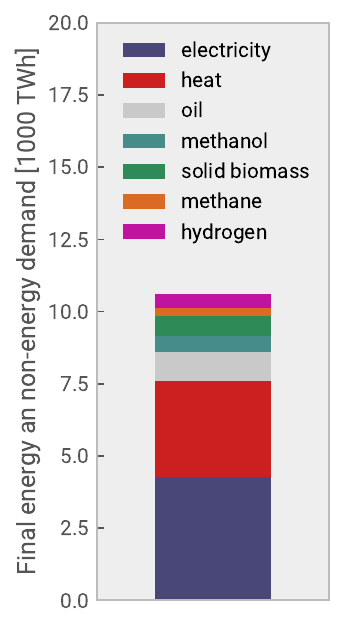}
    \includegraphics[height=0.27\textheight]{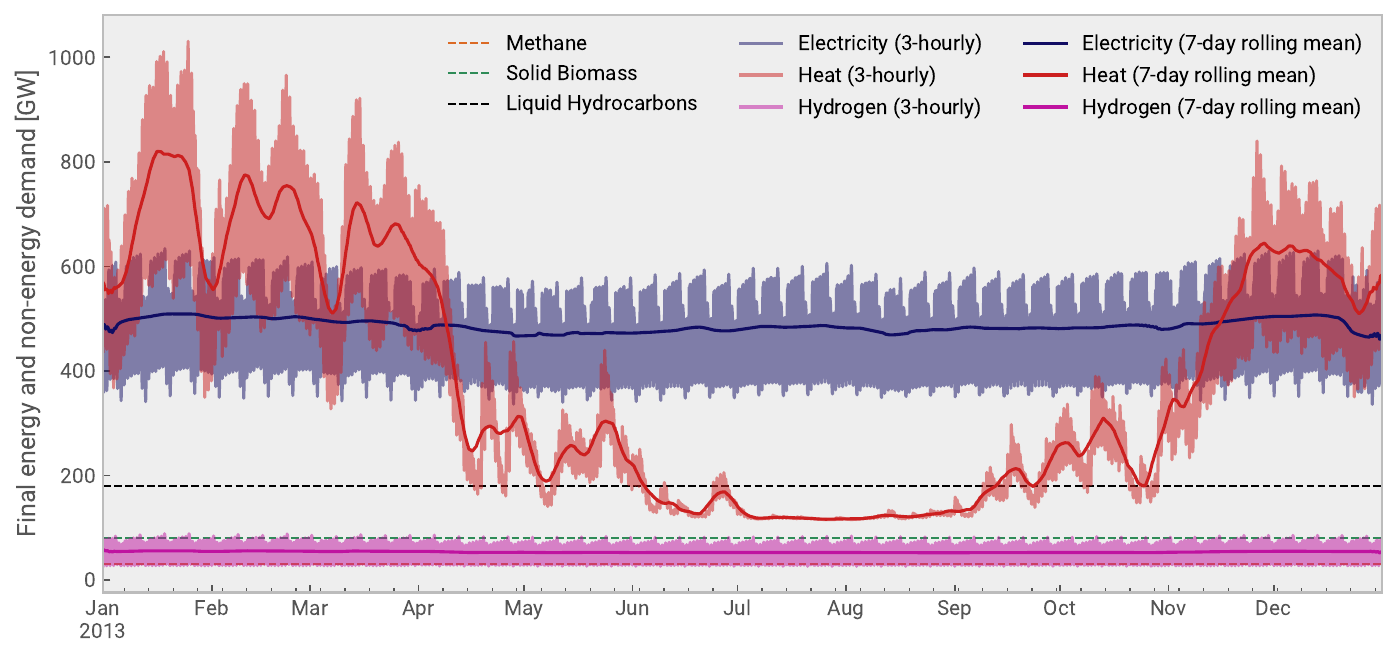}
    \caption{Annual final energy and non-energy demand (left) and system-level time series of demand by carrier (right).}
    \label{fig:demand-time}
\end{figure}

\begin{figure}[ht!]
    \centering
\begin{subfigure}[t]{0.49\textwidth}
    \centering
    \includegraphics[width=\textwidth]{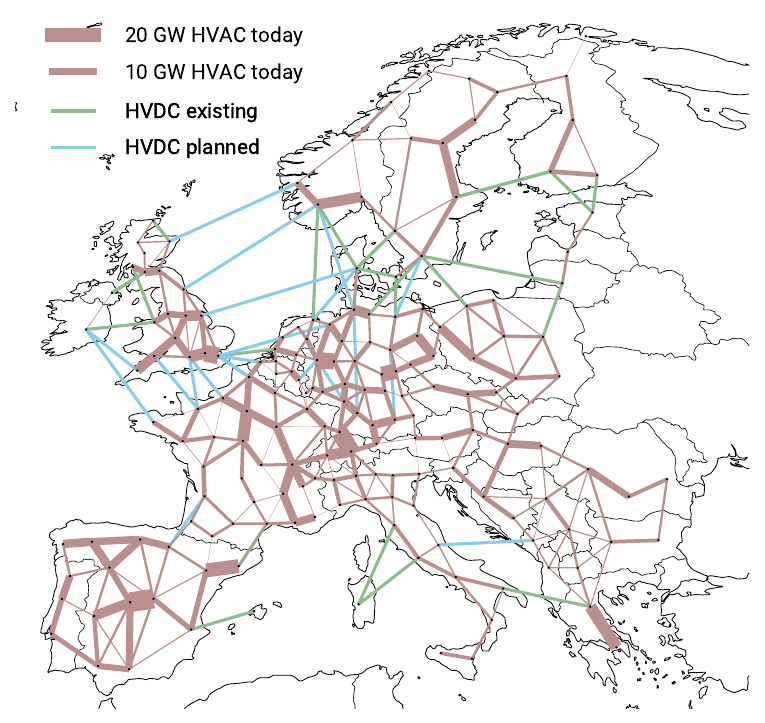}
\end{subfigure}
\begin{subfigure}[t]{0.49\textwidth}
    \centering
    \includegraphics[width=\textwidth]{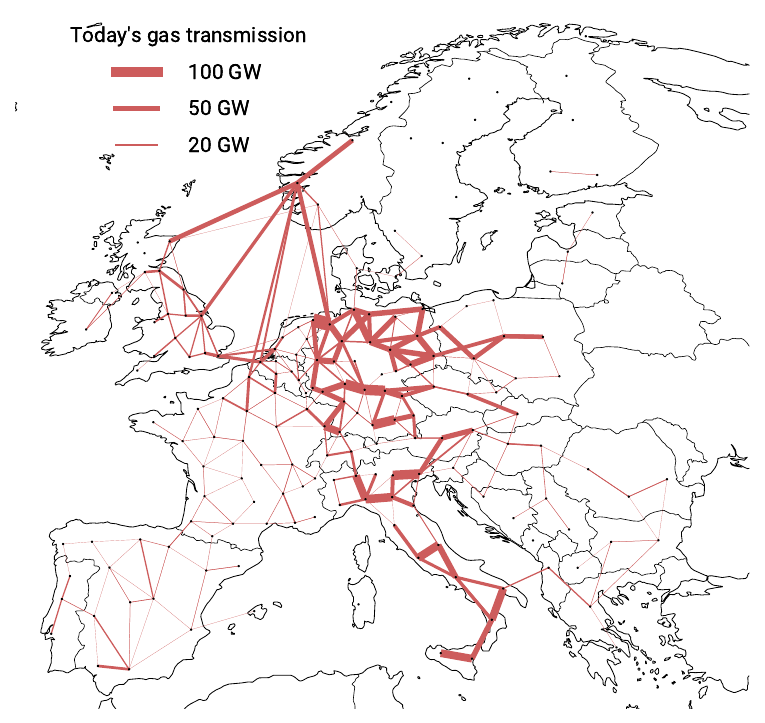}
\end{subfigure}
\caption{Clustered electricity and gas transmission networks before capacity expansion.}
\label{fig:clustered-networks}
\end{figure}

\begin{figure}
    \centering
    \begin{subfigure}[t]{0.49\textwidth}
        \centering
        \caption{electricity demand}
        \label{fig:demand-space:electricity}
        \vspace{-0.3cm}
        \includegraphics[width=\textwidth]{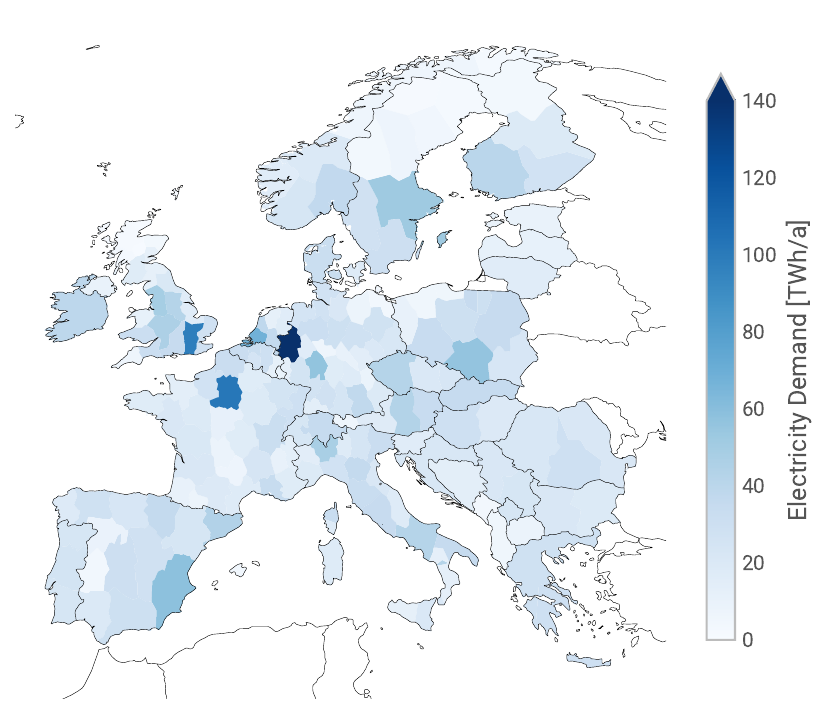}
    \end{subfigure}
    \begin{subfigure}[t]{0.49\textwidth}
        \centering
        \caption{hydrogen demand}
        \label{fig:demand-space:hydrogen}
        \vspace{-0.3cm}
        \includegraphics[width=\textwidth]{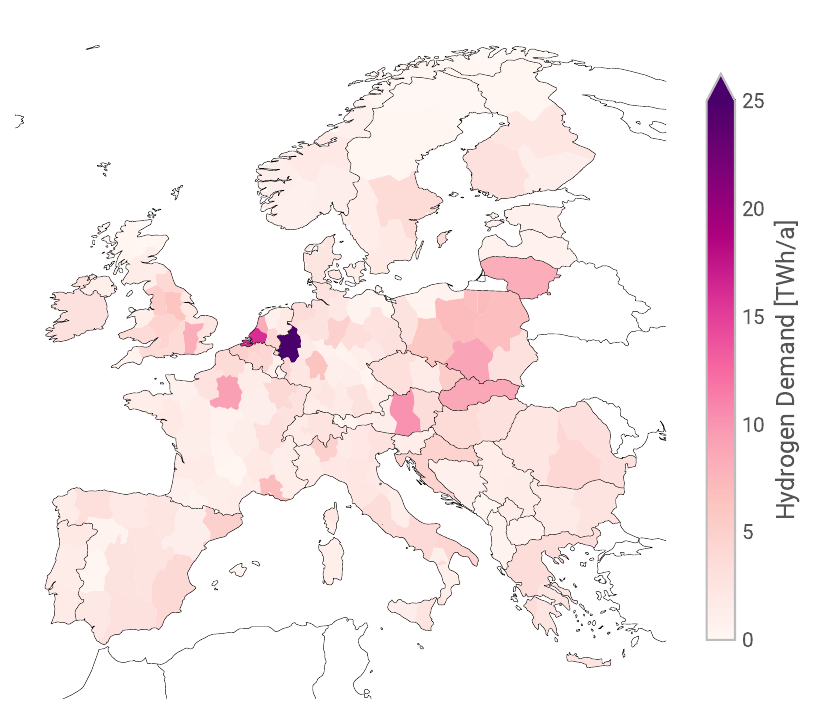}
    \end{subfigure}
    \begin{subfigure}[t]{0.49\textwidth}
        \centering
        \caption{methane demand}
        \label{fig:demand-space:methane}
        \vspace{-0.3cm}
        \includegraphics[width=\textwidth]{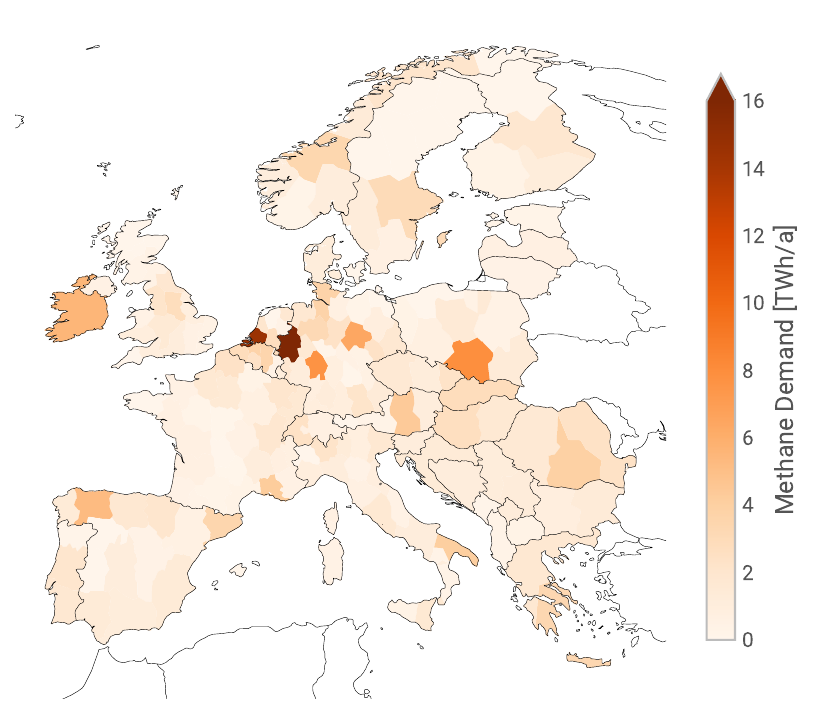}
    \end{subfigure}
    \begin{subfigure}[t]{0.49\textwidth}
        \centering
        \caption{heat demand}
        \label{fig:demand-space:heat}
        \vspace{-0.3cm}
        \includegraphics[width=\textwidth]{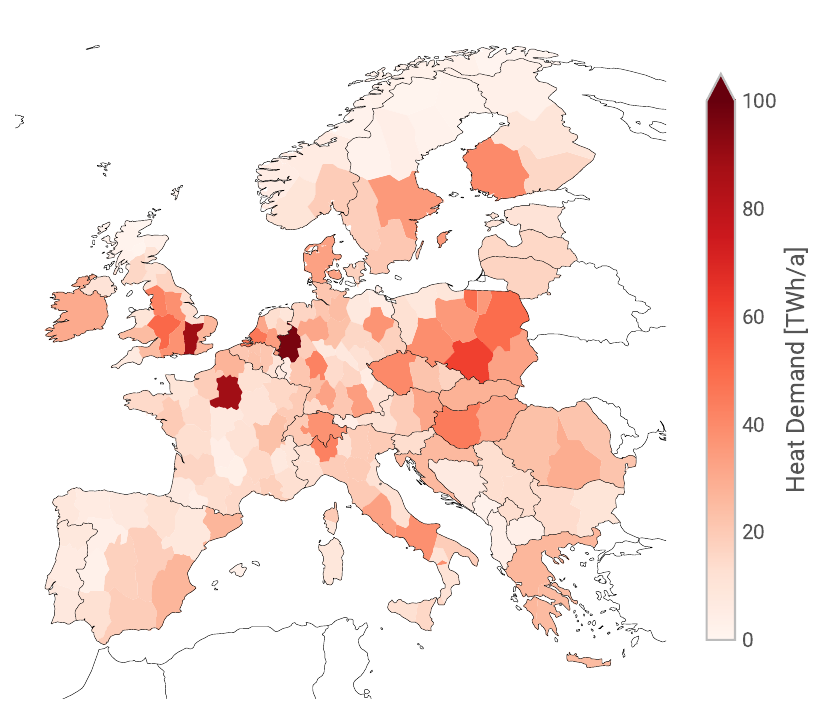}
    \end{subfigure}
    \begin{subfigure}[t]{0.49\textwidth}
        \centering
        \caption{oil-based product demand}
        \label{fig:demand-space:oil}
        \vspace{-0.3cm}
        \includegraphics[width=\textwidth]{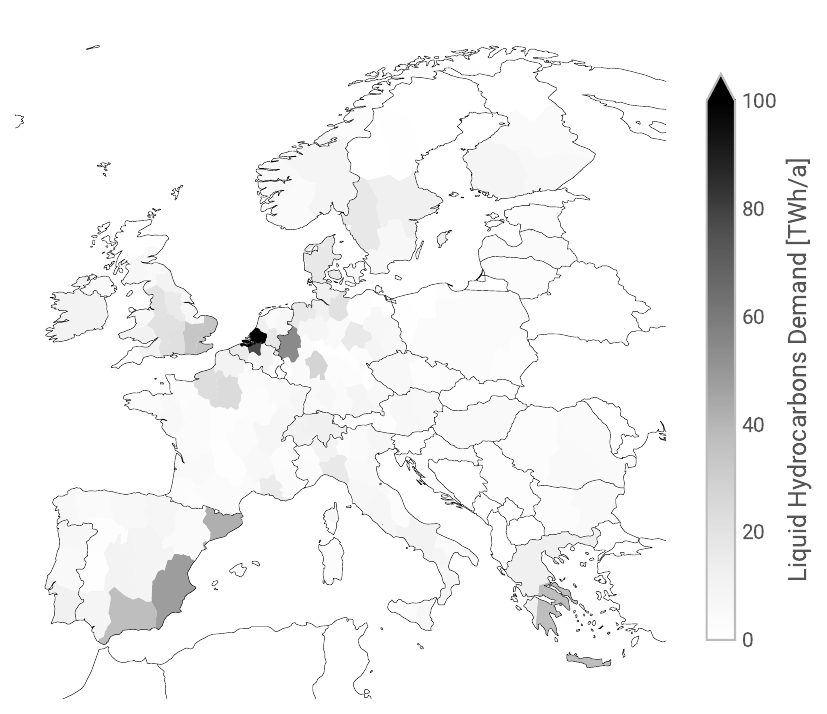}
    \end{subfigure}
    \begin{subfigure}[t]{0.49\textwidth}
        \centering
        \caption{solid biomass demand}
        \label{fig:demand-space:biomass}
        \vspace{-0.3cm}
        \includegraphics[width=\textwidth]{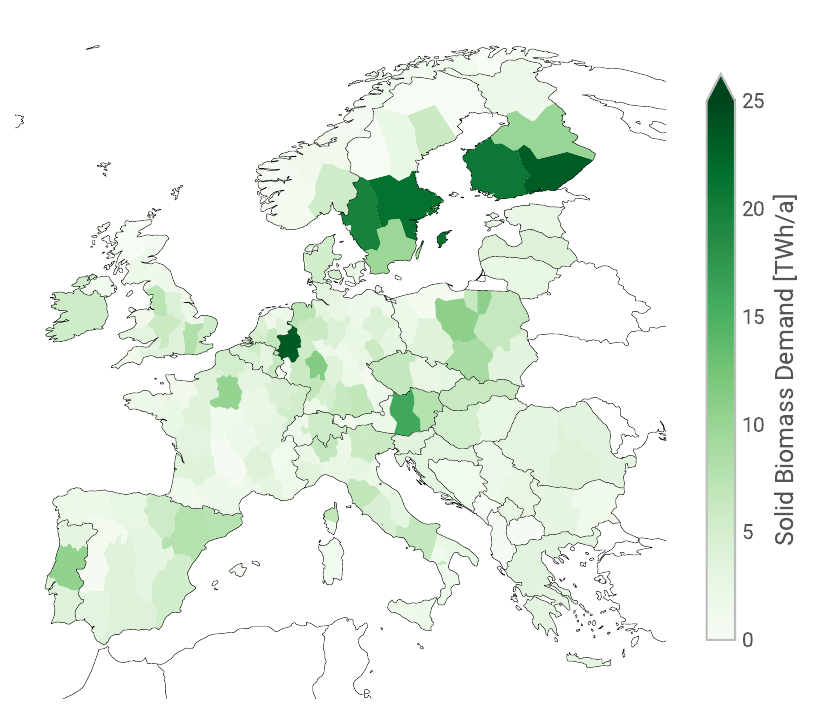}
    \end{subfigure}
    \caption{Spatial diversity of final energy and non-energy demand.}
    \label{fig:demand-space}
\end{figure}
\restoregeometry

\section{Model Overview}
\label{sec:si:model-overview}

PyPSA-Eur-Sec is an open model dataset of the European energy system at the
transmission network level that covers the electricity, heating, transport and
industry sectors. PyPSA-Eur-Sec builds a linear optimisation problem to plan
energy system infrastructure from various open data sources using the workflow
management tool Snakemake,\citeS{snakemake} which is then solved with the
commercial solver Gurobi.\citeS{gurobi} The overall circulation of energy and
carbon is shown in \cref{fig:multisector}. The modelling approaches for the
items listed there are described in detail in the following sections
\crefrange{sec:si:electricity}{sec:si:carbon-management}. A mathematical
formulation of the model is provided in \cref{sec:si:math}. The clustered model
resolution is shown in \cref{fig:clustered-networks} together with the existing
electricity and gas grid capacities. The carriers electricity, hydrogen and heat
nodally resolved, whereas other carriers like gas, oil, biomass and carbon
dioxide are copperplated in the current version to reduce the problem's
computational burden.
\section{Electricity Sector}
\label{sec:si:electricity}

Modelling of electricity supply and demand in Europe largely follows the open
electricity generation and transmission model PyPSA-Eur
\citeS{horschPyPSAEurOpen2018}. PyPSA-Eur processes publicly available data on
the topology of the power transmission network, historical time series of
weather observations and electricity consumption, conventional power plants, and
renewable potentials.

\subsection{Electricity Demand}
\label{sec:si:electricity:demand}

Hourly electricity demand at country-level for the reference year 2013 published
by ENTSO-E is retrieved via the interface of the Open Power System Data (OPSD)
initiative.\citeS{muehlenpfordtTimeSeries2020} Existing electrified heating is subtracted from this
demand, so that power-to-heat options can be optimised separately. Furthermore,
current industry electricity demand is subtracted and handled separately
considering further electrification in the industry sector (see \cref{sec:si:industry}).

For the distribution of electricity demand for industry we leverage geographical
data from the industrial database developed within the Hotmaps project.
\citeS{manzGeoreferencedIndustrial2018} The remaining
electricity demand for households and services is heuristically distributed
inside each country to 40\% proportional to population density and to 60\%
proportional to gross domestic product based on a regression performed by Hörsch
et al.~\citeS{horschPyPSAEurOpen2018}. The total spatial distribution of
electricity demands is shown in \cref{fig:demand-space:electricity}.

\subsection{Electricity Supply}

\begin{SCfigure}
    \caption{Existing conventional power plant capacities in Europe by technology. Marker size is proportional to nominal capacity.}
    \label{fig:powerplants}
    \includegraphics[width=0.7\textwidth]{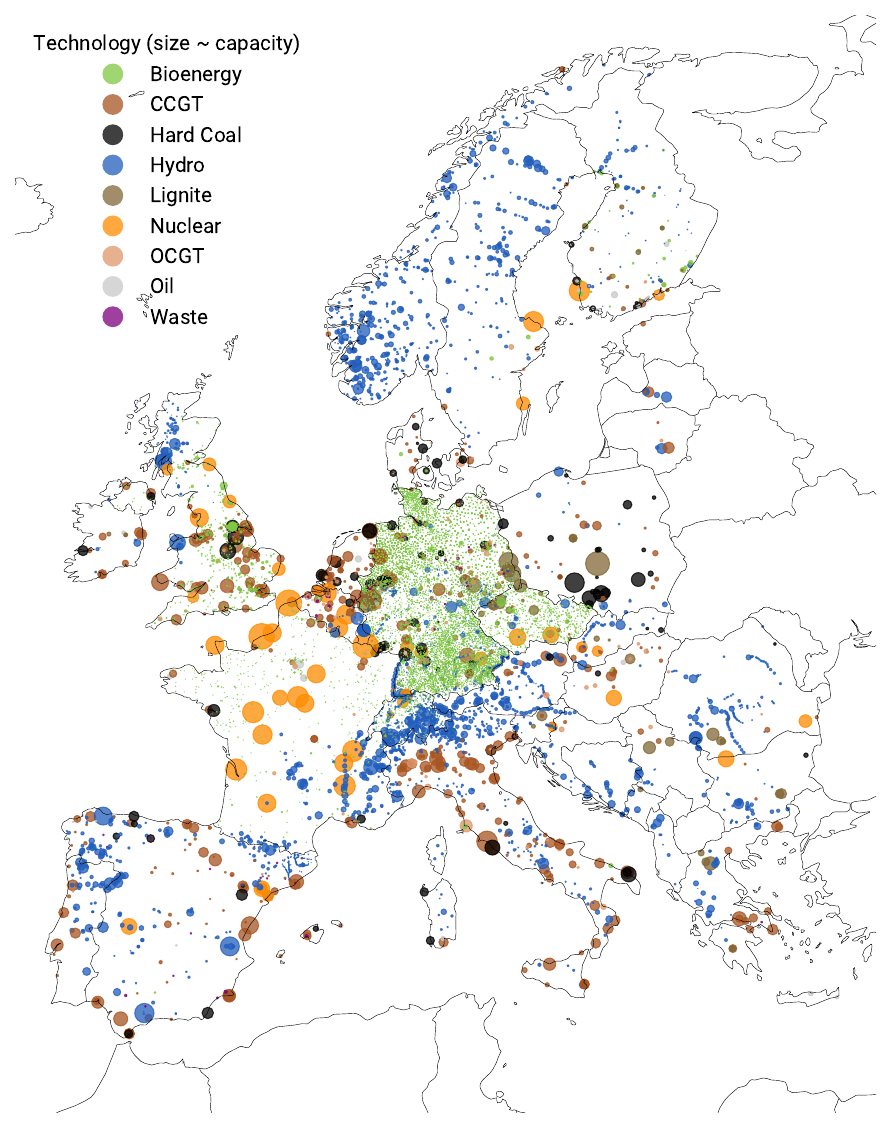}
\end{SCfigure}

For conventional electricity generators, PyPSA-Eur-Sec uses the open
\textit{powerplantmatching} tool, which merges datasets from a variety of
sources.\citeS{gotzensPerformingEnergy2019} As shown in \cref{fig:powerplants}, it
provides data on the power plants about their location, technology and fuel
type, age, and capacity, inlcuding hard coal, lignite, oil, open and combined
cycle gas turbines (OCGT and CCGT), and nuclear generators. Furthermore,
existing run-of-river, pumped-hydro storage plants, and hydro-electric dams, are
also part of the dataset, for which inflow is modelled based on runoff data from
reanalysis weather data and and scaled hydropower generation statistics (see
\cref{sec:si:renewable-ts}). In general, we suppose these to be non-extendable
due to assumed geographical constraints. The overnight scenarios in this study
only take into account existing hydro-electricity plants.

Expandable renewable generators include onshore and offshore wind, utility-scale
and rooftop solar photovoltaics, biomass from multiple feedstocks. The model
decides to build new capacities based on available land and on the weather
resource (see \cref{sec:si:renewable-potentials} and
\cref{sec:si:renewable-ts}). Because the continent-wide availability of data on
the locations of wind and solar installations is fragmentary, we disregard
already existing wind and solar capacities. Moreover, new OCGT and CCGT as well
as gas or biomass-fueled combined heat and power (CHP) generators may be built.
For CHP generators we assume back-pressure operation with heat production
proportional to electricity output. Specific techno-economic assumptions, like
costs, lifetimes and efficiencies are included in \cref{sec:si:costs}.

\subsection{Electricity Storage}
\label{sec:si:electricity:storage}

Electric energy can be stored in batteries (home, utility-scale, electric
vehicles), existing pumped-hydro storage (PHS), hydrogen storage and other
synthetically produced energy carriers (like methane, methanol and oil). For
stationary batteries we distinguish costs for inverters and for storage at home
or utility-scale. With these assumptions, home battery storage is about 40\%
more expensive than utility-scale battery storage (see \cref{sec:si:costs}). The
batteries' energy and power capacities can be independently sized.

To store electricity, hydrogen may be produced by water electrolysis (see
\cref{sec:si:h2:supply}), stored in overground steel tanks or underground salt
caverns (see see \cref{sec:si:h2:storage}), and re-electrified in a
utility-scale fuel cell. Synthetic methane can be re-electrified through an open
cycle gas turbine (OCGT) or a combined heat and power (CHP) plant.

\subsection{Electricity Transport}
\label{sec:si:electricity:transport}

\begin{figure}
    \includegraphics[width=\textwidth, center]{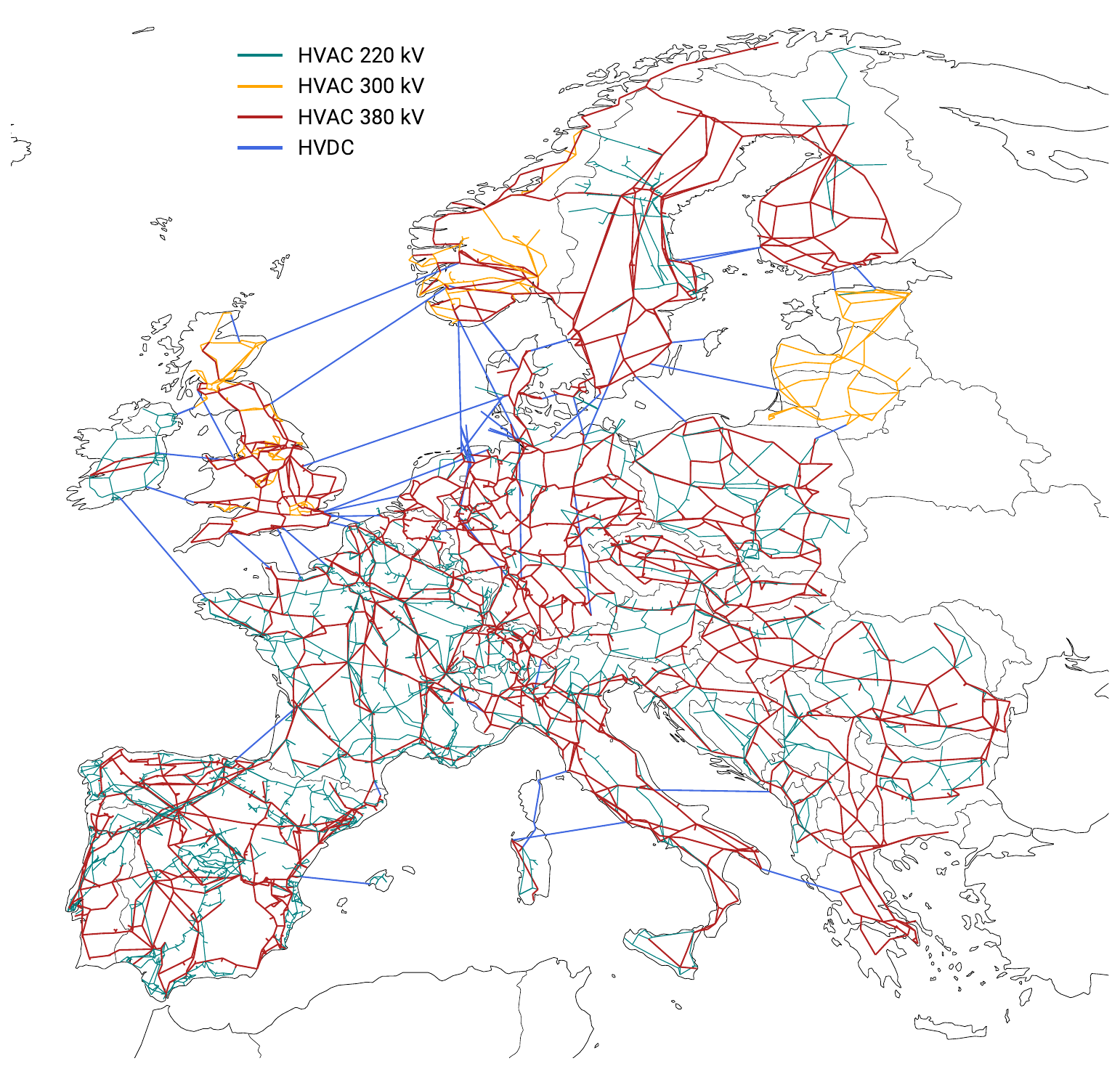}
    \caption{Unclustered European electricity transmission network by voltage level including planned TYNDP projects. Network data was retrieved from \href{https://www.entsoe.eu/data/map/}{entsoe.eu/data/map} and \href{https://tyndp.entsoe.eu/}{tyndp.entsoe.eu}.}
    \label{fig:base-network}
\end{figure}

\begin{SCfigure}
    \caption{Exemplary Voronoi cells of the transmission network's substations.}
    \includegraphics[width=0.55\textwidth]{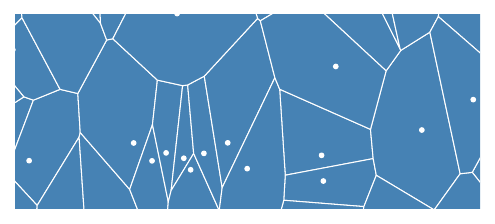}
    \label{fig:voronoi}
\end{SCfigure}

The topology of the European electricity transmission network is represented at
substation level based on maps released in the interactive \mbox{ENTSO-E} map\citeS{ENTSOE}
using a modified version of the GridKit tool.\citeS{gridkit} As displayed in
\cref{fig:base-network}, the dataset includes HVAC lines at and above 220 kV
across the mulitple synchronous zones of the \mbox{ENTSO-E} area, but excludes Turkey
and North-African countries which are also synchronised to the continental
European grid, interconnections to Russia, Belarus and Ukraine as well as small
island networks with less than four nodes at transmission level, such as Cyprus,
Crete and Malta. In total, the network encompasses around 3000 substations, 6600
HVAC lines and around 70 HVDC links, some of which are planned projects from the
Ten Year Network Development Plant (TYNDP) that are not yet in operation.
\citeS{tyndp2018}

The transmission network topology determines the basic regions of the
PyPSA-Eur-Sec model. Each substation has an associated Voronoi cell that
describes the region that is closer to the substation than to any other
substation except for country borders, which are kept to retain the integrity of
country totals. Exemplary Voronoi cells are illustrated in \cref{fig:voronoi}.
We use these as geographical catchment area for demands, renewable resource
potentials, and power plants, assuming that supply and demand always connect to
the closest substation. The Voronoi cells are also computed for offshore regions
based on the countries' Exclusive Economic Zones (EEZs) and the adjacent onshore
substations.

Capacities and electrical characteristics of transmission lines and substations,
such as impedances and thermal ratings, are
inferred from standard types for each voltage level from Oeding and Oswald.\citeS{oedingElektrischeKraftwerke2011} For each HVAC
line, we further restrict line loading to 70\% of the nominal rating to
approximate $N-1$ security, which protects the system against overloading if any
one transmission line fails. This conservative security margin is commonly
applied in the industry.\citeS{ackermannOptimisingEuropean2016} Dynamic line rating is not considered. Power
flow is modelled through lossless linearised power flow equations using an
efficient cycle-based formulation of Kirchhoff's voltage law.\citeS{horschLinearOptimal2018}

Solving the capacity expansion optimisation for the whole European energy system
at full network resolution is to large to be solved in reasonable time.
Therefore, we simplify the network topology by lowering the spatial resolution.
We initially remove the network's radial paths, i.e.~nodes with only one
connection, by linking remote resources to adjacent nodes and transforming the
network to a uniform voltage level of \SI{380}{\kilo\volt}. We also aggregate
generators of the same kind that connect to the same substation. Based on these
initial simplification, the network resolution is further reduced to a variable
number of nodes, in this case to 181 regions, by using a \textit{k-means}
clustering algorithm, which uses regional electricity consumption as weights.
\citeS{frysztackiStrongEffect2021, Hoersch2017} Only substations within the
same country can be aggregated. The equivalent lines connecting the clustered
regions are determined by the aggregated electro-technical characteristics of
original transmission lines. Their weighted cost takes into consideration the
underwater fraction of the lines and adds 25\% to the crow-fly distance to
approximate routing constraints. The clustered electricity network resolution
and associated model regions, as shown in \cref{fig:clustered-networks}, are
applied uniformly to the other nodally resolved energy carriers as well.

Contrary to the transmission level, the grid topology at the distribition level
(at and below \SI{110}{\kilo\volt}) is not included. Only the total power
exchange capacity between transmission and distribution level is co-optimised.
Costs of \SI{500}{\sieuro\per\kilo\watt} are assumed as well as lossless
distribution. Rooftop PV, heat pumps, resistive heaters, home batteries,
electric vehicles and electricity demands are connected to the low-voltage
level. All other remaining technologies connect directly to the transmission
grid. In this way, distribution grid capacity is developed if it is beneficial
to balance the local mismatch between supply and demand.
\section{Transport Sector}
\label{sec:si:transport}

Transport and mobility comprises light and heavy road, rail, shipping and
aviation transport. Annual energy demands for this sector are derived from the
JRC-IDEES database.\citeS{europeancommission.jointresearchcentre.JRCIDEESIntegrated2017}

\subsection{Land Transport}
\label{sec:si:transport:land}

\begin{figure}
    \centering
    \begin{subfigure}[t]{\textwidth}
        \centering
        \caption{demand (per-unit)}
        \includegraphics[width=.8\textwidth, trim=0cm 1cm 0cm 0cm, clip]{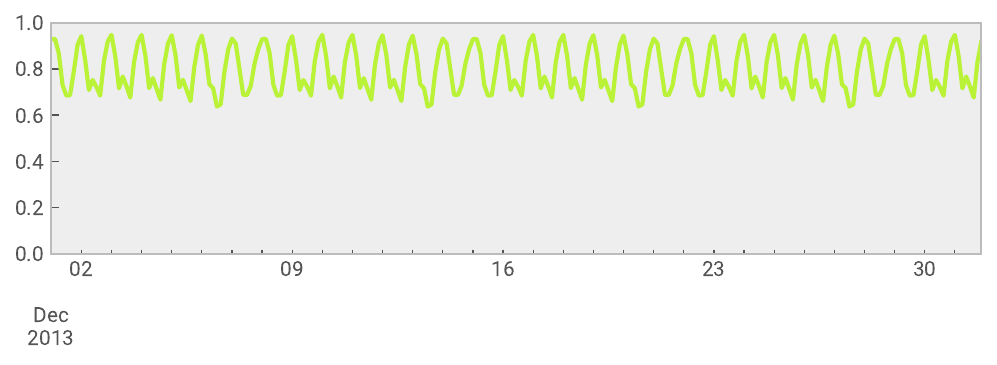}
    \end{subfigure}
    \vspace{-0.5cm}
    \begin{subfigure}[t]{\textwidth}
        \centering
        \caption{availability (per-unit)}
        \includegraphics[width=.8\textwidth, trim=0cm 1cm 0cm 0cm, clip]{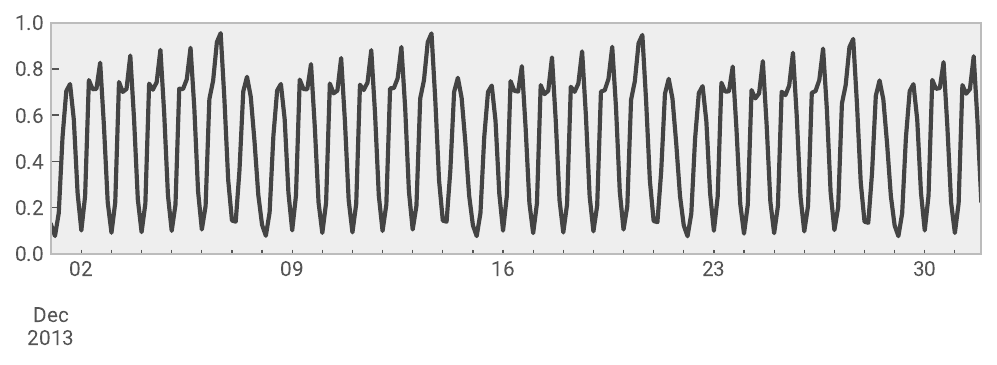}
    \end{subfigure}
    \caption{Normalised time series of battery electric vehicle demand and availability in December.}
    \label{fig:bev-demand-availability}
\end{figure}

The diffusion of battery electric vehicles (BEV) and fuel cell electric vehicles
(FCEV) in land transport is exogenously defined. For our mid-century scenarios,
we assume that 85\% of land transport is electrified and 15\% uses hydrogen fuel
cells. No more internal combustion engines exist.

The energy savings gained by electrifying road transport, are computed through
country-specific factors that compare the current final energy consumption of
cars per distance travelled (average for Europe
\SI{0.7}{\kwh\per\kilo\metre}\citeS{europeancommission.jointresearchcentre.JRCIDEESIntegrated2017}) to the
\SI{0.18}{\kilo\watt\hour\per\kilo\metre} assumed for the battery-to-wheel
efficiency of electric vehicles.

Weekly profiles of distances travelled published by the Germand Federal Highway
Research Institute (BASt)\citeS{bundesanstaltfurstrassenwesenAutomatischeZahlstellen2021} are used to generate hourly time series for
each European country taking into account their local time. Furthermore, a
temperature dependence is included in the time series to account for
heating/cooling demand in transport. For temperatures below \SI{15}{\celsius}
and above \SI{20}{\celsius} temperature coefficients of
\SI{0.98}{\percent\per\celsius} and \SI{0.63}{\percent\per\celsius} are assumed.
\citeS{brownSynergiesSector2018}

For battery electric vehicles, we assume a storage capacity of
\SI{50}{\kilo\watt\hour}, a charging capacity of \SI{11}{\kilo\watt} and a 90\%
charging efficiency. We assume that half of the BEV fleet can shift their
charging time and participate in vehicle-to-grid (V2G) services to facilitate
system operation. The BEV state of charge is forced to be higher than 75\% at
7am every day to ensure that the batteries are sufficiently charged for the peak
usage in the morning. This also restricts BEV demand to be shifted within a day
and prevent EV batteries from becoming seasonal storage. The percentage of BEV
connected to the grid at any time is inversely proportional to the transport
demand profile, which translates into an average/minimum availability of
80\%/62\% of the time. These values are conservative compared to most of the
literature, where average parking times of the European vehicle fleet is
estimated at 92\%. The battery cost of BEV is not included in the model
since it is assumed that BEV owners buy them to primarily satisfy their mobility
needs.

\subsection{Aviation}
\label{sec:si:transport:aviation}

The aviation sector consumes kerosene that is synthetically produced or of
fossil origin (see \cref{sec:si:oil:supply}). Biofuels are not considered.

\subsection{Shipping}
\label{sec:si:transport:shipping}

The shipping sector consumes synthetic methanol. In
\cref{sec:si:sensitivity-shipping} we also include a sensitivity analysis where
liquid hydrogen is used in shipping. For international shipping, the demand per
country is regionally distributed by port trade volumes taken from the World
Bank Data Catalog\citeS{worldbankWorldBank}. For domestic shipping, the demand is
distributed by population. Other fuel options like ammonia are currently not
considered.
\begin{SCfigure}
    \includegraphics[width=0.66\textwidth]{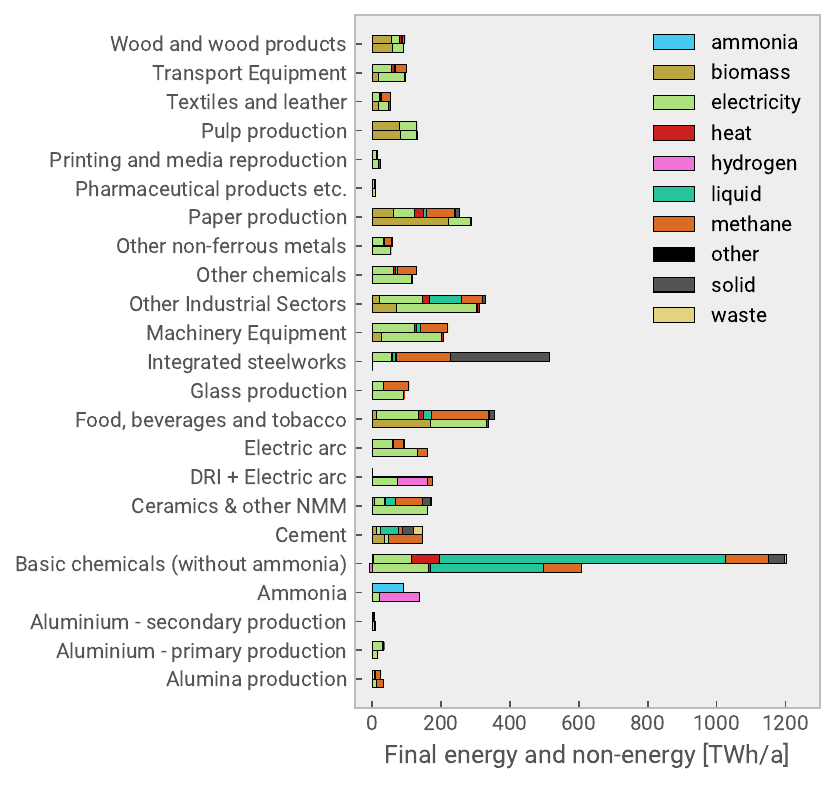}
    \caption{Final consumption of energy and non-energy feedstocks in industry today (top bar) and
    our scenario for net-zero emissions by mid-century (bottom bar)}
    \label{fig:fec-industry}
\end{SCfigure}

\begin{SCfigure}
    \includegraphics[width=0.66\textwidth]{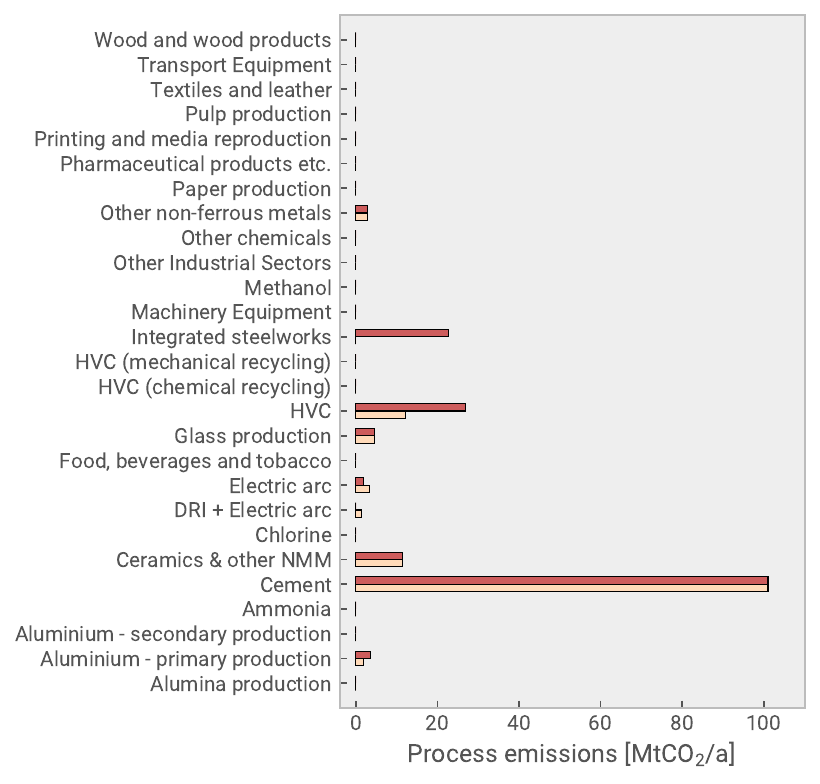}
    \caption{Process emissions in industry today (top bar) and mid-century without carbon capture (bottom bar)}
    \label{fig:process-emissions}
\end{SCfigure}

\begin{SCfigure}
    \includegraphics[width=0.7\textwidth]{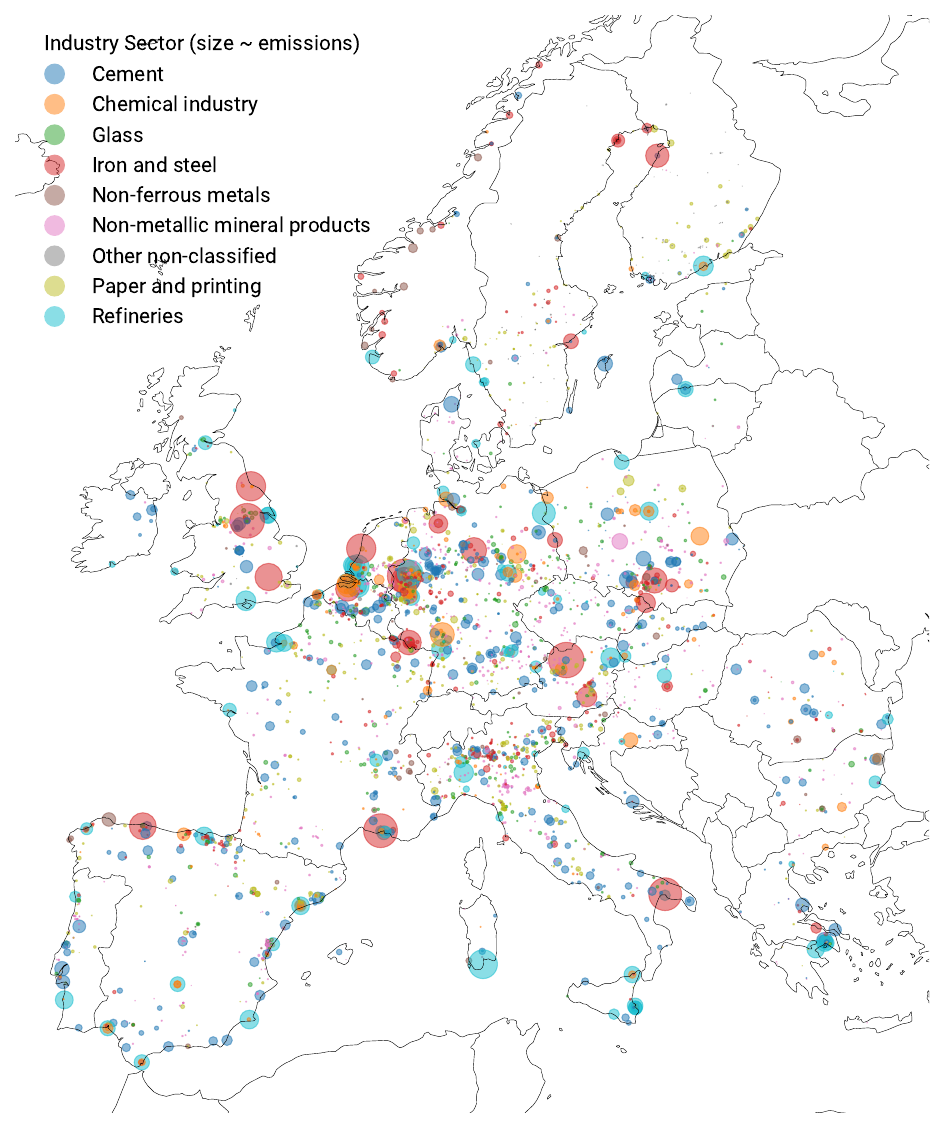}
    \caption{Distribution of industries according to emissions data from the Hotmaps industrial sites database. Marker size is proportional to the industrial site's reported emission levels.}
    \label{fig:hotmaps}
\end{SCfigure}

\begin{SCfigure}
    \includegraphics[width=0.6\textwidth]{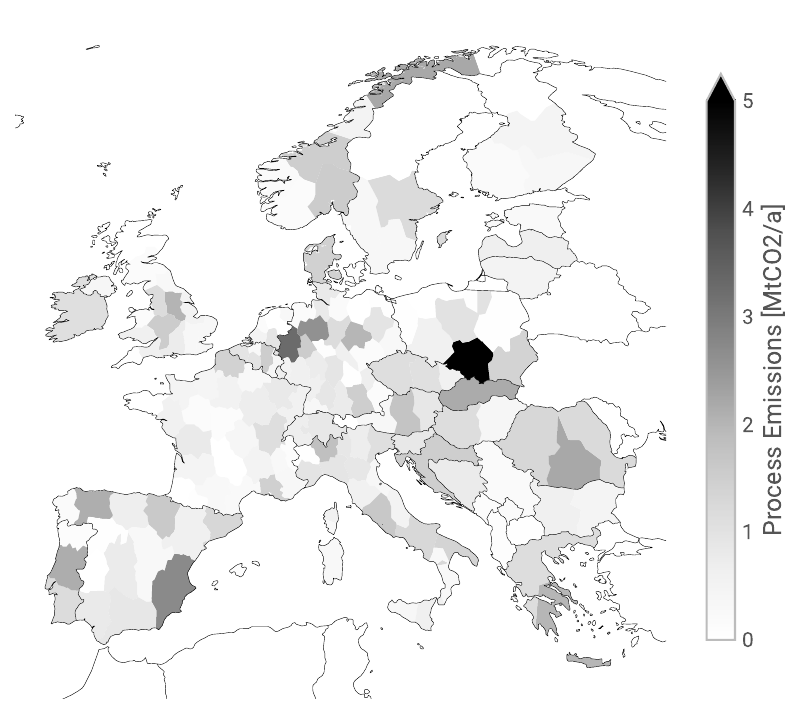}
    \caption{Spatial distribution of industrial process emissions.}
    \label{fig:spatial-process-emissions}
\end{SCfigure}

\section{Industry Sector}
\label{sec:si:industry}

Industry demand is split into a dozen different sectors with specific energy
demands, process emissions of carbon dioxide, as well as existing and
prospective mitigation strategies. \cref{sec:si:industry:overview} provides a
general description of the modelling approach for the industry sector in
PyPSA-Eur-Sec. The following subsections describe the current energy demands,
available mitigation strategies, and whether mitigation is exogenously fixed or
co-optimised with the other components of the model for each industry subsector
in more detail. In 2015, those subsectors with the larges final energy
consumption in Europe were iron and steel, chemicals industry, non-metallic
mineral products, pulp, paper and printing, food, beverages and tobacco, and
non-ferrous metals.\citeS{europeancommission.jointresearchcentre.JRCIDEESIntegrated2017}

\subsection{Overview}
\label{sec:si:industry:overview}

Greenhouse gas emissions associated with industry can be classified into
energy-related and process-related emissions (for the spatial distribution of
European process emissions see \cref{fig:spatial-process-emissions}). Today,
fossil fuels are used for process heat energy in the chemicals industry, but
also as a non-energy feedstock for chemicals like ammonia (\ce{NH3}), ethylene
(\ce{C2H4}) and methanol (\ce{CH3OH}). Energy-related emissions can be curbed by
using low-emission energy sources. The only option to reduce process-related
emissions is by using an alternative manufacturing process or by assuming a
certain rate of recyling so that a lower amount of virgin material is needed.

The overarching modelling procedure can be described as follows. First, the
energy demands and process emissions for every unit of material output are
estimated based on data from the JRC-IDEES
database\citeS{europeancommission.jointresearchcentre.JRCIDEESIntegrated2017}
and the fuel and process switching described in the subsequent sections. Second,
energy demands and process emissions for a climate-neutral Europe by mid-century
are calculated using the per-unit-of-material ratios based on the industry
transformations and the country-level material production in
2015,\citeS{europeancommission.jointresearchcentre.JRCIDEESIntegrated2017}
assuming constant material demand. Missing or too coarsely aggregated data in
the JRC-IDEES
database\citeS{europeancommission.jointresearchcentre.JRCIDEESIntegrated2017} is
supplemented with additional datasets: Eurostat energy
balances,\citeS{eurostatEnergyBalances2021} USGS for ammonia
production,\citeS{unitedstatesgeologicalsurveyAmmoniaProduction2021} DECHEMA for
methanol and chlorine,\citeS{bazzanellaLowCarbon2017} and national statistics from
Switzerland.\citeS{bundesamtfurenergieEnergieverbrauchIndustrie2021}

Where there are fossil and electrified alternatives for the same process (e.g.
in glass manufacture or drying) we assume that the process is completely
electrified. Current electricity demands (lighting, air compressors, motor
drives, fans, pumps) will remain electric. Where process heat is required our
approach depends on the temperature required.
\citeS{naeglerQuantificationEuropean2015,rehfeldtBottomupEstimation2018}
Processes that require temperatures below \SI{500}{\celsius} are supplied with
solid biomass, since we assume that residues and wastes are not suitable for
high-temperature applications (\cref{sec:si:heat:supply}). We see solid biomass
use primarily in the pulp and paper industry, where it is already widespread,
and in food, beverages and tobacco, where it replaces natural gas. Industries
which require high temperatures (above \SI{500}{\celsius}), such as metals,
chemicals and non-metalic minerals are either electified where suitable
processes already exist, or the heat is provided with synthetic methane.
Hydrogen for high-temperature process heat was not considered in our
scenarios.\citeS{neuwirthFuturePotential2022} For Europe, Rehfeldt et al.
\citeS{rehfeldtBottomupEstimation2018} estimated that, from 2015 industrial heat
demand, 45\% is above \SI{500}{\celsius}, 30\% within
\SIrange{100}{500}{\celsius}, 25\% below \SI{100}{\celsius}. Similarly, Naegler
et al. \citeS{naeglerQuantificationEuropean2015} estimate that 48\% is above
\SI{400}{\celsius}, 27\% within \SIrange{100}{400}{\celsius}, 25\% below
\SI{100}{\celsius}. Due to the high share of high-temperature process heat
demand, we disregard geothermal and solar thermal energy as source for process
heat. The final consumption of energy and non-energy feedstocks in industry
today in comparison to our scenarios for net-zero emissions by mid-century are
presented in \cref{fig:fec-industry}.

Inside each country the industrial demand is then distributed using the Hotmaps
Industrial Database, which is illustrated in \cref{fig:hotmaps}.\citeS{manzGeoreferencedIndustrial2018} This open database includes
georeferenced industrial sites of energy-intensive industry sectors in EU28,
including cement, basic chemicals, glass, iron and steel, non-ferrous metals,
non-metallic minerals, paper, refineries subsectors. The use of this spatial
dataset enables the calculation of regional and process specific energy demands.
This approach assumes that there will be no significant migration of
energy-intensive industries like, for instance, studied by Toktarova et
al.~\citeS{toktarovaInteractionElectrified2022} for the steel industry.

\subsection{Iron and Steel}
\label{sec:si:industry:steel}

Two alternative routes are used today to manufacture steel in Europe. The
primary route (integrated steelworks) represents 60\% of steel production, while
the secondary route (electric arc furnaces), represents the other 40\%.
\citeS{lechtenbohmerDecarbonisingEnergy2016}

The primary route uses blast furnaces in which coke is used to reduce iron ore
into molten iron.
\begin{align}
    \ce{CO2 + C &-> 2CO}, \\
    \ce{3Fe2O3 + CO &-> 2Fe3O4 + CO}, \\
    \ce{Fe3O4 + CO &-> 3FeO + CO2}, \\
    \ce{FeO + CO &-> Fe + CO2}.
\end{align}
which is then converted to steel. The primary route of steelmaking implies large
process emissions of \SI{0.22}{\tco\per\tonne} of steel, amounting to 7\% of
global greenhouse gas emissions.\citeS{voglPhasingOut2021}

In the secondary route, electric arc furnaces (EAF) are used to melt scrap
metal. This limits the \co emissions to the burning of graphite electrodes,
\citeS{Friedrichsen_2018} and reduces process emissions to
\SI{0.03}{\tco\per\tonne} of steel.

Integrated steelworks can be replaced by direct reduced iron (DRI) and subsequent processing in an electric arc furnace (EAF)
\begin{align}
    \ce{3Fe2O3 + H2 &-> 2Fe3O4 + H2O}, \\
    \ce{Fe3O4 + H2 &-> 3FeO + H2O}, \\
    \ce{FeO + H2 &-> Fe + H2O}.
\end{align}
This circumvents the process emissions associated with the use of coke. For
hydrogen-based DRI we assume energy requirements of 1.7 MWh$_{H_2}$/t steel
\citeS{voglAssessmentHydrogen2018} and 0.322 MWh$_{\text{el}}$/t steel
\citeS{hybritSummaryFindings2021}.

The shares of steel produced via each of the three routes by mid-century is exogenously
set in the model. We assume that hydrogen-based DRI plus EAF replaces integrated
steelworks for primary production completely, representing 30\% of total steel
production (down from 60\%). The remaining 70\% (up from 40\%) are manufactured
through the secondary route using scrap metal in EAF. According to
a Material Economics report,\citeS{circular_economy} circular economy practices even have the potential to
expand the share of the secondary route to 85\% by increasing the amount and
quality of scrap metal collected. Bioenergy as alternative to coke in blast
furnaces has not been considered.\citeS{mandovaPossibilitiesCO22018,suopajarviUseBiomass2018}

For the remaining subprocesses in this sector, the following transformations are
assumed. Methane is used as energy source for the smelting process. Activities
associated with furnaces, refining and rolling, product finishing are
electrified assuming the current efficiency values for these cases.
These transformations result in changes in process emissions as outlined in \cref{fig:process-emissions}.

\subsection{Chemicals Industry}
\label{sec:si:industry:chemicals}

The chemicals industry includes a wide range of diverse industries ranging from
the production of basic organic compounds (olefins, alcohols, aromatics), basic
inorcanic compounds (ammonia, chlorine), polymers (plastics), end-user products
(cosmetics, pharmaceutics).

The chemicals industry consumes large amounts of fossil-fuel based feedstocks,
\citeS{leviMappingGlobal2018} which can also be produced from renewables as
outlined for hydrogen in \cref{sec:si:h2:supply}, for methane in
\cref{sec:si:methane:supply}, and for oil-based products in
\cref{sec:si:oil:supply}. The ratio between synthetic and fossil-based fuels
used in the industry is an endogenous result of the optimisation.

The basic chemicals consumption data from the JRC IDEES\citeS{europeancommission.jointresearchcentre.JRCIDEESIntegrated2017} database
comprises high-value chemicals (ethylene, propylene and BTX), chlorine, methanol
and ammonia. However, it is necessary to separate out these chemicals because
their current and future production routes are different.

Statistics for the production of ammonia, which is commonly used as a
fertiliser, are taken from the United States Geological Survey (USGS) for every
country.\citeS{unitedstatesgeologicalsurveyAmmoniaProduction2021} Ammonia can
be made from hydrogen and nitrogen using the Haber-Bosch process.
\citeS{leviMappingGlobal2018}
\begin{equation}
    \ce{N2 + 3H2 -> 2NH3}
\end{equation}
The Haber-Bosch process is not explicitly represented in the model, such that
demand for ammonia enters the model as a demand for hydrogen (6.5
MWh$_{\ce{H2}}$/t$_{\ce{NH3}}$) and electricity (1.17 MWh$_{\text{el}}$/t$_{\ce{NH3}}$).
\citeS{wangGreeningAmmonia2018} Today, natural gas dominates in Europe as the source for
the hydrogen used in the Haber-Bosch process, but the model can choose among the
various hydrogen supply options described in
\cref{sec:si:h2:supply}

The total production and specific energy consumption of chlorine and methanol is
taken from a DECHEMA report.\citeS{bazzanellaLowCarbon2017} According to this
source, the production of chlorine amounts to 9.58 Mt$_{\ce{Cl}}$/a, which is assumed
to require electricity at 3.6 MWh$_{\text{el}}$/t of chlorine and yield hydrogen at
0.937 MWh$_{\ce{H2}}$/t of chlorine in the chloralkali process. The production of
methanol adds up to 1.5 Mt$_{\ce{MeOH}}$/a, requiring electricity at 0.167 MWh$_{\text{el}}$/t
of methanol and methane at 10.25 MWh$_{\ce{CH4}}$/t of methanol.

The production of ammonia, methanol, and chlorine production is deducted from
the JRC IDEES basic chemicals, leaving the production totals of high-value
chemicals. For this, we assume that the liquid hydrocarbon feedstock comes from
synthetic or fossil-origin naphtha (14 MWh$_{\text{naphtha}}$/t of HVC, similar
to Lechtenböhmer et al.~\citeS{lechtenbohmerDecarbonisingEnergy2016}), ignoring
the methanol-to-olefin route. Furthermore, we assume the following
transformations of the energy-consuming processes in the production of plastics:
the final energy consumption in steam processing is converted to methane since
requires temperature above \SI{500}{\celsius} (4.1 MWh$_{\ce{CH4}}$/t of
HVC);\citeS{rehfeldtBottomupEstimation2018} and the remaining processes are
electrified using the current efficiency of microwave for high-enthalpy heat
processing, electric furnaces, electric process cooling and electric generic
processes (2.85 MWh$_{\text{el}}$/t of HVC).

The process emissions from feedstock in the chemical industry are as high as
\SI{0.369}{\tco\per\tonne} of ethylene equivalent. We consider process emissions
for all the material output, which is a conservative approach since it assumes
that all plastic-embedded \co will eventually be released into the atmosphere.
However, plastic disposal in landfilling will avoid, or at least delay,
associated \co emissions.

Circular economy practices drastically reduce the amount of primary feedstock
needed for the production of plastics in the model
\citeS{kullmannValueRecycling2022,meysAchievingNetzero2021,meysCircularEconomy2020,guWastePlastics2017} and, consequently, also the energy demands
and level of process emissions\citeS{nicholsonManufacturingEnergy2021} (see
\cref{fig:process-emissions}). We assume that 30\% of plastics are mechanically
recycled requiring 0.547 MWh$_{\text{el}}$/t of HVC,
\citeS{meysCircularEconomy2020} 15\% of plastics are chemically recycled
requiring 6.9 MWh$_{\text{el}}$/t of HVC based on pyrolysis and electric steam
cracking,\citeS{materialeconomicsIndustrialTransformation2019} and 10\% of
plastics are reused (equivalent to reduction in demand). The remaining 45\% need
to be produced from primary feedstock. In comparison, Material Economics
\citeS{circular_economy} presents a scenario with circular economy scenario with
27\% primary production, 18\% mechanical recycling, 28\% chemical recycling, and
27\% reuse. Another new-processes scenario has 33\% primary production, 14\%
mechanical recycling, 40\% chemical recycling, and 13\% reuse.

\subsection{Non-metallic Mineral Products}
\label{sec:si:industry:nmmp}

This subsector includes the manufacturing of cement, ceramics, and glass.

\subsubsection*{Cement}

Cement is used in construction to make concrete. The production of cement
involves high energy consumption and large process emissions. The calcination of
limestone to chemically reactive calcium oxide, also known as lime, involves
process emissions of \SI{0.54}{\tco\per\tonne} cement.\citeS{fennellDecarbonizingCement2021}
\begin{equation}
    \ce{CaCO3 -> CaO + \co}
\end{equation}
Additionally, \co is emitted from the combustion of fossil fuels to provide
process heat. Thereby, cement constitutes the biggest source of industry
process emissions in Europe (\cref{fig:process-emissions}).

Cement process emissions can be captured assuming a capture rate of 90\%.
\citeS{DEA} Whether emissions are captured is decided by the model taking
into account the capital costs of carbon capture modules. The electricity and
heat demand of process emission carbon capture is currently ignored. For
net-zero emission scenarios, the remaining process emissions need to be
compansated by negative emissions.

With the exception of electricity demand and biomass demand for low-temperature
heat (0.06~MWh/t and 0.2~MWh/t), the final energy consumption of this subsector is assumed to be supplied
by methane (0.52 MWh/t), which is capable of delivering the required high-temperature heat.
This implies a switch from burning solid fuels to burning gas which will require
adjustments of the kilns.\citeS{akhtarCoalNatural2013}

Other mitigation strategies to reduce energy consumption or process emissions
(using new raw materials, recovering unused cement from concrete at end of life,
oxyfuel cement production to facilitate carbon
sequestration, electric kilns for heat
provision) are at a early development stage and have therefore not been
considered.\citeS{kuramochiComparativeAssessment2012}

\subsubsection*{Ceramics}

The ceramics sector is assumed to be fully electrified based on the current
efficiency of already electrified processes which include microwave drying and
sintering of raw materials, electric kilns for primary production processes,
electric furnaces for the product finishing.\citeS{europeancommission.jointresearchcentre.JRCIDEESIntegrated2017} In total, the final
electricity consumption is 0.44 MWh/t of ceramic. The manufacturing of ceramics
includes process emissions of \SI{0.03}{\tco\per\tonne} of ceramic. For a
detailed overview of the ceramics industry sector see Furszyfer Del Rio et
al.\citeS{furszyferdelrioDecarbonizingCeramics2022a}

\subsubsection*{Glass}

The production of glass is assumed to be fully electrified based on the current
efficiency of electric melting tanks and electric annealing which adds up to an
electricity demand of 2.07 MWh\el/t of glass
\citeS{lechtenbohmerDecarbonisingEnergy2016}. The manufacturing of glass incurs
process emissions of \SI{0.1}{\tco\per\tonne} of glass. Potential efficiency
improvements, which according to Lechtenböhmer et al.~\citeS{lechtenbohmerDecarbonisingEnergy2016}
could reduce energy demands to 0.85~MWh\el/t of glass, have not been considered.
For a detailed overview of the glass industry sector see Furszyfer Del Rio et
al.~\citeS{furszyferdelrioDecarbonizingGlass2022}

\subsection{Non-ferrous Metals}
\label{sec:si:industry:nfm}

The non-ferrous metal subsector includes the manufacturing of base metals
(aluminium, copper, lead, zink), precious metals (gold, silver), and technology
metals (molybdenum, cobalt, silicon).

The manufacturing of aluminium accounts for more than half of the final energy
consumption of this subsector. Two alternative processing routes are
used today to manufacture aluminium in Europe. The primary route represents 40\%
of the aluminium production, while the secondary route represents the remaining
60\%.

The primary route involves two energy-intensive processes: the production of
alumina from bauxite (aluminium ore) and the electrolysis to transform alumina
into aluminium via the  Hall-H\'{e}roult process
\begin{equation}
    \ce{2Al2O3 + 3C -> 4Al + 3\co}.
\end{equation}
The primary route requires high-enthalpy heat (2.3 MWh/t) to produce alumina
which is supplied by methane and causes process emissions of
\SI{1.5}{\tco\per\tonne} aluminium. According to Friedrichsen et al.,\citeS{Friedrichsen_2018}
inert anodes might become commercially available by 2030 that would eliminate
the process emissions. However, they have not been considered in this study.
Assuming all subprocesses are electrified, the primary route requires 15.4
MWh$_{\text{el}}$/t of aluminium.

In the secondary route, scrap aluminium is remelted. The energy demand for this
process is only 10\% of the primary route and there are no associated process
emissions. Assuming all subprocesses are electrified, the secondary route
requires 1.7 MWh/t of aluminium. Following Friedrichsen et al.,\citeS{Friedrichsen_2018} we assume
a share of recycled aluminium of 80\% by mid-century.

For the other non-ferrous metals, we assume the electrification of the entire
manufacturing process with an average electricity demand of 3.2 MWh\el/t lead
equivalent.

\subsection{Other Industry Subsectors}
\label{sec:si:industry:other}

The remaining industry subsectors include (a) pulp, paper, printing, (b) food,
beverages, tobacco, (c) textiles and leather, (d) machinery equipment, (e)
transport equipment, (f) wood and wood products, (g) others. Low- and
mid-temperature process heat in these industries is assumed to be supplied by
biomass,\citeS{sovacoolDecarbonizingFood2021} while the remaining processes are
electrified. None of the subsectors involve process emissions.

Energy demands for the agriculture, forestry and fishing sector per country are
taken from the JRC IDEES database.\citeS{europeancommission.jointresearchcentre.JRCIDEESIntegrated2017} Missing countries are filled
with eurostat data.\citeS{eurostatEnergyBalances2021} Agricultural energy
demands are split into electricity (lighting, ventilation, specific electricity
uses, electric pumping devices), heat (specific heat uses, low enthalpy heat)
machinery oil (motor drives, farming machine drives, diesel-fueled pumping
devices). Heat demand is for this sector is classified as services rural heat.
Time series for demands are assumed to be constant and distributed inside
countries in proportion to population.

\section{Heating Sector}
\label{sec:si:heat}

\subsection{Heat Demand}
\label{sec:si:heat:demand}

Building heating considering space and water heating in the residential and
services sectors is resolved for each region, both for individual buildings and
district heating systems, which include different supply options.

Annual heat demands per country are retrieved from JRC-IDEES\citeS{europeancommission.jointresearchcentre.JRCIDEESIntegrated2017} for
the year 2011 and split into space and water heating. The space heating demand
is reduced by retrofitting measures that improve the buildings' thermal
envelopes. This reduction is exogenously fixed at 29\%.\citeS{zeyenMitigatingHeat2021} For space
heating, the annual demands are converted to daily values based on the
population-weighted Heating Degree Day (HDD) using the \textit{atlite} tool,
\citeS{hofmannAtliteLightweight2021} where space heat demand is proportional to the difference between the
daily average ambient temperature (read from ERA5\citeS{ecmwf}) and a threshold
temperature above which space heat demand is zero. A threshold temperature of
\SI{15}{\celsius} is assumed. The daily space heat demand is distributed to the
hours of the day following heat demand profiles from BDEW.\citeS{bdewBDEWHeat2021} These differ
for weekdays and weekends/holidays and between residential and services demand.
Hot water demand is assumed to be constant throughout the year.

For every country, heat demand is split between low and high population density
areas. These country-level totals are then distributed to each region in
proportion to their rural and urban populations respectively. Urban areas with
dense heat demand can be supplied with large-scale district heating systems. We
assume that by mid-century, 60\% of urban heat demand is supplied by district heating
networks. Lump-sum losses of 15\% are assumed in district heating systems.
Cooling demand is supplied by electricity and included in the electricity
demand. Cooling demand is assumed to remain at current levels.

The regional distribution of the total heat demand is depicted in
\cref{fig:demand-space:heat}. As \cref{fig:demand-time} reveals, the total heat
demand is similar to the total electricity demand but features much more
pronounced seasonal variations. The total building heating demand adds up to
\SI{3084}{\twh\per\year} of which 78\% occurs in urban areas.

\subsection{Heat Supply}
\label{sec:si:heat:supply}

Different supply options are available depending on whether demand is met
centrally through district heating systems or decentrally through appliances in
individual buidlings. Supply options in individual buildings include gas and oil
boilers, air- and ground-sourced heat pumps, resistive heaters, and solar
thermal collectors. For large-scale district heating systems more options are
available: combined heat and power (CHP) plants consuming gas or biomass from
waste and residues with and without carbon capture (CC), large-scale air-sourced
heat pumps, gas and oil boilers, resistive heaters and fuel cell CHPs.
Additionally, waste heat from the Fischer-Tropsch and Sabatier processes for the
production of synthetic hydrocarbons can supply district heating systems.
Ground-source heat pumps are only allowed in rural areas because of space constraints.
Thus, only air-source heat pumps are allowed in urban areas. This is a conservative
assumption, since there are many possible sources of low-temperature heat that
could be tapped in cities (e.g.~waste water, ground water, or natural bodies of water).
Costs, lifetimes and efficiencies for these technologies are listed in \cref{sec:si:costs}.

CHPs are based on back pressure plants operating with a fixed ratio of
electricity to heat output. The efficiencies of each are given on the back
pressure line, where the back pressure coefficient $c_b$ is the electricity
output divided by the heat output. For biomass CHP, we assume $c_b=0.46$,
whereas for gas CHP, we assume $c_b=1$.

The coefficient of performance (COP) of air- and ground-sourced heat pumps
depends on the ambient or soil temperature respectively. Hence, the COP is a
time-varying parameter. Generally, the COP will be lower during winter when
temperatures are low. Because the ambient temperature is more volatile than the
soil temperature, the COP of ground-sourced heat pumps is less variable.
Moreover, the COP depends on the difference between the source and sink
temperatures
\begin{equation}
    \Delta T = T_{sink} - T_{source}.
\end{equation}
For the sink water temperature $T_{sink}$ we assume \SI{55}{\celsius} For the
time- and location-dependent source temperatures $T_{source}$, we rely on the
ERA5 reanalysis weather data.\citeS{ecmwf} The temperature differences are
converted into COP time series using results from a regression analysis
performed in.\citeS{staffellReviewDomestic2012} For air-sourced heat pumps
(ASHP), we use the function
\begin{equation}
    COP(\Delta T) = 6.81 + 0.121 \Delta T + 0.000630 \Delta T^2;
\end{equation}
for ground-sourced heat pumps (GSHP), we use the function
\begin{equation}
    COP(\Delta T) = 8.77 + 0.150 \Delta T + 0.000734 \Delta T^2.
\end{equation}
The resulting time series are displayed in \cref{fig:cfs-ts}.
The spatial diversity of heat pump coefficients is shown in \cref{fig:cfs-maps}.

\subsection{Heat Storage}
\label{sec:si:heat:storage}

Thermal energy storage (TES) is available in large water pits associated with
district heating networks for seasonal storage and small water tanks for
decentral short-term storage. A thermal energy density
46.8~kWh$_{\text{th}}$/m$^3$ is assumed, corresponding to temperature difference
of \SI{40}{\kelvin}. The decay of thermal energy $1-\exp(-\sfrac{1}{24\tau})$ is
assumed to have a time constant of $\tau=180$ days for central TES and $\tau=3$
days for individual TES. The charging and discharging efficiencies are 90\% due
to pipe losses.

\section{Renewables}
\label{sec:si:renewables}

\subsection{Potentials}
\label{sec:si:renewable-potentials}

\begin{figure}
    \centering
    \begin{subfigure}[t]{0.48\textwidth}
            \centering
        \caption{solar land eligibility}
        \includegraphics[width=\textwidth]{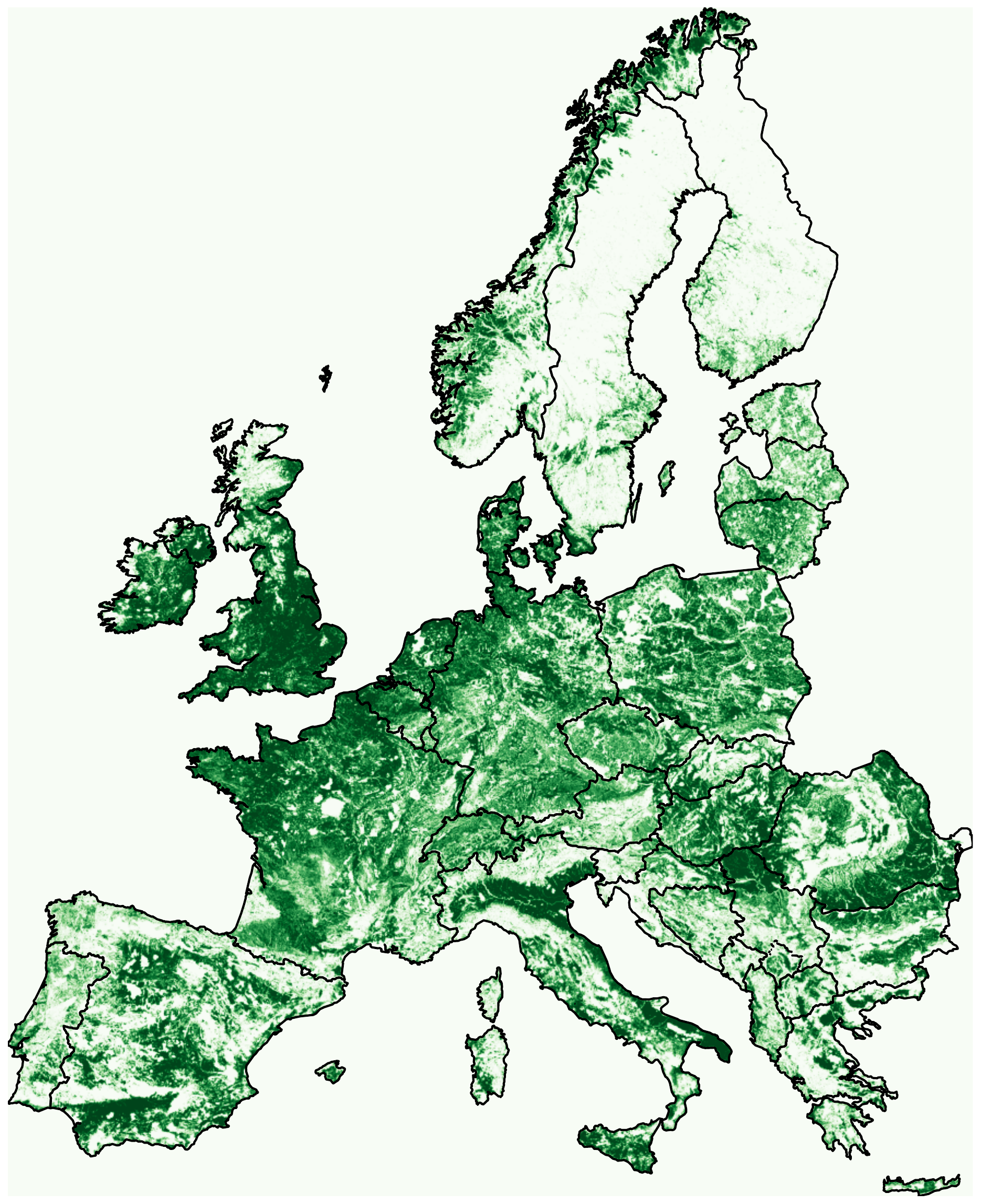}
    \end{subfigure}
    \begin{subfigure}[t]{0.48\textwidth}
        \centering
        \caption{onshore wind land eligibility}
        \includegraphics[width=\textwidth]{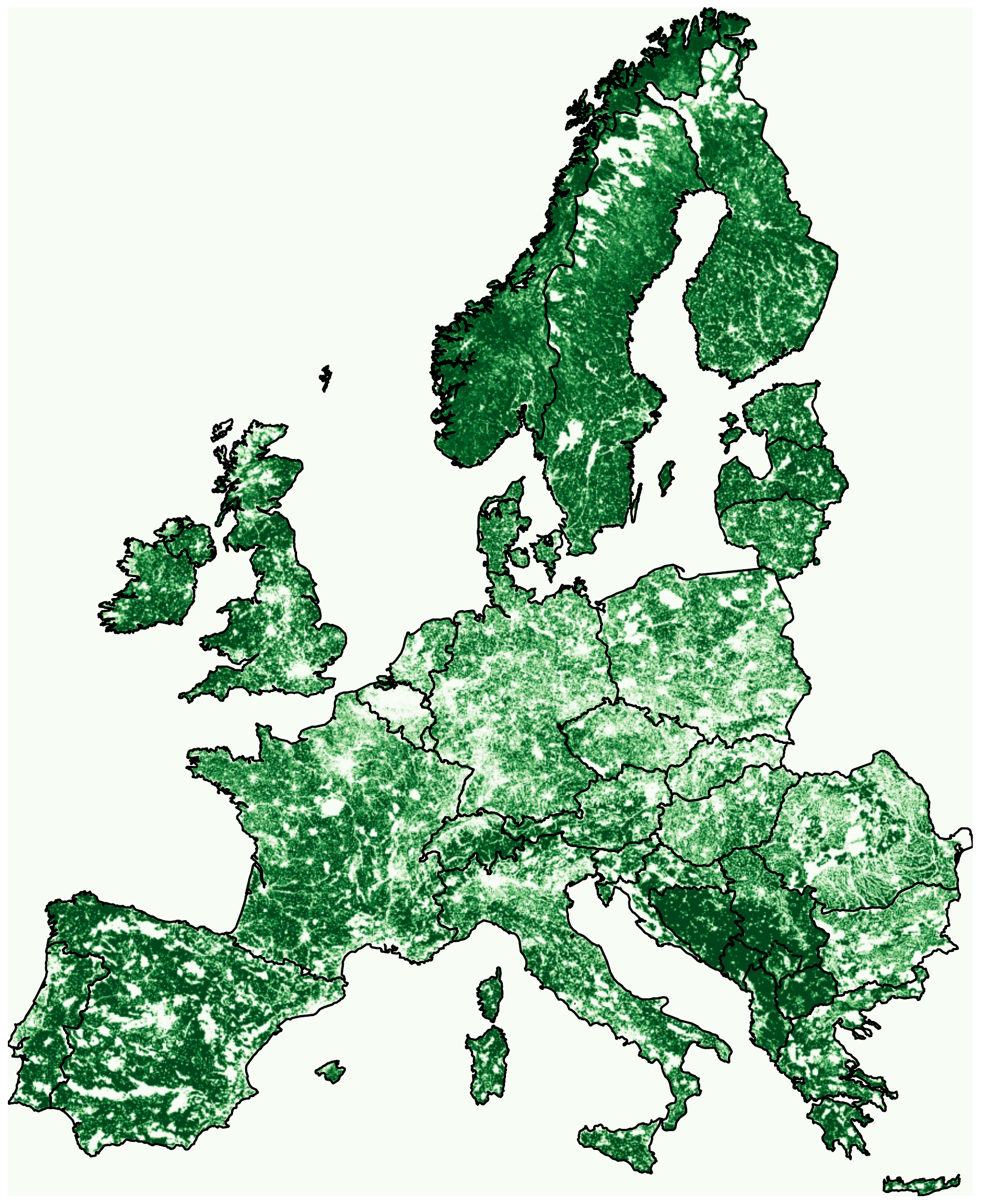}
    \end{subfigure}
    \begin{subfigure}[t]{0.48\textwidth}
        \centering
        \vspace{.5cm}
        \caption{offshore wind land eligibility}
        \includegraphics[width=\textwidth]{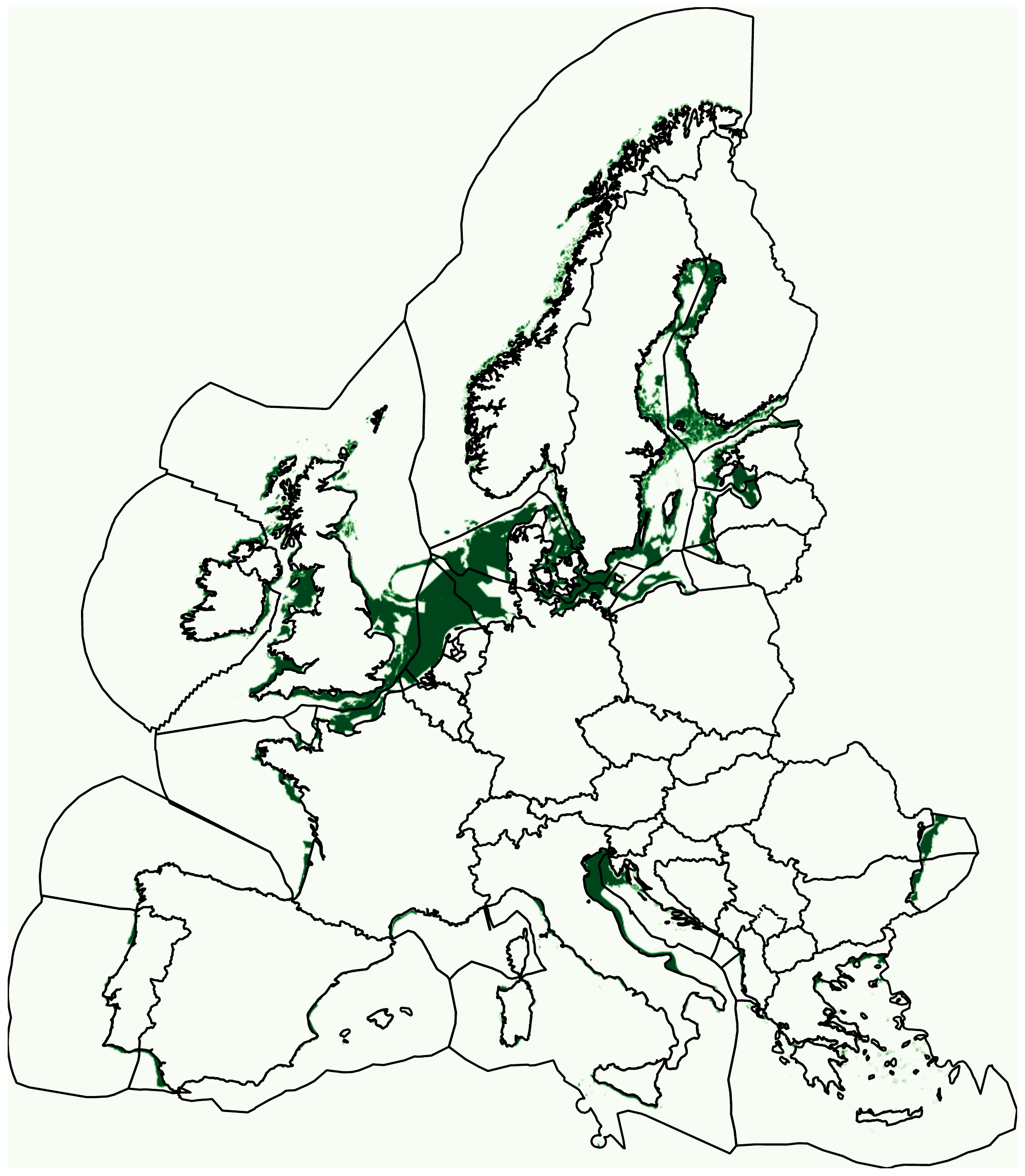}
    \end{subfigure}
    \caption{Land eligibility for the development of renewable generation capacities. Green color indicates
    areas eligible to build wind or utility-scale solar parks based on suitable land types, natural protection areas, and water depths.}
    \label{fig:eligibility}
\end{figure}

\begin{table}
    \caption{Land types considered suitable for every technology from Corine Land Cover database. Land type codes are referenced in brackets.}
    \small
    \begin{tabularx}{\textwidth}{lX}
        \toprule
        Solar PV & artificial surfaces (1-11), agriculture land except for those
        areas already occupied by agriculture with significant natural
        vegetation and agro-forestry areas (12-20), natural grasslands (26), bare rocks (31),
        sparsely vegetated areas (32) \\ \midrule
        Onshore wind & agriculture areas (12-22), forests (23-25), scrubs and herbaceous vegetation associations (26-29), bare rocks (31), sparsely vegetated areas (32) \\ \midrule
        Offshore wind & sea and ocean (44) \\ \bottomrule
    \end{tabularx}
    \label{tab:eligibility}
\end{table}

Eligibile areas for developing renewable infrastructure are calculated per
technology and substation's Voronoi cell using the
\textit{atlite}\citeS{hofmannAtliteLightweight2021} tool and shown in
\cref{fig:eligibility}.

The land available for wind and utility-scale solar PV capacities in a
particular region is constrained by eligible codes of the
CORINE\citeS{europeanenvironmentagencyeeaCorineLand} land use database (100m
resolution)  and is further restricted by distance criteria and the natural
protection areas specified in the Natura
2000\citeS{europeanenvironmentagencyeeaNatura2000} dataset. These criteria are
summarised in \cref{tab:eligibility}. The installable potentials for rooftop PV
are included with an assumption of 1 kWp per person (0.1 kW/m$^2$ and 10
m$^2$/person). A more sophisticated potential estimate can be found in Bódis et
al.~\citeS{bodisHighresolutionGeospatial2019}. Moreover, offshore wind farms may
not be built at sea depths exceeding \SI{50}{\metre}, as indicated by the GEBCO\citeS{gebcoGEBCO2014}
bathymetry dataset. This currently disreagards the
possibility of floating wind turbines.
\citeS{lerchSensitivityAnalysis2018,lauraLifecycleCost2014,myhrLevelisedCost2014,kauscheFloatingOffshore2018,castro-santosEconomicFeasibility2016}
For near-shore locations (less than 30 km off the shore) AC connections are
considered, whereas for far-shore locations, DC connections including AC-DC
converter costs are assumed. Reservoir hydropower and run-of-river capacities
are exogenously fixed at current values and not expandable.

To express the potential in terms of installable capacities, the available areas
are multiplied with allowed deployment densities, which we consider to be a
fraction of the technology's technical deployment density to preempt public
acceptance issues. These densities are 3 MW/m$^ 2$ for onshore wind, 2 MW/m$^2$
for offshore wind, 5.1 MW/m$^2$ for utility-scale solar. For a review of
alternative potential wind potential assessments, see McKenna et
al.~\citeS{mckennaHighresolutionLargescale2022} and Ryberg et
al.~\citeS{rybergFutureEuropean2019}.

\subsection{Time Series}
\label{sec:si:renewable-ts}

\begin{figure}
    \centering
    \begin{subfigure}[t]{0.49\textwidth}
        \centering
        \caption{solar}
        \includegraphics[width=\textwidth]{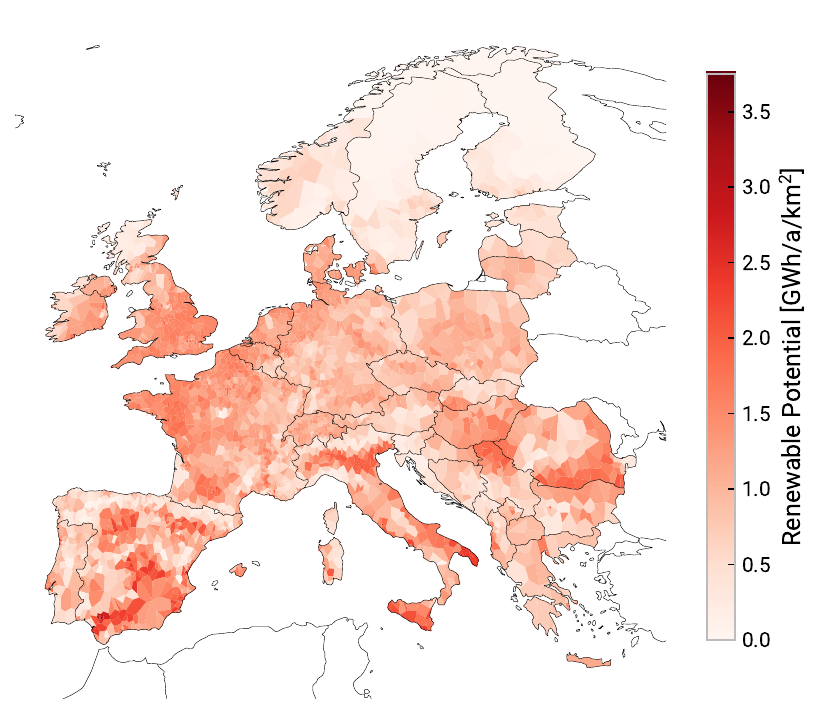}
    \end{subfigure}
    \begin{subfigure}[t]{0.49\textwidth}
        \centering
        \caption{wind}
        \includegraphics[width=\textwidth]{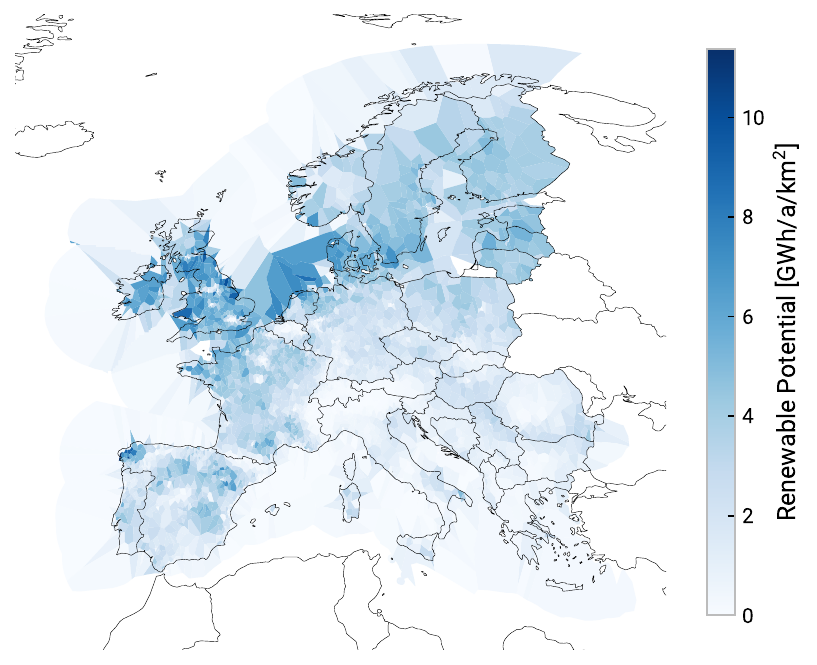}
    \end{subfigure}
    \caption{Available energy density for wind and utility-scale solar PV power generation.}
    \label{fig:energy-density}
\end{figure}

\begin{figure}
    \centering
        \begin{subfigure}[t]{0.49\textwidth}
            \centering
        \includegraphics[width=\textwidth]{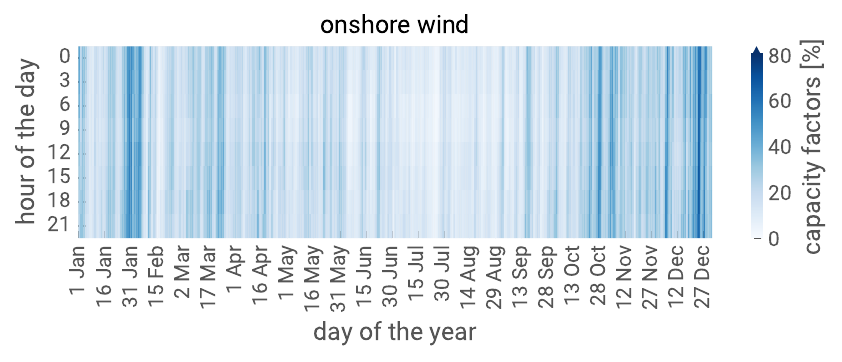}
    \end{subfigure}
    \begin{subfigure}[t]{0.49\textwidth}
        \centering
        \includegraphics[width=\textwidth]{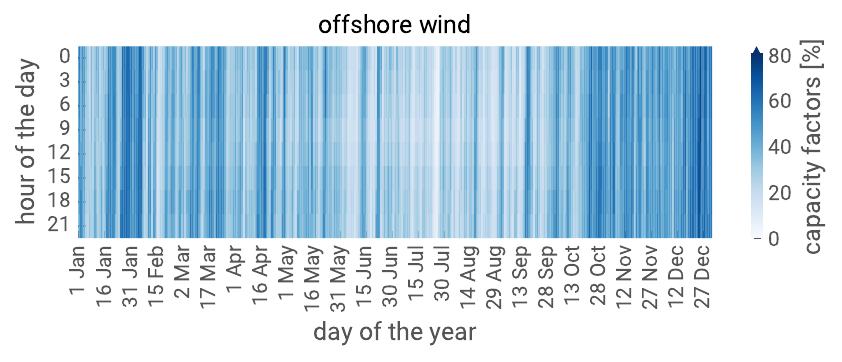}
    \end{subfigure}
    \begin{subfigure}[t]{0.49\textwidth}
        \centering
        \includegraphics[width=\textwidth]{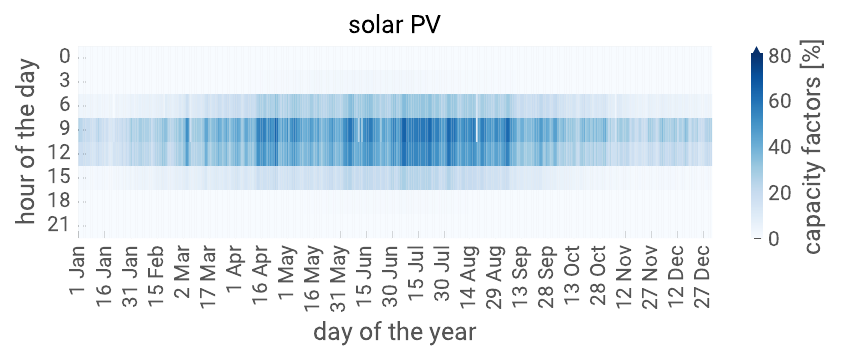}
    \end{subfigure}
    \begin{subfigure}[t]{0.49\textwidth}
        \centering
        \includegraphics[width=\textwidth]{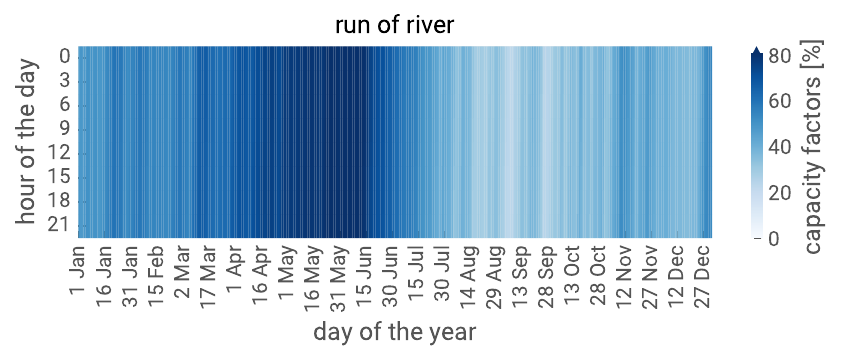}
    \end{subfigure}
    \begin{subfigure}[t]{0.49\textwidth}
        \centering
        \includegraphics[width=\textwidth]{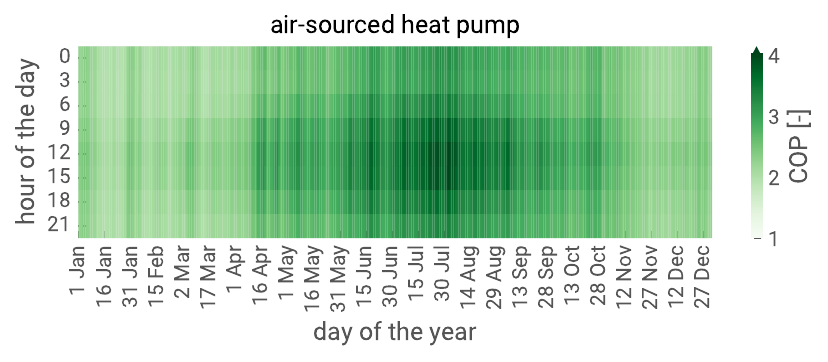}
    \end{subfigure}
    \begin{subfigure}[t]{0.49\textwidth}
        \centering
        \includegraphics[width=\textwidth]{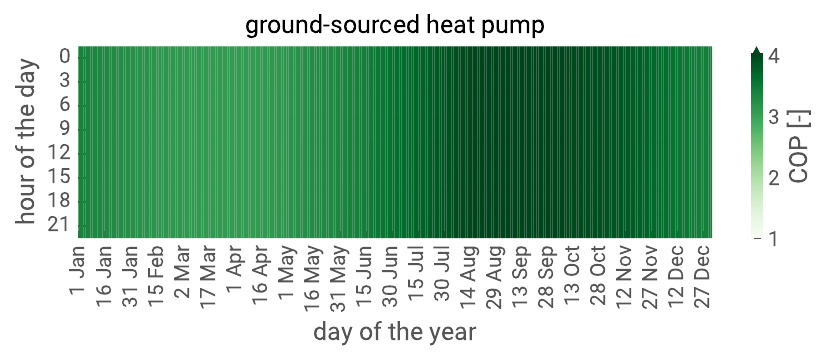}
    \end{subfigure}
    \caption{Spatially aggregated capacity factor time series of renewable energy sources.}
    \label{fig:cfs-ts}
\end{figure}

\begin{figure}
    \centering
    \begin{subfigure}[t]{0.49\textwidth}
        \centering
        \caption{onshore wind}
        \includegraphics[width=\textwidth]{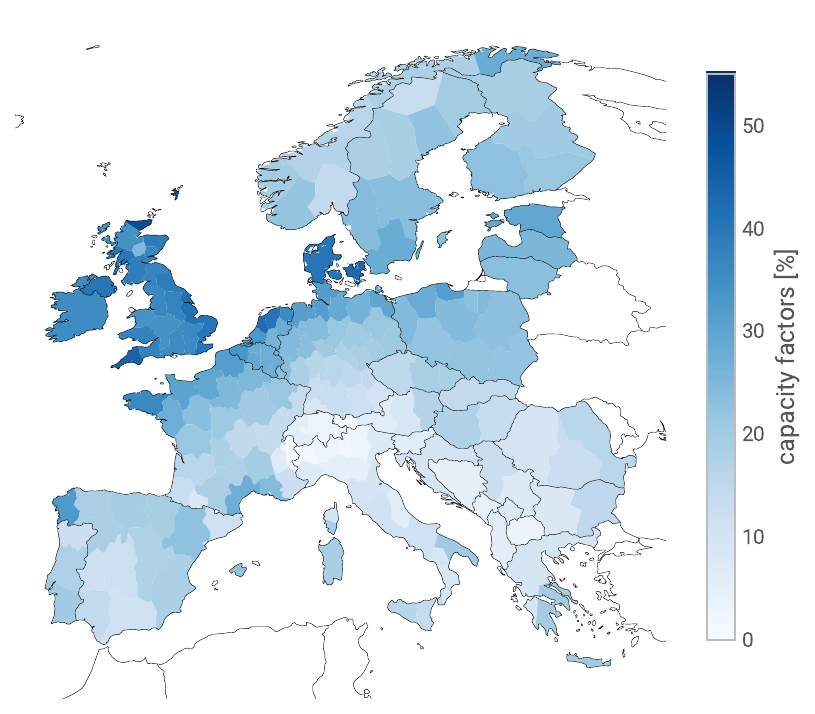}
    \end{subfigure}
    \begin{subfigure}[t]{0.49\textwidth}
        \centering
        \caption{solar photovoltaics}
        \includegraphics[width=\textwidth]{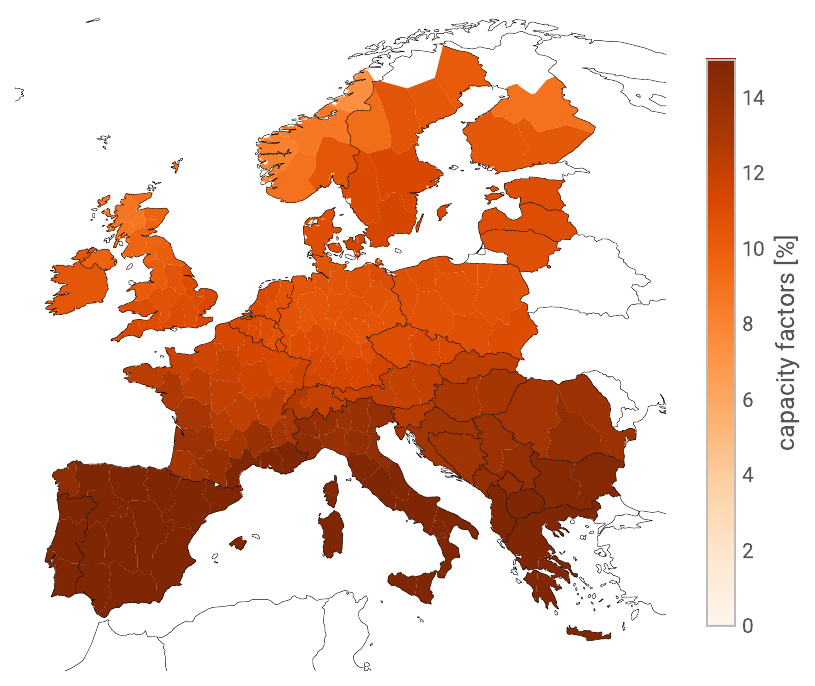}
    \end{subfigure}
    \begin{subfigure}[t]{0.49\textwidth}
        \centering
        \caption{offshore wind (DC-connected)}
        \includegraphics[width=\textwidth]{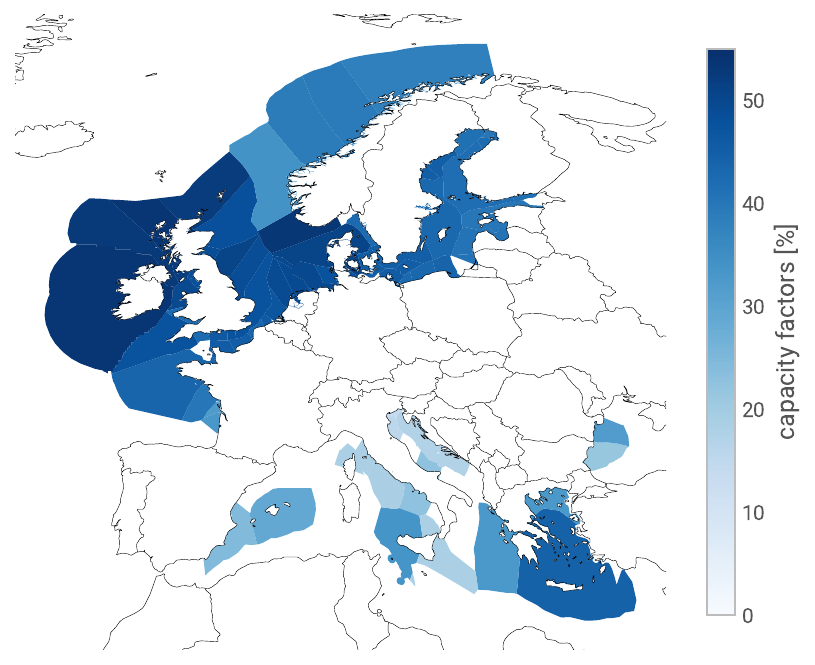}
    \end{subfigure}
    \begin{subfigure}[t]{0.49\textwidth}
        \centering
        \caption{offshore wind (AC-connected)}
        \includegraphics[width=\textwidth]{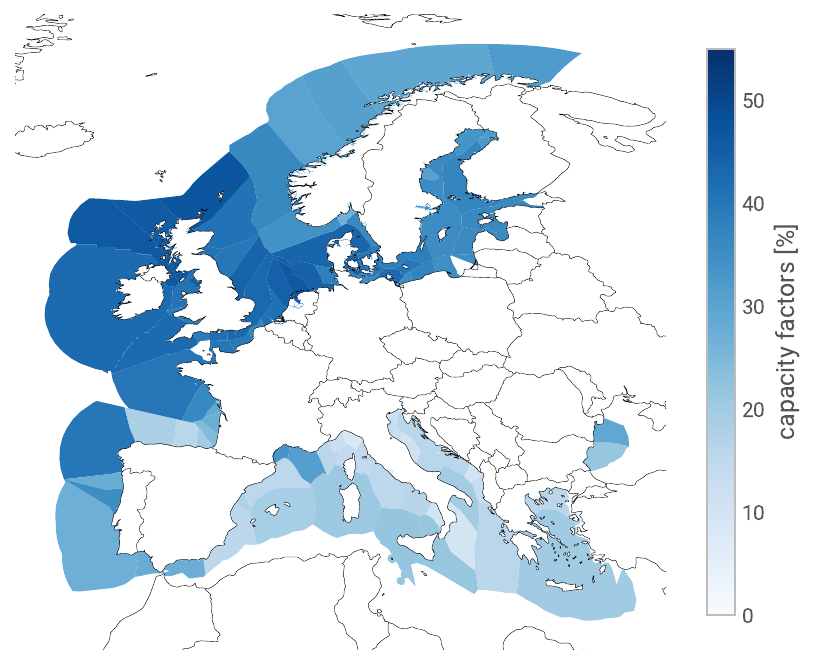}
    \end{subfigure}
    \begin{subfigure}[t]{0.49\textwidth}
        \centering
        \caption{air-sourced heat pump}
        \includegraphics[width=\textwidth]{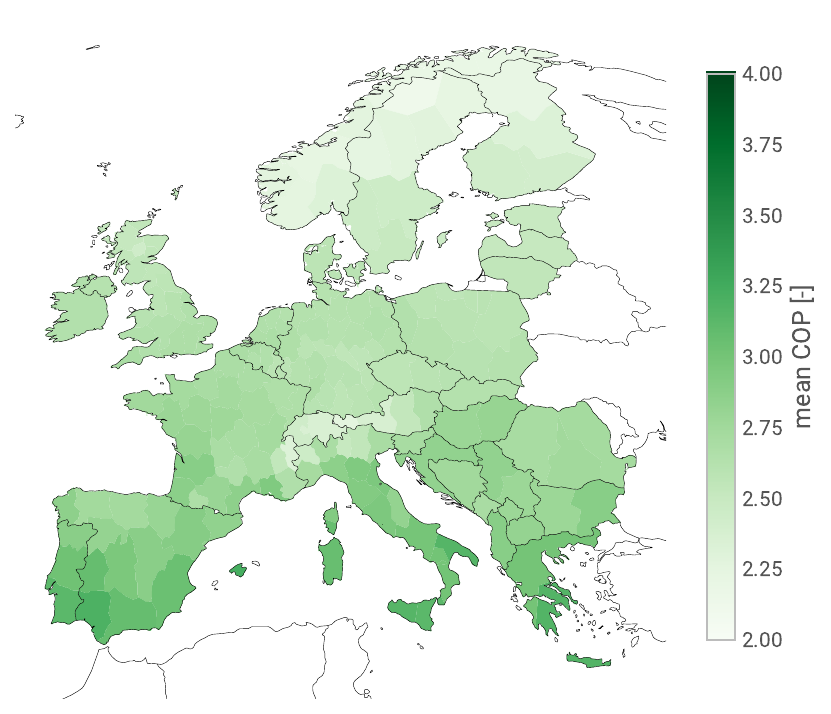}
    \end{subfigure}
    \begin{subfigure}[t]{0.49\textwidth}
        \centering
        \caption{ground-sourced heat pump}
        \includegraphics[width=\textwidth]{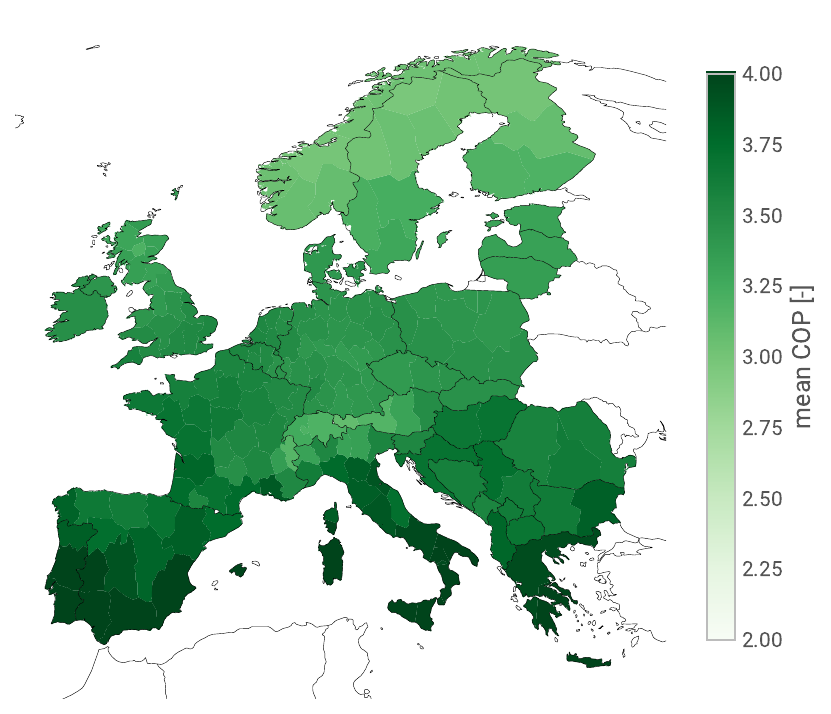}
    \end{subfigure}
    \caption{Regional distribution of average capacity factors of renewable energy sources.}
    \label{fig:cfs-maps}
\end{figure}

The location-dependent renewables availability time series are generated based
on two gridded historical weather datasets. We
retrieve wind sepeeds at \SI{100}{\metre}, surface roughness, soil and air
temperatures, and surface run-off from rainfall or melting snow from the global
ERA5 reanalysis dataset provided by the ECMWF \citeS{ecmwf}. It provides hourly
values for each of these parameters since 1950 on a $0.25^{\circ} \times
0.25^{\circ}$ grid. In Germany, such a weather cell expands approximately
\SI{20}{\kilo\metre} from east to west and \SI{31}{km} from north to south. For
the direct and diffuse solar irradiance, we use the satellite-aided SARAH-2
dataset \citeS{SARAH}, which assesses cloud cover in more detail than the ERA5 dataset. It
features values from 1983 to 2015 at an even higher resolution with a
$0.05^{\circ} \times 0.05^{\circ}$ grid and 30-minute intervals \citeS{SARAH}. In
general, the reference weather year can be freely chosen for the optimisation,
but in this contribution all analyses are based on the year 2013, which is
regarded as characteristic year for both wind and solar resources (e.g.
\citeS{cokerInterannualWeather2020}).

Models for wind turbines, solar panels, heat pumps and the inflow into hydro
basins convert the weather data to hourly time series for capacity factors and
performance coefficients. Using power curves of selected wind turbines types
(Vestas V112 for onshore, NREL 5MW for offshore), wind speeds scaled to the
according hub height are mapped to power outputs. For offshore wind, we
additionally take into account wake effects by applying a uniform correction
factor of 88.55\% to the capacity factors \citeS{boschTemporallyExplicit2018}.
The solar photovoltaic panels' output is calculated based on the incidence angle
of solar irradiation, the panel's tilt angle, and conversion efficiency.
Similarly, solar thermal generation is determined based on collector orientation
and a clear-sky model based on \citeS{henningComprehensiveModel2014a}. The
creation of heat pump time series follows regression analyses that map soil or
air temperatures to the coefficient of performance (COP)
\citeS{nouvelEuropeanMapping2015, staffellReviewDomestic2012}. Hydroelectric
inflow time series are derived from run-off data from ERA5 and scaled using EIA
annual hydropower generation statistics
\citeS{u.s.energyinformationadministrationHydroelectricityNet2022}. The
open-source library \textit{atlite} \citeS{hofmannAtliteLightweight2021}
provides functionality to perform all these calculations efficiently. Finally,
the obtained time series are aggregated to each region heuristically in
proportion to each grid cell's mean capacity factor. This assumes a capacity
layout proportional to mean capacity factors. The resulting spatial and temporal
variability of capacity factors are shown in
\cref{fig:cfs-ts,fig:cfs-maps}.

In combination with the capacity potentials derived from the assumed land use
restrictions, the time-averaged capacity factors are used to display in
\cref{fig:energy-density} the energy that could be produced from wind and solar
energy in the different regions of Europe.

\section{Hydrogen}
\label{sec:si:h2}

\subsection{Hydrogen Demand}
\label{sec:si:h2:demand}

Hydrogen is consumed in the industry sector to produce ammonia and direct
reduced iron (DRI) (see \cref{sec:si:industry:steel}). Hydrogen is also consumed
to produce synthetic methane and liquid hydrocarbons (see
\cref{sec:si:methane:supply} and \cref{sec:si:oil:supply}) which have multiple
uses in industry and other sectors. For transport applications, the consumption
of hydrogen is exogenously fixed. It is used in heavy-duty land transport (see
\cref{sec:si:transport:land}). Furthermore, stationary fuel cells may
re-electrify hydrogen (with waste heat as a byproduct) to balance renewable
fluctuations. The regional distribution of spatially-fixed final hydrogen
demands is shown in \cref{fig:demand-space:hydrogen}.

\subsection{Hydrogen Supply}
\label{sec:si:h2:supply}

Today, most hydrogen is produced from natural gas by steam methane reforming
(SMR)
\begin{equation}
    \ce{ CH4 + H2O -> CO + 3H2 }
\end{equation}
combined with a water-gas shift reaction
\begin{equation}
    \ce{CO + H2O -> CO2 + H2}.
\end{equation}
We consider this route of production with and without carbon capture (CC),
assuming a capture rate of 90\%. These routes are also referred to as blue and
grey hydrogen. The methane input can be of fossil, biogenic, or synthetic origin.

Furthermore, we consider water electrolysis (green hydrogen) which uses electric
energy to split water into hydrogen and oxygen
\begin{equation}
    \ce{2H2O -> 2 H2 + O2}.
\end{equation}
For the electrolysis, we assume alkaline electrolysers since they have lower
cost\citeS{DEA} and higher cumulative installed
capacity\citeS{staffellRoleHydrogen2019a} than polymer electrolyte membrane
(PEM) electrolysers. Waste heat from electrolysis is not leveraged in the model.

The split between these three different technology
options and their installed capacities are a result of the optimisation
depending on the techno-economic assumptions listed in \cref{sec:si:costs}.

\subsection{Hydrogen Transport}
\label{sec:si:h2:transport}

Hydrogen can be transported in pipelines. These can be retrofitted natural gas
pipelines or completely new pipelines. The cost of retrofitting a gas pipeline
is about half that of building a new hydrogen pipeline. These costs include the
cost for new compressors but neglect the energy demand for compression.

The endogenous retrofitting of gas pipelines to hydrogen pipelines is
implemented in a way, such that for every unit of gas pipeline decommissioned,
60\% of its nominal capacity are available for hydrogen transport on the
respective route, following assumptions from the European Hydrogen Backbone
report.\citeS{gasforclimateEuropeanHydrogen2020} When the gas network is not
resolved, this value denotes the potential for repurposed hydrogen pipelines.

New pipelines can be built additionally on all routes where there currently is a
gas or electricity network connection. These new pipelines will be built where
no sufficient retrofitting options are available. The capacities of new and
repurposed pipelines are a result of the optimisation.

\subsection{Hydrogen Storage}
\label{sec:si:h2:storage}

\begin{figure}
    \centering
    \makebox[\textwidth][c]{
    \begin{subfigure}[t]{0.6\textwidth}
        \centering
        \includegraphics[width=\textwidth]{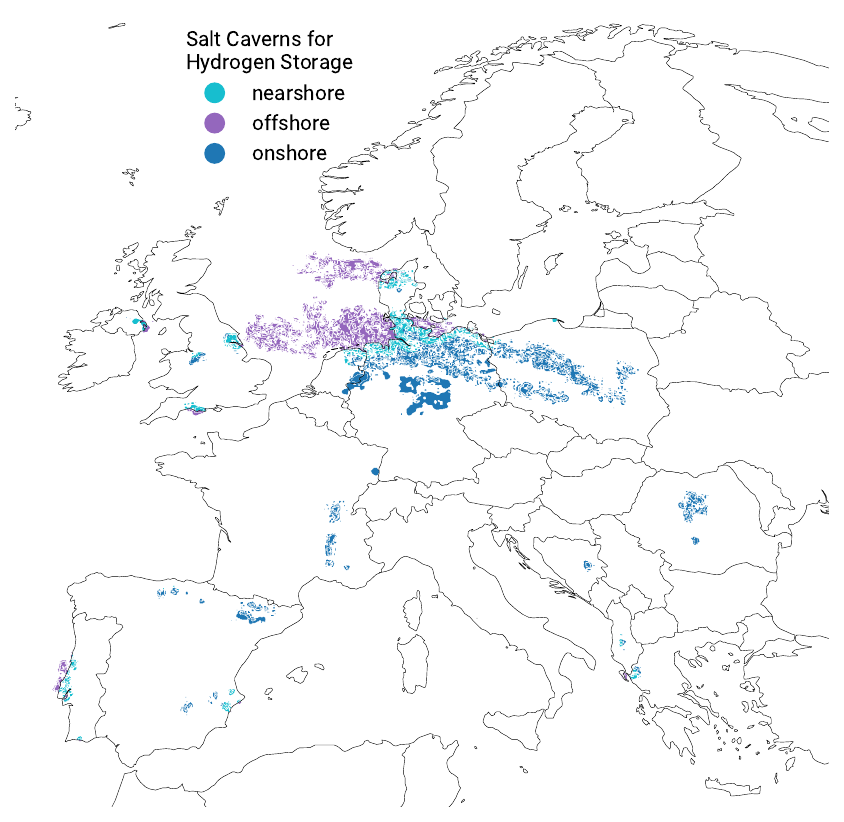}
    \end{subfigure}
    \begin{subfigure}[t]{0.6\textwidth}
        \centering
        \includegraphics[width=\textwidth]{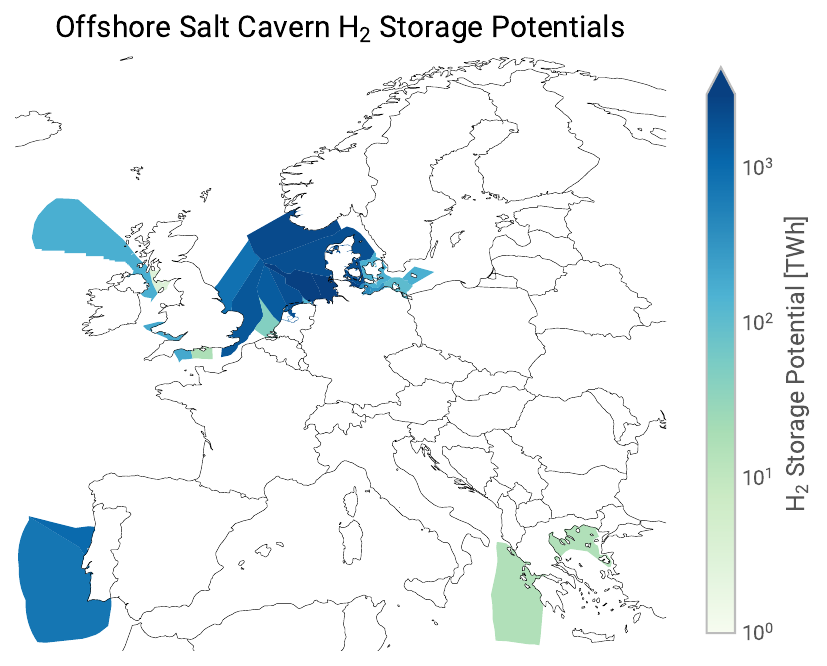}
    \end{subfigure}
    }
    \makebox[\textwidth][c]{
    \begin{subfigure}[t]{0.6\textwidth}
        \centering
        \includegraphics[width=\textwidth]{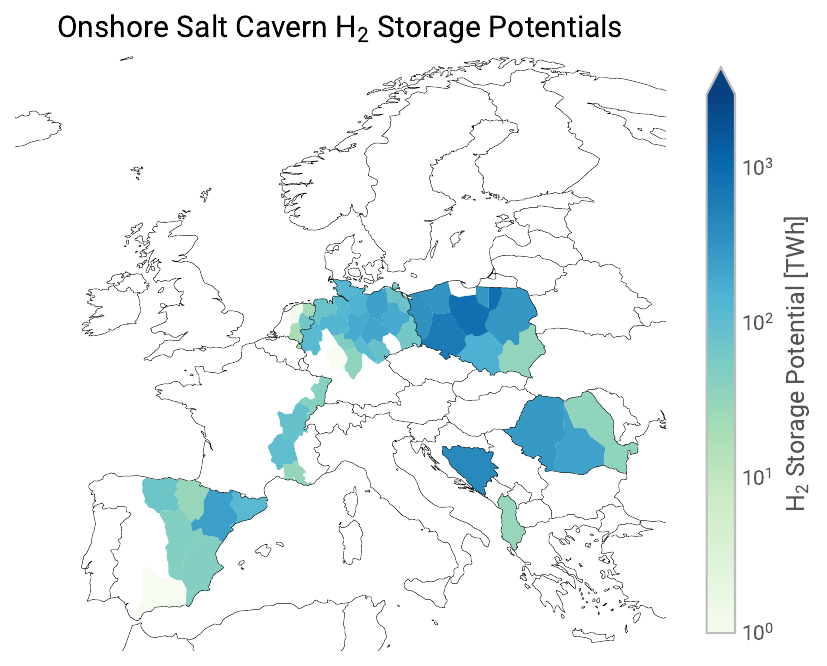}
    \end{subfigure}
    \begin{subfigure}[t]{0.6\textwidth}
        \centering
        \includegraphics[width=\textwidth]{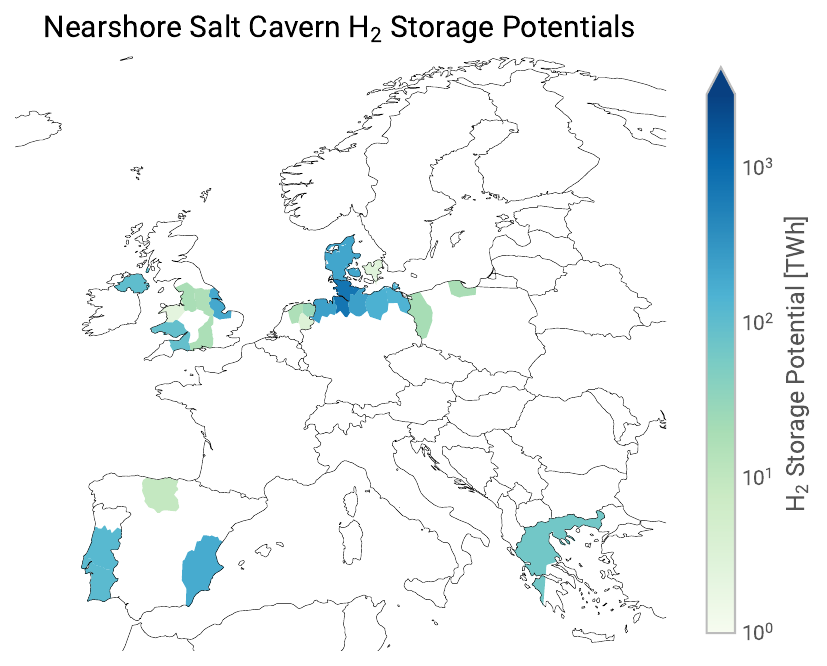}
    \end{subfigure}
    }
    \caption{Potentials for hydrogen underground storage in salt caverns. Potentials are separated into offshore, onshore and near-shore (within 50km of the coast) potentials.}
    \label{fig:clustered-caverns}
\end{figure}

Hydrogen can be stored in overground steel tanks or underground salt caverns.
The annuitised cost for cavern storage is around 30 times lower than for storage
in steel tanks including compression. For underground storage potentials for
hydrogen in European salt caverns we take data from Caglayan et
al.~\citeS{caglayanTechnicalPotential2020} and map it to
each of the 181 model regions (\cref{fig:clustered-caverns}). We include only
those caverns that are located on land and within 50 km of the shore
(nearshore). We impose this restriction to circumvent environmental problems
associated with brine water disposal.\citeS{caglayanTechnicalPotential2020} The
storage potential is abundant and the constraining factor is more where they
exist and less how large the energy storage potentials are.

\section{Methane}
\label{sec:si:methane}

\subsection{Methane Demand}
\label{sec:si:methane:demand}

Methane is used in individual and large-scale gas boilers, in CHP plants with
and without carbon capture, in OCGT and CCGT power plants, and in some industry
subsectors for the provision of high temperature heat (see
\cref{sec:si:industry}) Methane is not used in the transport sector because of
engine slippage. The regional distribution of methane demands is shown in
\cref{fig:demand-space:hydrogen}.

\subsection{Methane Supply}
\label{sec:si:methane:supply}

Besides methane from fossil origins, the model also considers biogenic and
synthetic sources. Fossil gas can enter the European system at existing and
planned LNG terminals, pipeline entry-points, and intra-European gas extraction
sites (see \cref{fig:gas-raw}), which are retrieved from the SciGRID Gas
IGGIELGN dataset\citeS{plutaSciGRIDGas2022a} and the GEM
Wiki.\citeS{gemwikiLNGTerminals2021} Biogas can be upgraded to methane (see
\cref{sec:si:bio:potentials}). Synthetic methane can be produced by processing
hydrogen and captures \co in the Sabatier reaction
\begin{equation}
    \ce{CO2 + 4H2 -> CH4 + 2H2O}.
\end{equation}
The share of synthetic, biogenic and fossil methane is an optimisation result
depending on the techno-economic assumptions listed in \cref{sec:si:costs}.

\subsection{Methane Transport}
\label{sec:si:methane:transport}

\begin{figure}
    \includegraphics[width=1\textwidth,center]{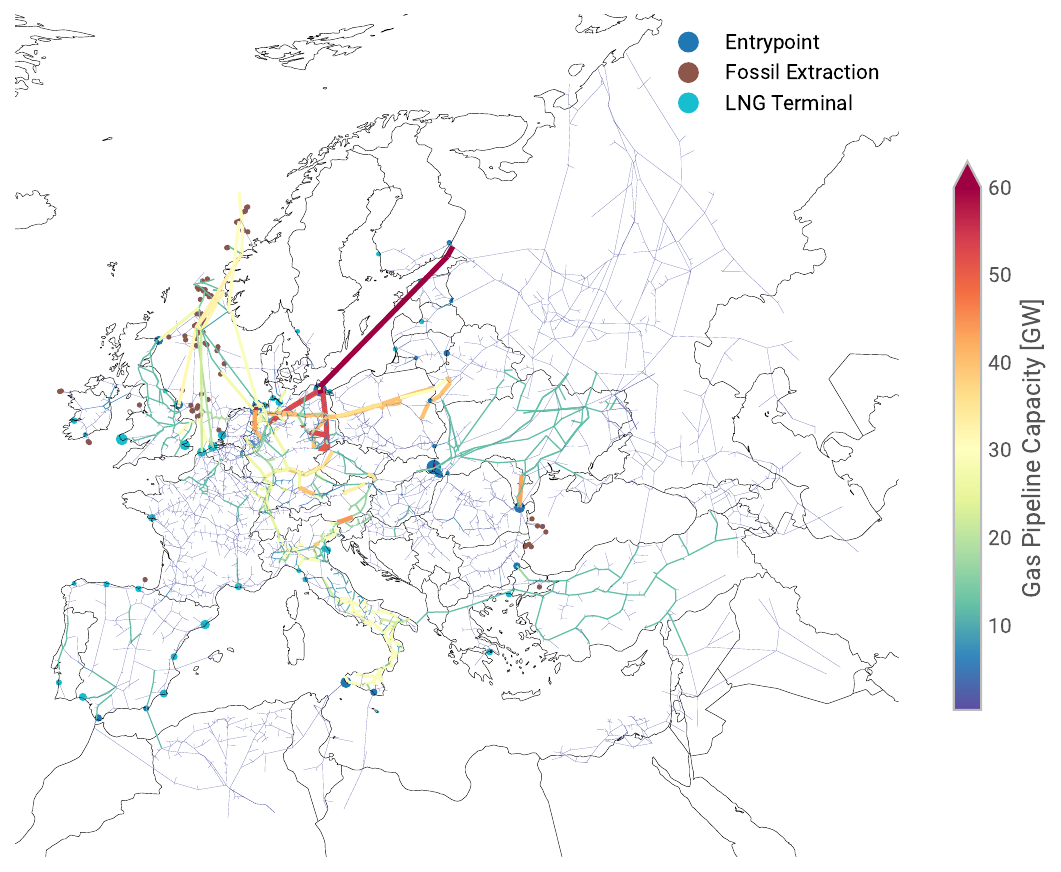}
    \label{fig:gas-raw}
    \caption{Unclustered European gas transmission network based on the
    SciGRID Gas IGGIELGN dataset. The pipelines are color-coded by estimated capacities. Markers indicate entry-points, sites of fossil resource extraction, and LNG terminals.}
\end{figure}

The existing European gas transmission network is represented based on the
SciGRID Gas IGGIELGN dataset,\citeS{plutaSciGRIDGas2022a} as shown in
\cref{fig:gas-raw}. This dataset is based on compiled and merged data from the
ENTSOG maps\citeS{entsogTransmissionCapacity2021} and other publicly available
data sources. It includes data on the capacity, diameter, pressure, length, and
directionality of pipelines. Missing capacity data is conservatively inferred
from the pipe diameter following conversion factors derived from an EHB report
\citeS{gasforclimateEuropeanHydrogen2021}. The gas network is clustered to the
model's 181 regions (see \cref{fig:clustered-networks}). Gas pipelines can be
endogenously expanded or repurposed for hydrogen transport (see
\cref{sec:si:h2:transport}). Gas flows are represented by a lossless transport
model.

The results shown regard the gas transmission network only to determine the
retofitting potentials for hydrogen pipelines. These assume methane to be
transported without cost or capacity constraints, since future demand is
predicted to be low compared to available transport capacities even if a certain
share is repurposed for hydrogen transport such that no bottlenecks are
expected. This assumption has been verified in selected runs with
spatially-resolved gas network infrastructure.

\section{Oil-based Products}
\label{sec:si:oil}

\subsection{Oil-based Product Demand}
\label{sec:si:demand}

Naphtha is used as a feedstock in the chemicals industry (see
\cref{sec:si:industry:chemicals}). Furthermore, kerosene is used as transport
fuel in the aviation sector (see \cref{sec:si:transport:aviation}). International
and domestic shipping uses methanol as transport fuel.
Non-electrified agriculture machinery also consumes gasoline. The regional
distribution of the demand for oil-based products is shown in
\cref{fig:demand-space:oil}. However, this carrier is copperplated in the model,
which means that transport costs and constraints are neglected.

\subsection{Oil-based Product Supply}
\label{sec:si:oil:supply}

In addition to fossil origins, oil-based products can be synthetically produced
by processing hydrogen and captured \co in Fischer-Tropsch plants
\begin{equation}
    \ce{$n$CO + ($2n$ + 1)H2 -> C_$n$H_{2n+2} + $n$H2O}.
\end{equation}
with costs as included in \cref{sec:si:costs}. The waste heat from the
Fischer-Tropsch process is supplied to district heating networks. Likewise,
methanol can be synthesized from captured \co and hydrogen
\begin{equation}
    \ce{CO2 + 3H2 -> CH3OH + H2O}
\end{equation}
with an assumed consumption of 1.14 MWh hydrogen, 0.27 MWh electricity and
\SI{0.25}{\tco} per MWh of methanol produced and costs as listed in
\cref{sec:si:costs}.

\subsection{Oil-based Product Transport}
\label{sec:si:oil:transport}

Liquid hydrocarbons are assumed to be transported freely among the model region
since future demand is predicted to be low, transport costs for liquids are low
and no bottlenecks are expected.

\section{Biomass}
\label{sec:si:bio}

\subsection{Biomass Supply and Potentials}
\label{sec:si:bio:potentials}

\begin{figure}
    \centering
    \begin{subfigure}[t]{0.49\textwidth}
        \centering
        \caption{solid biomass potentials}
        \includegraphics[width=\textwidth]{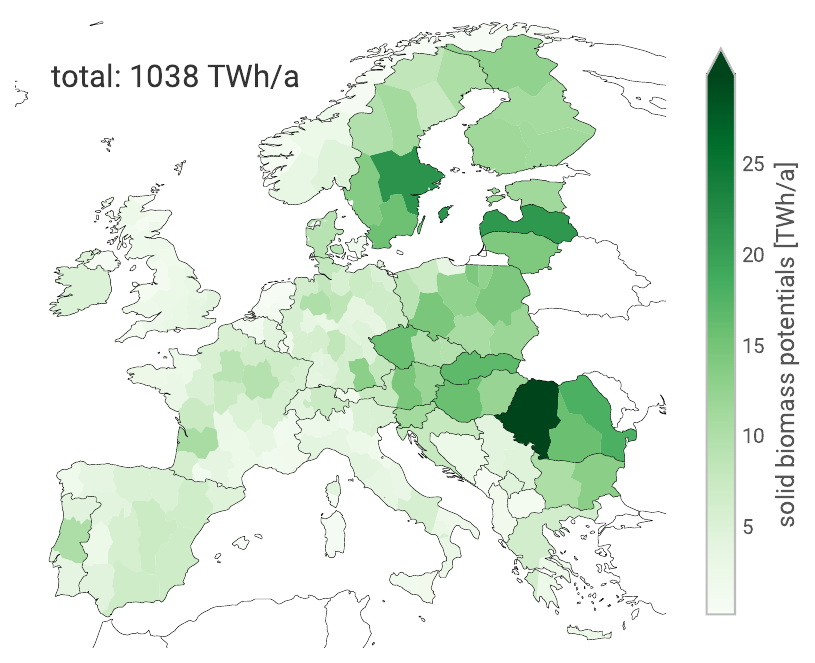}
    \end{subfigure}
    \begin{subfigure}[t]{0.49\textwidth}
        \centering
        \caption{biogas potentials}
        \includegraphics[width=\textwidth]{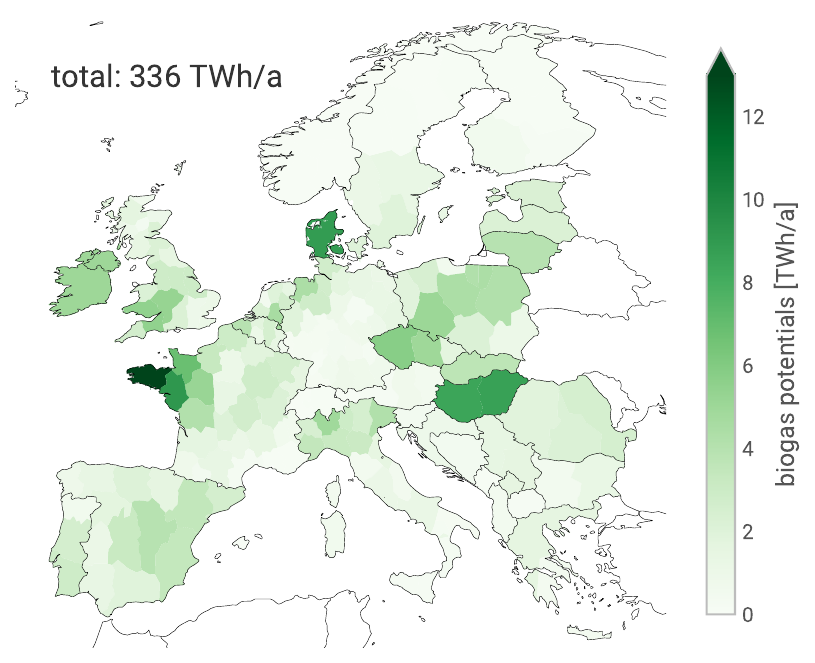}
    \end{subfigure}
    \caption{Regional distribution of biomass potentials separated by solid biomass and biogas. Only residual biomass feedstocks are included. Potentials are based on the medium availability scenario for 2030 from the JRC ENSPRESO database.}
    \label{fig:biomass-potentials}
\end{figure}

Regional biomass supply potentials are taken from the JRC ENSPRESO database
\citeS{ruizENSPRESOOpen2019}. This dataset includes various biomass feedstocks at
NUTS2 resolution for low, medium and high availability scenarios.  We use the
medium availability scenario for 2030, assuming no biomass import from outside
Europe. The data for NUTS2 regions is mapped to PyPSA-Eur-Sec model regions in
proportion to the area overlap.

Only residues from agriculture, forestry, and biodegradable municipal
waste are considered as energy feedstocks. Fuel crops are avoided because they
compete with scarce land for food production, while primary wood as well as wood
chips and pellets are avoided because of concerns about sustainability.
\citeS{bentsenCarbonDebt2017}
Manure and sludge waste are available to the model as biogas, whereas other
wastes and residues are classified as solid biomass. Solid biomass resources are
available for combustion in combined-heat-and-power (CHP) plants and for medium
temperature heat (below \SI{500}{\celsius}) applications in industry.
The technical characteristics for the solid biomass CHP are taken from the
Danish Energy Agency Technology Database\citeS{DEA} assumptions for a
medium-sized back pressure CHP with wood pellet feedstock; this has very similar
costs and efficiencies to CHPs with feedstocks of straw and wood chips.

A summary of which feedstocks are used in the model is shown in
\cref{tab:biomass}; the respective regional distribution of potentials is
included in \cref{fig:biomass-potentials}. In 2015, the EU28 biomass energy
consumption consisted of \SI{180}{\twh} of biogas, \SI{1063}{\twh} of solid
biofuels, \SI{109}{\twh} renewable municipal waste and \SI{159}{\twh} of liquid
biofuels.\citeS{ruizENSPRESOOpen2019} In comparison, PyPSA-Eur-Sec implies a doubling of biogas
consumption and similar amounts of solid biofuels, but a shift from energy crops
and primary wood to residues and wastes. Zappa et al.
\citeS{zappa100Renewable2019} additionally allowed the use of roundwood chips
and pellets, and grassy, willow and poplar energy crops.

\begin{table}
    \centering
    \small
    \begin{tabularx}{\textwidth}{lXr}
        \toprule
        Application & Source & Potential [\si{\twh\per\year}] \\
        \midrule
        solid biomass & primary agricultural residues; forest energy residue; secondary forestry residues: woodchips, sawdust; forestry residues from landscape care; biodegradable municipal waste & 1037 \\
        biogas & wet and dry manure; biodegradable sludge & 336\\
        not used & energy crops: sugar beet bioethanol, rape seed and other oil crops, starchy crops, grassy, willow, poplar; roundwood fuelwood; roundwood chips and pellets & 1661 \\
        \bottomrule
    \end{tabularx}
    \caption{Use of biomass potentials according to classifications from the JRC ENSPRESO database in the medium availability scenario for 2030.}
    \label{tab:biomass}
\end{table}

\subsection{Biomass Demand}
\label{sec:si:bio:demand}

Solid biomass provides process heat up to \SI{500}{\celsius} in industry and can
also feed CHP plants in district heating networks. As noted in
\cref{sec:si:industry}, solid biomass is used as heat supply in the paper and
pulp and food, beverages and tobacco industries, where required temperatures are
lower.\citeS{naeglerQuantificationEuropean2015,rehfeldtBottomupEstimation2018}
The regional distribution of solid biomass demand is shown in
\cref{fig:demand-space:biomass}.

\subsection{Biomass Transport}
\label{sec:si:bio:transport}

Solid biomass is assumed to be transported freely among the modelled regions.
Biogas can be upgraded and then transported via the methane network.
\section{Carbon dioxide capture, usage and sequestration (CCU/S)}
\label{sec:si:carbon-management}

Carbon management becomes important in net-zero scenarios.
\citeS{martin-robertsCarbonCapture2021} PyPSA-Eur-Sec includes carbon capture
from air, electricity generators and industrial facilities, carbon dioxide
storage and transport, the usage of carbon dioxide in synthetic hydrocarbons, as
well as the ultimate sequestration of carbon dioxide underground.

\subsection{Carbon Capture}

Carbon dioxide can be captured from industry process emissions, steam methane
reforming, methane or biomass used for process heat in the industry, combined
heat and power plants (CHP using biomass or methane), and directly from the air
using direct air capture (DAC). The capacities of each carbon capture technology
are co-optimised.

As shown in \cref{fig:process-emissions}, the model includes industrial process
emissions with fossil-origin totalling 127 Mt$_{\ce{CO2}}$/a based on
the JRC-IDEES database.\citeS{europeancommission.jointresearchcentre.JRCIDEESIntegrated2017} Process emissions originate, for instance,
from limestone in cement production. These emissions need to be captured and
sequestered or offset to achieve net-zero emissions. Industry process emissions
are captured assuming a capture rate of 90\% and assuming costs of \co capturing
like in the cement industry.\citeS{DEA} The electricity and heat demand of process
emission carbon capture is currently ignored.

For steam methane reforming (SMR), CHP units, and biomass and methane demand in
industry the model can decide between two options (with and without carbon
capture) with different costs. Here, we also apply a capture rate of 90\%.

DAC includes the energy requirements of the adsorption phase with inputs
electricity and heat to assist adsorption process and regenerate adsorbent, as
well as the compression of \co prior to storage which consumes electricity and
rejects heat. We assume a net energy consumption of 1.8 MWh/t$_{\ce{CO2}}$ heat and 0.47
MWh/t$_{\ce{CO2}}$ electricity based on DEA data.\citeS{DEA} These values are a bit higher
compared to Breyer et al.,~\citeS{breyerCarbonDioxide2020} who assume
requirements of 1.2 MWh/t$_{\ce{CO2}}$ heat at \SI{100}{\celsius} and 0.2 MWh\el/t$_{\ce{CO2}}$
electricity.

\subsection{Carbon Usage}

Captured \co can be used to produce synthetic methane and liquid hydrocarbons
(e.g. naphtha, methanol, Fischer-Tropsch fuels). See
\cref{sec:si:methane:supply} and \cref{sec:si:oil:supply}. If carbon captured
from biomass is used, the \co emissions of the synthetic fuels are net-neutral.

\subsection{Carbon Transport and Sequestration}

Captured \co can also be stored underground up to an annual sequestration limit
of 200 Mt$_{\ce{CO2}}$/a. Compared to other studies, this is a conservative assumption but
sufficient to capture and sequester process emissions. The sequestration of
captured \co from bioenergy results in net negative emissions. As stored carbon
dioxide is modelled as a single node for Europe, transport constraints are
neglected. For for \co transport and sequestration we assume a cost of
\SI{20}{\sieuro\per\tco} based on IEA data.\citeS{internationalenergyagencyCarbonCapture}

\section{Mathematical Model Formulation}
\label{sec:si:math}

The objective is to minimise the total annual energy system costs of the energy system
that comprises both investment costs and operational expenditures of generation,
storage, transmission and conversion infrastructure. To express both as annual
costs, we use the annuity factor $(1-(1+\tau)^{-n}) / \tau$ that, like a
mortgage, converts the upfront investment of an asset to annual payments
considering its lifetime $n$ and cost of capital $\tau$. Thus, the objective
includes on one hand the annualised capital costs $c_*$ for investments at bus
$i$ in generator capacity $G_{i,r}\in\R^+$ of technology $r$, storage energy
capacity $E_{i,s}\in\R^+$ of technology $s$, electricity transmission line
capacities $P_{\ell}\in\R^+$, and energy conversion and transport capacities
$F_k\in\R^+$ (links), as well as the variable operating costs $o_*$ for
generator dispatch $g_{i,r,t}\in\R^+$ and link dispatch $f_{k,t}\in\R^+$ on the
other:
\begin{align}
  \label{eq:objective}
  \min_{G,E,P,F,g} \quad &\left[\sum_{i,r} c_{i,r}\cdot G_{i,r} + \sum_{i,s} c_{i,s}\cdot E_{i,s} + \sum_{\ell}c_{\ell}\cdot P_{\ell}+ \sum_{k}c_{k}\cdot F_k +\right. \\
  & \left.  \sum_{t} w_t \cdot \left( \sum_{i,r} o_{i,r} \cdot g_{i,r,t} + \sum_k o_k \cdot f_{k,t} \right) \right].
\end{align}
Thereby, the representative time snapshots $t$ are weighted by the time span
$w_t$ such that their total duration adds up to one year; \mbox{$\sum_{t\in \cT}
w_t=365\cdot 24\text{h}=8760\text{h}$}. A bus $i$ represents both a regional
scope and an energy carrier. Represented carriers include electricity, heat
(various subdivisions), hydrogen, methane, oil and carbon dioxide.

In addition to the cost-minimising objective function, we further impose a set
of linear constraints that define limits on (i) the capacities of generation,
storage, conversion and transmission infrastructure from geographical and
technical potentials, (ii) the availability of variable renewable energy sources
for each location and point in time (iii) the limit for \co emissions or transmission expansion, (iv)
storage consistency equations, and (v) a multi-period linearised optimal power
flow (LOPF) formulation. Overall, this results in a large linear problem (LP).

The capacities of generation, storage, conversion and transmission
infrastructure are constrained from above by their installable potentials and
from below by any existing components:
\begin{align}
  \underline{G}_{i,r}  &  & \leq &  & G_{i,r}  &  & \leq &  & \overline{G}_{i,r}  & \qquad\forall i, r \label{eq:genlimit} \\
  \underline{E}_{i,s}  &  & \leq &  & E_{i,s}  &  & \leq &  & \overline{E}_{i,s}  & \qquad\forall i, s \\
  \underline{P}_{\ell} &  & \leq &  & P_{\ell} &  & \leq &  & \overline{P}_{\ell} & \qquad\forall \ell \\
  \underline{F}_{k} &  & \leq &  & F_{k} &  & \leq &  & \overline{F}_{k} & \qquad\forall k
\end{align}

Moreover, the dispatch of generators and links may not only be constrained by their rated capacity but also by the weather-dependent availability of
variable renewable energy or must-run conditions.
This can be expressed as a time- and location-dependent availability
factor $\overline{g}_{i,r,t}$/$\overline{f}_{k,t}$  and must-run factor $\underline{g}_{i,r,t}$/$\underline{f}_{k,t}$, given per unit of the nominal capacity:
\begin{align}
    \underline{g}_{i,r,t}  G_{i,r} &  & \leq &  & g_{i,r,t} &  & \leq &  & \overline{g}_{i,r,t} G_{i,r} & \qquad\forall i, r, t \\
    \underline{f}_{k,t}  F_{k} &  & \leq &  & f_{k,t} &  & \leq &  & \overline{f}_{k,t} F_{i,r} & \qquad\forall k, t
\end{align}
The parameter $\underline{f}_{k,t}$ can also be used to define whether a link is
bidirectional or unidirectional. For instance, for HVDC links
$\underline{f}_{k,t}=-1$ allows power flows in either direction. On the other
hand, a heat resistor has $\underline{f}_{k,t}=0$ since it can only convert
electricity to heat.

The energy levels $e_{i,s,t}$ of all stores are constrained by their energy capacity
\begin{align}
  0 &  & \leq &  & e_{i,s,t} &  & \leq &  & E_{i,s} & \qquad\forall i, s, t,
\end{align}
and have to be consistent with the dispatch variable $h_{i,s,t}\in\R$ in all
hours
\begin{align}
  e_{i,s,t} =\: & \eta_{i,s,0}^{w_t} \cdot e_{i,s,t-1} + w_t \cdot h_{i,s,t}, \label{eq:stoe}
\end{align}
where $\eta_{i,s,0}$ denotes the standing loss . Furthermore, the storage energy
levels are either assumed to be cyclic or given an initial state of charge,
\begin{align}
  e_{i,s,0} = e_{i,s,\abs{\mathcal{T}}} \qquad\forall i, s, or\\
  e_{i,s,0} = e_{i,s,\text{initial}} \qquad\forall i, s.
\end{align}

The modelling of hydroelectricity storage deviates from regular storage to
additionally account for natural inflow and spillage of water. We also assume
fixed power ratings $H_{i,s}$ for hydroelectricity storage. The dispatch of
hydroelectricity storage units is split into two positive variables; one each
for charging $h_{i,s,t}^+$ and discharging $h_{i,s,t}^-$, and limited by
$H_{i,s}$.
\begin{align}
  0 &  & \leq &  & h_{i,s,t}^+ &  & \leq &  & H_{i,s} & \qquad\forall i, s, t \label{eq:sto1} \\
  0 &  & \leq &  & h_{i,s,t}^- &  & \leq &  & H_{i,s} & \qquad\forall i, s, t \label{eq:sto2}
\end{align}
The energy levels $e_{i,s,t}$ of all hydroelectric storage also have to match
the dispatch across all hours
\begin{align}
  e_{i,s,t} =\: & \eta_{i,s,0}^{w_t} \cdot e_{i,s,t-1} + w_t \cdot h_{i,s,t}^\text{inflow} - w_t \cdot h_{i,s,t}^\text{spillage} & \quad\forall i, s, t \nonumber \\
                & + \eta_{i,s,+} \cdot w_t \cdot h_{i,s,t}^+ - \eta_{i,s,-}^{-1} \cdot w_t \cdot h_{i,s,t}^-, \label{eq:stoe}
\end{align}
whereby hydropower storage units can additionally have a charging efficiency
$\eta_{i,s,+}$, a discharging efficiency $\eta_{i,s,-}$, natural inflow
$h_{i,s,t}^\text{inflow}$ and spillage $h_{i,s,t}^\text{spillage}$, besides the
standing loss $\eta_{i,s,0}$.

The nodal balance constraint for supply and demand (Kirchoff's current law for electricity
buses) requires local generators and storage units as well as incoming or
outgoing energy flows $f_{\ell,t}$ of incident transmission lines $\ell$ to
balance the perfectly inelastic electricity demand $d_{i,t}$ at each location
$i$ and snapshot $t$
\begin{align}
    \sum_r g_{i,r,t} + \sum_s \left(h_{i,s,t}^- - h_{i,s,t}^+ \right) + \sum_s h_{i,s,t} + \sum_\ell K_{i\ell} f_{\ell,t} + \sum_k L_{ikt} f_{k,t} = d_{i,t}  \quad \leftrightarrow \quad \lambda_{i,t} \quad \forall i,t,
\end{align}
where $K_{i\ell}$ is the incidence matrix of the electricity network with
non-zero values $-1$ if line $\ell$ starts at node $i$ and $1$ if it ends at
node $i$. $L_{ikt}$ is the lossy incidence matrix of the network with
non-zero values $-1$ if link $k$ starts at node $i$ and $\eta_{i,k,t}$ if one of
its terminal buses is node $i$. For a link with more than two outputs (e.g. CHP
converts gas to heat and electricity in a fixed ratio), the respective column of
the lossy incidence matrix has more than two non-zero entries (hypergraph). The
efficiency may be time-dependent and greater than one for certain technologies
(e.g. for heat pumps converting electricity and ambient heat to hot water).

The Lagrange multiplier (KKT multiplier) $\lambda_{i,t}$ associated with the
nodal balance constraint indicates the marginal price of the respective energy
carrier and location of bus $i$ at time $t$, e.g. the local marginal price (LMP)
of electricity at the electricity bus.

The power flows $p_{\ell,t}$ are limited by their nominal capacities $P_\ell$
\begin{align}
	|p_{\ell,t}| \leq \overline{p}_{\ell} P_{\ell} & \qquad\forall \ell, t,
	\label{eq:cap}
\end{align}
where $\overline{p}_\ell$ acts as an additional per-unit security margin on the line capacity
to allow a buffer for the failure of single circuits ($N-1$ criterion) and reactive power flows.

Kirchoff's voltage law (KVL) imposes further constraints on the flow of AC
transmission lines and there are several ways to formulate KVL with large
impacts on performance. Here, we use linearised load flow assumptions, where the
voltage angle difference around every closed cycle in the electricity
transmission network must add up to zero. Using a cycle basis $C_{\ell c}$ of
the network graph where the independent cycles $c$ are expressed as directed
linear combinations of lines $\ell$,\citeS{horschLinearOptimal2018} we can write
KVL as
\begin{align}
    \sum_\ell C_{\ell c} \cdot x_\ell \cdot p_{\ell,t} = 0 \qquad\forall c,t
    \label{eq:kvl}
\end{align}
where $x_\ell$ is the series inductive reactance of line $\ell$.

We may further regard a constraint on the total annual CO$_2$ emissions $\Gamma_{\text{CO}_2}$
to achieve sustainability goals.
The emissions are determined from the time-weighted generator dispatch $ w_t \cdot g_{i,r,t}$ using the specific emissions $\rho_r$ of technology $r$
and the generator efficiencies $\eta_{i,r}$
\begin{align}
	\sum_{i,r,t}  \rho_r \cdot \eta_{i,r}^{-1} \cdot w_t \cdot g_{i,r,t} + \sum_{i,s} \rho_s \left(e_{i,s,t=0} - e_{i,s,t=\abs{\mathcal{T}}}\right) \leq \Gamma_{\text{CO}_2}  \quad \leftrightarrow \quad \mu_{\text{CO}_2}.
\end{align}
In this case, the Lagrange multiplier (KKT multiplier) $\mu_{\text{CO}_2}$ denotes the shadow price of emitting an additional tonne of \co, i.e. the \co price necessary to achieve the respective \co emission reduction target.

Additionally, another global constraint may be set on the volume of electricity transmission network expansion
\begin{align}
    \sum_\ell l_\ell \cdot P_\ell \leq \Gamma_{LV} \quad \leftrightarrow \quad \mu_{LV},
\end{align}
where the sum of transmission capacities $P_\ell$ multiplied by their lengths $l_\ell$ is bounded by a transmission volume cap
$\Gamma_{LV}$. In this case, the Lagrange multiplier (KKT multiplier) $\mu_{LV}$ denotes the shadow price of a marginal increase in transmission volume.

This formulation does not include pathway optimisation (i.e.~no
sequences of investments), but searches for a cost-optimal layout corresponding
to a given \co emission reduction level and assumes perfect foresight for
the reference year based on which capacities are optimised. This optimisation
problem is implemented in the open-source Python-based modelling framework
PyPSA.\citeS{brownPyPSAPython2018}

\section{Sensitivity Analysis}
\label{sec:si:sensitivity}

\subsection{Electricity Grid Reinforcement Restrictions}
\label{sec:si:lv}

\begin{figure}
    \centering
    \makebox[\textwidth][c]{
    \begin{subfigure}[t]{1.2\textwidth}
        \centering
        \caption{Sensitivity of total system cost towards electricity transmission grid expansion limits}
        \includegraphics[width=\textwidth]{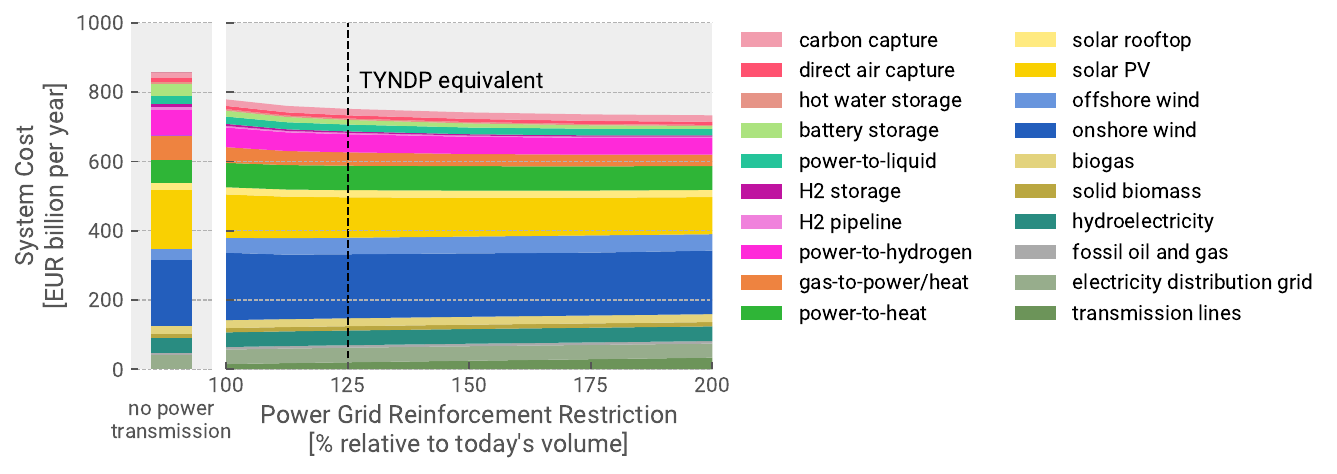}
        \label{fig:lv-restriction}
    \end{subfigure}
    }
    \makebox[\textwidth][c]{
    \begin{subfigure}[t]{1.2\textwidth}
        \centering
        \caption{Sensitivity of total system cost towards onshore wind expansion limits}
        \includegraphics[width=\textwidth]{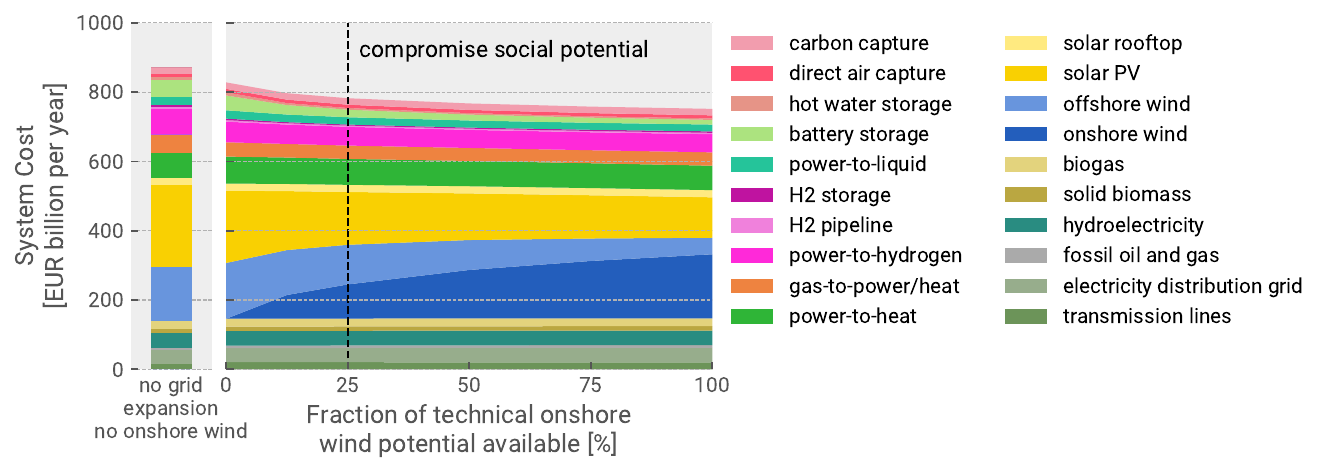}
        \label{fig:onw-restriction}
    \end{subfigure}
    }
    \caption{Sensitivity of total system cost towards electricity transmission grid expansion limits and onshore wind restrictions.
    The sweep for grid expansion restrictions allows full onshore wind potentials.
    The extreme case (a, left) removes all existing power transmission lines as well.
    The sweep for onshore wind potential restrictions allows power grid reinforcements by up to 25\% of today's transmission capacities.
    The extreme case (b, left) combines no grid expansion with no onshore wind potentials.}

    \label{fig:lv-onw-restriction}
\end{figure}

In the following sensitivity runs, the model is allowed to build new electricity
transmission infrastructure wherever is cost-optimal, but the total volume of
new transmission capacity (sum of line length times capacity, TWkm) is
successively limited. The volume limit is given in fractions of today's grid
volume: a line volume limit of 100\% means no new capacity is allowed beyond
today's grid (since the model cannot remove existing lines); a limit of 125\%
means the total grid capacity can grow by 25\% (25\% is similar to the planned
extra capacity in the European network operators' Ten Year Network Development
Plan (TYNDP)\citeS{tyndp2018}). For this investigation, a hydrogen network
could be built.

\cref{fig:lv-restriction} shows the composition of total annual energy system costs
(including all investment and operational costs) as we vary the allowed power
grid expansion, from no expansion (only today's grid) to a doubling of today's
grid capacities (the model optimises where new capacity is placed). As the grid
is expanded, total costs decrease only slightly, despite the increasing costs of
the grid. The total cost benefit of a doubling of grid capacity is around
46~bn\euro/a (6\%) corresponding to an expansion of 715 TWkm. However, over half
of the benefit (27~bn\euro/a, 3.5\%) is available already at a 25\% expansion
corresponding to an expansion of 447 TWkm.

\cref{fig:lv-restriction} also includes a scenario where today's electricity
transmission infrastructure is completely removed from the model, similar to an
electricity system study on geographic trade-offs by Tröndle et
al.\citeS{trondleTradeOffsGeographic2020} While doubling the transmission grid
yields a benefit of 46~bn\euro/a, removing what exists incurs a cost of
108~bn\euro/a. The lack of electricity grid is mostly compensated by more solar
PV generation, battery storage and re-electrified hydrogen.

\subsection{Onshore Wind Potential Elimination}
\label{sec:si:onw}

\begin{figure}
    \centering
    \makebox[\textwidth][c]{
        \begin{subfigure}[t]{0.6\textwidth}
            \centering
            \caption{hydrogen network}
            \includegraphics[height=0.39\textheight]{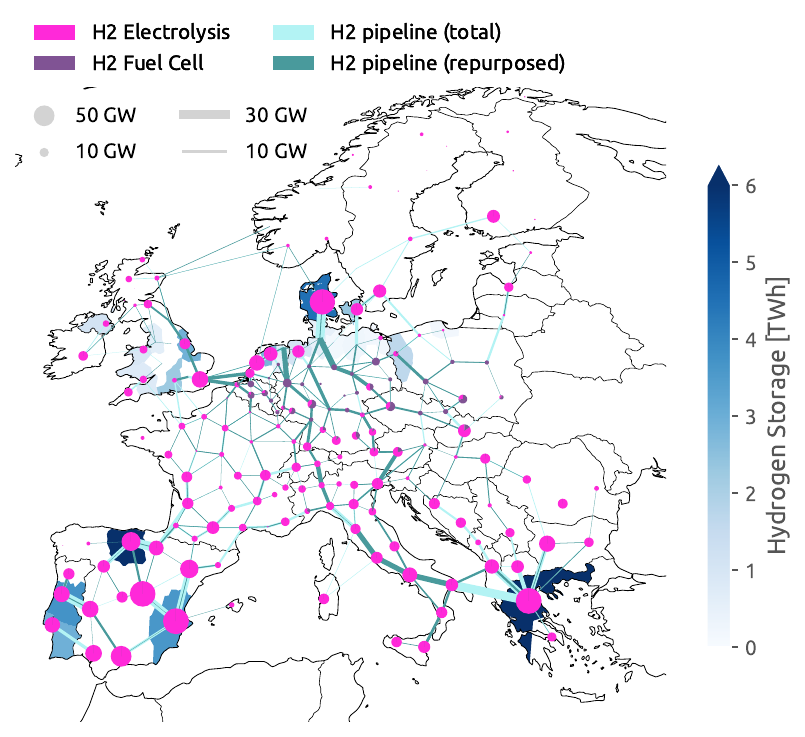}
            \label{fig:no-onw:h2}
        \end{subfigure}
        \begin{subfigure}[t]{0.6\textwidth}
            \centering
            \caption{energy balance}
            \includegraphics[height=0.4\textheight]{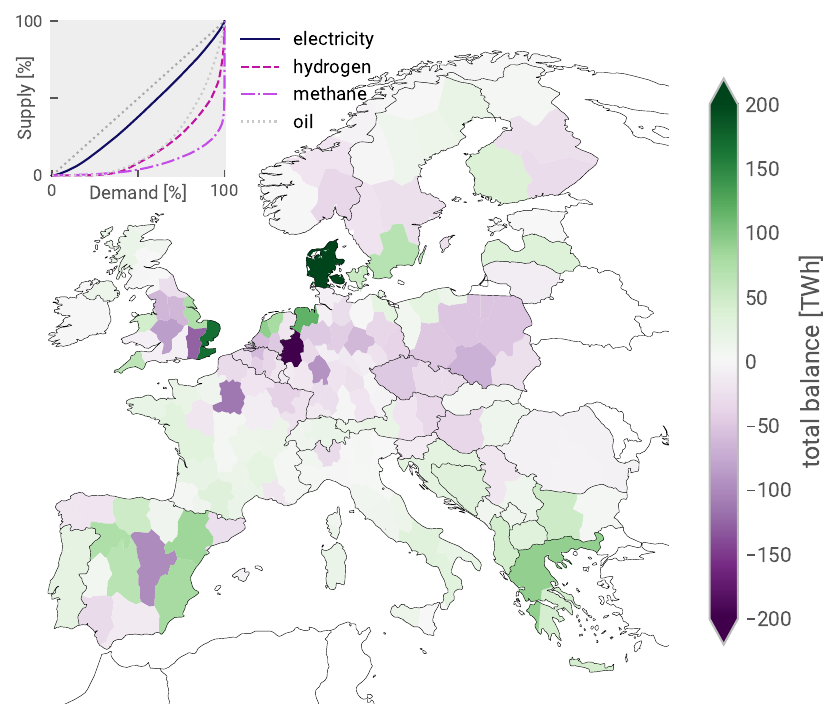}
            \label{fig:no-onw:io}
        \end{subfigure}
        }
        \makebox[\textwidth][c]{
            \begin{subfigure}[t]{0.67\textwidth}
                \centering
                \caption{system cost}
                \includegraphics[height=0.38\textheight]{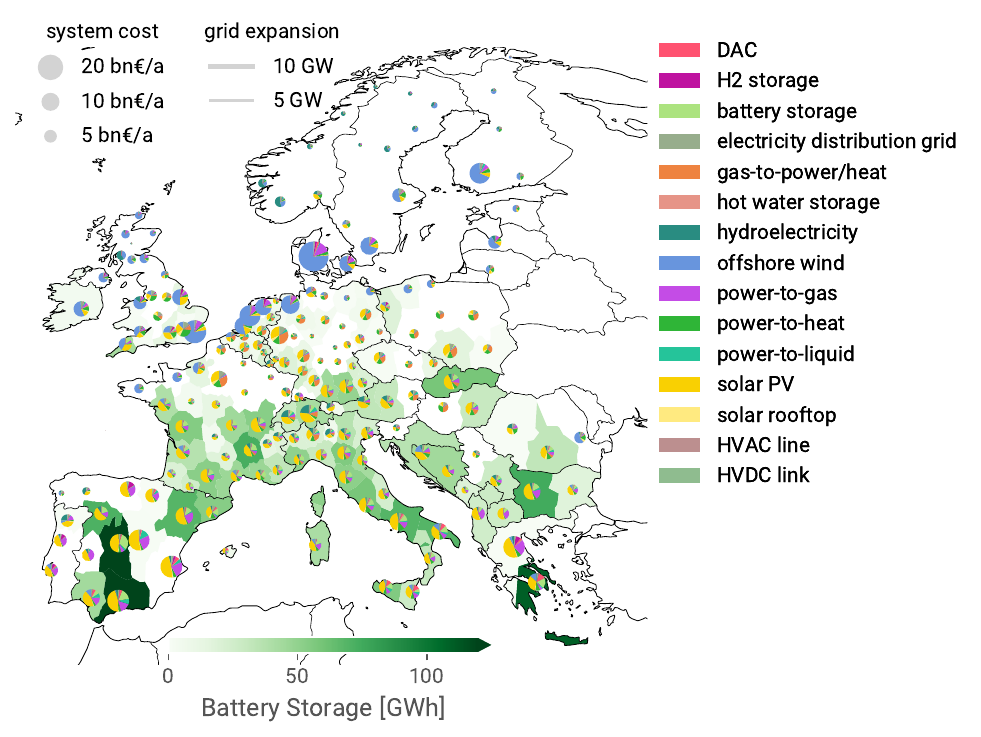}
                \label{fig:no-onw:tsc}
            \end{subfigure}
            \begin{subfigure}[t]{0.53\textwidth}
                \centering
                \caption{hydrogen network flows}
                \includegraphics[height=0.33\textheight]{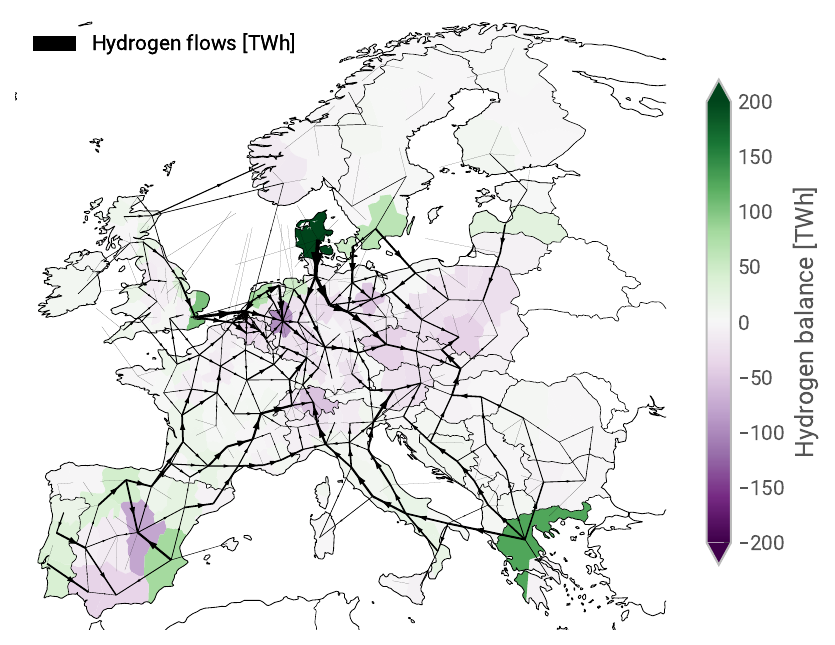}
                \label{fig:no-onw:flow}
            \end{subfigure}
            }

    \caption{ Maps of regional energy balance, hydrogen network and production
        sites, the spatial and technological distribution of total energy system
        costs, and hydrogen flows for a scenario without onshore wind and
        without power grid expansion. }
    \label{fig:no-onw}
\end{figure}

Like building new power transmission lines, the deployment of onshore wind may
not always be socially accepted, such that it may not be possible to leverage
its full potential.\citeS{mckennaScenicnessAssessment2021,weinandImpactPublic2021,weinandExploringTrilemma2021} In the following additional sensitivity analysis, we explore
the hypothetical impact of restricting the installable potentials of onshore
wind down to zero (\cref{fig:no-onw}).

We find that as onshore wind is eliminated, costs rise by \euro~92 bn/a (12\%)
when the electricity grid is fixed to today's capacities, but a hydrogen network
can still be developed. In comparison to the least-cost solution with full
network expansion, this solution is 19\% more expensive. A solution in which
neither a hydrogen network could be developed would be 23\% more expensive.
\cref{sec:si:onw-compromise} presents further intermediate results between full
and no onshore wind expansion for scenarios with hydrogen network expansion and
TYNDP-equivalent power grid reinforcements. The model substitutes onshore wind,
particularly in the British Isles, for higher investment in offshore wind in the
continental shores of the North Sea and solar generators plus batteries in
Southern and Central Europe (\cref{fig:no-onw:tsc}).  Without onshore wind, the
potentials for rooftop solar PV and fixed-pole offshore wind in Europe are
largely exhausted, such that in this self-sufficient scenario for Europe, the
effect of installable potentials becomes critical.

Whereas with onshore wind, we observe both wind-backed electrolysis in
Northwestern Europe and solar-backed hydrogen production in Southern Europe, the
latter becomes the dominant producer of hydrogen if the development of onshore
wind capacities is restricted (\cref{fig:no-onw:h2,fig:no-onw:io}). This shift
in hydrogen infrastructure also impacts the share of gas pipelines being
retrofitted for hydrogen transport. As the Iberian Peninsula becomes a preferred
region for hydrogen production but has a more sparse gas transmission network
today, the rate of retrofitted pipeline capacity reduces from 65\% to 58\%. Many
new hydrogen pipelines are built to connect Spain with France, but also to
connect Denmark to Germany and Greece to Italy. Gas pipeline retrofitting is
then concentrated in Germany, Austria and Italy.

\begin{figure}
    \centering
    \includegraphics[width=.9\textwidth]{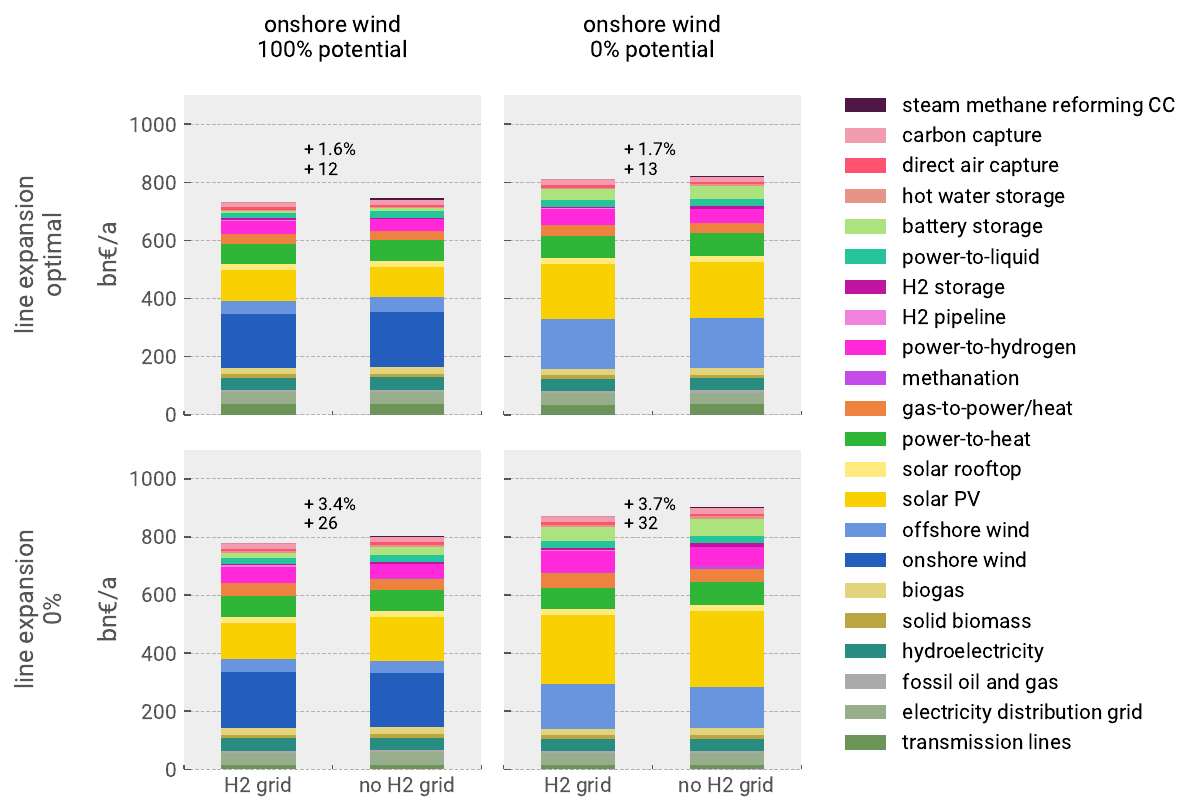}
    \caption{Varying cost benefits of hydrogen network infrastructure and changes in system composition as power grid and onshore wind expansion options are altered. The cost benefit of a hydrogen network varies between 1.6\% and 3.7\% across all scenarios shown.}
    \label{fig:h2-restriction-w-onw}
\end{figure}

The cost benefit of a hydrogen network is similar whether or not onshore wind
capacities are built in Europe, even though the hydrogen network topology is
then built around supply from solar PV from Southern Europe and offshore wind in
the North Sea rather than from onshore wind in Northwestern Europe. As
\cref{fig:h2-restriction-w-onw} illustrates, the net benefit is again strongest
when power grid expansion is restricted. If both onshore wind and power grid
expansion are excluded, costs for a system without a hydrogen network option
were by 32~bn\euro/a (3.7\%) higher. With cost-optimal electricity grid
reinforcement, the net benefit of a hydrogen network is lower with
13~bn\euro/a (1.7\%).

\subsection{Compromises on Onshore Wind Potential Restrictions}
\label{sec:si:onw-compromise}

In the following sensitivity runs, the maximum installable capacity of onshore
wind is successively restricted down to zero at each node. The upper limit is
derived from land use restriction and yields a maximum technical potential
corresponding to about \SI{481}{\giga\watt} for Germany. For this investigation,
a compromise electricity grid expansion by 25\% compared to today and no limits
on hydrogen network infrastructure are assumed.

In this case, system costs rise by 77~bn\euro/a (10\%)
by restricting the installable potentials of onshore down to zero. Just as in
the case of restricted line volumes, \cref{fig:onw-restriction} reveals a
nonlinear rise in system costs: if we constrain the model to 25\% of the onshore
potential (around 120~GW for Germany), costs rise by only 46~bn\euro/a (6\%).
Thereby, 25\% of the onshore wind potential may represent a social compromise
between total system cost, and social concerns about onshore wind development.

In comparison, Schlachtberger et al.~\citeS{schlachtbergerCostOptimal2018} found
a similar change between 9\% and 12\% in system costs in an electricity-only
model when onshore wind potentials were restricted across various grid expansion
limitations. The the biggest change was observed when the power grid could not
be reinforced. Onshore wind was largely replaced with offshore wind in that
model. Unlike that model, here we have a higher grid resolution (181 versus 30
regions) which allows us to better assess the grid integration costs of offshore
wind. Our results show that moderate power grid expansion is particularly
important when onshore wind development is severely limited. For the extreme
case where no onshore wind capacities would be built, reducing power grid
expansion from 25\% to none incurs another rise in system cost of an additional
43~bn\euro/a (6\%).

\subsection{Using Technology and Cost Projections for 2050}
\label{sec:si:sensitivity-costs}

In this sensitivity analysis, we investigate the impact of using more
progressive technology cost projections.\citeS{wayEmpiricallyGrounded2022}
Rather than using assumptions for the year 2030, we use cost assumptions for
2050 as outlined in \cref{tab:si:costs}. These assumptions include cost
reductions of solar photovoltaics and power-to-liquid processes by 25\% beyond
2030, as well as a reduction by 33\% for direct air capture and 45\% to 60\% for
battery storage and electrolysers.

\begin{SCfigure}
    \centering
    \includegraphics[width=0.75\textwidth]{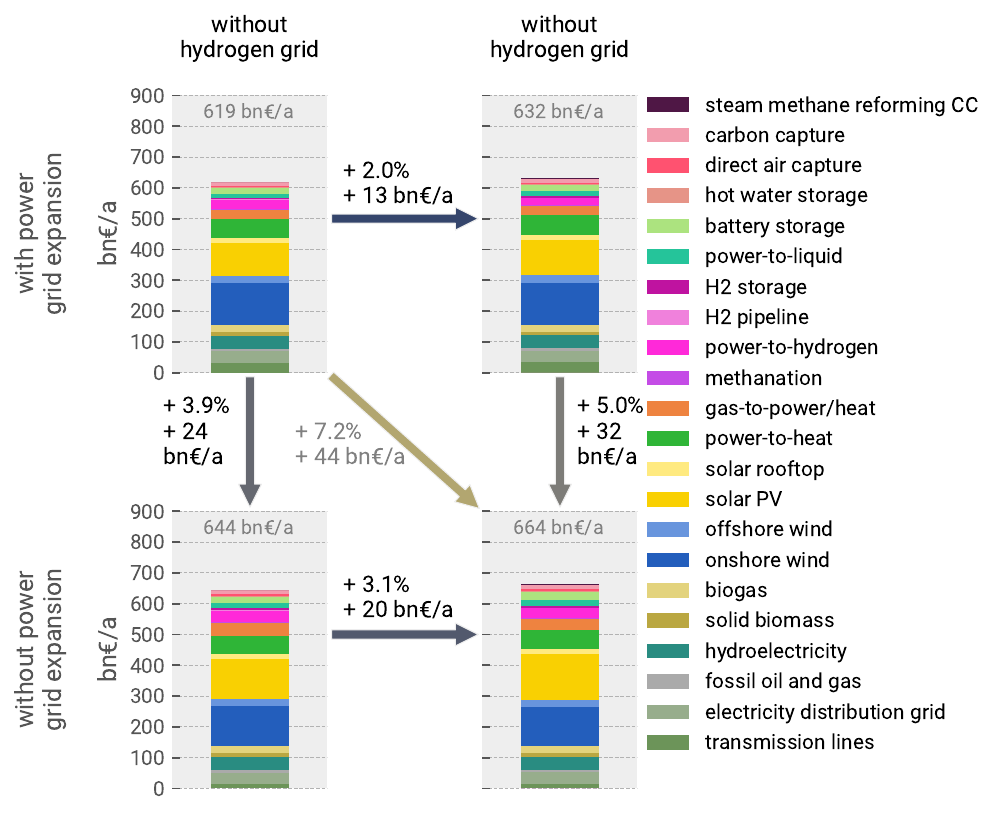}
    \caption{Cost benefits of electricity and hydrogen network infrastructure with cost projections for 2050.}
    \label{fig:sensitivity-costs}
\end{SCfigure}

\begin{figure}
    \centering
    \begin{subfigure}[t]{\textwidth}
        \centering
        \caption{differences in system cost compared to 2030 cost projections}
        \includegraphics[width=\textwidth]{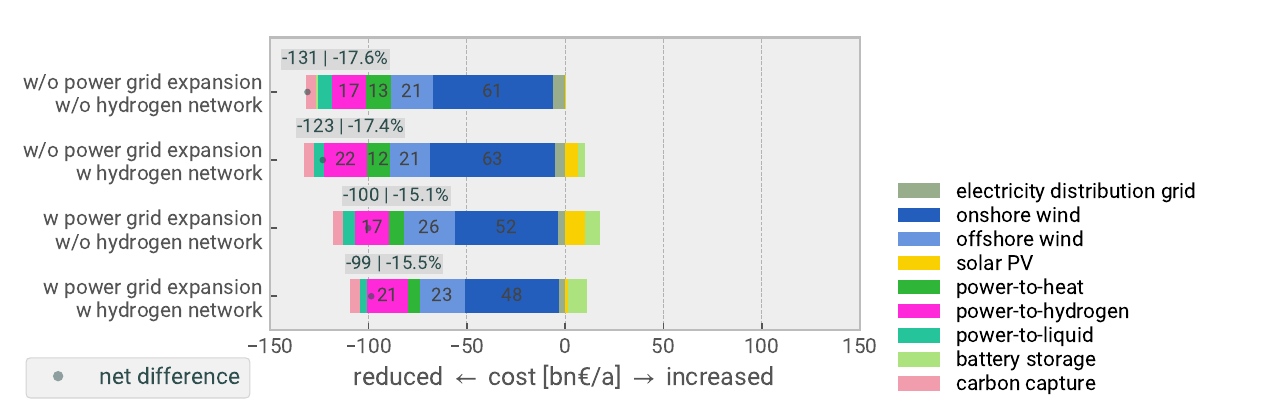}
        \label{fig:sensitivity-costs-cost}
    \end{subfigure}
    \begin{subfigure}[t]{\textwidth}
        \centering
        \caption{differences in generation and conversion capacities compared to 2030 cost projections}
        \includegraphics[width=\textwidth]{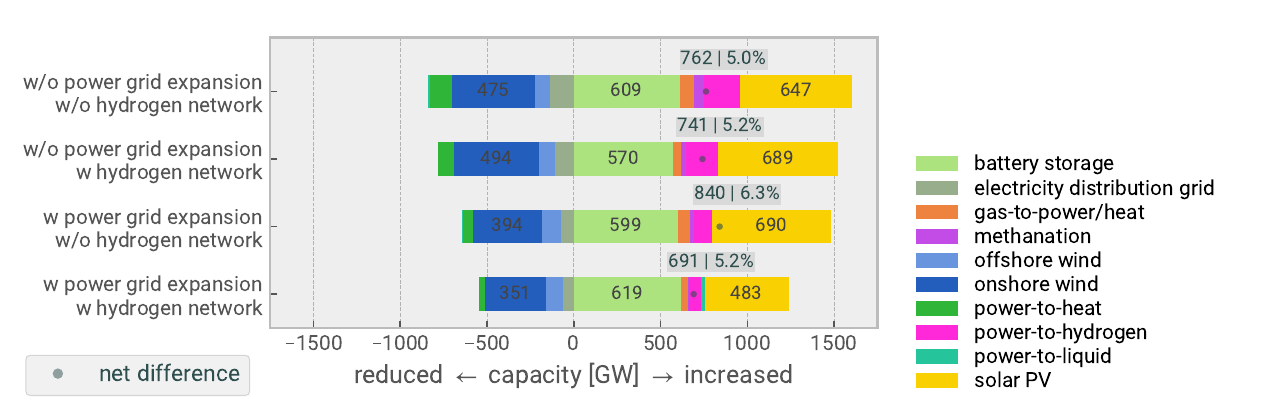}
        \label{fig:sensitivity-costs-cap}
    \end{subfigure}
    \begin{subfigure}[t]{\textwidth}
        \centering
        \caption{differences in storage capacities compared to 2030 cost projections}
        \includegraphics[width=\textwidth]{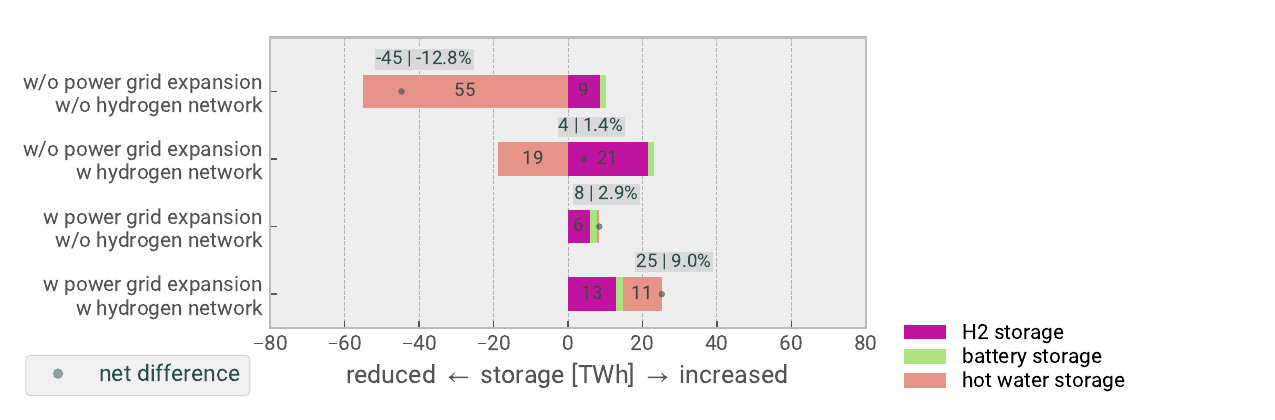}
        \label{fig:sensitivity-costs-sto}
    \end{subfigure}
    \caption{Differences in total system cost and optimised capacities for more progressive 2050 cost projections compared to more conservative 2030 cost projections.}
    \label{fig:sensitivity-costs-diff}
\end{figure}

Using more progressive assumptions diminishes the cost benefit of power grid
reinforcements from 6-8\% to 4-5\% (\cref{fig:sensitivity-costs}). However, the
cost benefit of the hydrogen network is robust against variations in cost and
technology projections, changing merely from 1.6-3.4\% to 2.0-3.1\%. With cost
projections for 2050, we see a total cost reduction between 15\% and 18\% and a
shift towards more distributed and decentral solutions
(\cref{fig:sensitivity-costs-diff}). This includes significantly more solar and
battery deployment, more electrolysers with more flexible operation supported by
additional hydrogen storage for buffering, and less wind generation and
distribution grid capacities. Owing to plummeting costs of solar photovoltaics,
\cref{fig:hydrogen-inframaps:costs} reveals a shift towards more hydrogen
production in sunny Southern Europe. This leads to more hydrogen storage and a
stronger hydrogen network buildout in this region. The consequence is a more
balanced production of solar-based hydrogen in Southern Europe and wind-based
hydrogen in the North Sea region.

\subsection{Importing all Liquid Hydrocarbons}
\label{sec:si:sensitivity-imports}

In this sensitivity analysis, we explore the cost benefit of a hydrogen network
if all liquid hydrocarbons were imported from outside of Europe. We chose this
case as it constitutes an import scenario that should reduce the
benefits of a hydrogen network. We assume uniform import costs of 115
\euro/MWh\citeS{staissOptionenFuer2022,schornMethanolRenewable2021} for
methanol and Fischer-Tropsch fuels, for a total import volume of 1573 TWh
(roughly one-third methanol and two-thirds Fischer-Tropsch fuels). By replacing the
domestic production of electrofuels with imports, 1903 TWh (80\%) of domestic
hydrogen demand fall away, which is much more compared to the 333 TWh for
domestic and foreign hydrogen supply each mentioned in the REPowerEU plans for
2030.\citeS{europeancommissionRepowerEUPlan}

\begin{SCfigure}
    \centering
    \includegraphics[width=0.75\textwidth]{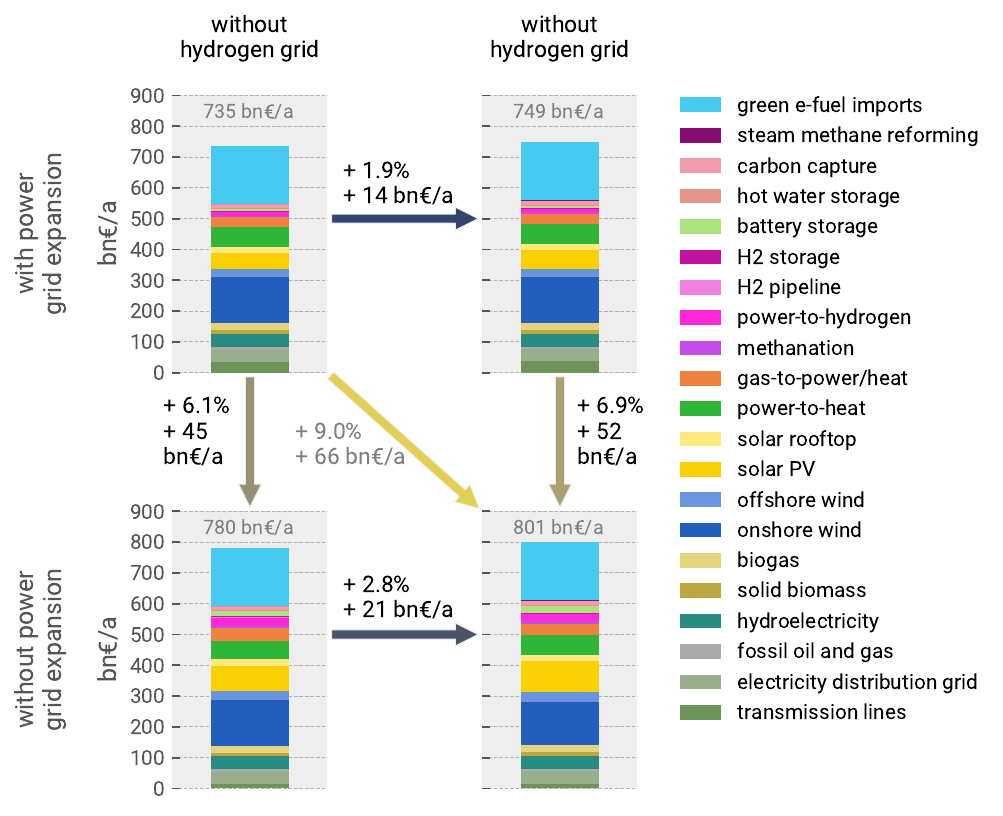}
    \caption{Cost benefits of electricity and hydrogen network infrastructure if all liquid hydrocarbons are imported.}
    \label{fig:sensitivity-imports}
\end{SCfigure}

\begin{figure}
    \centering
    \begin{subfigure}[t]{\textwidth}
        \centering
        \caption{differences in system cost compared to scenarios without imports}
        \includegraphics[width=\textwidth]{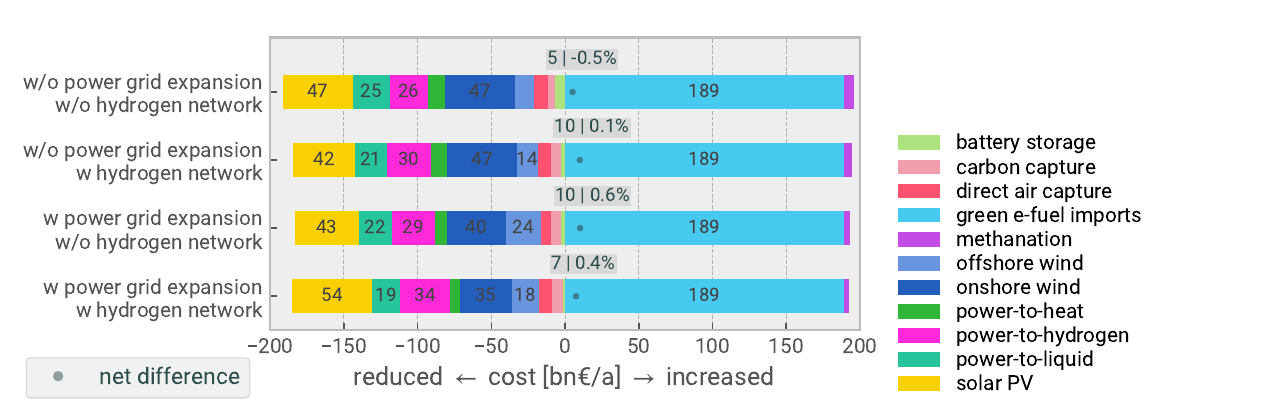}
        \label{fig:sensitivity-import-cost}
    \end{subfigure}
    \begin{subfigure}[t]{\textwidth}
        \centering
        \caption{differences in generation and conversion capacities compared to scenarios without imports}
        \includegraphics[width=\textwidth]{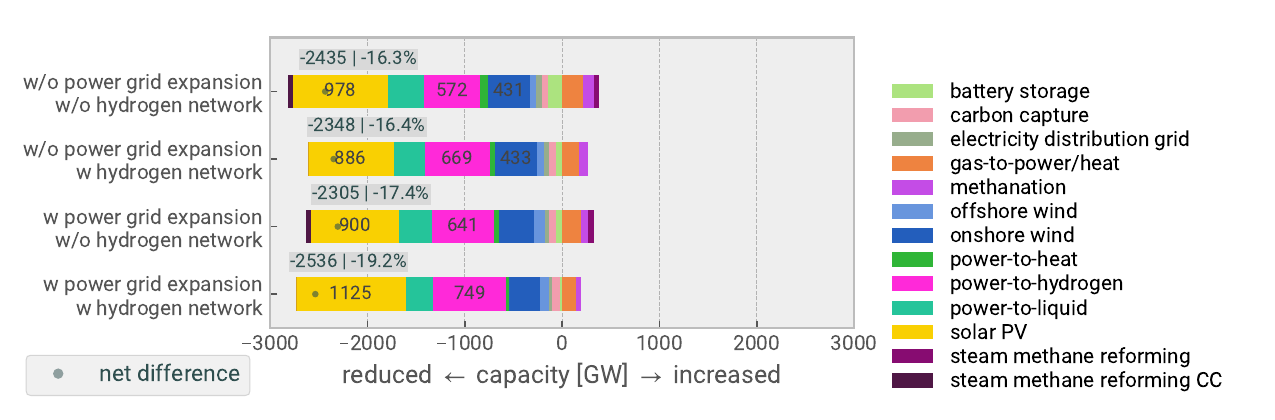}
        \label{fig:sensitivity-import-cap}
    \end{subfigure}
    \begin{subfigure}[t]{\textwidth}
        \centering
        \caption{differences in storage capacities compared to scenarios without imports}
        \includegraphics[width=\textwidth]{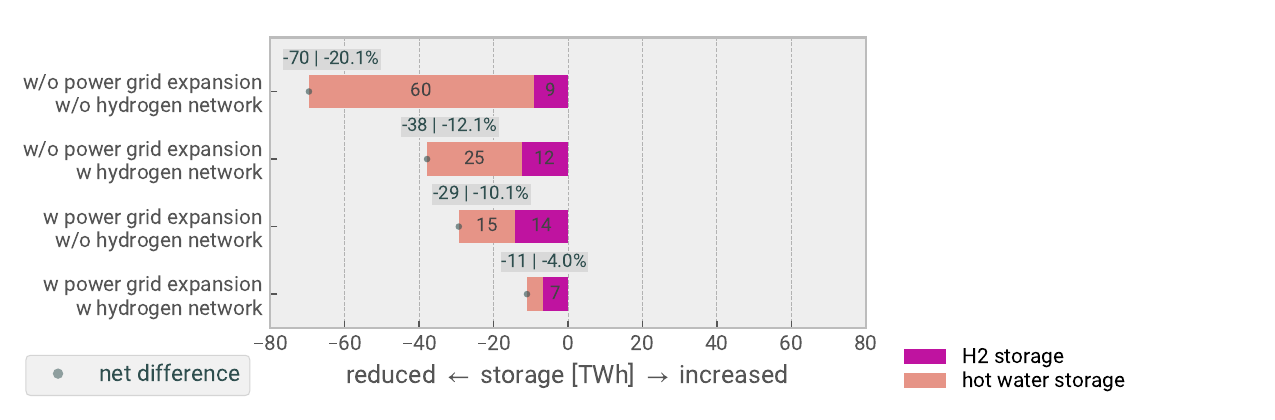}
        \label{fig:sensitivity-import-sto}
    \end{subfigure}
    \caption{Differences in total system cost and optimised capacities for scenarios with all liquid hydrocarbons imported compared to scenarios without imports.}
    \label{fig:sensitivity-import-diff}
\end{figure}

As \cref{fig:sensitivity-imports} shows, the relative cost benefits of network
expansion do not change much (from 1.6-3.4\% to 1.9-2.8\%), while the overall
benefit of network expansion is slightly reduced from 10\% to 9\%. Moreover,
there is practically no change in total system costs
(\cref{fig:sensitivity-import-diff}). The costs of 189 bn\euro/a for electrofuel
imports (23.6-25.7\% of system costs) displace almost equal costs for the
domestic supply chain for fuel synthesis comprising wind and solar electricity
generation, direct air capture, hydrogen storage and power-to-X processes
(electrolysis, methanolisation, Fischer-Topsch). Since the domestic electrofuels
can mostly use captured carbon dioxide from point-sources, whereas imported
fuels rely on direct air capture as a carbon source, the higher costs for direct
air capture cancel out the savings from utilising better renewable resources
abroad.

Regarding the spatial deployment of hydrogen infrastructure, as shown in
\cref{fig:hydrogen-inframaps:imports}, we see fewer hydrogen pipelines built
overall, reduced to a total network volume of 103-180 TWkm compared to 204-307
TWkm in scenarios without imports. However, the reduced network achieves higher
retrofitting shares of 78\%. Hydrogen production hubs in Southern Europe
disappear such that the remaining hubs are located in the broader North Sea region.

It is necessary to underline that this sensitivity analysis explores the impact
of importing the majority of hydrogen derivatives and does not analyse the
impact of direct hydrogen imports on network development. For instance, if
hydrogen were imported via pipelines from the MENA region to supply hydrogen to
domestic synthetic fuel production sites, much of the buildout of the European
hydrogen network would likely be diverted to Italy and Spain.

\subsection{Liquid Hydrogen in Shipping}
\label{sec:si:sensitivity-shipping}

In this sensitivity analysis, we examined the impact of changing the primary
fuel used in shipping from methanol to liquid hydrogen. The hydrogen demand for
international shipping was geographically distributed based on trade volumes of
international ports,\citeS{worldbankWorldBank} while the demand for domestic
shipping was distributed by population. The costs for hydrogen liquefaction were
also included in our analysis (see \cref{tab:si:costs}).

\begin{SCfigure}
    \centering
    \includegraphics[width=0.75\textwidth]{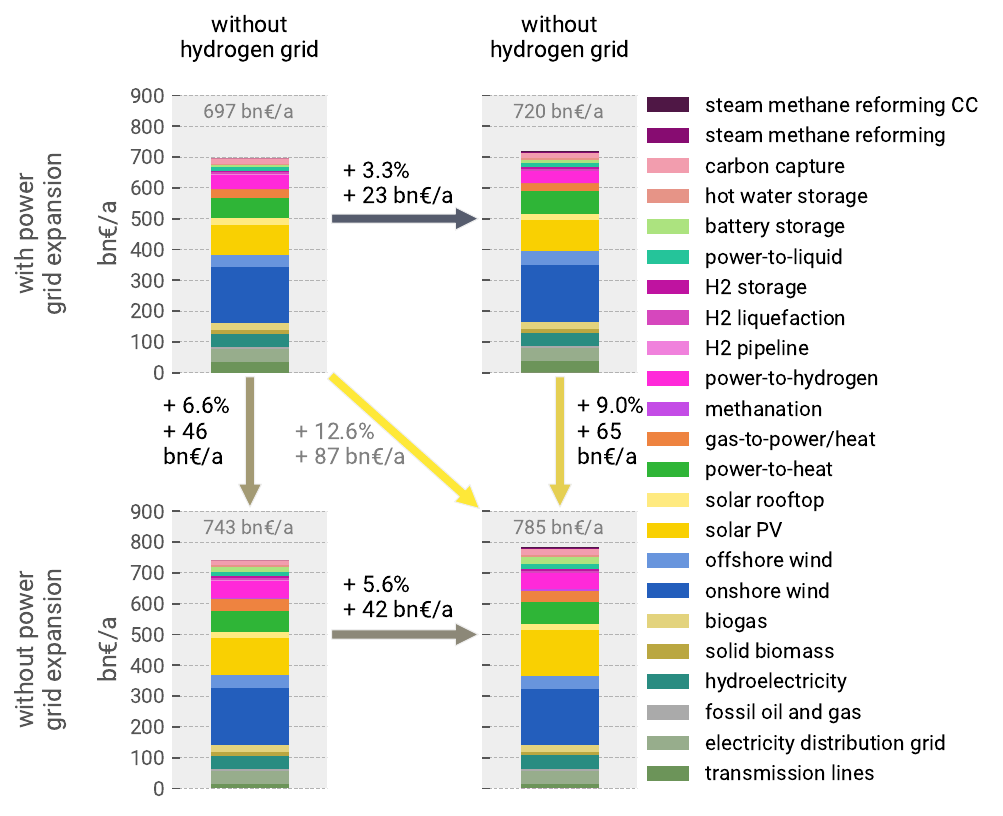}
    \caption{Cost benefits of electricity and hydrogen network infrastructure with use of liquid hydrogen in shipping instead of methanol.}
    \label{fig:sensitivity-shipping}
\end{SCfigure}

The results of this analysis are shown in \cref{fig:sensitivity-shipping} and
indicate that while the cost benefit of electricity grid reinforcements is
similar (6.6-9.0\%), the cost benefit of a hydrogen network almost doubles (from
1.6-3.4\% to 3.3-5.6\%). The overall cost benefit of network expansion rose from
9.9\% to 12.6\%. This difference can be attributed to the added need to
transport the hydrogen for the shipping sector from the most cost-effective
hydrogen production sites to the ports, whereas previously methanol offered
low-cost transport allowing for methanolisation directly where hydrogen was
produced.

\begin{figure}
    \centering
    \includegraphics[width=\textwidth]{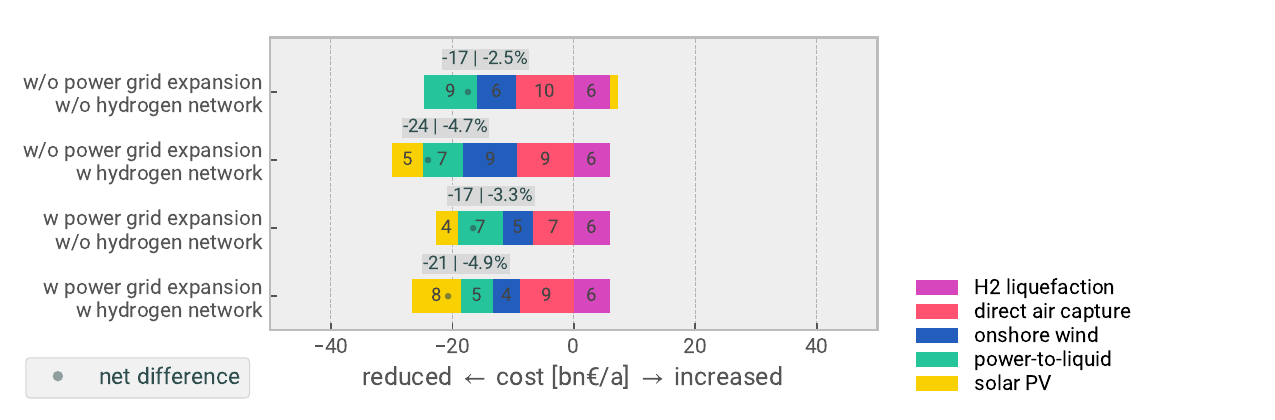}
    \caption{Differences in total system cost for usage of liquid hydrogen in shipping compared to methanol.}
    \label{fig:sensitivity-shipping-diff}
\end{figure}

\begin{figure}
    \centering
    \makebox[\textwidth][c]{
        \begin{subfigure}[t]{0.6\textwidth}
            \centering
            \caption{hydrogen infrastructure with 2030 costs, methanol in shipping, no imports}
            \includegraphics[width=\textwidth]{\hyrun/maps/elec_s_181_lv1.0__Co2L0-3H-T-H-B-I-A-solar+p3-linemaxext10-h2_network_2050.pdf}
            \label{fig:hydrogen-inframaps:base}
        \end{subfigure}
        \begin{subfigure}[t]{0.6\textwidth}
            \centering
            \caption{hydrogen infrastructure with 2050 cost assumptions}
            \includegraphics[width=\textwidth]{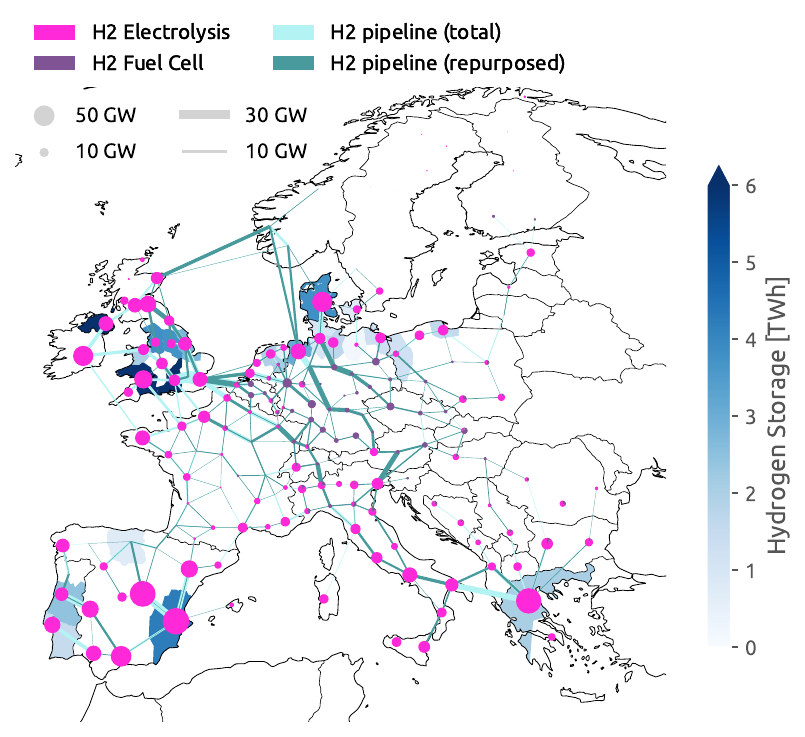}
            \label{fig:hydrogen-inframaps:costs}
        \end{subfigure}
    }
    \makebox[\textwidth][c]{
        \begin{subfigure}[t]{0.6\textwidth}
            \centering
            \caption{hydrogen infrastructure with liquid hydrogen in shipping}
            \includegraphics[width=\textwidth]{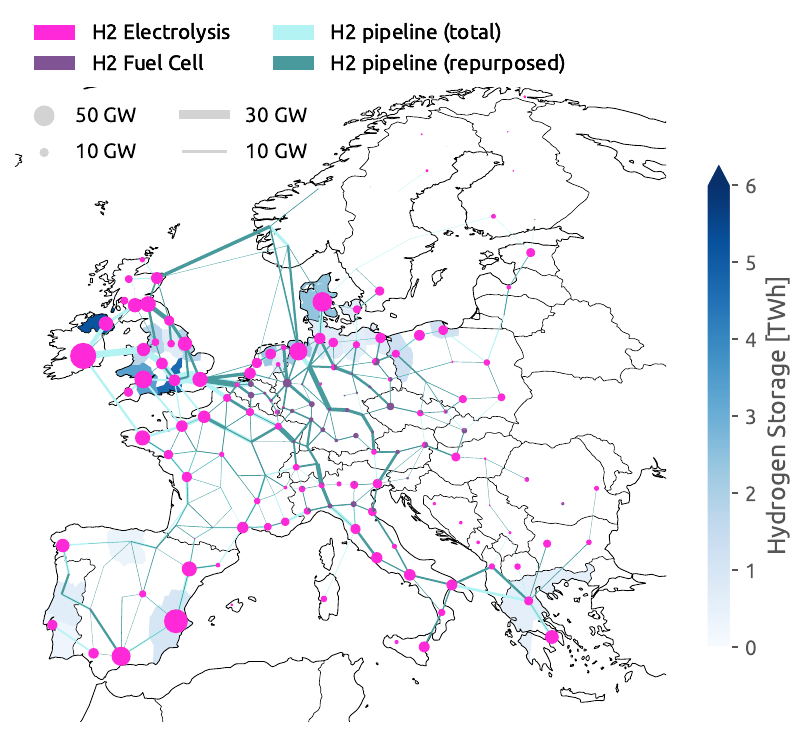}
            \label{fig:hydrogen-inframaps:lh2}
        \end{subfigure}
        \begin{subfigure}[t]{0.6\textwidth}
            \centering
            \caption{hydrogen infrastructure with all liquid hydrocarbons imported}
            \includegraphics[width=\textwidth]{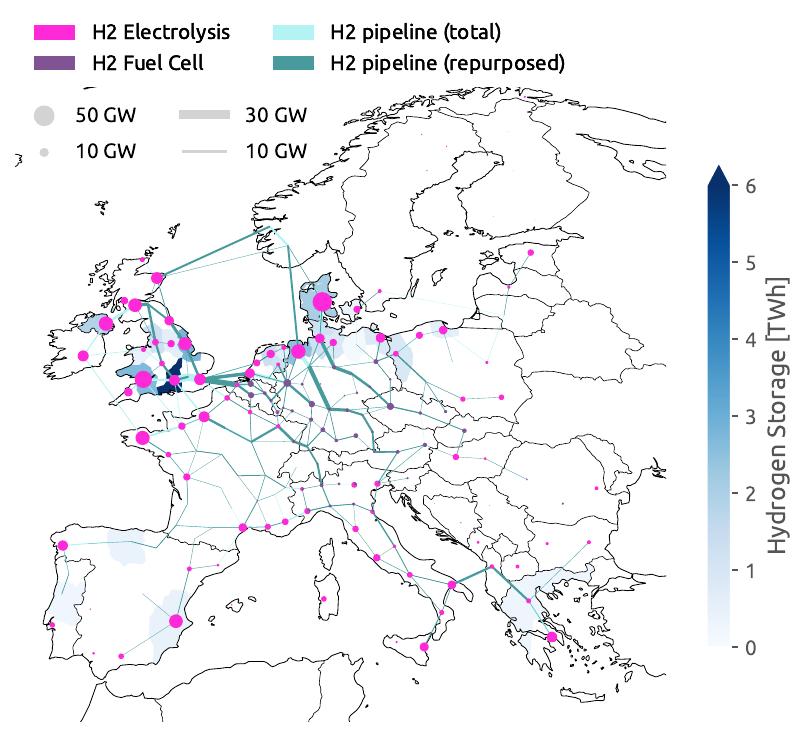}
            \label{fig:hydrogen-inframaps:imports}
        \end{subfigure}
        }
    \caption{Hydrogen infrastructure buildout with different cost assumptions (\cref{fig:hydrogen-inframaps:costs}), shipping fuel (\cref{fig:hydrogen-inframaps:lh2}) or import levels (\cref{fig:hydrogen-inframaps:imports}) in scenarios without electricity grid reinforcements.}
    \label{fig:hydrogen-inframaps}
\end{figure}

As shown in \cref{fig:sensitivity-shipping-diff}, energy system costs are
between 2.5\% and 4.9\% cheaper when exchanging methanol with liquid hydrogen in
ships. Methanolisation plants are substituted by hydrogen liquefaction plants,
and because less carbon needs to be handled in the system, direct air capture is
no longer required. The higher energy efficiency of liquid hydrogen in shipping
also lowers the requirements for wind and solar buildout. However, the cost
differences should be viewed in the context that the costs of ships fueled by
liquid hydrogen are likely to be considerably higher than those fueled by
methanol.\citeS{johnstonShippingSunshine2022} The spatial patterns of hydrogen
infrastructure buildout remain largely unchanged
(\cref{fig:hydrogen-inframaps:lh2}).

\subsection{Temporal Resolution}
\label{sec:si:sensitivity-time}

\begin{figure}
    \centering
    \begin{subfigure}[t]{\textwidth}
        \centering
        \caption{differences in total energy system cost compared to hourly resolution}
        \includegraphics[width=.9\textwidth]{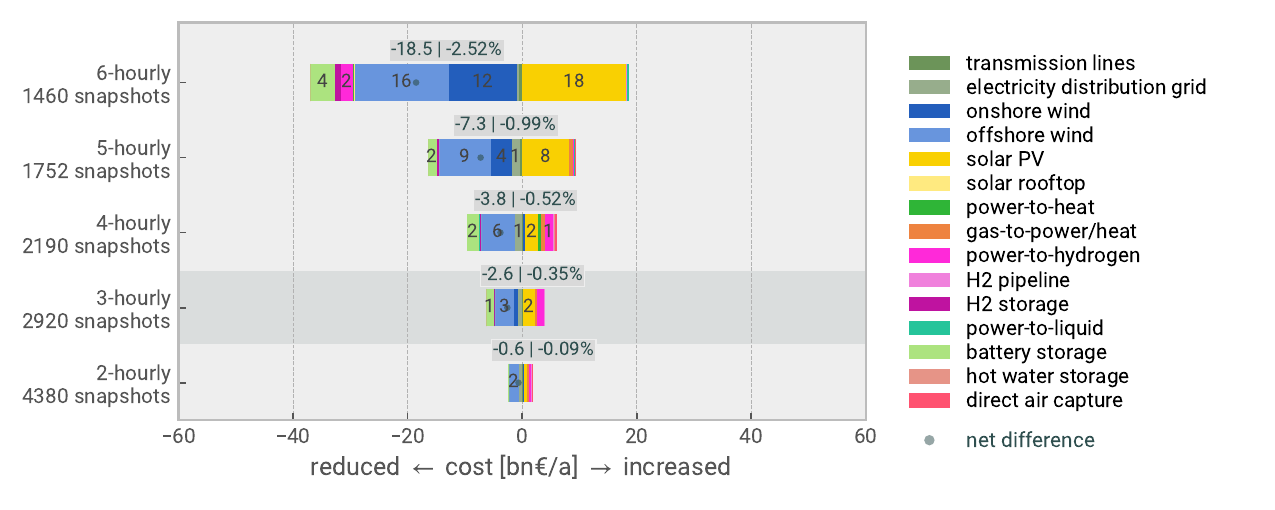}
        \label{fig:sensitivity-time-cost}
    \end{subfigure}
    \begin{subfigure}[t]{\textwidth}
        \centering
        \caption{differences in generation and conversion capacities compared to hourly resolution}
        \includegraphics[width=.9\textwidth]{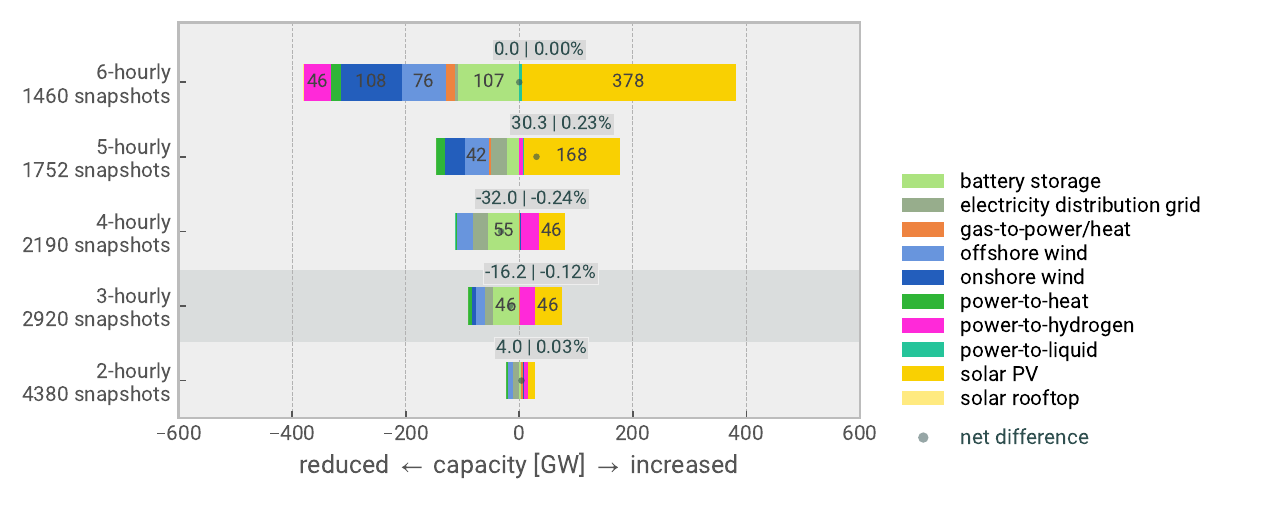}
        \label{fig:sensitivity-time-cap}
    \end{subfigure}
    \caption{ Total system costs and optimised capacities for varying temporal
    resolutions relative to hourly resolution. The comparison refers to scenario
    with both power grid reinforcements and hydrogen network expansion. }
    \label{fig:sensitivity-time}
\end{figure}

In \cref{fig:sensitivity-time}, we varied the temporal resolution for the
scenario with both power and hydrogen network expansion with a reduced spatial
resolution of 90 regions. In this way, we were computationally able to sweep the
time resolution from a 6-hourly model up to an hourly model. Total energy system
costs and optimised capacities are shown relative to the hourly model.

Overall, with a system cost difference of -0.35\% the error induced by
resampling the model from an hourly to a 3-hourly resolution is small and
justifies a model size reduction by factor 3. The temporal aggregation causes a
minor underestimation of short-term battery storage and offshore wind as well as
a minor overestimation of solar photovoltaics and hydrogen storage. This trend
intensifies with coarser temporal resolution, such that with a 6-hourly
resolution the system cost deviation exceeds 2.5\% since balancing needs for
solar electricity are discounted, which makes this technology more attractive.

\subsection{Spatial Resolution}
\label{sec:si:sensitivity-space}

\begin{figure}
    \centering
    \begin{subfigure}[t]{\textwidth}
        \centering
        \caption{differences in system cost compared to 181-region model}
        \includegraphics[width=.9\textwidth]{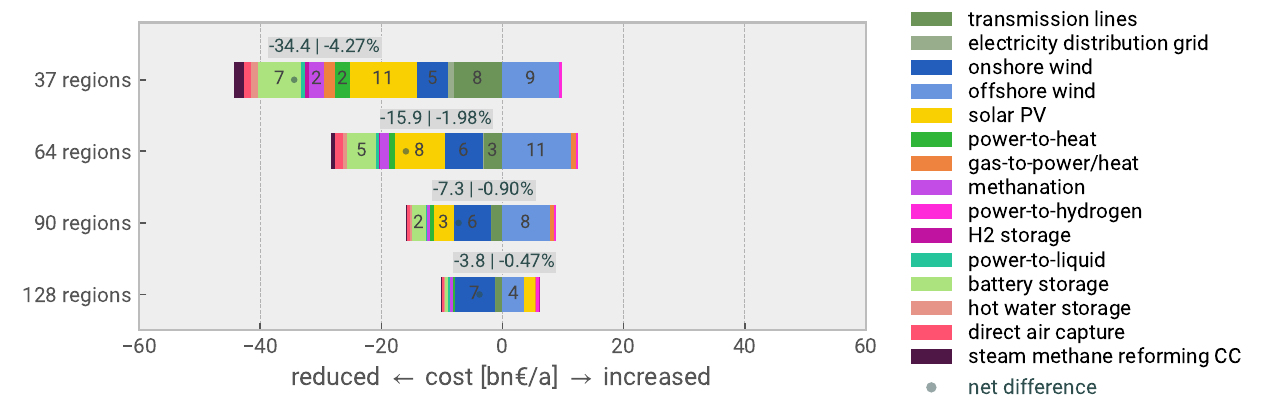}
        \label{fig:sensitivity-space-cost}
    \end{subfigure}
    \begin{subfigure}[t]{\textwidth}
        \centering
        \caption{differences in generation and conversion capacities compared to 181-region model}
        \includegraphics[width=.9\textwidth]{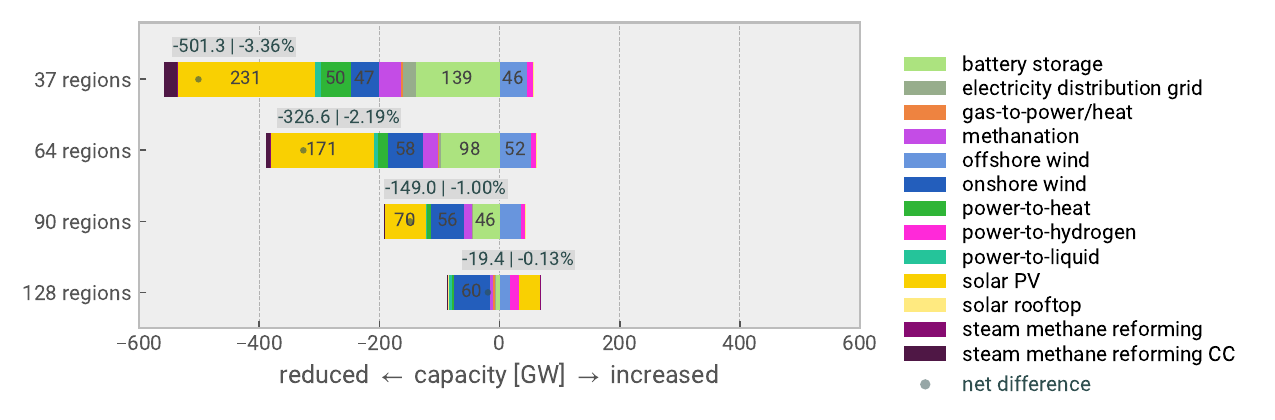}
        \label{fig:sensitivity-space-cap}
    \end{subfigure}
    \caption{ Total system costs and optimised capacities for varying spatial
    resolutions relative to 181-regions model. The case with 37 regions corresponds to
    a single node per country and synchronous zone. The comparison refers to the scenario
    with neither power grid reinforcements nor hydrogen network expansion. }
    \label{fig:sensitivity-space}
\end{figure}

In \cref{fig:sensitivity-space}, we also varied the spatial resolution of the
model from a one-node-per-country version (37 regions) to 181 regions for the
scenario without power or hydrogen network expansion and 3-hourly resolution.
Total energy system costs and optimised capacities are shown relative to the
181-region model.

Compared to the model with 181 regions, a one-node-per-country resolution
underestimates system cost by 4.3\%, favouring remote offshore wind over more
localised production with solar photovoltaics and batteries. The differences can
be explained by a combination of two opposing
effects.\citeS{frysztackiStrongEffect2021} The aggregation of the transmission
networks lifts bottlenecks within clustered regions, lowering system costs. On
the other hand, the aggregation of wind and solar capacity factors blur the most
productive sites, increasing costs. In terms of system costs, the error induced
by reducing the spatial resolution from 181 regions to 128 regions (-0.47\%) is
comparable to the error caused by choosing 3-hourly over hourly time resolution
(-0.35\%).

\section{Supplementary Results for Network Expansion Scenarios}
\label{sec:si:results-network-expansion}

\begin{figure}
    \begin{subfigure}{0.49\textwidth}
        \centering
        \caption{With power grid expansion, with hydrogen network}
        \includegraphics[width=\textwidth]{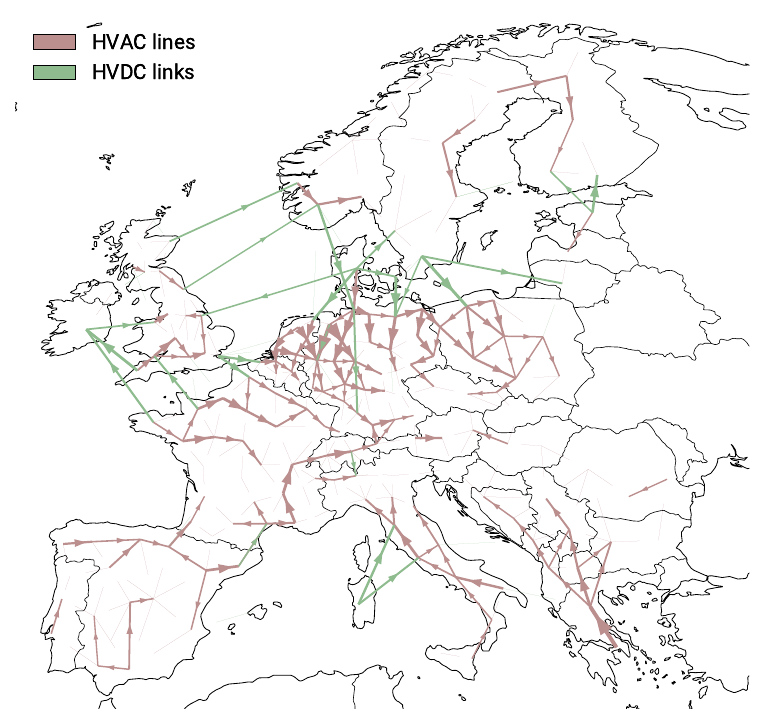}
    \end{subfigure}
    \begin{subfigure}{0.49\textwidth}
        \centering
        \caption{With power grid expansion, without hydrogen network}
        \includegraphics[width=\textwidth]{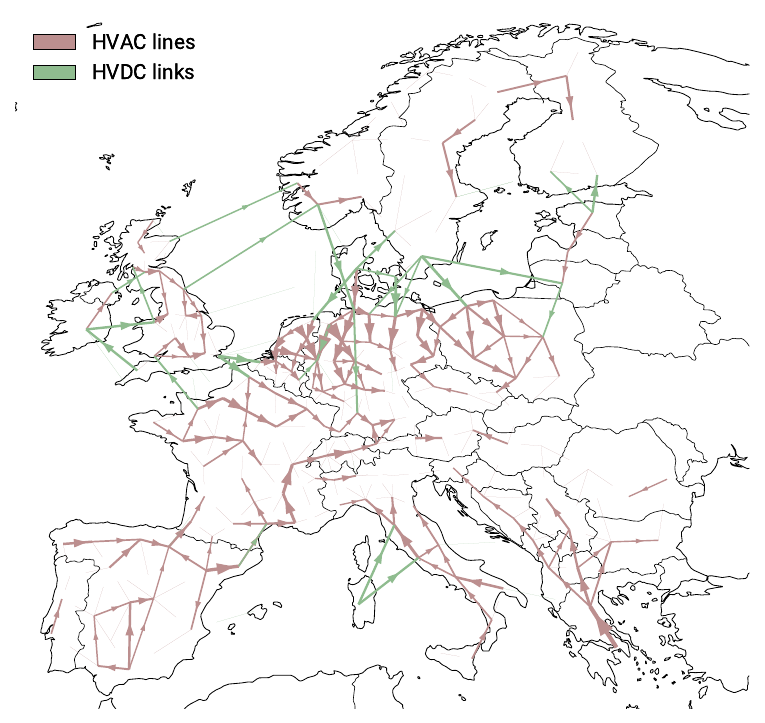}
    \end{subfigure}
    \begin{subfigure}{0.49\textwidth}
        \centering
        \caption{Without power grid expansion, with hydrogen network}
        \includegraphics[width=\textwidth]{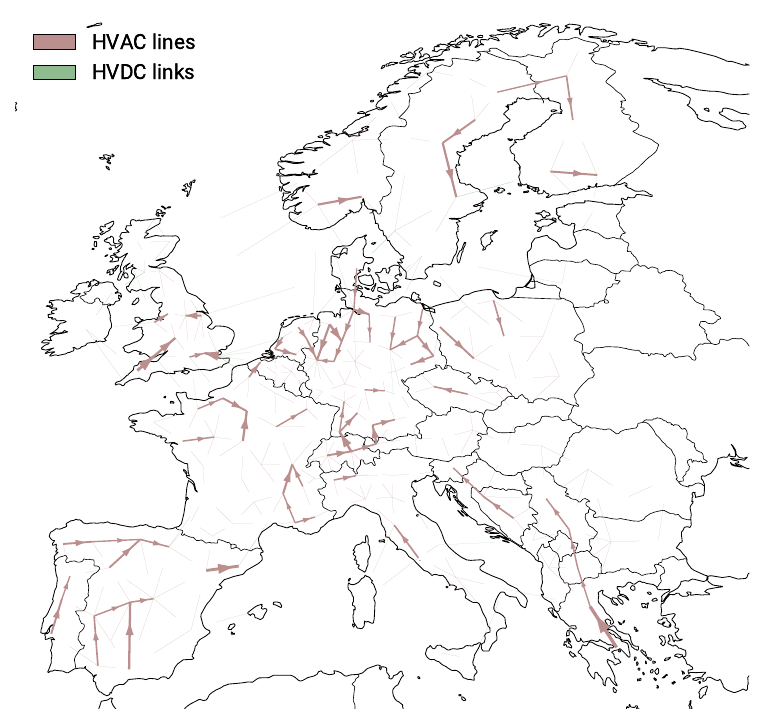}
    \end{subfigure}
    \begin{subfigure}{0.49\textwidth}
        \centering
        \caption{Without power grid expansion, without hydrogen network}
        \includegraphics[width=\textwidth]{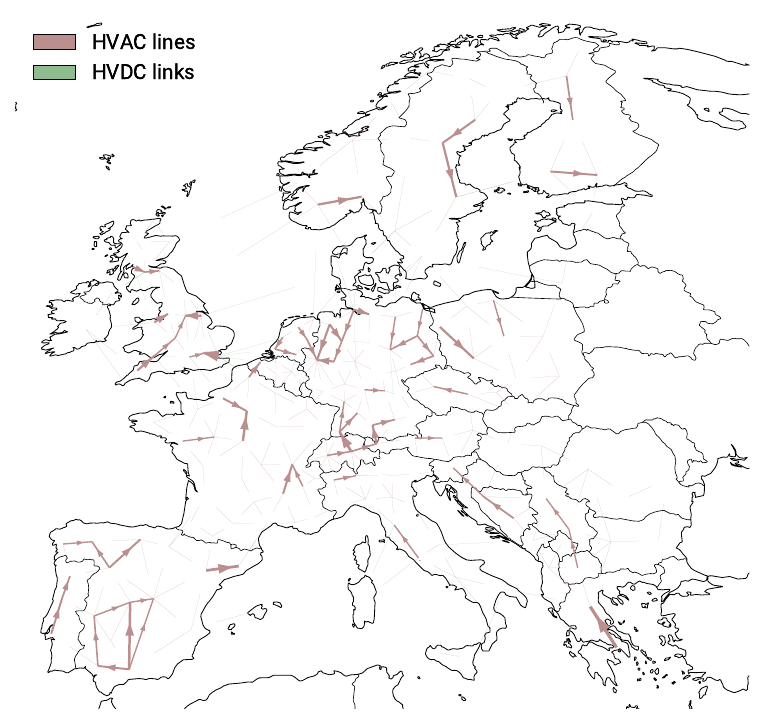}
    \end{subfigure}
    \caption{Net flow of electricity in the network. The maps shows net flows larger than 10 TWh with arrow sizes proportional to net flow volume. Only power grid expansion enables bulk energy transport in form of electricity. With the existing transmission network, net flows are limited and the transmission infrastructure is rather used for synoptic balancing as weather systems pass the continent.}
    \label{fig:si:flow-ac}
\end{figure}

\begin{figure}
    \begin{subfigure}{0.49\textwidth}
        \centering
        \caption{With power grid expansion, with hydrogen network}
        \includegraphics[width=\textwidth]{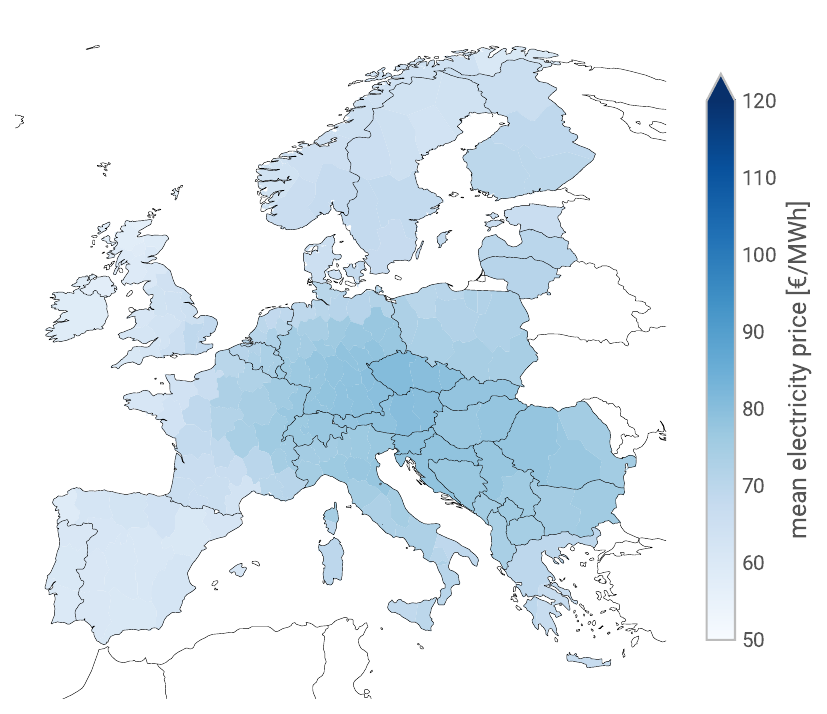}
    \end{subfigure}
    \begin{subfigure}{0.49\textwidth}
        \centering
        \caption{With power grid expansion, without hydrogen network}
        \includegraphics[width=\textwidth]{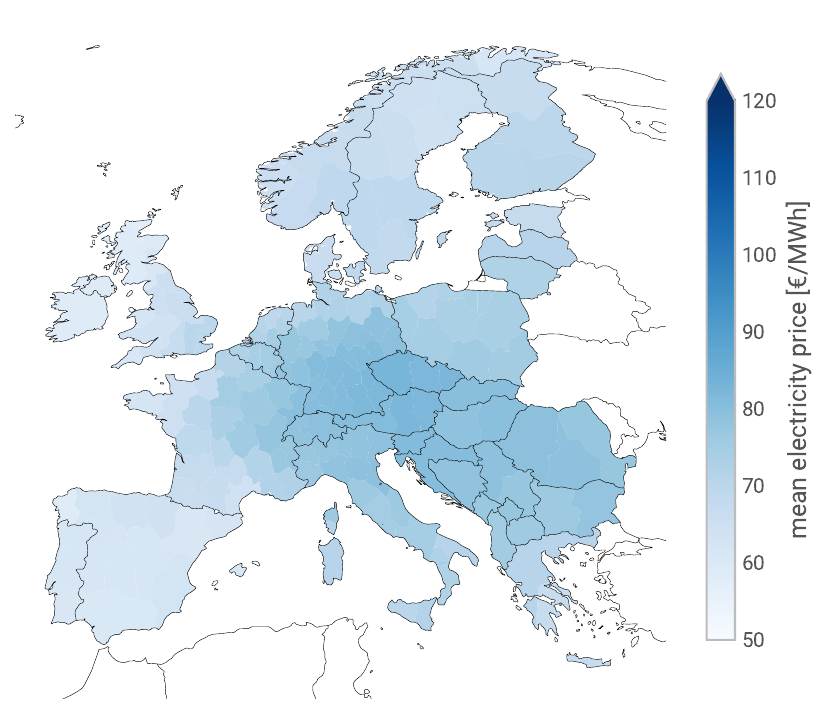}
    \end{subfigure}
    \begin{subfigure}{0.49\textwidth}
        \centering
        \caption{Without power grid expansion, with hydrogen network}
        \includegraphics[width=\textwidth]{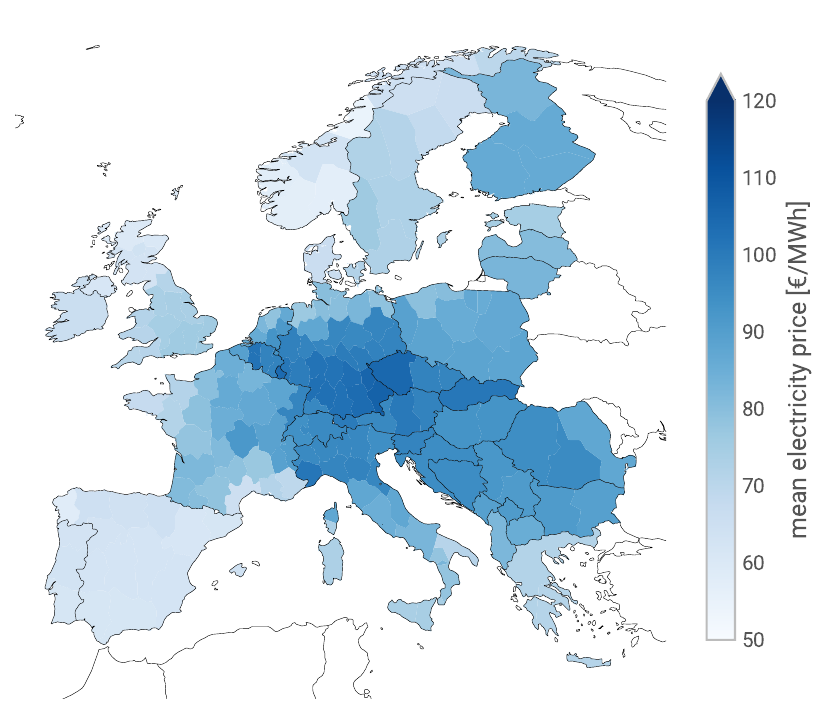}
    \end{subfigure}
    \begin{subfigure}{0.49\textwidth}
        \centering
        \caption{Without power grid expansion, without hydrogen network}
        \includegraphics[width=\textwidth]{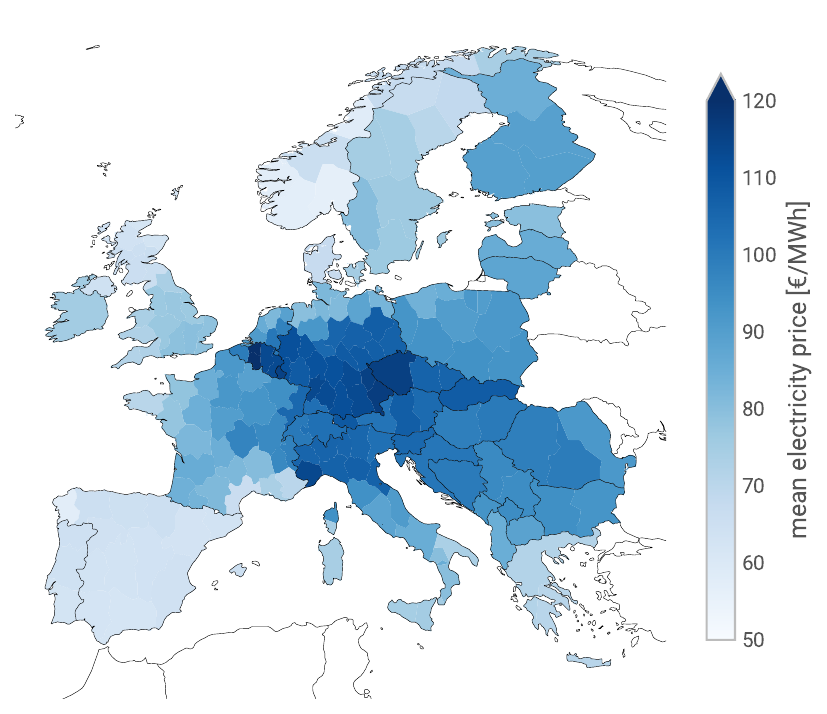}
    \end{subfigure}
    \caption{Regional distribution of average nodal electricity prices. The reinforcement of the electricity grid mitigates regional price differences. Some price differences persist because of expansion constraints on individual lines.}
    \label{fig:si:lmp-ac}
\end{figure}

\begin{figure}
    \begin{subfigure}{0.49\textwidth}
        \centering
        \caption{With power grid expansion, with hydrogen network}
        \includegraphics[width=\textwidth]{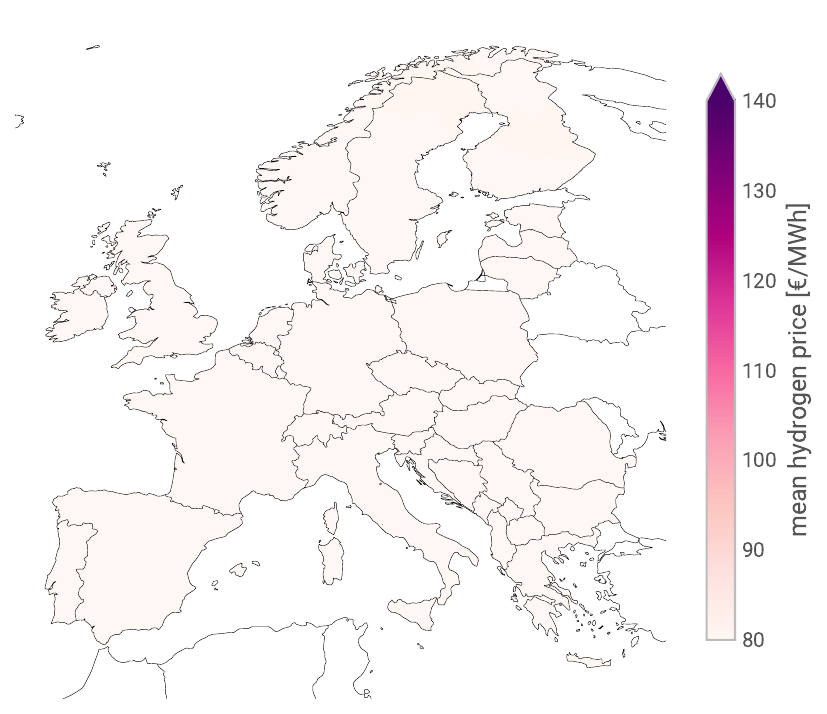}
    \end{subfigure}
    \begin{subfigure}{0.49\textwidth}
        \centering
        \caption{With power grid expansion, without hydrogen network}
        \includegraphics[width=\textwidth]{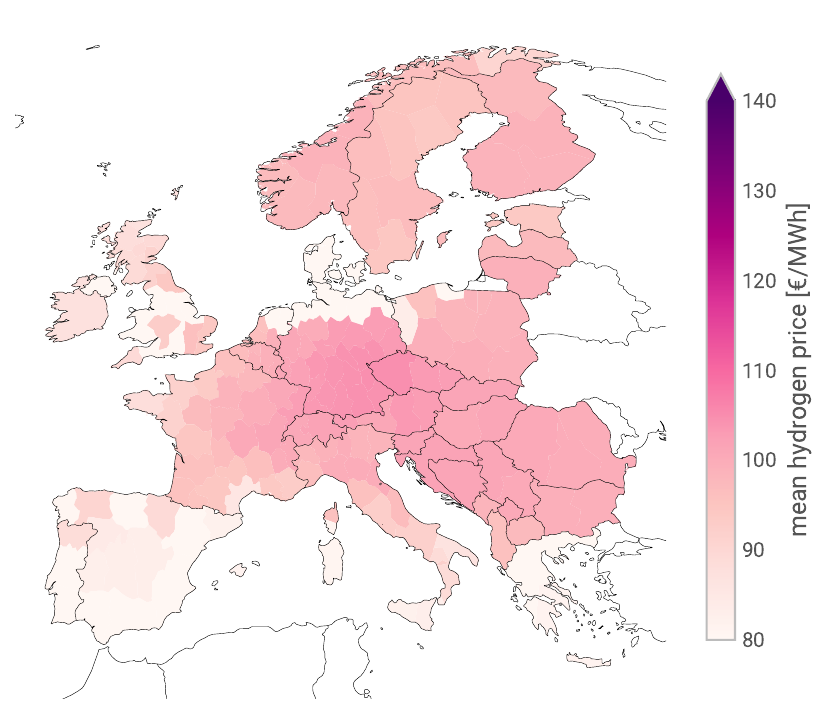}
    \end{subfigure}
    \begin{subfigure}{0.49\textwidth}
        \centering
        \caption{Without power grid expansion, with hydrogen network}
        \includegraphics[width=\textwidth]{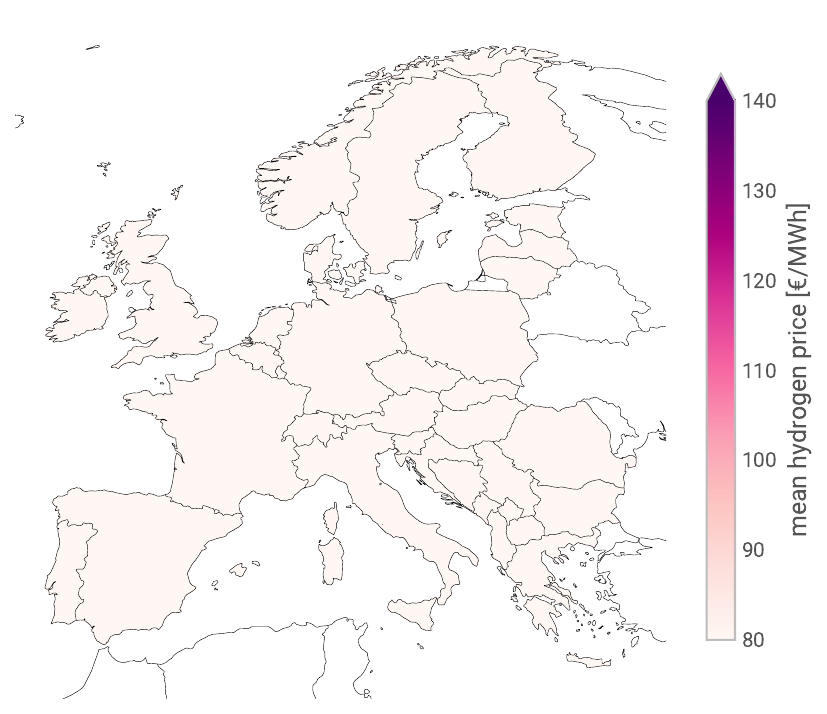}
    \end{subfigure}
    \begin{subfigure}{0.49\textwidth}
        \centering
        \caption{Without power grid expansion, without hydrogen network}
        \includegraphics[width=\textwidth]{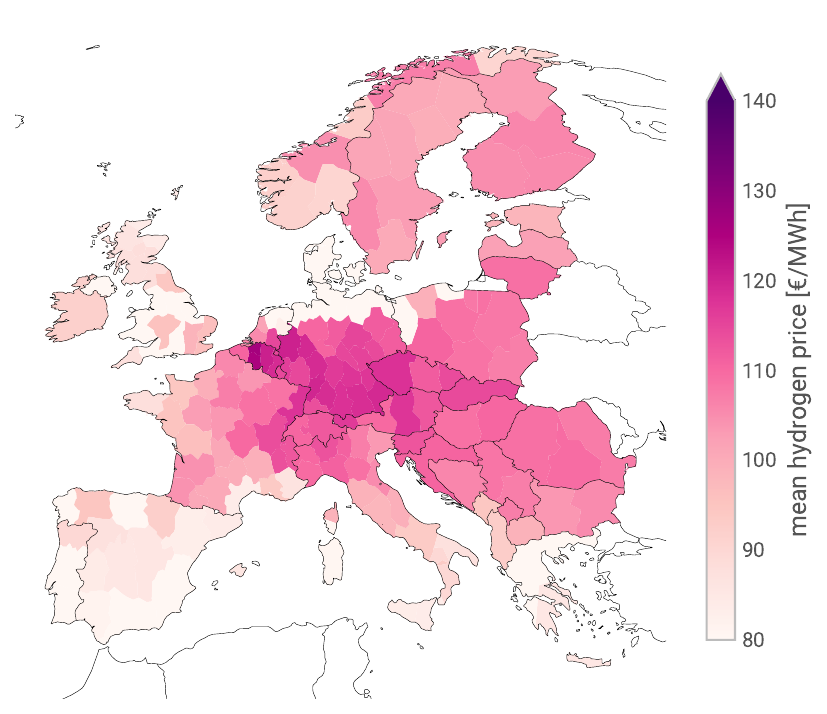}
    \end{subfigure}
    \caption{Regional distribution of average nodal hydrogen prices. The
    development of a hydrogen network evens out regional price differences. With
    limited hydrogen network expansion prices are almost twice as high in
    Europe's industrial clusters than the most cost-effective hydrogen
    production sites.}
    \label{fig:si:lmp-h2}
\end{figure}

\begin{figure}
    \begin{subfigure}{0.49\textwidth}
        \centering
        \caption{With power grid expansion, with hydrogen network}
        \includegraphics[width=\textwidth]{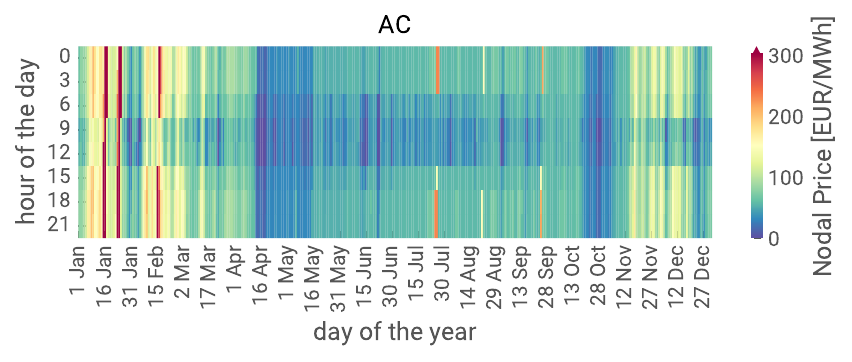}
    \end{subfigure}
    \begin{subfigure}{0.49\textwidth}
        \centering
        \caption{With power grid expansion, without hydrogen network}
        \includegraphics[width=\textwidth]{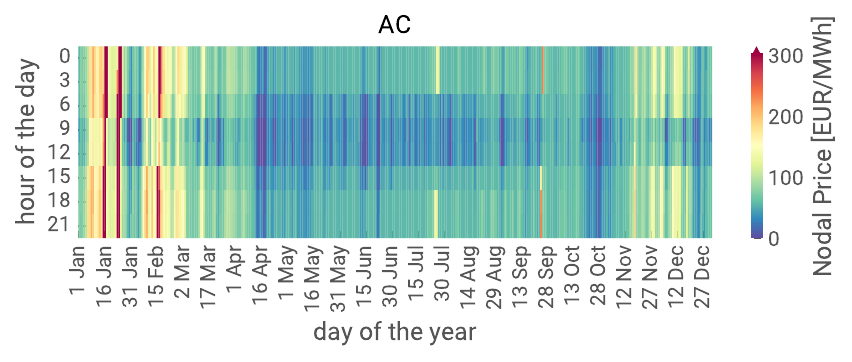}
    \end{subfigure}
    \begin{subfigure}{0.49\textwidth}
        \centering
        \caption{Without power grid expansion, with hydrogen network}
        \includegraphics[width=\textwidth]{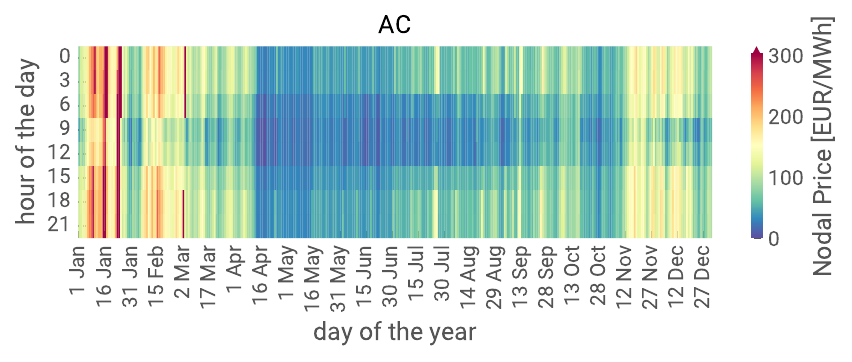}
    \end{subfigure}
    \begin{subfigure}{0.49\textwidth}
        \centering
        \caption{Without power grid expansion, without hydrogen network}
        \includegraphics[width=\textwidth]{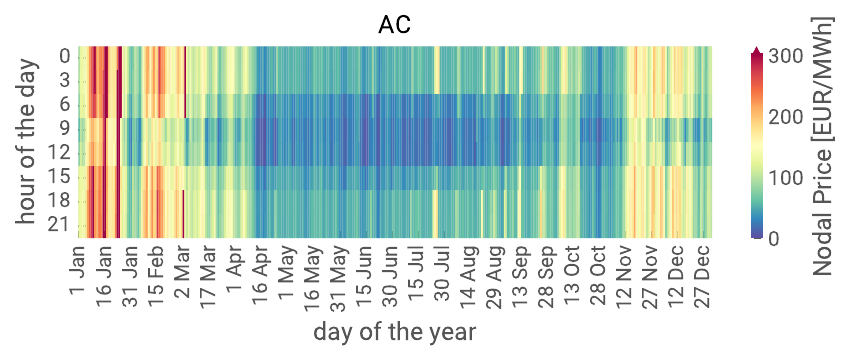}
    \end{subfigure}
    \caption{Temporal distribution of average nodal electricity prices. The graphs show daily patterns with price troughs during the day, especially in summer, as well as seasonal patterns with higher prices in winter than in the summer. A few periods in January and February are particularly challenging to the system, resulting in very high electricity prices.}
    \label{fig:si:lmp-ts-ac}
\end{figure}

\begin{figure}
    \begin{subfigure}{0.49\textwidth}
        \centering
        \caption{With power grid expansion, with hydrogen network}
        \includegraphics[width=\textwidth]{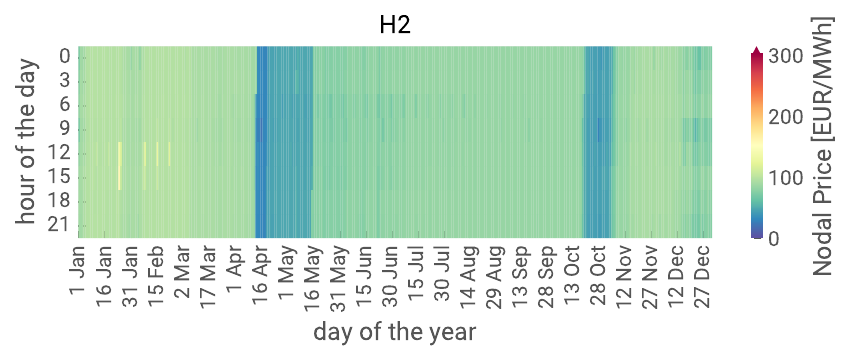}
    \end{subfigure}
    \begin{subfigure}{0.49\textwidth}
        \centering
        \caption{With power grid expansion, without hydrogen network}
        \includegraphics[width=\textwidth]{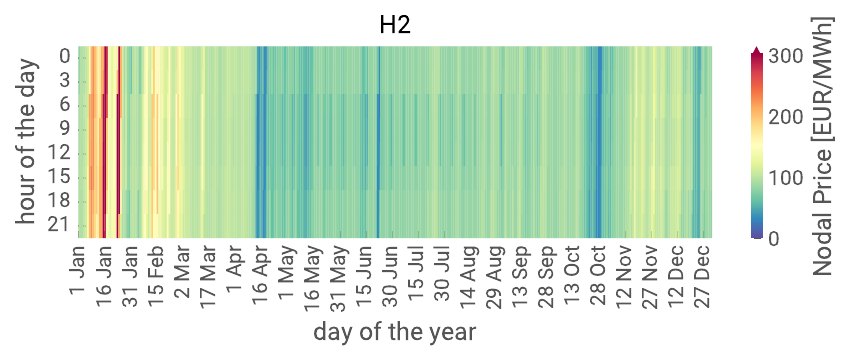}
    \end{subfigure}
    \begin{subfigure}{0.49\textwidth}
        \centering
        \caption{Without power grid expansion, with hydrogen network}
        \includegraphics[width=\textwidth]{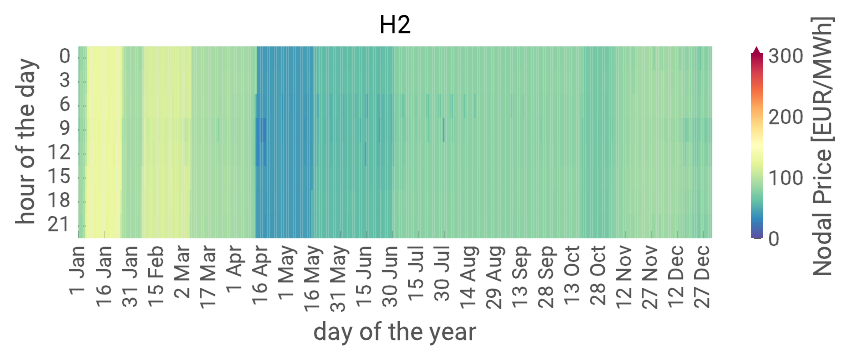}
    \end{subfigure}
    \begin{subfigure}{0.49\textwidth}
        \centering
        \caption{Without power grid expansion, without hydrogen network}
        \includegraphics[width=\textwidth]{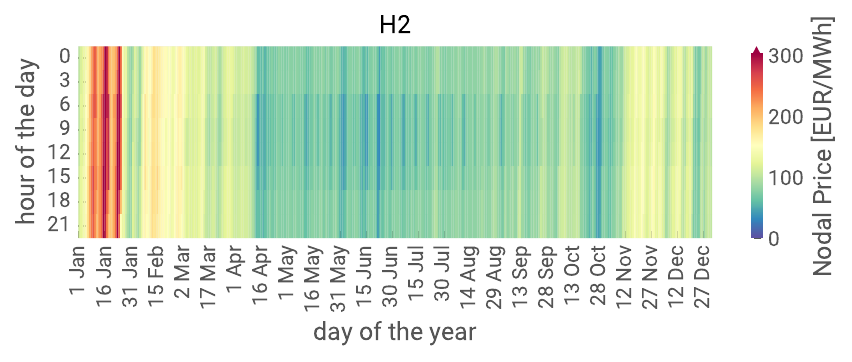}
    \end{subfigure}
    \caption{Temporal distribution of average nodal hydrogen prices. Compared to
    electricity prices, the seasonal component dominates daily patterns. Price
    spikes occur with limited hydrogen network expansion in winter periods that
    are challenging to the system.}
    \label{fig:si:lmp-ts-h2}
\end{figure}

\begin{figure}
    \begin{subfigure}{0.49\textwidth}
        \centering
        \caption{With power grid expansion, with hydrogen network}
        \includegraphics[width=.8\textwidth]{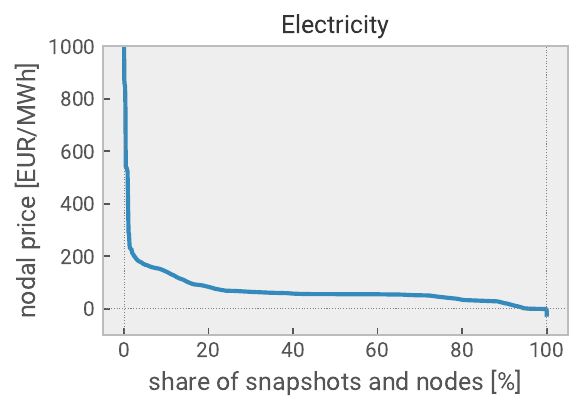}
    \end{subfigure}
    \begin{subfigure}{0.49\textwidth}
        \centering
        \caption{With power grid expansion, without hydrogen network}
        \includegraphics[width=.8\textwidth]{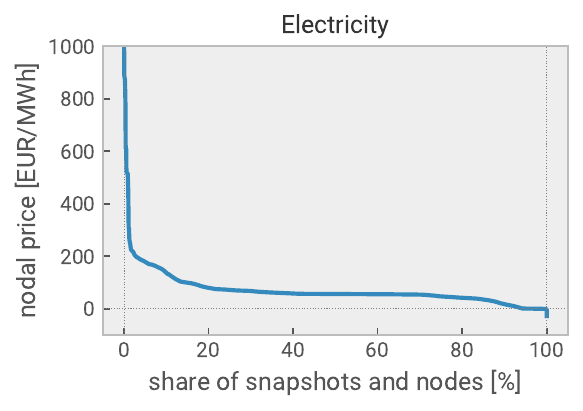}
    \end{subfigure}
    \begin{subfigure}{0.49\textwidth}
        \centering
        \includegraphics[width=.8\textwidth]{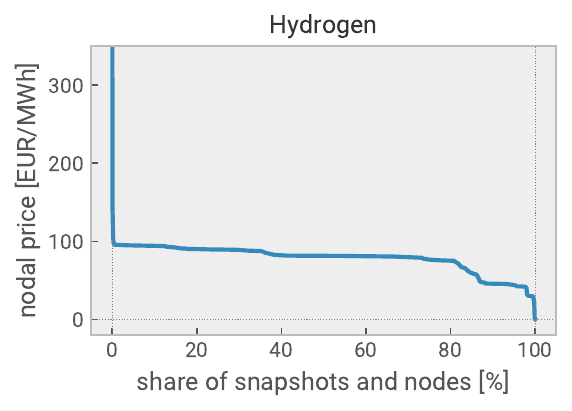}
    \end{subfigure}
    \begin{subfigure}{0.49\textwidth}
        \centering
        \includegraphics[width=.8\textwidth]{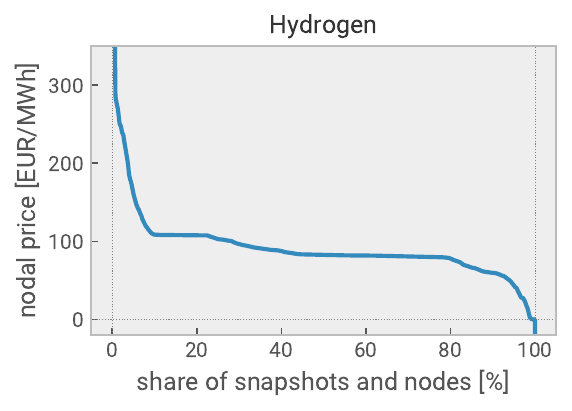}
    \end{subfigure}
    \begin{subfigure}{0.49\textwidth}
        \centering
        \caption{Without power grid expansion, with hydrogen network}
        \includegraphics[width=.8\textwidth]{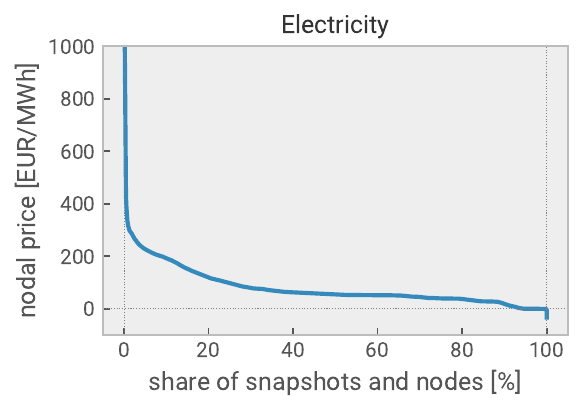}
    \end{subfigure}
    \begin{subfigure}{0.49\textwidth}
        \centering
        \caption{Without power grid expansion, without hydrogen network}
        \includegraphics[width=.8\textwidth]{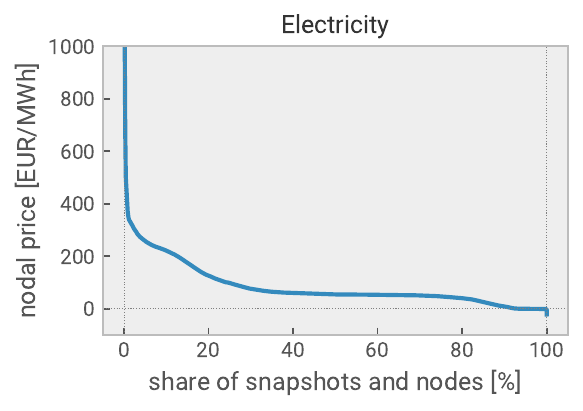}
    \end{subfigure}
    \begin{subfigure}{0.49\textwidth}
        \centering
        \includegraphics[width=.8\textwidth]{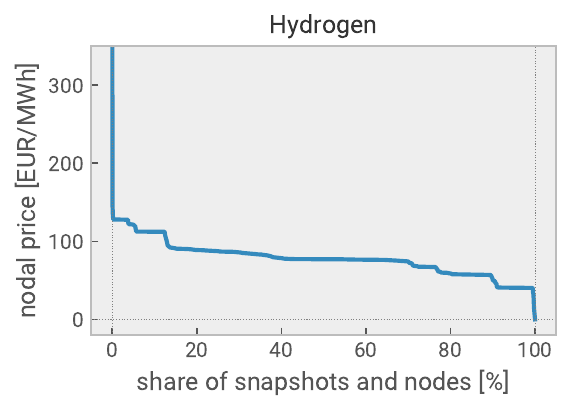}
    \end{subfigure}
    \begin{subfigure}{0.49\textwidth}
        \centering
        \includegraphics[width=.8\textwidth]{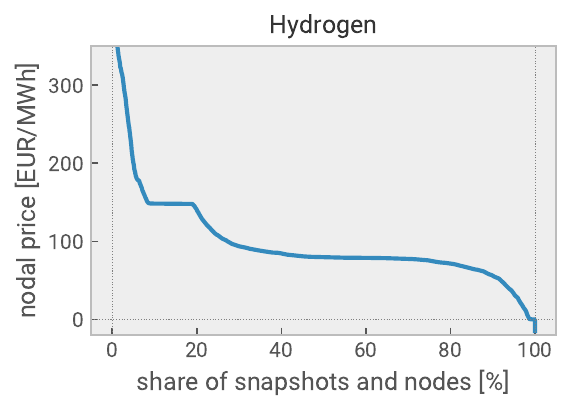}
    \end{subfigure}
    \caption{Duration curve of nodal electricity and hydrogen prices.}
    \label{fig:si:lmp-dc}
\end{figure}

\begin{figure}
    \makebox[\textwidth][c]{
    \begin{subfigure}{0.6\textwidth}
        \centering
        \caption{With power grid expansion, with hydrogen network}
        \includegraphics[trim=1.5cm 2cm 1.5cm 2.5cm, clip, width=\textwidth]{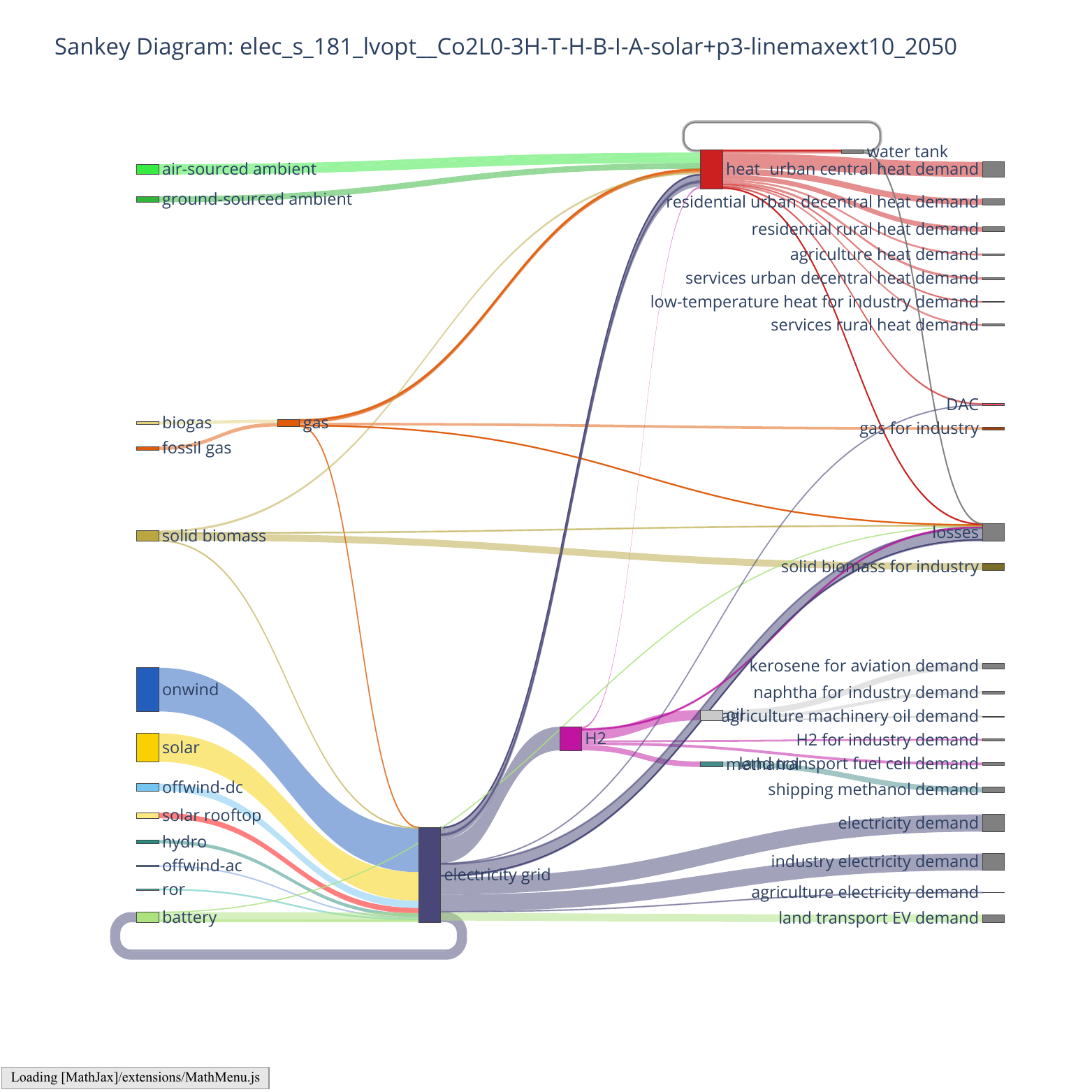}
    \end{subfigure}
    \begin{subfigure}{0.6\textwidth}
        \centering
        \caption{With power grid expansion, without hydrogen network}
        \includegraphics[trim=1.5cm 2cm 1.5cm 2.5cm, clip, width=\textwidth]{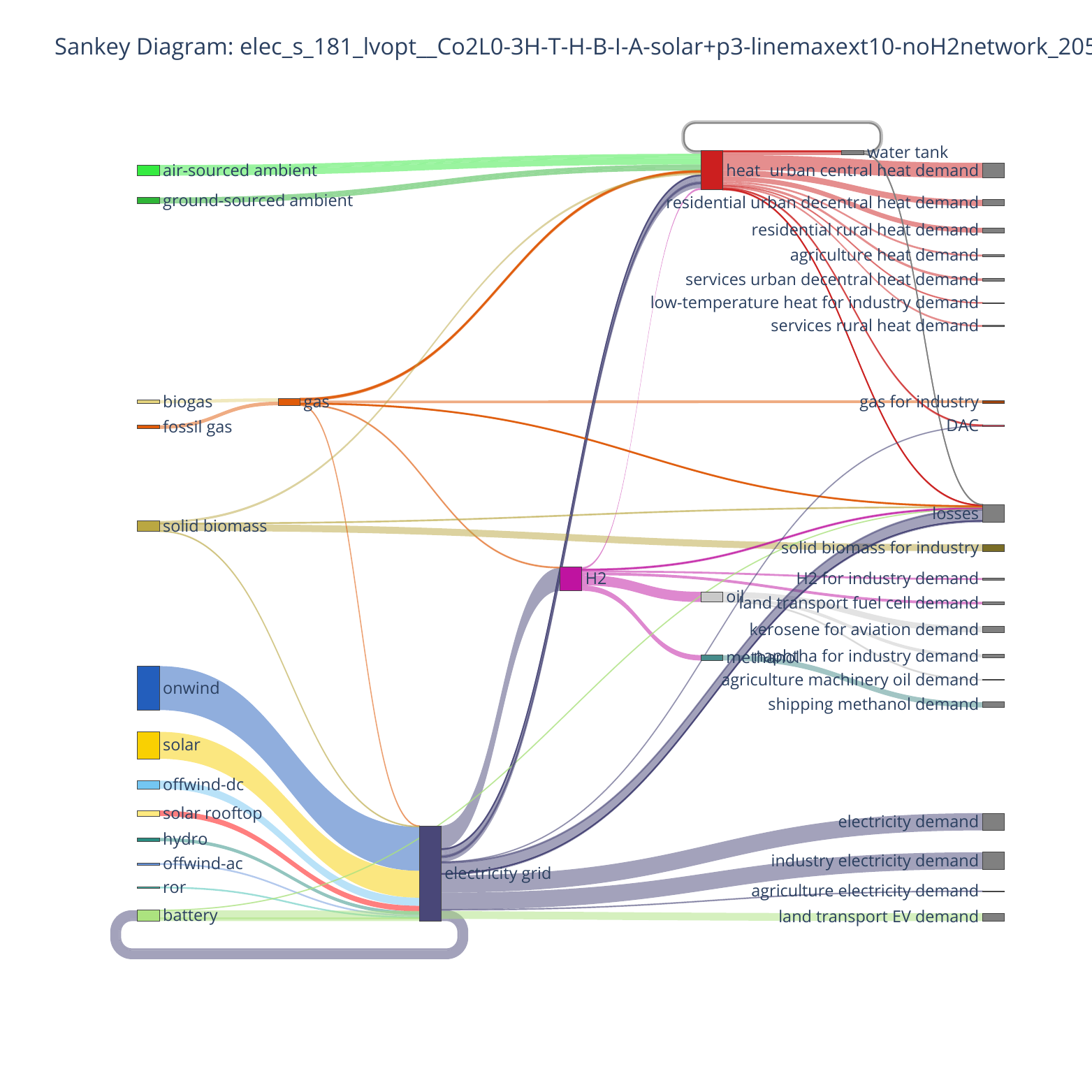}
    \end{subfigure}
    }
    \makebox[\textwidth][c]{
    \begin{subfigure}{0.6\textwidth}
        \centering
        \vspace{1cm}
        \caption{Without power grid expansion, with hydrogen network}
        \includegraphics[trim=1.5cm 2cm 1.5cm 2.5cm, clip, width=\textwidth]{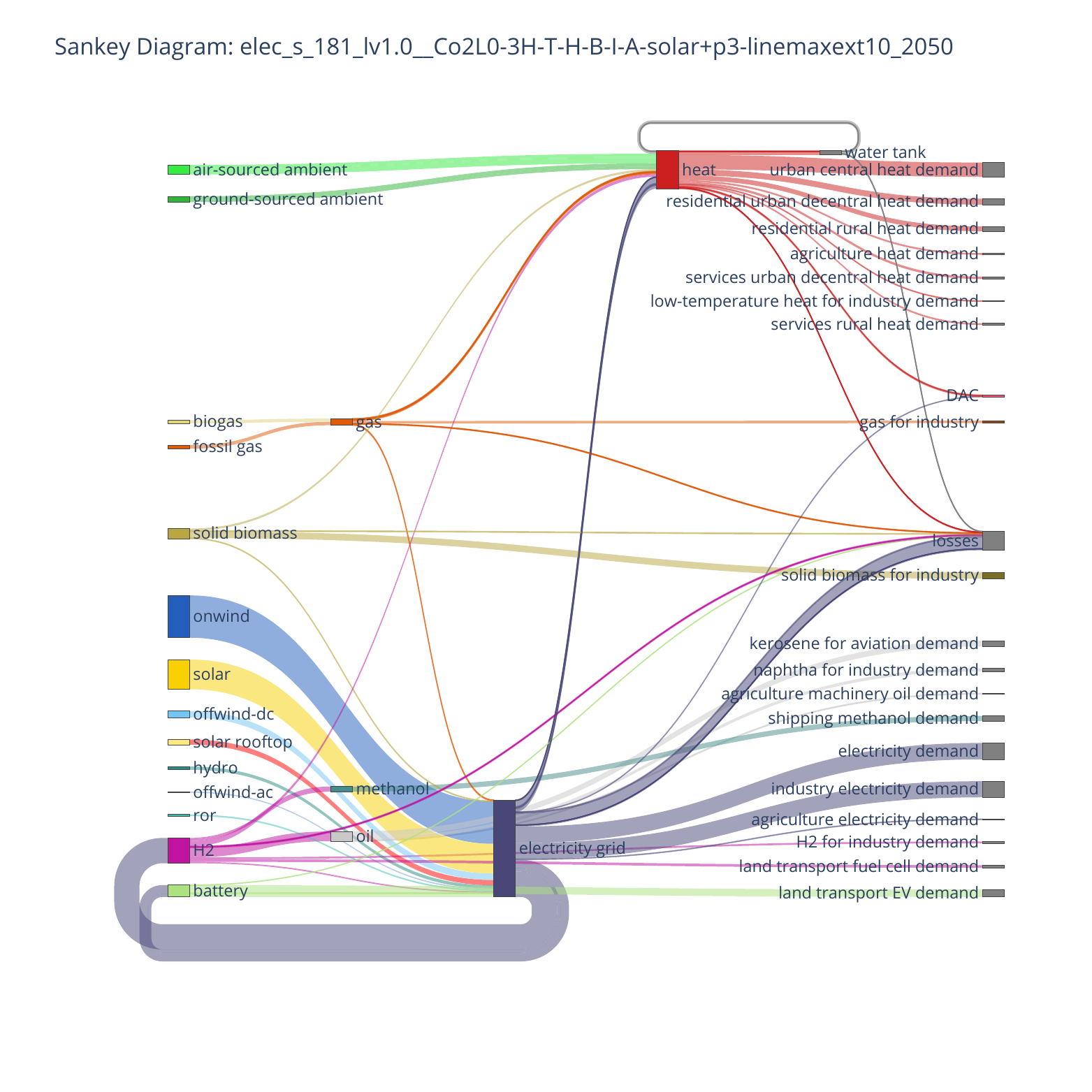}
    \end{subfigure}
    \begin{subfigure}{0.6\textwidth}
        \centering
        \vspace{1cm}
        \caption{Without power grid expansion, without hydrogen network}
        \includegraphics[trim=1.5cm 2cm 1.5cm 2.5cm, clip, width=\textwidth]{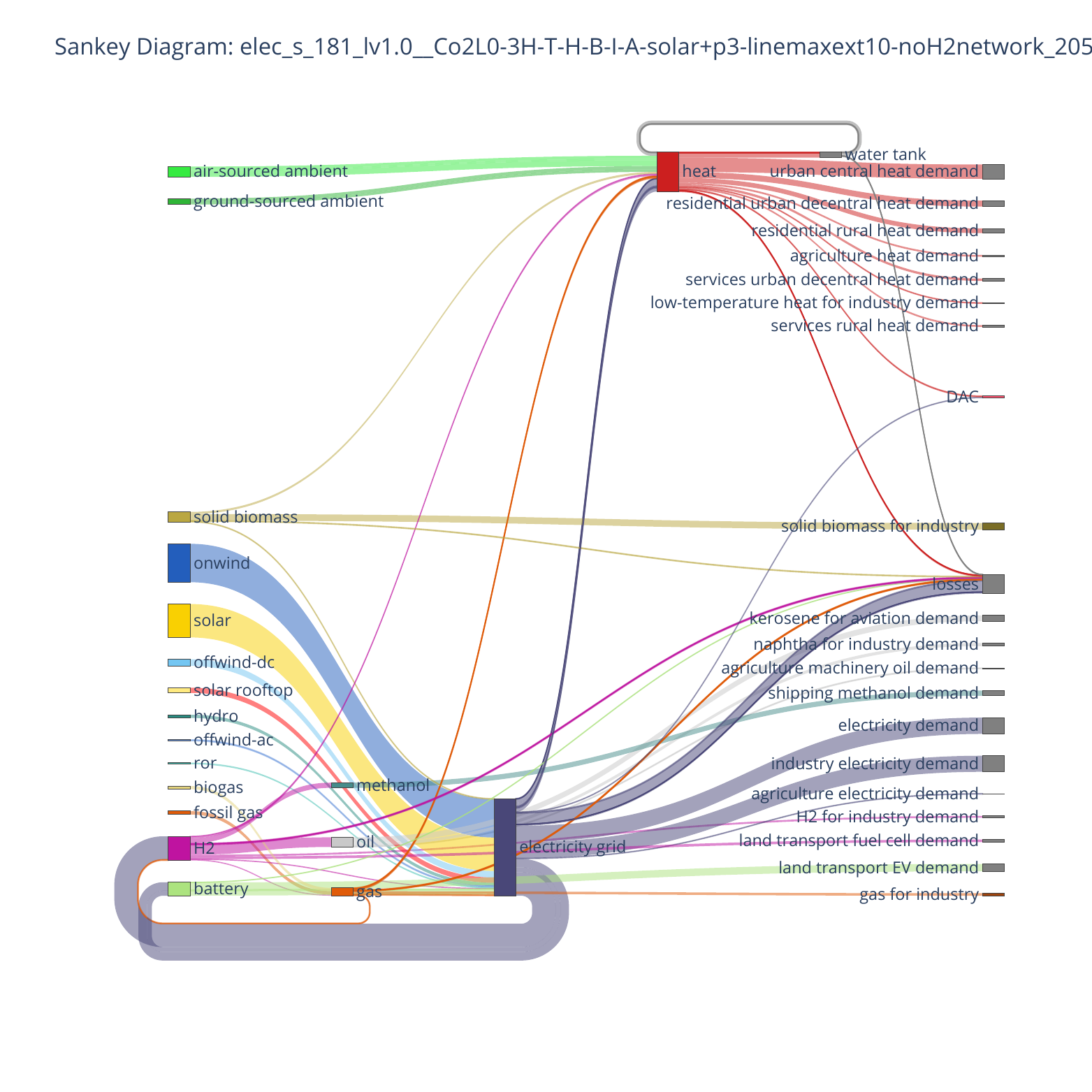}
    \end{subfigure}
    }
    \caption{Sankey diagrams of energy flows in the European system. An interactive version of these plots can be explored at \href{https://h2-network.streamlit.app}{h2-network.streamlit.app}.}
    \label{fig:si:sankey}
\end{figure}

\begin{figure}
    \centering
    \includegraphics[width= \textwidth]{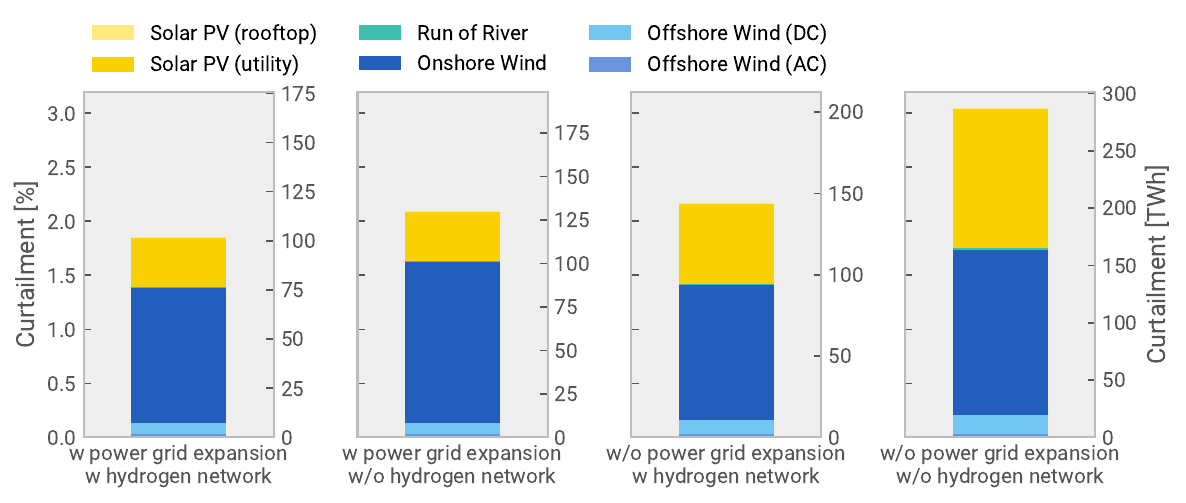}
    \caption{Curtailment of variable renewables by technology and scenario in
    relative and absolute terms. Using small exogenously set marginal costs for
    renewables, the model prioritises the curtailment of wind over solar
    electricity.}
    \label{fig:si:curtailment}
\end{figure}

In this section, supplementary results for the different network expansion
scenarios are presented. \cref{fig:si:flow-ac} displays net electricity in their
respective transmission networks analogous to the hydrogen flows presented in
\cref{fig:h2-network}. Further figures show the variation of average nodal
prices of electricity and hydrogen in space
(\cref{fig:si:lmp-ac,fig:si:lmp-h2}), in time
(\cref{fig:si:lmp-ts-ac,fig:si:lmp-ts-h2}), and as duration curves
(\cref{fig:si:lmp-dc}). Sankey diagrams in \cref{fig:si:sankey} illustrate
energy flows in the system. Across the scenarios, we infer that a carbon price
between \SI{385}{\sieuro\per\tco} with transmission infrastructure expansion and
\SI{579}{\sieuro\per\tco} without would be required to achieve both climate
neutrality and self-sufficiency in Europe. Curtailment of renewables varies between 1.8\% and 3\% (\cref{fig:si:curtailment}).

\begin{figure}
    \makebox[\textwidth][c]{
    \begin{subfigure}{1.2\textwidth}
        \centering
        \caption{With power grid expansion, with hydrogen network}
        \includegraphics[trim=0cm 0cm 0cm 0cm, clip, width=1\textwidth]{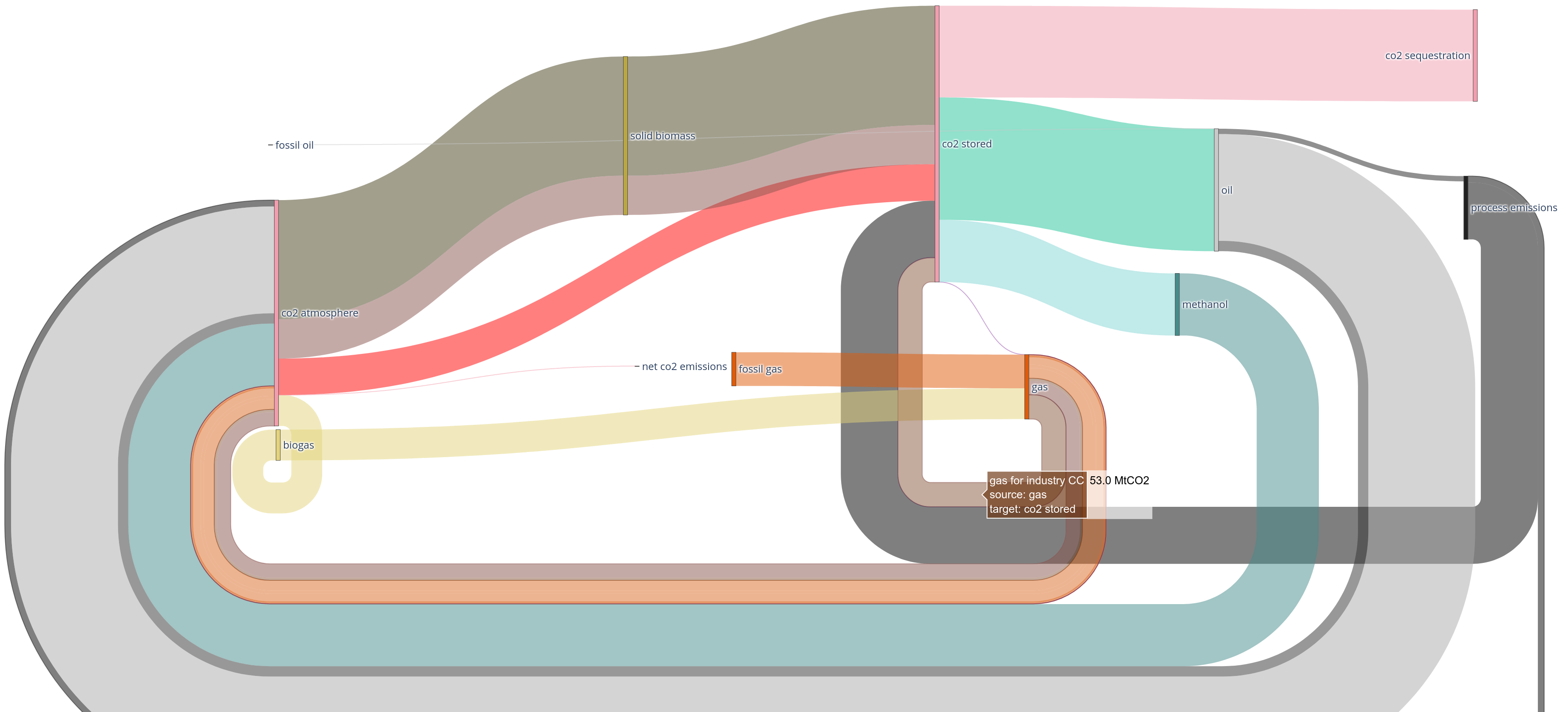}
    \end{subfigure}
    }
    \makebox[\textwidth][c]{
    \begin{subfigure}{1.2\textwidth}
        \centering
        \vspace{1cm}
        \caption{Without power grid expansion, with hydrogen network}
        \includegraphics[trim=0cm 0cm 0cm 0cm, clip, width=1\textwidth]{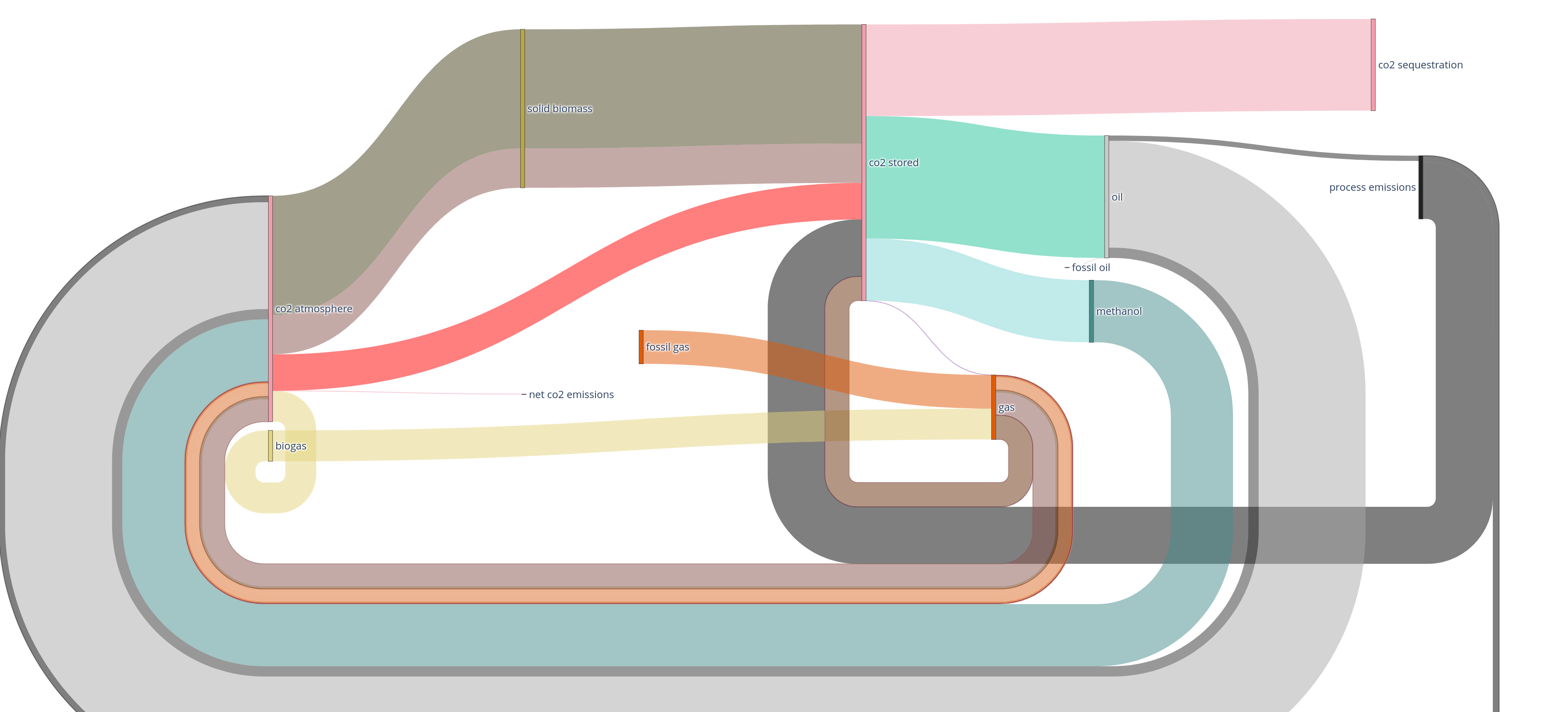}
    \end{subfigure}
    }
    \caption{Sankey diagrams of carbon flows in the European system with hydrogen network expansion. An interactive version of these plots can be explored at \href{https://h2-network.streamlit.app}{h2-network.streamlit.app}.}
    \label{fig:si:carbon-sankey}
\end{figure}

\begin{figure}
    \makebox[\textwidth][c]{
    \begin{subfigure}{1.2\textwidth}
        \centering
        \caption{With power grid expansion, without hydrogen network}
        \includegraphics[trim=0cm 0cm 0cm 0cm, clip, width=1\textwidth]{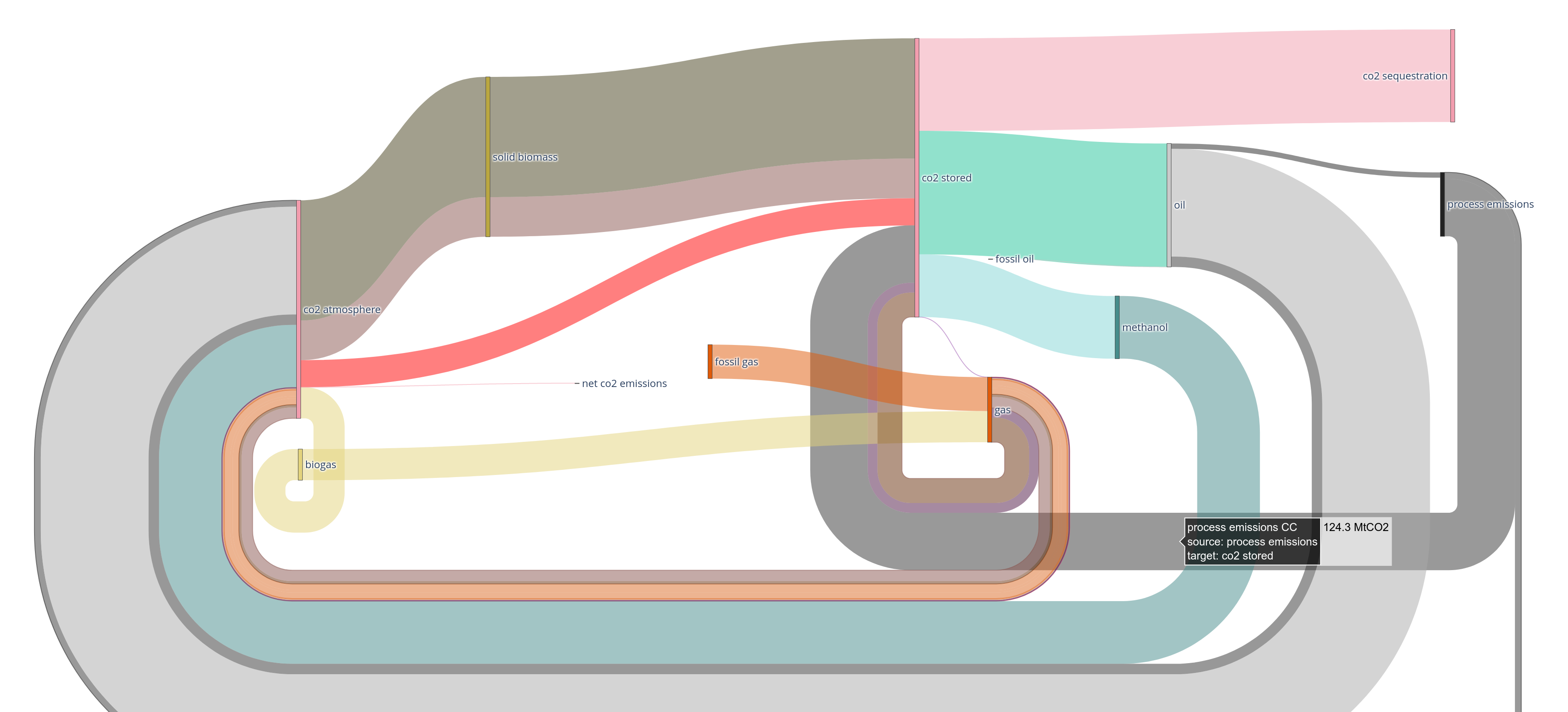}
    \end{subfigure}
    }
    \makebox[\textwidth][c]{
    \begin{subfigure}{1.2\textwidth}
        \centering
        \vspace{1cm}
        \caption{Without power grid expansion, without hydrogen network}
        \includegraphics[trim=0cm 0cm 0cm 0cm, clip, width=1\textwidth]{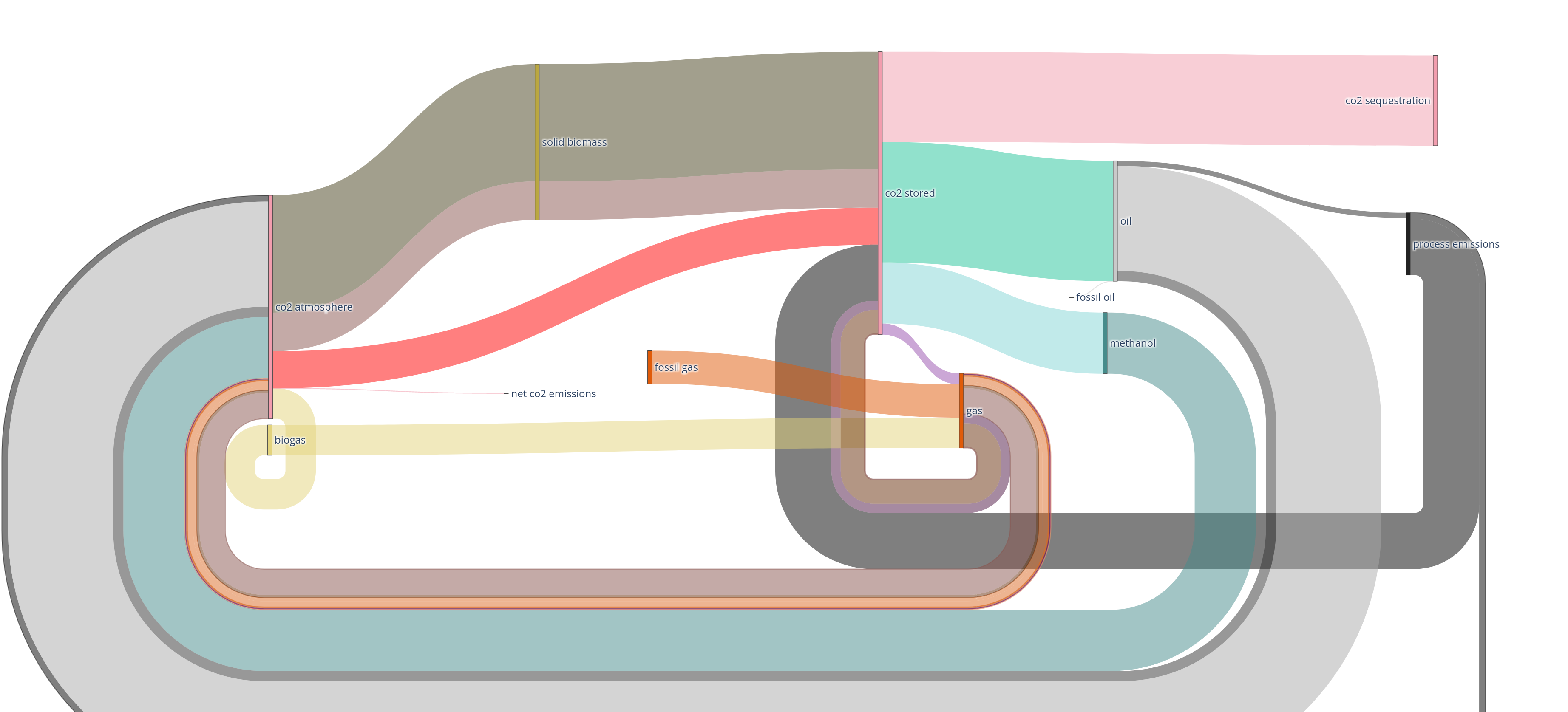}
    \end{subfigure}
    }
    \caption{Sankey diagrams of carbon flows in the European system without hydrogen network expansion. An interactive version of these plots can be explored at \href{https://h2-network.streamlit.app}{h2-network.streamlit.app}.}
    \label{fig:si:carbon-sankey-2}
\end{figure}

\begin{figure}
    \centering
    \includegraphics[width=\textwidth]{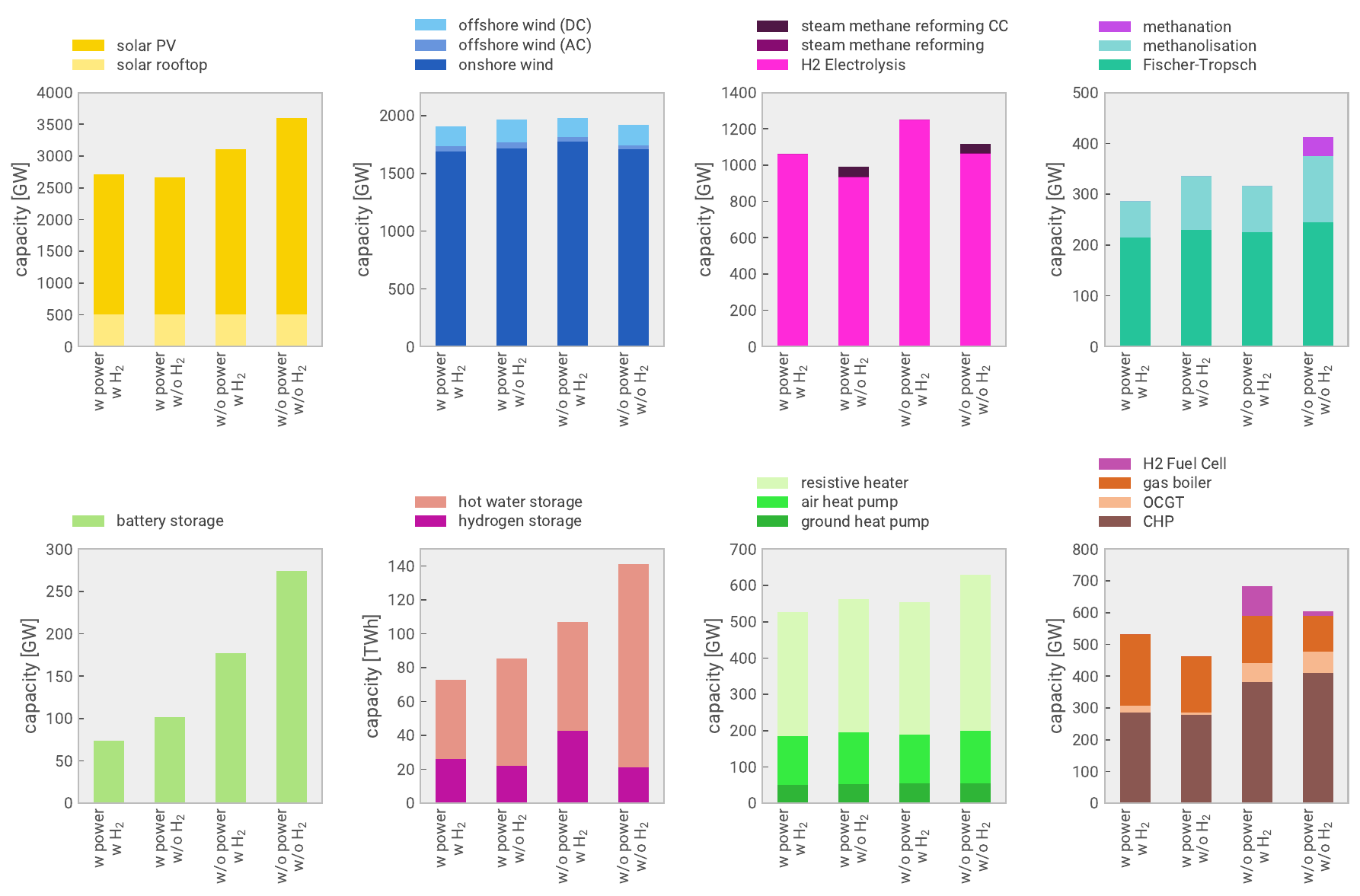}
    \caption{Installed capacities per scenario and technology group.}
    \label{fig:si:capacities}
\end{figure}

\section{Detailed Results of Least-Cost Solution with Full Grid Expansion}
\label{sec:si:detailed}

\begin{figure}
    \centering

    \begin{subfigure}[t]{\textwidth}
        \centering
        \caption{electricity}
        \includegraphics[width=\textwidth]{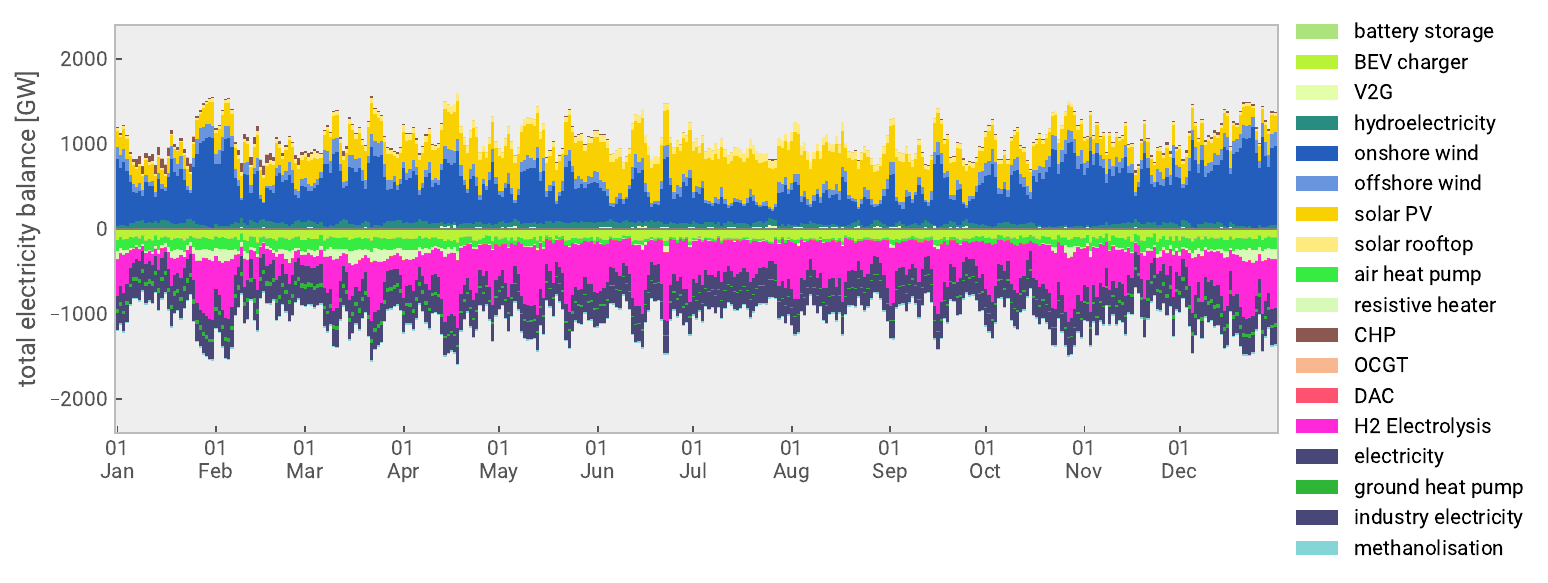}
    \end{subfigure}
    \begin{subfigure}[t]{\textwidth}
        \centering
        \caption{space and water heating}
        \includegraphics[width=\textwidth]{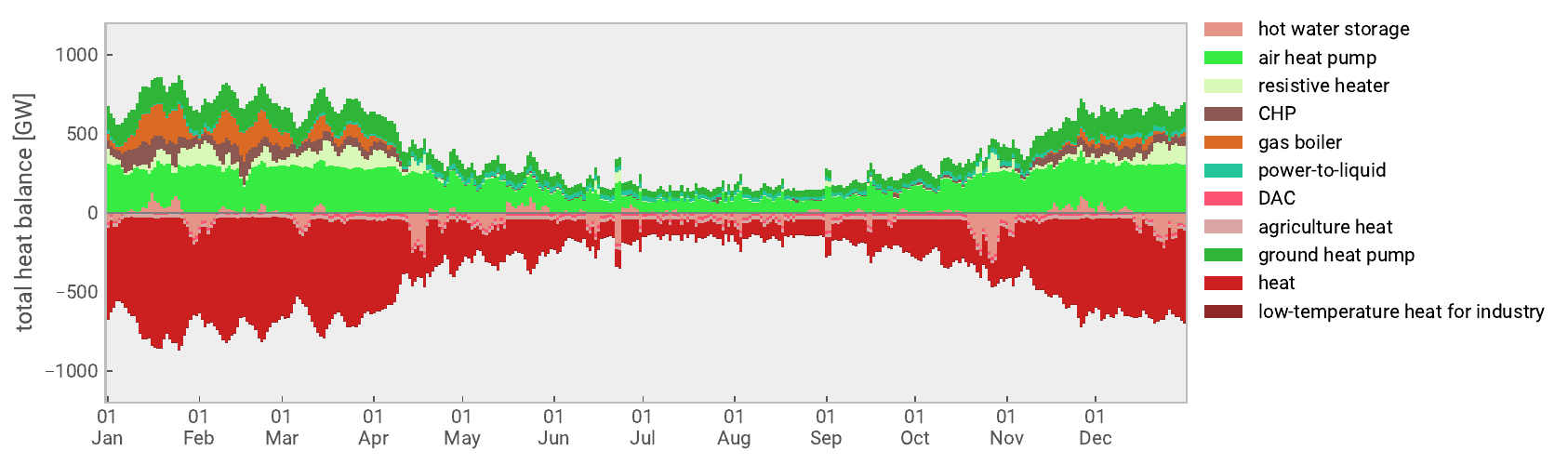}
    \end{subfigure}
    \begin{subfigure}[t]{\textwidth}
        \centering
        \caption{hydrogen}
        \includegraphics[width=\textwidth]{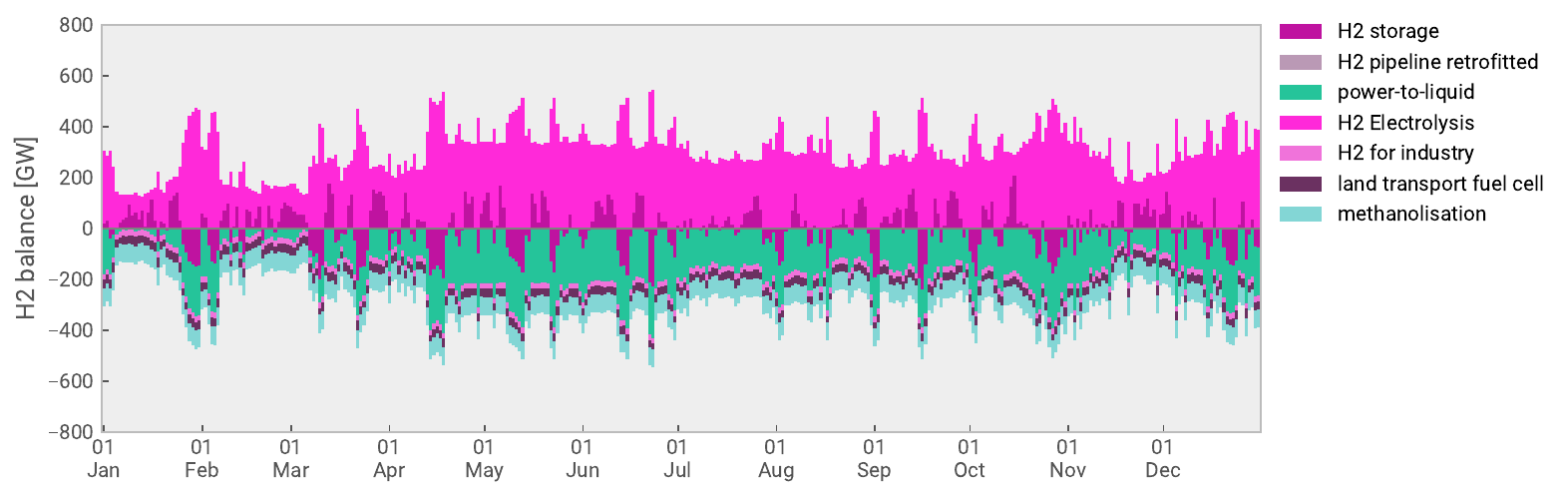}
    \end{subfigure}

    \caption{Daily sampled time series for (a) electricity, (b) heat, and (c) hydrogen supply (above zero) and consumption (below zero) composition. Supply and consumption balance for each bar by definition.}
    \label{fig:output-ts-1}
\end{figure}

\begin{figure}
    \centering

    \begin{subfigure}[t]{\textwidth}
        \centering
        \caption{methane}
        \includegraphics[width=\textwidth]{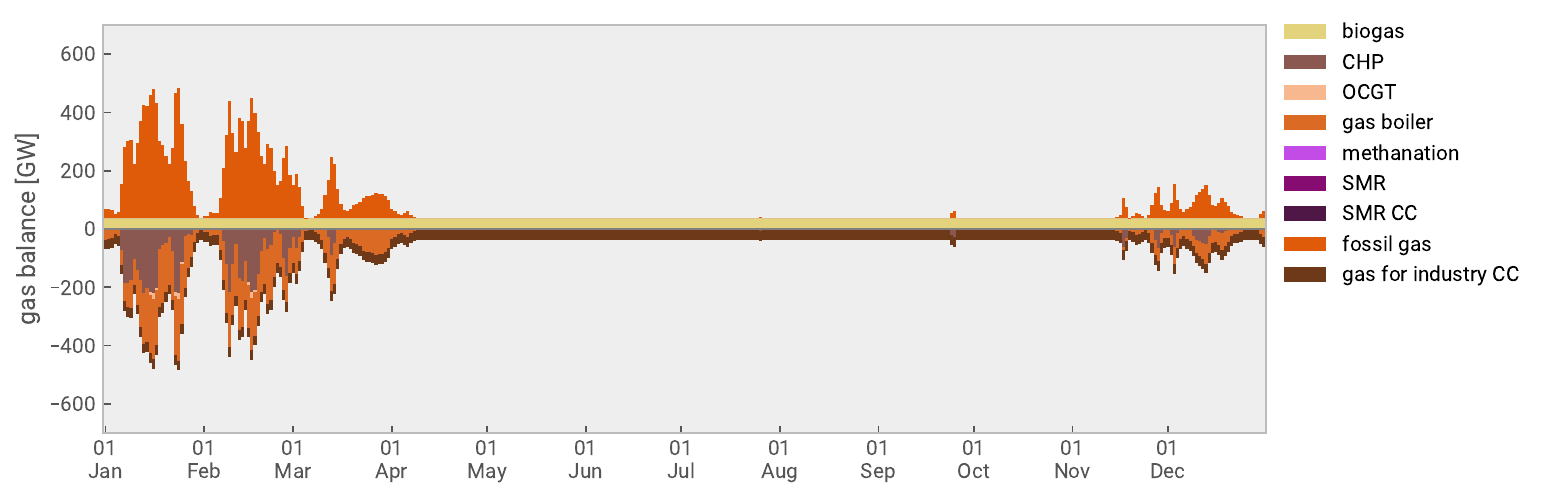}
    \end{subfigure}
    \begin{subfigure}[t]{\textwidth}
        \centering
        \caption{oil-based products}
        \includegraphics[width=\textwidth]{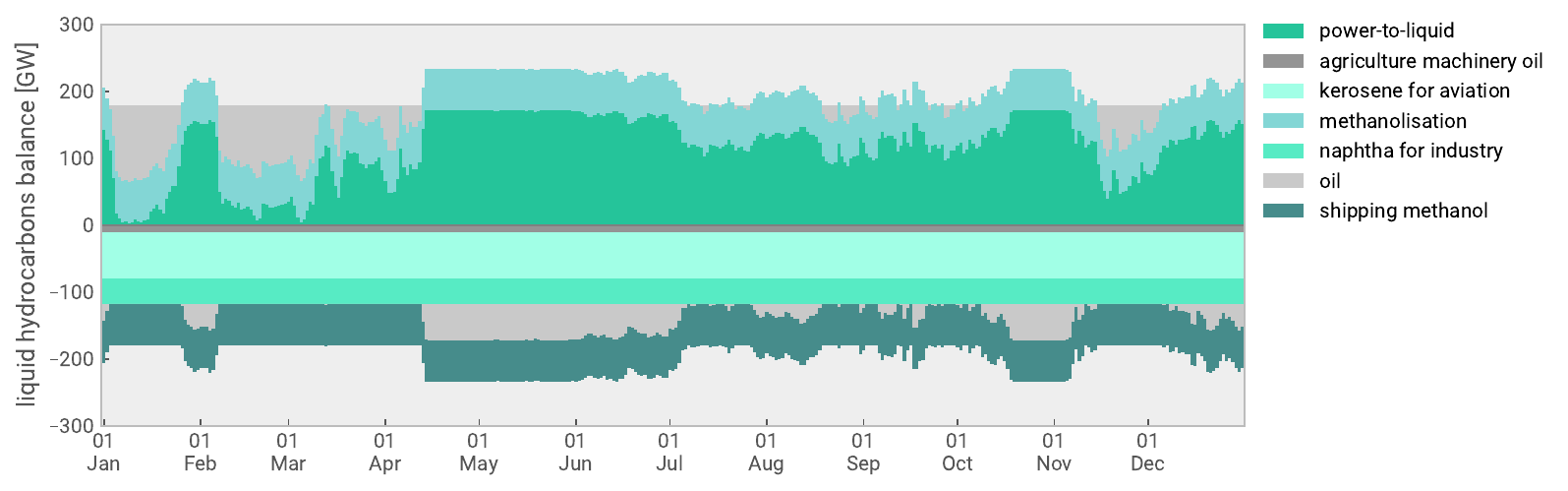}
    \end{subfigure}
    \begin{subfigure}[t]{\textwidth}
        \centering
        \caption{stored \co}
        \includegraphics[width=\textwidth]{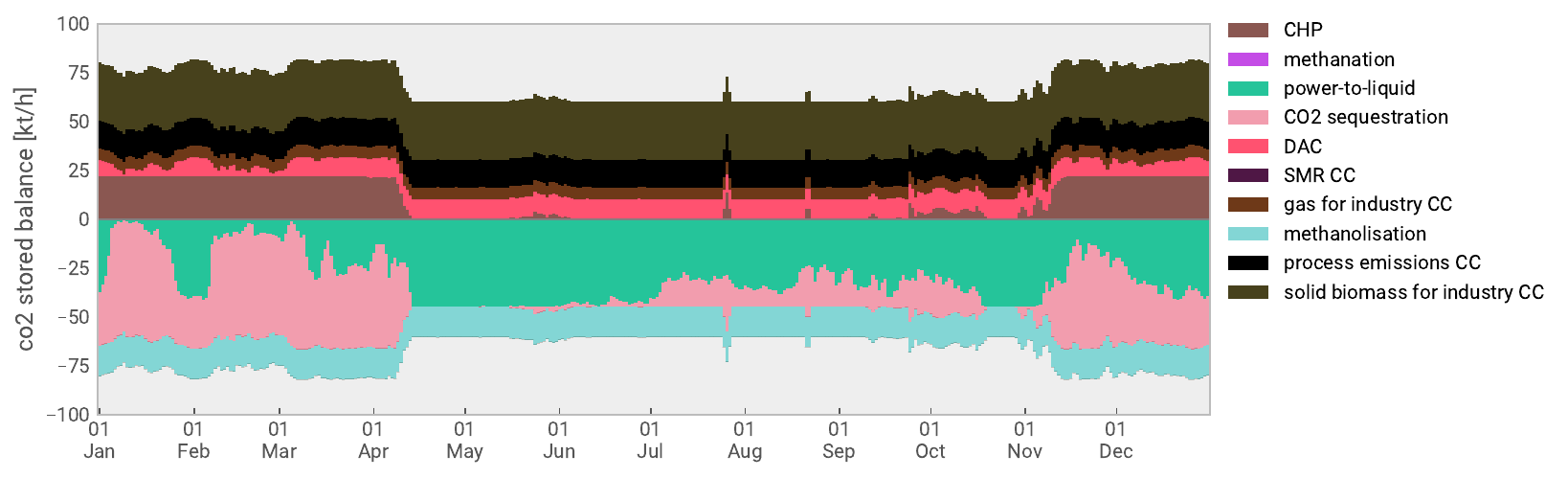}
    \end{subfigure}

    \caption{Daily sampled time series for (a) methane, (b) oil-based products, and (c) carbon dioxide supply (above zero) and consumption (below zero) composition. Supply and consumption balance for each bar.}
    \label{fig:output-ts-2}
\end{figure}

\begin{figure}
    \centering

    \begin{subfigure}[t]{\textwidth}
        \centering
        \caption{electricity}
        \includegraphics[width=\textwidth]{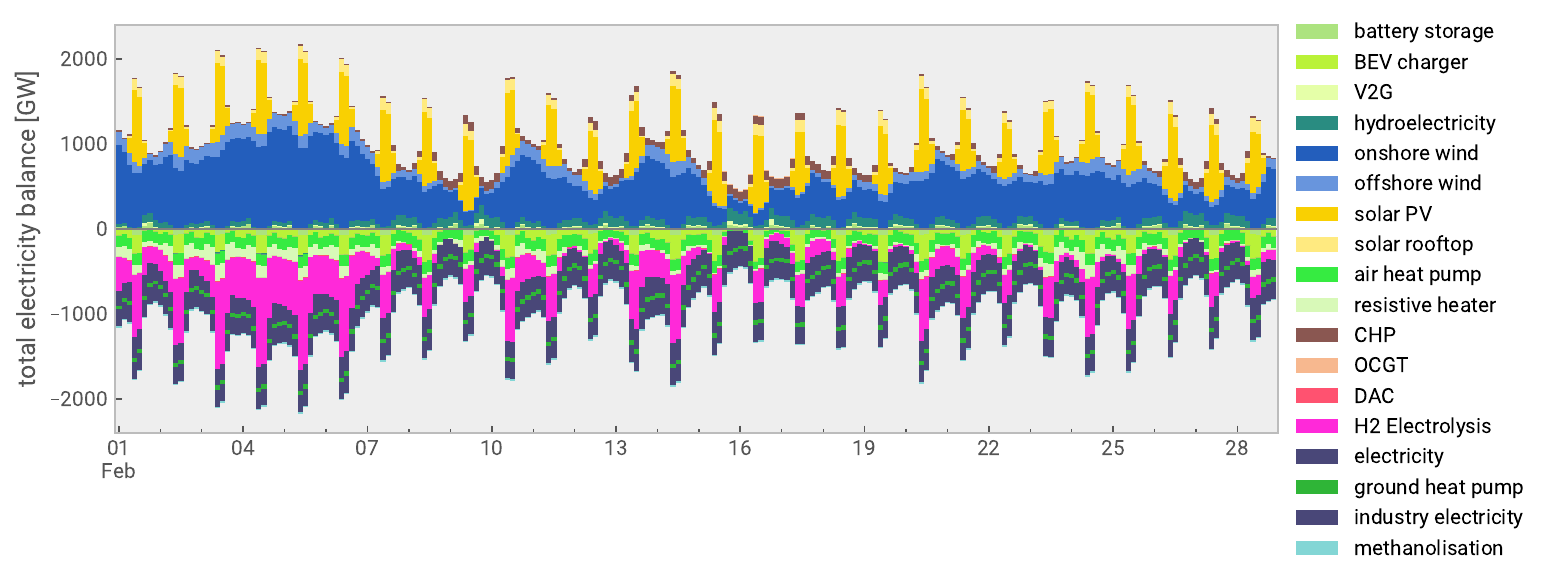}
    \end{subfigure}
    \begin{subfigure}[t]{\textwidth}
        \centering
        \caption{space and water heating}
        \includegraphics[width=\textwidth]{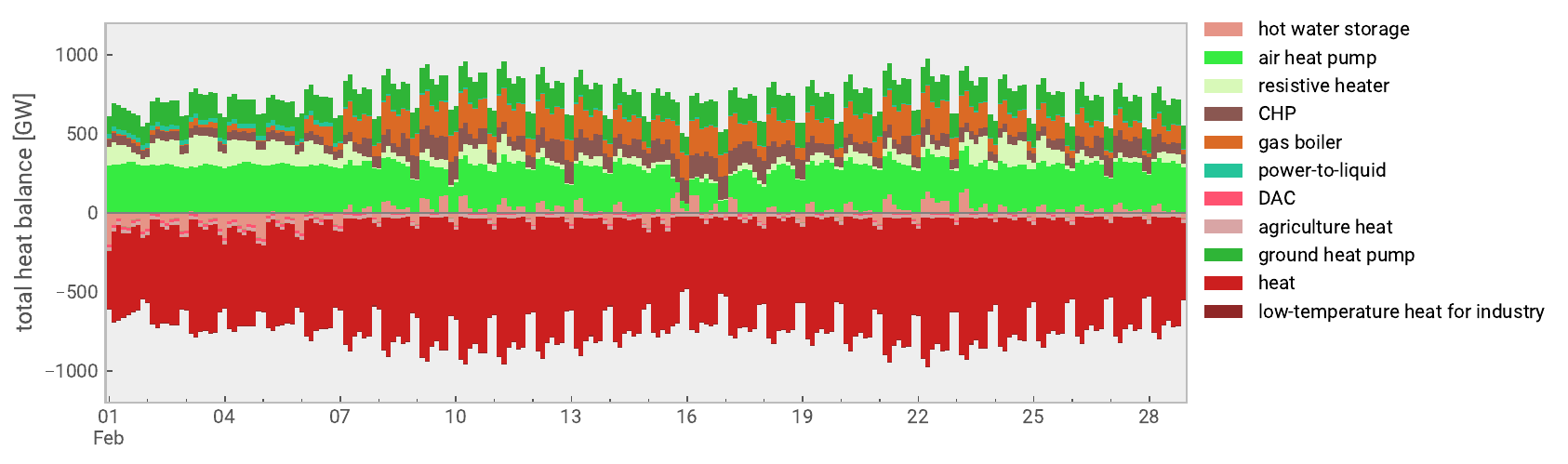}
    \end{subfigure}
    \begin{subfigure}[t]{\textwidth}
        \centering
        \caption{hydrogen}
        \includegraphics[width=\textwidth]{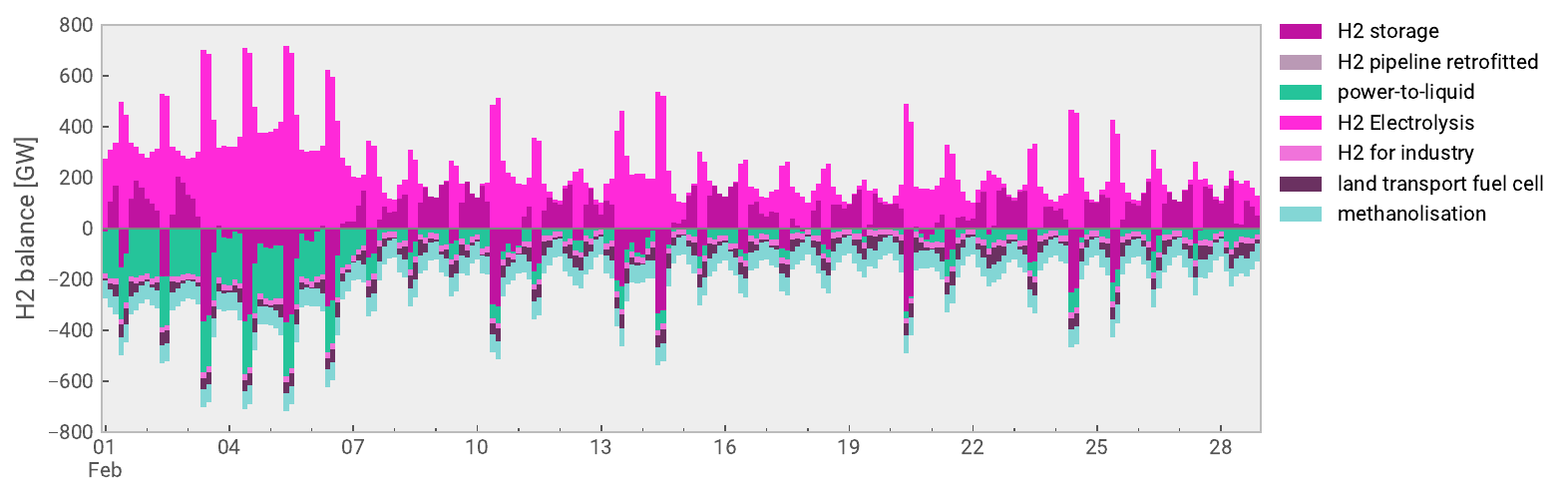}
    \end{subfigure}

    \caption{Hourly sampled time series of February for (a) electricity, (b) heat, and (c) hydrogen supply (above zero) and consumption (below zero) composition. Supply and consumption balance for each bar.}
    \label{fig:output-ts-3}
\end{figure}

\begin{figure}
    \centering

    \begin{subfigure}[t]{\textwidth}
        \centering
        \caption{methane}
        \includegraphics[width=\textwidth]{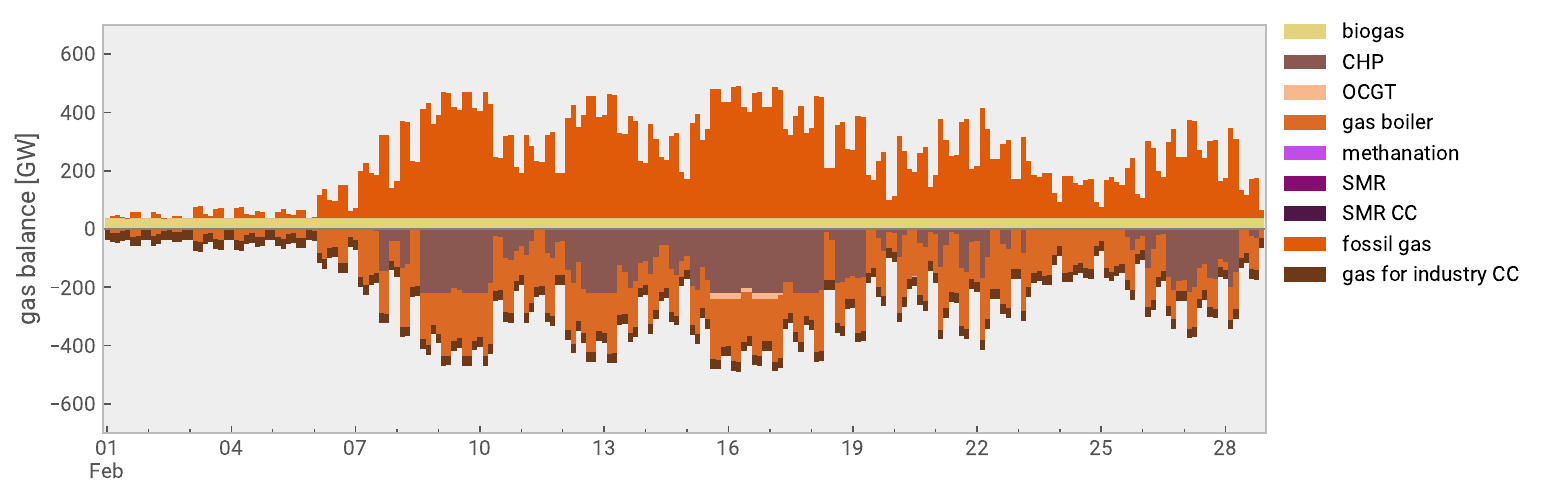}
    \end{subfigure}
    \begin{subfigure}[t]{\textwidth}
        \centering
        \caption{oil-based products}
        \includegraphics[width=\textwidth]{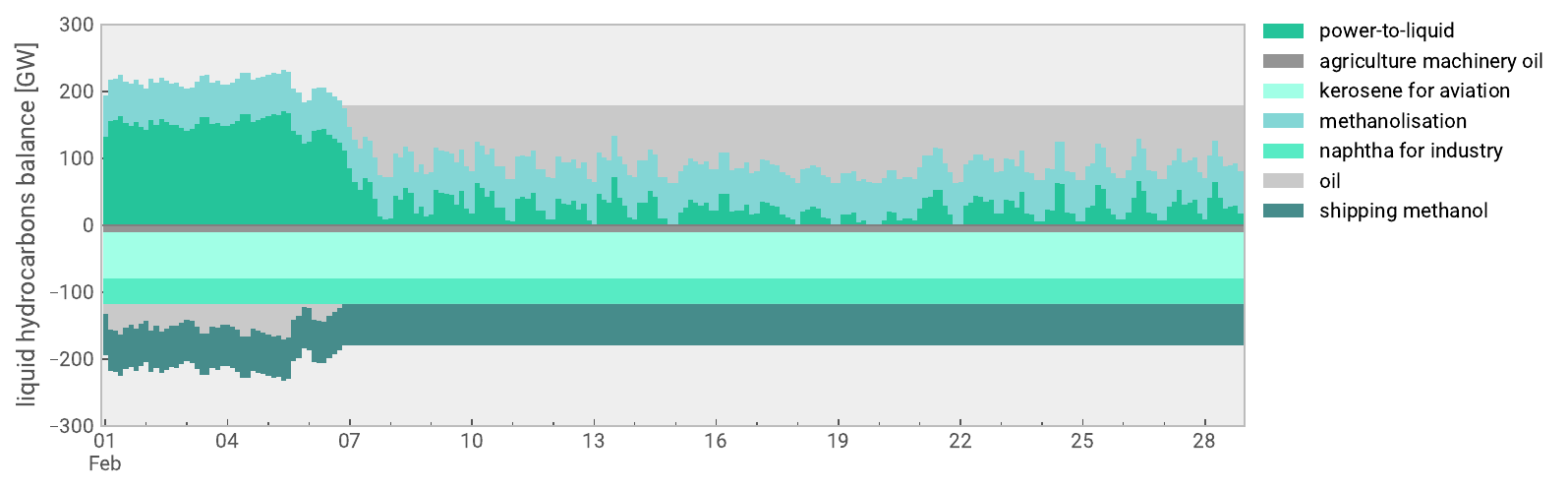}
    \end{subfigure}
    \begin{subfigure}[t]{\textwidth}
        \centering
        \caption{stored \co}
        \includegraphics[width=\textwidth]{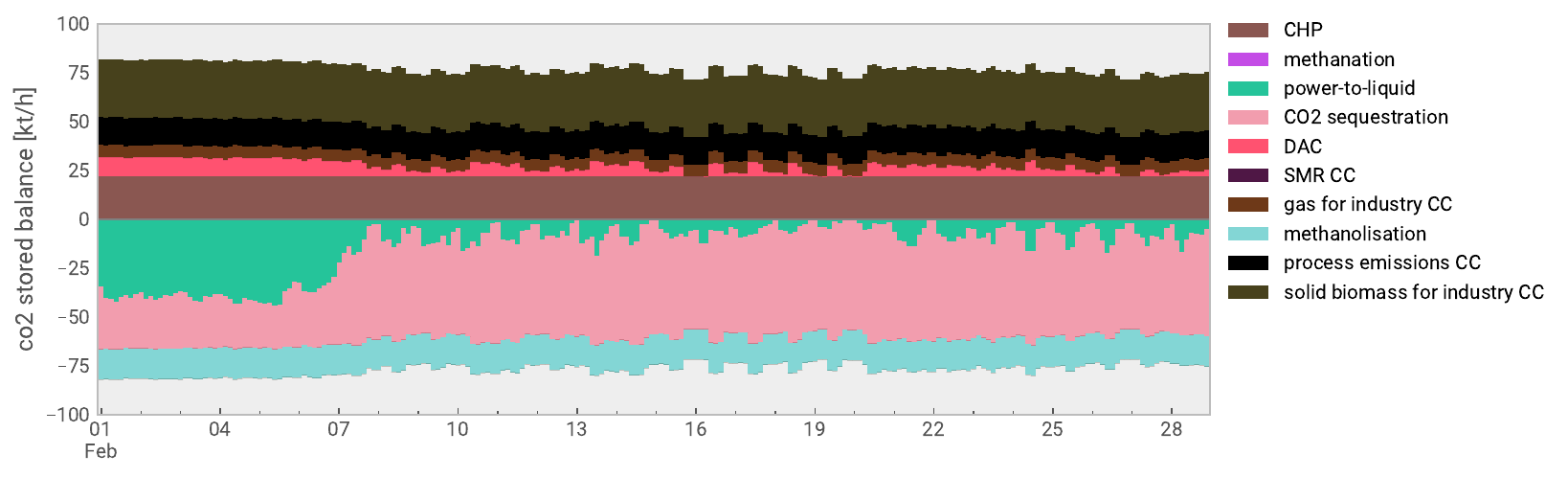}
    \end{subfigure}

    \caption{Hourly sampled time series of February for (a) methane, (b) oil-based products, and (c) carbon dioxide supply (above zero) and consumption (below zero) composition. Supply and consumption balance for each bar.}
    \label{fig:output-ts-4}
\end{figure}

\begin{figure}
    \begin{subfigure}{0.49\textwidth}
        \centering
        \includegraphics[width=\textwidth]{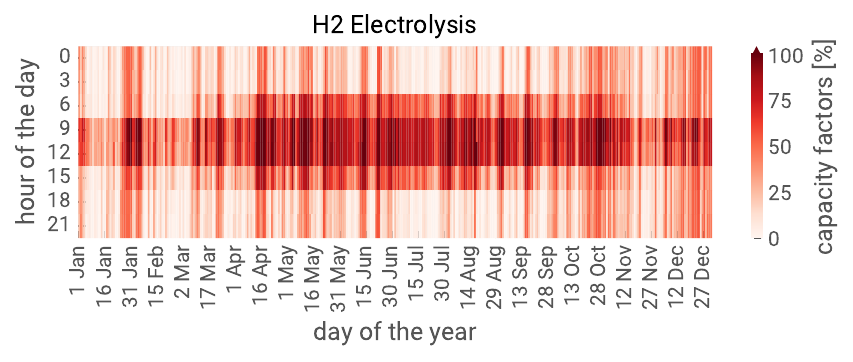}
    \end{subfigure}
    \begin{subfigure}{0.49\textwidth}
        \centering
        \includegraphics[width=\textwidth]{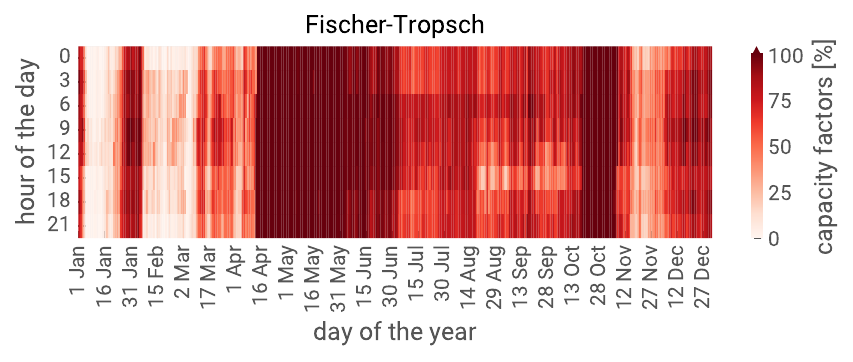}
    \end{subfigure}
    \begin{subfigure}{0.49\textwidth}
        \centering
        \includegraphics[width=\textwidth]{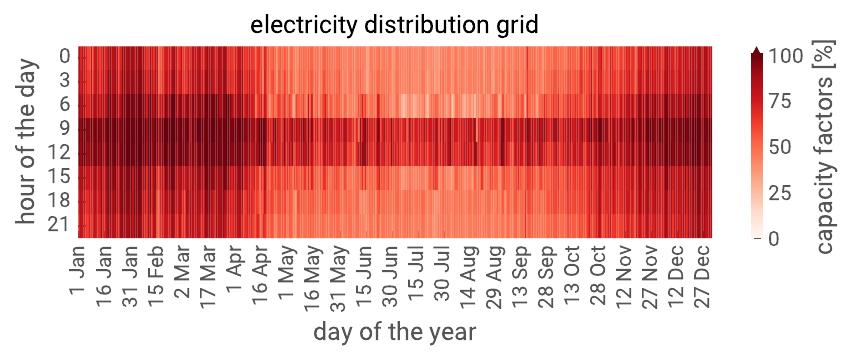}
    \end{subfigure}
    \begin{subfigure}{0.49\textwidth}
        \centering
        \includegraphics[width=\textwidth]{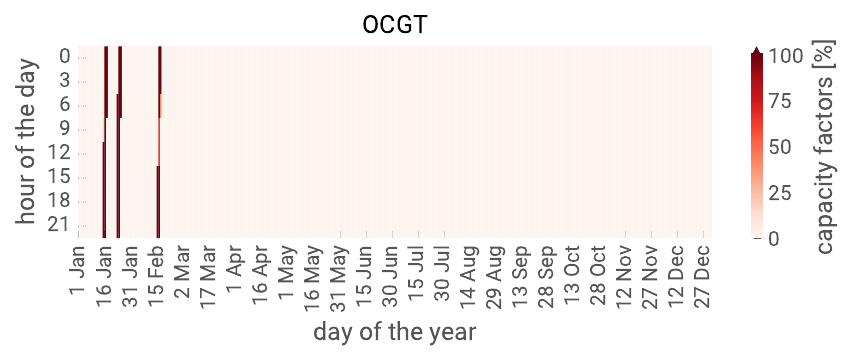}
    \end{subfigure}
    \begin{subfigure}{0.49\textwidth}
        \centering
        \includegraphics[width=\textwidth]{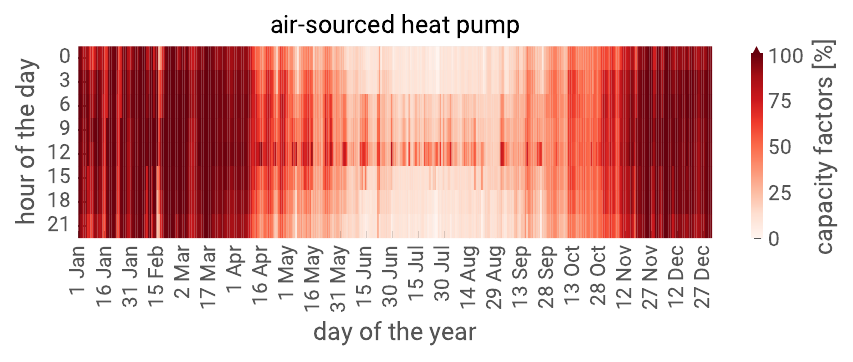}
    \end{subfigure}
    \begin{subfigure}{0.49\textwidth}
        \centering
        \includegraphics[width=\textwidth]{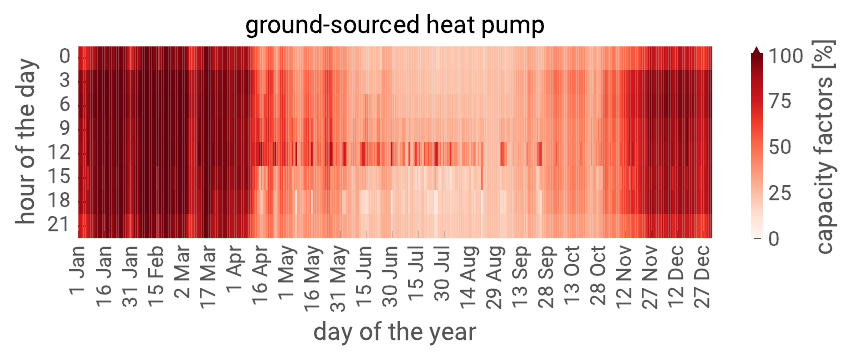}
    \end{subfigure}
    \begin{subfigure}{0.49\textwidth}
        \centering
        \includegraphics[width=\textwidth]{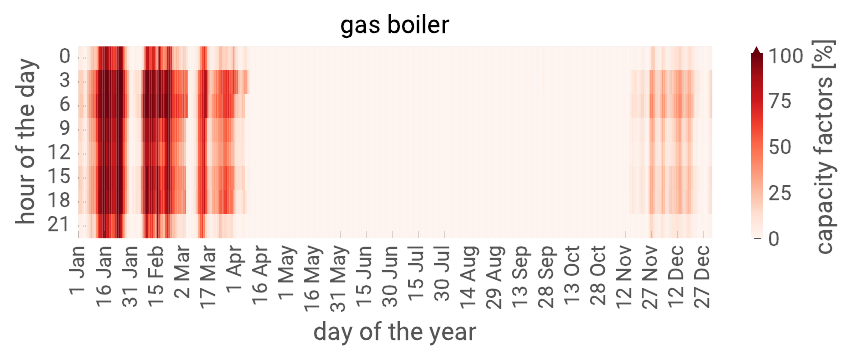}
    \end{subfigure}
    \begin{subfigure}{0.49\textwidth}
        \centering
        \includegraphics[width=\textwidth]{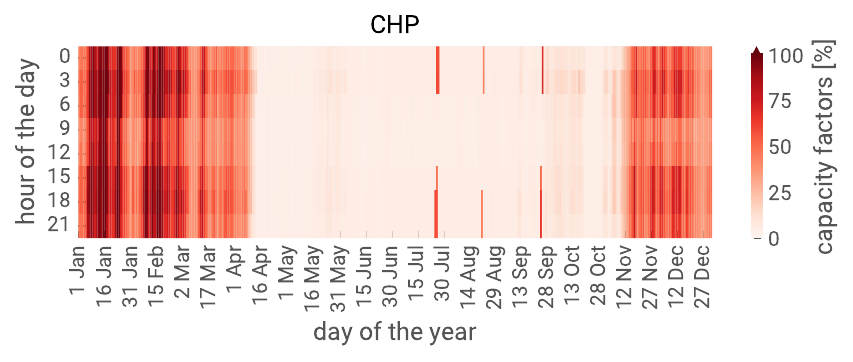}
    \end{subfigure}
    \begin{subfigure}{0.49\textwidth}
        \centering
        \includegraphics[width=\textwidth]{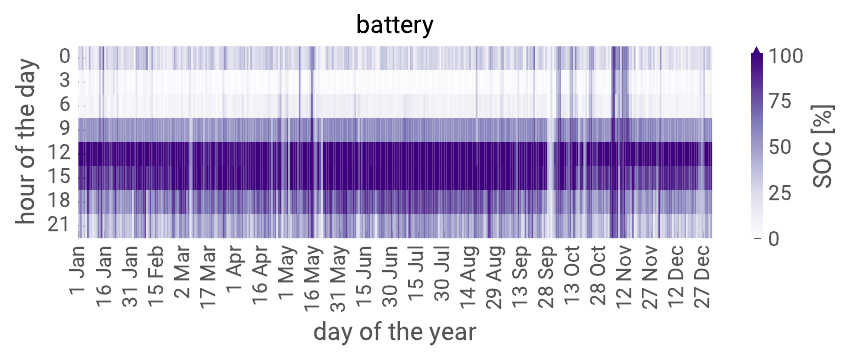}
    \end{subfigure}
    \begin{subfigure}{0.49\textwidth}
        \centering
        \includegraphics[width=\textwidth]{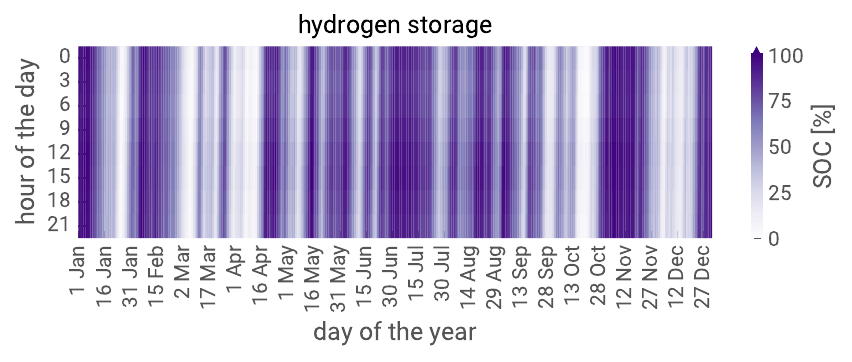}
    \end{subfigure}
    \caption{Operations and storage filling levels of selected energy system components. The figure outlines
    the flexible operation of electrolysers (both weeky due to wind-based and daily due to solar-based production)
    the operation of synthetic fuel production
    the backup role of gas power plants (OCGT),
    the seasonal operation of heat pumps, gas boilers, CHP, and hydrogen storage,
    the daily pattern of battery storage filling levels, and
    periods of peak loading of the power distribution grid.
    }
    \label{fig:si:utilisation-rate-ts}
\end{figure}

\begin{figure}
    \centering
    \begin{subfigure}{0.49\textwidth}
        \centering
        \caption{electricity network loading}
        \includegraphics[width=\textwidth]{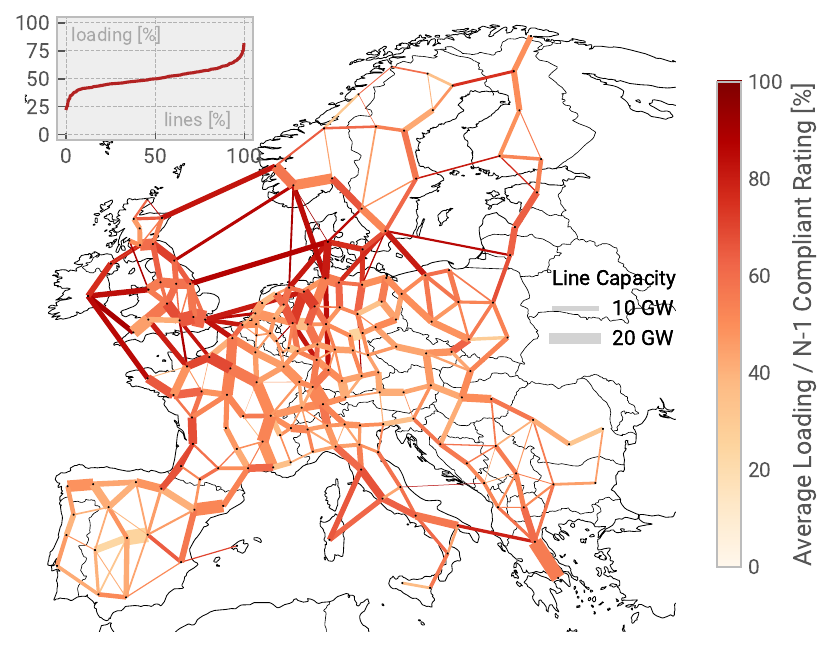}
    \end{subfigure}
    \begin{subfigure}{0.49\textwidth}
        \centering
        \caption{hydrogen network loading}
        \includegraphics[width=\textwidth]{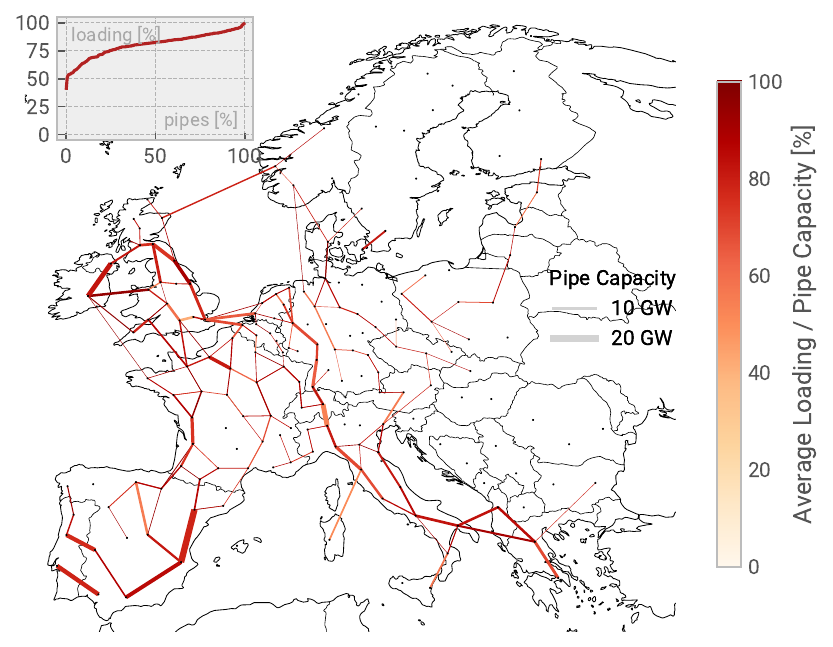}
    \end{subfigure}
    \begin{subfigure}{0.65\textwidth}
        \vspace{1cm}
        \centering
        \caption{electricity grid congestion and curtailment}
        \includegraphics[width=\textwidth]{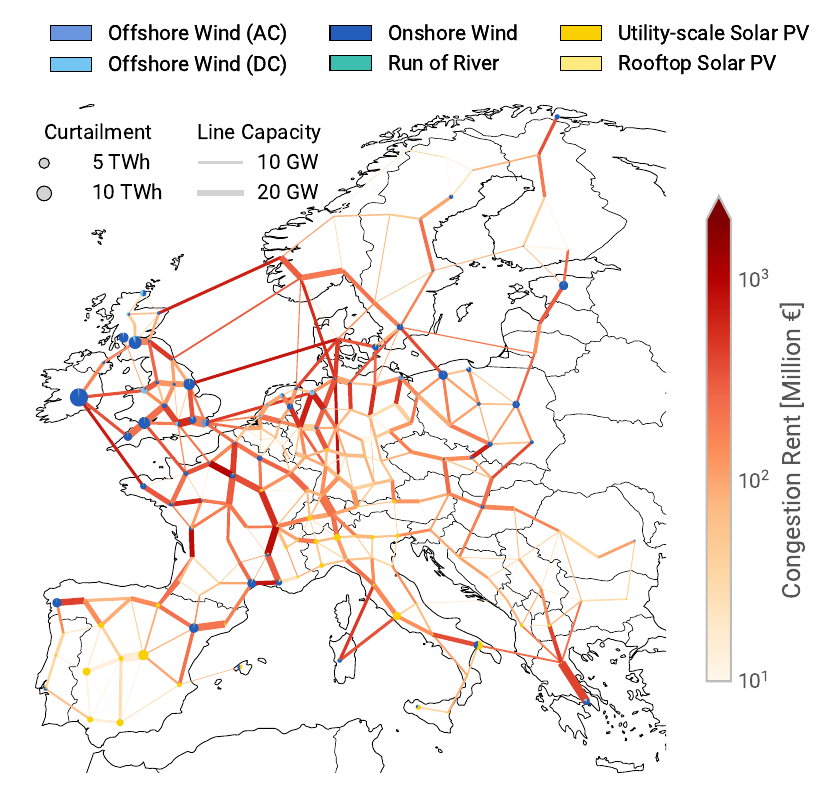}
    \end{subfigure}
    \caption{Utilisation rate of electricity and hydrogen network, curtailment and congestion. Subplot (a) shows average electricity network loading relative to $N-1$ compliant line rating (70\% of nominal rating) and the corresponding duration curve of line loadings. Subplot (b) shows the average hydrogen pipeline loading relative to the nominal pipeline capacity and also the corresponding duration curve of pipeline loadings. Subplot (c) shows the regional and technological distribution of curtailment in the system as well as realised congestion rents in the electricity network.}
    \label{fig:si:grid-utilisation}
\end{figure}

In the following section we present more detailed results from the scenario
where both the hydrogen and electricity grid could be expanded. Among the
scenarios we investigated, this represents the least-cost solution.
\crefrange{fig:output-ts-1}{fig:output-ts-4} show temporally resolved energy
balances for different carriers: electricity, hydrogen, heat, methane, oil-based
products, and carbon dioxide. These are daily sampled time series for a year and
3-hourly sampled time series for the month February, and indicate how different
technologies are operated both seasonally and daily. How selected energy system
components are operated throughout the year is shown in
\cref{fig:si:utilisation-rate-ts}. The utilisation of electricity and hydrogen
network assets are presented in \cref{fig:si:grid-utilisation}, alongside
information about where energy is curtailed and what congestion rents are
incurred.

\begin{figure}
    \centering
    \begin{subfigure}{0.66\textwidth}
        \centering
        \caption{Electrolyser capacities}
        \includegraphics[width=\textwidth]{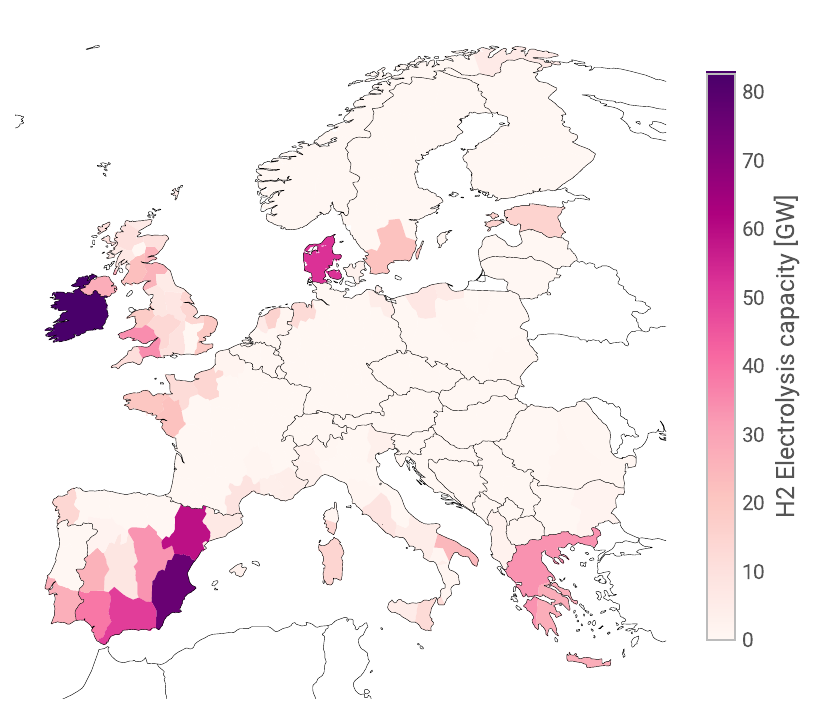}
    \end{subfigure}
    \begin{subfigure}{0.66\textwidth}
        \centering
        \caption{Fischer-Tropsch conversion plant capacities}
        \includegraphics[width=\textwidth]{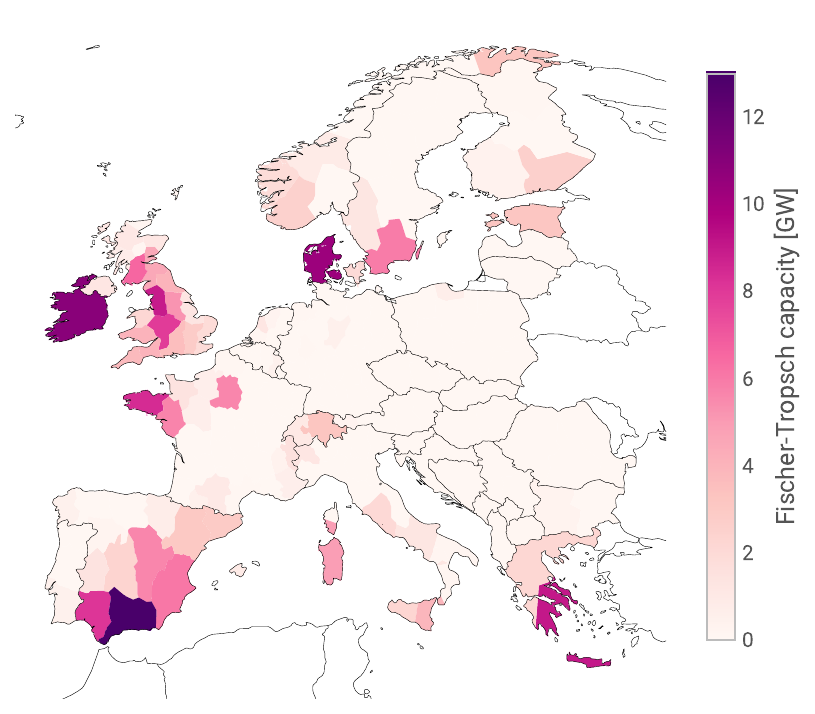}
    \end{subfigure}
    \caption{Pairing of electrolysers and Fischer-Tropsch fuel production sites. }
    \label{fig:si:colocation}
\end{figure}

\begin{figure}
    \centering
    \includegraphics[width=\textwidth]{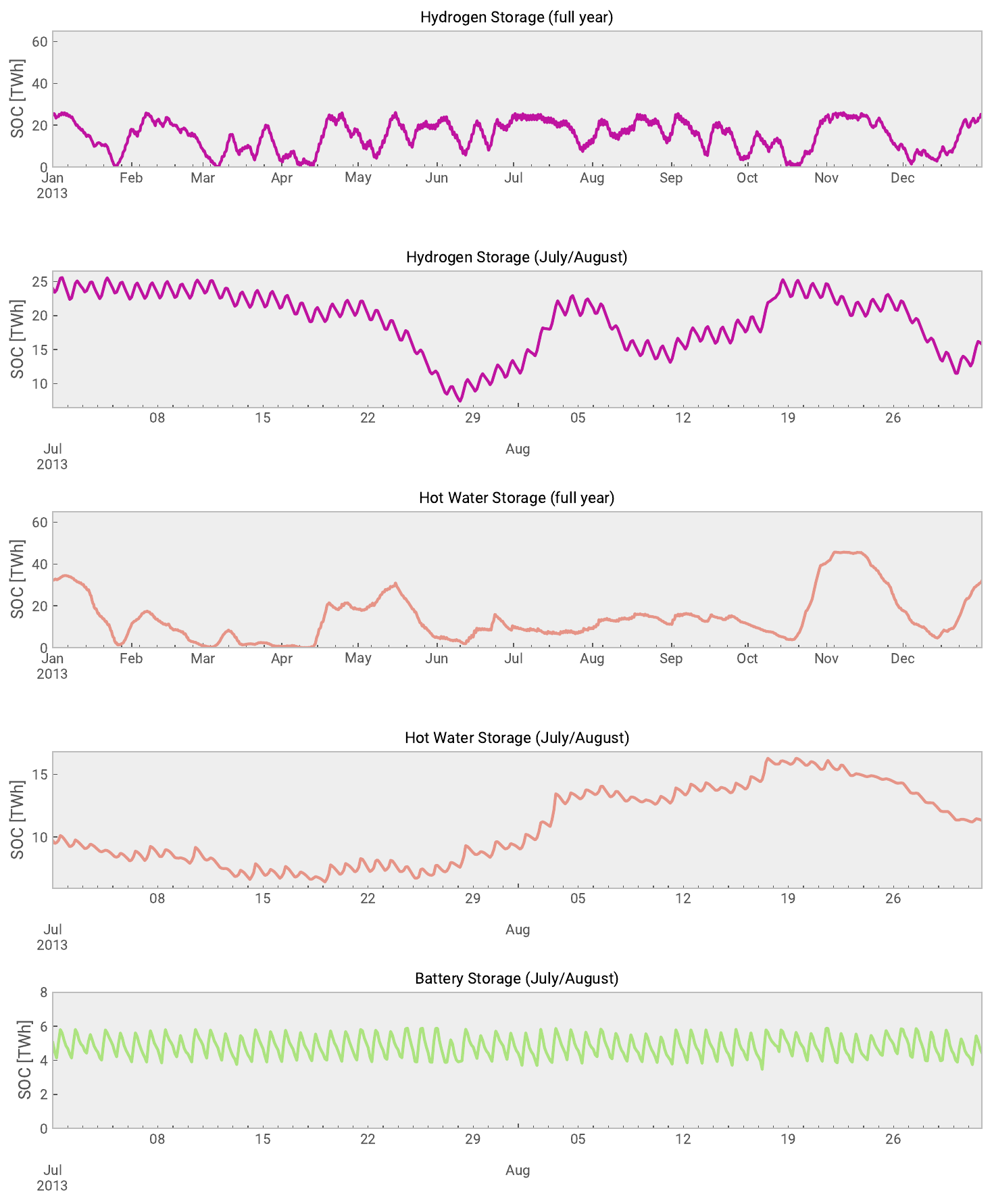}
    \caption{Patterns of storage filling levels for hydrogen storage, hot water storage and battery storage (including electric vehicles). The figures show daily patterns for battery storage, and daily as well as synoptic patterns for hydrogen and hot water storage. Neither hydrogen nor hot water storage have a dominant seasonal pattern.}
    \label{fig:si:soc}
\end{figure}

\section{Techno-Economic Assumptions}
\label{sec:si:costs}

For the technology assumptions, we take estimates for the year 2030 for the main
scenarios and run a sensitivity analysis with more progressive 2050 cost
assumptions in \cref{sec:si:sensitivity-costs}.  Many of those come from a
database published by the Danish Energy Agency (DEA).\citeS{DEA} We take 2030
technology assumptions for the main scenarios to account for expected technology
cost reductions in the near-term while acknowledging that the gradual transition
to climate neutrality implies that much of the infrastructure must be built well
in advance of reaching net-zero emissions. A complete list is compiled in
\cref{tab:si:costs}. Assumptions are maintained at
\href{https://github.com/pypsa/technology-data}{github.com/pypsa/technology-data}
and were taken from version 0.4.0.

\newgeometry{margin=2cm}
\begin{landscape}

\begin{footnotesize}
\begin{longtable}{p{6cm}p{3.8cm}rrp{2cm}p{5.4cm}}
\caption{Overview of technology assumptions for respective projection year.}
\label{tab:si:costs}\\
\toprule
                      &            &  value (2030) &  value (2050) &                          unit &                                                                                                                                         source \\
technology & parameter &               &               &                               &                                                                                                                                                \\
\midrule
\endfirsthead
\caption[]{Overview of technology assumptions for respective projection year.} \\
\toprule
                      &            &  value (2030) &  value (2050) &                          unit &                                                                                                                                         source \\
technology & parameter &               &               &                               &                                                                                                                                                \\
\midrule
\endhead
\midrule
\multicolumn{6}{r}{{Continued on next page}} \\
\midrule
\endfoot

\bottomrule
\endlastfoot
AC grid connection (station) & overnight investment &        250.00 &        250.00 &               \euro/kW$_{el}$ &                                                                                                                                 DEA\citeS{DEA} \\
AC grid connection (submarine) & overnight investment &      2,685.00 &      2,685.00 &                   \euro/MW/km &                                                                                                                                 DEA\citeS{DEA} \\
AC grid connection (underground) & overnight investment &      1,342.00 &      1,342.00 &                   \euro/MW/km &                                                                                                                                 DEA\citeS{DEA} \\
CCGT & Cb coefficient &          2.00 &          2.20 &  50$^{\circ}$C/100$^{\circ}$C &                                                                                                DEA\citeS{danishenergyagencyTechnologyData2018} \\
                      & Cv coefficient &          0.15 &          0.15 &  50$^{\circ}$C/100$^{\circ}$C &                                                                                                DEA\citeS{danishenergyagencyTechnologyData2018} \\
                      & FOM &          3.35 &          3.25 &                       \%/year &                                                                                                DEA\citeS{danishenergyagencyTechnologyData2018} \\
                      & VOM &          4.20 &          4.00 &                     \euro/MWh &                                                                                                DEA\citeS{danishenergyagencyTechnologyData2018} \\
\end{longtable}

\end{footnotesize}

\end{landscape}

\restoregeometry

\addcontentsline{toc}{section}{Supplementary References}
\renewcommand{\ttdefault}{\sfdefault}
\bibliographyS{/home/fneum/zotero}

\newpage
\begin{small}
	\tableofcontents
\end{small}

\end{document}